\documentclass[twoside,11pt]{article}

\usepackage{amsmath,amsfonts}
\usepackage{amssymb,latexsym}
\usepackage{amsxtra}
\usepackage{upref}
\usepackage{exscale}
\usepackage[mathscr]{eucal}

\newcommand{\be}{\begin{equation}}
\newcommand{\ee}{\end{equation}}
\newcommand{\beq}{\begin{eqnarray}}
\newcommand{\eeq}{\end{eqnarray}}
\newcommand{\no}{\nonumber}
\newcommand{\bea}{\begin{array}}
\newcommand{\eea}{\end{array}}
\newcommand{\lb}{\label}
\newcommand{\mcal}{\mathcal}
\newcommand{\mscr}{\mathscr}
\newcommand{\mfrak}{\mathfrak}
\newcommand{\ve}{\varepsilon}

\newcommand{\ts}{\textstyle}
\newcommand{\pp}{\partial}
\newcommand{\im}{\imath}
\newcommand{\ppr}{^{\boldsymbol{\prime}}}
\newcommand{\pppr}{^{\boldsymbol{\prime\prime}}}
\newcommand{\bpr}{\boldsymbol{\prime}}

\newcommand{\pdag}{^{\dagger}}
\newcommand{\wt}{\widetilde}
\newcommand{\ovv}{\overline}
\newcommand{\TRAB}{\raisebox{-4pt}{$\mbox{Tr}\atop {\scriptstyle a,b}$}}
\newcommand{\ph}{\phantom}
\newcommand{\scr}{\scriptstyle}
\newcommand{\scrscr}{\scriptscriptstyle}
\newcommand{\scz}{\scriptsize}

\newcommand{\Teta}[1]{\mscr{T}_{{\scrscr#1}}^{{\scrscr(\eta_{#1})}}}
\newcommand{\Tetax}[1]{\mscr{T}_{j_{{\scrscr#1}},\vec{x}_{{\scrscr#1}}}^{{\scrscr(\eta_{j_{#1}})}}}
\newcommand{\Tetaxx}[2]{\mscr{T}_{{\scrscr#1},{\scrscr#2}}^{{\scrscr(\eta_{#1})}}}
\newcommand{\TT}{\mscr{T}}
\newcommand{\etan}[1]{\eta_{{\scrscr#1}}}
\newcommand{\deltaN}{\delta^{\scrscr(\mcal{N}_{x})}}
\newcommand{\nn}[1]{\mathsf{{\scr'#1'}}}
\newcommand{\sdelta}{{\scr\Delta}}
\newcommand{\sdeltaT}{\hat{\pp}_{\hspace*{-0.8pt}\hat{\mathrm{T}}}}
\numberwithin{equation}{section}
\allowdisplaybreaks[4]

\begin{document}

\begin{center}
{\large\bf Fermionic coherent state path integral for ultrashort laser pulses} \vspace*{0.1cm}\\
{\large\bf and transformation to a field theory of coset matrices} \vspace*{0.1cm}\\
(Classical field theory for the self-energy matrices from various Hubbard-Stratonovich transformations) \vspace*{0.3cm}\\
{\bf Bernhard Mieck}\footnote{e-mail: "bjmeppstein@arcor.de"; freelance
activity; current location : Zum Kohlwaldfeld 16, 65817 Eppstein, Germany.} \\  {\sf\small 12th April 2010}
\end{center}

\begin{abstract}
\noindent A coherent state path integral of anti-commuting fields is considered for a two-band, semiconductor-related solid
which is driven by a ultrashort, classical laser field. We describe the generation of exciton quasi-particles from the driving
laser field as anomalous pairings of the fundamental, fermionic fields. This gives rise to Hubbard-Stratonovich transformations
from the quartic, fermionic interaction to various Gaussian terms of self-energy matrices; the latter self-energy matrices
are solely coupled to bilinear terms of anomalous-doubled, anti-commuting fields which are subsequently removed by
integration and which create the determinant with the one-particle operator and the prevailing self-energy. We accomplish
path integrals of even-valued self-energy matrices with Euclidean integration measure where three cases of increasing
complexity are classified (scalar self-energy variable, density-related self-energy matrix and also a self-energy including anomalous
doubled terms). According to the driving, anomalous-doubled Hamiltonian part, we also specify the case of a SSB with
'hinge' fields which factorizes the total self-energy matrix by a coset decomposition into density-related, block diagonal
self-energy matrices of a background functional and into coset matrices with off-diagonal block generators for the anomalous
pairings of fermions. In particular we investigate the transformation from the coset fields of a curved coset space,
as the independent field degrees of freedom, to locally 'flat' fields with Euclidean integration measure.
This allows to reduce the final path integral to solely
'Nambu'-doubled fields after a saddle point approximation for the density-related self-energy matrices and also allows to
derive classical field equations for exciton quasi-particles from various kinds of gradient expansions of the determinant.
Despite classical evolution equations for 'Nambu'-doubled terms, one thus incorporates a semi-classical notion with
quantum properties, due to the transformation and reduction to self-energy matrices which represent the irreducible parts
of a diagrammatic propagation. The derived equations with various kinds of classical self-energy matrices also allow
to examine self-induced transparency effects and the area theorem for ultrashort coherent transients of matter
in combination with holography and selective Fourier optics.\newline
\vspace*{0.1cm}

\noindent {\bf Keywords} : 'Nambu' doubling of fields, Hubbard-Stratonovich transformations to self-energies,
coset integration measure, coset decomposition for pair condensates, coherent state path integral, self-induced transparency, 
ultrashort coherent transients\newline
\vspace*{0.1cm}

\noindent 1st {\bf PACS} : 03.65.Db , 71.10.-w , 02.20.Qs ;\newline
\noindent 2nd{\bf PACS} : 42.50.Md , 42.65.Tg , 78.47.jb .
\end{abstract}

\newpage

\tableofcontents

\section{Introduction} \lb{s1}

\subsection{Hubbard-Stratonovich transformation of quartic fields to self-energies} \lb{s11}

Phenomena of many body physics can be mapped to different kinds of theories which can start from very direct approaches as in
computational physics \cite{Suzuki1}-\cite{Thijs} or which can be based on various approximations. 
Apart from the YBE-related, integrable systems \cite{QiMa}, there is a general lack of constant integrals of motion so that simple solutions of functions do usually not appear in many particle systems. In consequence, one has to rely on approximations, as by choosing a dependence of the self-energy on the irreducible diagrams or Green function parts or as by taking various manners of correlated field operator terms. The chosen approximation of the self-energy or correlated operators has a satisfactory, correct form if the prevailing physical phenomenon is considered within its computation; therefore, different manners of calculations can occur, each approximation being regarded as a particular theory aside from the original, beginning formulation with a fundamental Hamiltonian, Green function and Dyson equation or the original definition of the correlations 
with the {\it 'full'} operators \cite{Negele}-\cite{Sadovs}. 
We prefer those "approximating theories" as the more appealing kind in many particle physics where one
only reduces to a few terms, but still regards the essentials of the corresponding phenomenon.

In the present paper we investigate a coherent state path integral of fermions or anti-commuting fields, by way of example
for a two-band, semiconductor-related solid which is driven by an external, classical electric field \cite{Negele,pop1,pop2}. 
We apply various kinds of Gaussian transformations from the quartic interaction of Fermi fields to various self-energy matrices.
The latter operation or Hubbard-Stratonovich transformation ('HST') maps the quartic Fermi-fields of the original coherent state path integral
to a bilinear term of anti-commuting fields with linear coupling to a self-energy matrix and a separate Gaussian factor
of the corresponding self-energy. In section \ref{s2} we define the total Hamiltonian from second quantized, fermionic operators in normal ordering whose time development is determined by a coherent state path integral \cite{pop1,pop2}. 
Since the driving interaction with the classical electric field creates electron-hole pairs, one has to perform an anomalous doubling of anti-commuting fields for the presence of exciton quasi-particles within the original coherent state path integral.
In section \ref{s3} subsequent HSTs lead to quadratic self-energy terms and to the linear coupling of the anomalous-doubled, bilinear Fermi fields
with the self-energy. Three different kinds of HSTs are regarded within sections \ref{s31} to \ref{s33} which we abbreviate symbolically in advance :
\begin{align}
\mbox{Section \ref{s31}} &: \mbox{Auxiliary variables }\sigma(\Teta{j},\vec{x})\;\mbox{ as scalar, real-valued self-energy} \\
\notag & \sum_{\mu,s}
\chi_{\mu,s}^{*}(\Teta{j},\vec{x})\;q_{\mu}\;\chi_{\mu,s}(\Teta{j},\vec{x})
\stackrel{\mbox{\scz 'HST' section \ref{s31}}}{\Longrightarrow}\;\sigma(\Teta{j},\vec{x})\;\;
(\mbox{cf. (\ref{s3_1a}-\ref{s3_5c})}) \;;  \\
\mbox{Section \ref{s32}} &: \mbox{Hermitian self-energy matrix }
\hat{\Sigma}_{\nu,s\ppr;\mu,s}(\Teta{j_{2}},\vec{x}_{2};\Teta{j_{1}},\vec{x}_{1})\;\mbox{of densities} \\
\notag & \chi_{\nu,s\ppr}(\Teta{j_{2}},\vec{x}_{2}){\bf\otimes}
\chi^{*}_{\mu,s}(\Teta{j_{1}},\vec{x}_{1}) \stackrel{\mbox{\scz 'HST' section \ref{s32}}}{\Longrightarrow}\;
\hat{\Sigma}_{\nu,s\ppr;\mu,s}(\Teta{j_{2}},\vec{x}_{2};\Teta{j_{1}},\vec{x}_{1})
(\mbox{cf. (\ref{s3_9a}-\ref{s3_13c})}) \;;  \\
\mbox{Section \ref{s33}} &: \mbox{Self-energy matrix }
\hat{\Sigma}_{\nu,s\ppr;\mu,s}^{ba}(\Teta{j_{2}},\vec{x}_{2};\Teta{j_{1}},\vec{x}_{1})\;
\mbox{for densities and anomalous pairs} \\
\notag & \left(\bea{c} \chi_{\nu,s\ppr}(\Teta{j_{2}},\vec{x}_{2}) \\
\chi_{\nu,s\ppr}^{*}(\Teta{j_{2}},\vec{x}_{2}) \eea\right)^{b(=1,2)}\otimes
\Big(\chi_{\mu,s}^{*}(\Teta{j_{1}},\vec{x}_{1})\;\boldsymbol{,}\;
\chi_{\mu,s}(\Teta{j_{1}},\vec{x}_{1})\Big)^{a(=1,2)}\;
\stackrel{\mbox{\scz 'HST' section \ref{s33}}}{\Longrightarrow}\;  \\ \notag &
\stackrel{\mbox{\scz 'HST' section \ref{s33}}}{\Longrightarrow}\;
\hat{\Sigma}_{\nu,s\ppr;\mu,s}^{ba}(\Teta{j_{2}},\vec{x}_{2};\Teta{j_{1}},\vec{x}_{1})\;\;
(\mbox{cf. (\ref{s3_16}-\ref{s3_22b})}) \;.
\end{align}
In section \ref{s31} we only transform to scalar self-energy variables with a contour time and spatial dependence, but without regard of any spin or electron-hole band labels. In section \ref{s32} we extend the HST to a self-energy {\it matrix} which replaces dyadic products of Fermi fields
for pure density-related terms of field operators. Section \ref{s33} contains self-energy matrices, which follow from
{\it dyadic products of anomalous-doubled}, anti-commuting fields and which thus consist of density-related, block
diagonal matrix parts \(\hat{\Sigma}_{\nu,s\ppr;\mu,s}^{b=a}(\Teta{j_{2}},\vec{x}_{2};\Teta{j_{1}},\vec{x}_{1})\) and also off-diagonal matrix parts
\(\hat{\Sigma}_{\nu,s\ppr;\mu,s}^{b\neq a}(\Teta{j_{2}},\vec{x}_{2};\Teta{j_{1}},\vec{x}_{1})\);
the latter substitute anomalous-doubled pairings of Fermi operators \cite{pop1}-\cite{Gold}. 
After the various HSTs in sections \ref{s31} to \ref{s33} and the removal of bilinear Fermi fields by integration, we obtain path integrals with the corresponding self-energy variables and matrices as the independent field degrees of freedom whose first order and second order variation straightforwardly result in saddle point equations and corresponding solutions for fluctuation terms.
We emphasize the precise time step order of the time evolution for a normal ordered Hamiltonian throughout the presented article and follow along the lines of the article
\cite{mie3} for a proper time step separation of field operators within coherent state path integrals.
Section \ref{s4} has a more profound kind of HST where one transforms separately density-related and anomalous-doubled terms to corresponding self-energy matrices in combination with a spontaneous symmetry breaking ('SSB') and with 'hinge' fields \cite{pop1}; this leads to a factorization of the total self-energy matrix into density-related, block diagonal self-energy matrices and into anomalous-doubled, off-diagonal blocks. The latter are taken into account by coset matrices within
a coset decomposition \(\mbox{SO}(\mfrak{N},\mfrak{N})\,/\,\mbox{U}(\mfrak{N})\otimes\mbox{U}(\mfrak{N})\)
of the total self-energy matrix which itself has the detailed structure of a \(\mbox{so}(\mfrak{N},\mfrak{N})\) generator
within the orthogonal group \(\mbox{SO}(\mfrak{N},\mfrak{N})\)
(dimension \(\mfrak{N}=(\mu=\mbox{'e','h'})\times(s=\uparrow,\downarrow)\times(\eta_{j}=\pm)\times(N=t_{N}/\Delta t)\times
(\mcal{N}_{x}=(L/\Delta x)^{d})=2\times 2\times 2\times N\times \mcal{N}_{x}\); a factor of two for the electron-hole, spin and contour time metric degrees of
freedom, respectively, times the number of time-like and discrete, spatial points defines the relevant dimension
\(\mfrak{N}\) within \(\mbox{SO}(\mfrak{N},\mfrak{N})\,/\,\mbox{U}(\mfrak{N})\otimes\mbox{U}(\mfrak{N})\;\)).
The coset decomposition combined with the factorization of the total self-energy
allows for a projection onto the anomalous-doubled field degrees of freedom with the coset matrices whose path integral consists of the density-related self-energy as a background functional or as a solution from a saddle point equation. The coset decomposition involves various transformations, as the appropriate, invariant integration measure and as the transformation to Euclidean integration variables for first, second or higher order variations with classical coset matrices \cite{miesinh}. These involved appearing, but straightforward transformations are described in sections \ref{s4} to \ref{s5} and are important for the appropriate consideration of the invariant integration measure, 
following from the square root of the coset metric tensor \cite{pop1,miesinh}.

\subsection{The importance of the appropriate, invariant integration measure}\lb{s12}

In section \ref{s4} we attain a coherent state path integral whose remaining field degrees of freedom, adapted by
coset matrices for exciton-related parts, possess a nontrivial, non-Euclidean integration measure determined by the
coset metric tensor of \(\mbox{SO}(\mfrak{N},\mfrak{N})\,/\,\mbox{U}(\mfrak{N})\otimes\mbox{U}(\mfrak{N})\).
In analogy we illustrate this problem for a multidimensional integral \(Z[\vec{x}]\) (\ref{s1_4a}-\ref{s1_5c}),
(\(\vec{x}=\{x^{1},\ldots,x^{N}\}\)), with an action \(\mscr{A}[\vec{x}]\) where the Euclidean integration measure
\(d[\vec{x}]\) is modified by the square root of a metric tensor \(\hat{g}_{ij}(\vec{x})\), similarly to the square
root of the coset metric tensor in section \ref{s4} (cf. eqs. (\ref{s4_11a}-\ref{s4_12})) \cite{Hobson,Burg2}
\begin{subequations}
\begin{align}\lb{s1_4a}
Z[\vec{x}]&=\int d[\vec{x}]\;\sqrt{\mbox{det}\big(\hat{g}(\vec{x})\big)}\;\;
\exp\big\{\im\;\mscr{A}[\vec{x}]\big\}\;;  \\ \lb{s1_4b}
d[\vec{x}]&:\;\mbox{Euclidean integration measure }\;d[\vec{x}]=dx^{1}\wedge\,\ldots\,
\wedge dx^{N} \;;
\end{align}
\end{subequations}
\begin{subequations}
\begin{align}\lb{s1_5a}
\big(ds\big)^{2}&:=dx^{i}\;\;\hat{g}_{ij}(\vec{x})\;\;dx^{j}\;; \\ \lb{s1_5b}
\hat{g}(\vec{x}) &:\stackrel{\wedge}{=} \;\hat{g}_{ij}(\vec{x}) =\;\mbox{metric tensor with covariant indices}\;; \\
\lb{s1_5c} \hat{g}^{-1}(\vec{x}) &:\stackrel{\wedge}{=}\; \hat{g}^{ij}(\vec{x}) =\;\mbox{inverted metric tensor with
contravariant indices}\;.
\end{align}
\end{subequations}
The transformation to Euclidean variables \(dy^{j}=dy_{j}\) (\ref{s1_6}-\ref{s1_8}) is related to the inverse
square root of the metric tensor \(\hat{g}^{-1/2}(\vec{x})\) as the appropriate Jacobi matrix where
the symmetric property of the metric tensor allows a decomposition into orthogonal
matrices \(\hat{O}_{ij}(\vec{x})\) with \(\det(\hat{O}(\vec{x})\,)=1\)
and real eigenvalues \(\hat{\lambda}^{k}(\vec{x})\) (\(i,j,k,l=1,\ldots,N\))
\beq \lb{s1_6}
\big(ds\big)^{2}&=&dx^{i}\;\hat{g}_{ij}(\vec{x})\;dx^{j}=
dx^{i}\;\hat{O}_{ik}^{T}(\vec{x})\;\hat{\lambda}^{k}(\vec{x})\;
\hat{O}_{kj}(\vec{x})\;dx^{j}=  \\ \no &=&
\underbrace{dx^{i}\;\Big(\hat{O}^{T}(\vec{x})\cdot\hat{\lambda}^{1/2}(\vec{x})\Big)_{i}^{\ph{i}k}}_{dy^{k}}\;
\underbrace{\Big(\hat{\lambda}^{1/2}(\vec{x})\cdot\hat{O}(\vec{x})\Big)_{kj}\;dx^{j}}_{dy_{k}}=
dy^{k}\;dy_{k}=dy^{k}\;dy^{k}\;;  \\ \lb{s1_7}
dy^{j}=dy_{j}&=& \Big(\hat{\lambda}^{1/2}(\vec{x})\cdot\hat{O}(\vec{x})\Big)^{j}_{\ph{j}i}\;dx^{i} \;;
 \\ \lb{s1_8}
\hat{O}_{ji}(\vec{x})\;dx^{i}&=&  \Big(\hat{\lambda}^{-1/2}(\vec{x})\cdot d\vec{y}\Big)_{j} \;;
\;\;\Longrightarrow y^{j}=y^{j}(\vec{x})\;\;\Longleftrightarrow x^{i}=x^{i}(\vec{y})\;.
\eeq
One obtains from (\ref{s1_8}) an arbitrariness of further orthogonal transformations,
e.g. to \(dy^{\bpr j}\) (\ref{s1_9a}), which are given by locally rotated Euclidean differentials, maintaining the
local volume elements (\ref{s1_9b}). These transformations correspond to gauge transformations and to the gauge
invariance of the coset metric integration measure, determined in section \ref{s4} for the anomalous-doubled
field degrees of freedom
\begin{subequations}
\begin{align}\lb{s1_9a}
dy^{\bpr j}=dy_{j}\ppr &= \bigl(\hat{g}^{1/2}(\vec{x})\bigr)_{ji}\;dx^{i} =
\Bigl(\hat{O}^{T}(\vec{x})\cdot\hat{\lambda}^{1/2}(\vec{x})\cdot\hat{O}(\vec{x})\Bigr)_{ji}\;dx^{i}
= \hat{O}_{jk}^{T}(\vec{x})\;\Bigl(\hat{\lambda}^{1/2}(\vec{x})\cdot\hat{O}(\vec{x})\Bigr)_{\ph{k}i}^{k}\;dx^{i} \\ \notag
&= \hat{O}_{jk}^{T}(\vec{x})\;dy^{k}=\hat{O}^{T}_{jk}(\vec{x})\;dy_{k}\;;\;\;
(dy^{\bpr j}=dy_{j}\ppr\;;\;\;\;\mbox{locally rotated Euclidean differentials}); \\ \lb{s1_9b}\Longrightarrow\;
d[\vec{y}\ppr] &= d[\vec{y}]\;;\;\;\;(\mbox{invariance of volume elements under 'gauge' transformations}) \;.
\end{align}
\end{subequations}
The transformation to Euclidean differentials \(d\vec{y}\) (or to equivalently rotated versions \(d\vec{y}\ppr\))
involves the additional Jacobi matrix
\(\hat{J}^{i}_{\ph{i}k}=(\pp x^{i}/\pp y^{k})=(\,\hat{O}^{T}(\vec{x})\cdot\hat{\lambda}^{-1/2}(\vec{x})\,)^{i}_{\ph{i}k}\)
whose determinant leads in combination with the original square root of the metric tensor to Euclidean integration
variables. The simplified, Euclidean integration measure is accompanied by a transformation of field variables within the
action \(\mscr{A}[\vec{x}]\) to \(\mscr{A}\ppr[\vec{y}]=\mscr{A}[\vec{x}(\vec{y})]\) which alters the dependence
of the field degrees of freedom within the classical action of the exponential
\beq \lb{s1_10}
\hat{J}^{i}_{\ph{i}k}&=&\frac{\pp x^{i}}{\pp y^{k}}=\Big(\hat{O}^{T}(\vec{x})\cdot
\hat{\lambda}^{-1/2}(\vec{x})\Big)^{i}_{\ph{i}k} \;; \\  \lb{s1_11}
\mbox{det}\big(\hat{J}^{i}_{\ph{i}k}\big)&=&
\mbox{det}\Big[\big(\,\hat{O}^{T}(\vec{x})\cdot\hat{\lambda}^{-1/2}(\vec{x})\,\big)^{i}_{\ph{i}k}\Big]  =
\mbox{det}\big[\hat{g}^{-1/2}(\vec{x})\big]=\Big(\mbox{det}\big[\hat{g}(\vec{x})\big]\Big)^{-1/2} \;;  \\  \lb{s1_12}
Z[\vec{x}(\vec{y})]&=&\int d[\vec{y}]\;
\underbrace{\mbox{det}\Big[\hat{g}^{\boldsymbol{-1/2}}(\vec{x})\Big]\;
\sqrt{\mbox{det}\big(\hat{g}(\vec{x})\big)}}_{\equiv 1}\;\exp\big\{\im\;\mscr{A}[\vec{x}(\vec{y})]\big\} =
Z\ppr[\vec{y}]=\int d[\vec{y}]\;\exp\{\im\;\mscr{A}\ppr[\vec{y}]\}\;; \\  \lb{s1_13}
\mscr{A}\ppr[\vec{y}] &=& \mscr{A}[\vec{x}(\vec{y})]\;;\hspace*{0.5cm}
Z\ppr[\vec{y}]=Z[\vec{x}(\vec{y})]\;;\hspace*{0.5cm}
d[\vec{y}]=dy^{1}\wedge\,\ldots\,\wedge dy^{N}\;.
\eeq
In the following we compare the two cases 'I' and 'II' of variations for classical equations and their quadratic
fluctuation terms 'without' (case 'I') and under inclusion of the transformation to Euclidean integration variables (case 'II').
Case 'I' disregards the nontrivial integration measure \((\mbox{det}(\hat{g}(\vec{x})\,)\,)^{1/2}\) (\ref{s1_4a}-\ref{s1_5c})
in the first and second order variations so that one can simply expand the action \(\mscr{A}[\vec{x}]\). We thus attain
from the first order variation the classical equations \(\pp\mscr{A}[\vec{x}]/\pp x^{i}\equiv0\) ('in a transferred sense') whose
solution (or 'solutions') determine the quadratic fluctuation term (\ref{s1_14d}) for the prevailing 'field' configuration
\(\vec{x}=\vec{x}_{0}\)
\begin{subequations}
\begin{align}\lb{s1_14a}
\mbox{\bf Case 'I' : }Z[\vec{x}] &= \int d[\vec{x}]\;\;\sqrt{\mbox{det}\big(\hat{g}(\vec{x})\big)}\;\;
\exp\{\im\;\mscr{A}[\vec{x}]\} \;;  \\ \notag \mbox{neglect} & \hspace*{0.15cm}\sqrt{\mbox{det}\big(\hat{g}(\vec{x})\big)}\;
\mbox{and expand }\mscr{A}[\vec{x}]\;\mbox{for }\vec{x}=\vec{x}_{0}\;\mbox{following from :} \\ \lb{s1_14b}
\frac{\pp \mscr{A}[\vec{x}]}{\pp x^{i}} &\equiv 0 \;;\;\;\;(i=1,\ldots,N)\Longrightarrow\;
\vec{x}=\vec{x}_{0}\;;\;\;(\mbox{extremum}); \\ \lb{s1_14c}
\mscr{A}[\vec{x}] &= \mscr{A}[\vec{x}_{0}]+(x^{i}-x_{0}^{i})\;
\frac{\pp \mscr{A}[\vec{x}]}{\pp x^{i}}\bigg|_{\vec{x}=\vec{x}_{0}} +\frac{1}{2!}(x^{k}-x^{k}_{0})\;
\frac{\pp^{2}\mscr{A}[\vec{x}]}{\pp x^{k}\;\pp x^{l}}\bigg|_{\vec{x}=\vec{x}_{0}}\;(x^{l}-x^{l}_{0})+\ldots\;; \\ \lb{s1_14d}
\frac{\pp^{2}\mscr{A}[\vec{x}]}{\pp x^{k}\;\pp x^{l}}\bigg|_{\vec{x}=\vec{x}_{0}} &\Longrightarrow
\;\mbox{quadratic fluctuation term}\;.
\end{align}
\end{subequations}
In case 'II' with the Euclidean integration variables we can similarly expand the action \(\mscr{A}\ppr[\vec{y}]\) around a
classical solution \(\vec{y}_{0}\) up to second order, but have to take into account the chain rule compared to case 'I'.
The first order variation (\ref{s1_15b}) contains the multiplicative Jacobi matrix, compared to (\ref{s1_14b}) of case 'I',
so that the classical solution \(\vec{y}_{0}\) directly follows from the transformation \(\vec{y}_{0}=\vec{y}(\vec{x}_{0})\)
according to a necessary 'rank $N$'-property of the Jacobian for proper transformations (except the Jacobi determinant
vanishes for particular points which causes separate ranges of integration and transformation intervals).
Nevertheless, this implies a different, second order fluctuation term (\ref{s1_15d}) which is modified by a derivative and two multiplications
of Jacobi matrices compared to case 'I'
\begin{subequations}
\begin{align}\lb{s1_15a}
\mbox{\bf Case 'II' : }Z[\vec{x}(\vec{y})] &= Z\ppr[\vec{y}]=\int d[\vec{y}]\;\;\exp\bigl\{\im\;\mscr{A}\ppr[\vec{y}]\bigr\}=
\int d[\vec{y}]\;\;\exp\bigl\{\im\;\mscr{A}[\vec{x}(\vec{y})]\bigr\} \;;   \\ \lb{s1_15b}
\frac{\pp\mscr{A}\ppr[\vec{y}]}{\pp y^{k}}\bigg|_{\vec{y}=\vec{y}_{0}} &=\frac{\pp x^{i}}{\pp y^{k}}\;
\frac{\pp\mscr{A}[\vec{x}]}{\pp x^{i}}\equiv 0 \;\Longleftrightarrow\;
\frac{\pp\mscr{A}[\vec{x}]}{\pp x^{i}}\equiv 0\;;\;(\mbox{i=1,\ldots,N})\;\Longrightarrow\;
\vec{x}=\vec{x}_{0}=\vec{x}(\vec{y}_{0})\;, \\  \notag \mbox{because }
\hat{J}_{\ph{i}k}^{i}&=\bigl(\tfrac{\pp x^{i}}{\pp y^{k}}\bigr)_{\vec{y}=\vec{y}_{0}}\;
\mbox{has to be of rank }N\;\mbox{for a valid transformation
to Euclidean coordinates}\;; \\  \lb{s1_15c}
\mscr{A}\ppr[\vec{y}] = \mscr{A}\ppr[\vec{y}_{0}]&+(y^{i}-y^{i}_{0})\;
\frac{\pp\mscr{A}\ppr[\vec{y}]}{\pp y^{i}}\bigg|_{\vec{y}=\vec{y}_{0}}+
\frac{1}{2!}\;(y^{k}-y^{k}_{0})\;\frac{\pp^{2}\mscr{A}\ppr[\vec{y}]}{\pp y^{k}\;\pp y^{l}}\bigg|_{\vec{y}=\vec{y}_{0}}\;
(y^{l}-y^{l}_{0})+\ldots = \\ \notag =\mscr{A}[\vec{x}(\vec{y})]&=\mscr{A}[\vec{x}(\vec{y}_{0})]+
(y^{k}-y^{k}_{0})\;\frac{\pp x^{i}}{\pp y^{k}}\bigg|_{\vec{y}=\vec{y}_{0}}\;
\frac{\pp\mscr{A}[\vec{x}]}{\pp x^{i}}\bigg|_{\vec{x}_{0}=\vec{x}(\vec{y}_{0})}+   \\ \notag +\frac{1}{2!}\;
(y^{k}-y^{k}_{0})&\bigg(\frac{\pp^{2}x^{i}}{\pp y^{k} \pp y^{l}}\bigg|_{\vec{y}=\vec{y}_{0}}
\frac{\pp\mscr{A}[\vec{x}]}{\pp x^{i}}\bigg|_{\vec{x}_{0}=\vec{x}(\vec{y}_{0})}+\frac{\pp x^{i}}{\pp y^{k}}\bigg|_{\vec{y}=\vec{y}_{0}}\;
\frac{\pp^{2}\mscr{A}[\vec{x}]}{\pp x^{i}\;\pp x^{j}}\bigg|_{\vec{x}_{0}=\vec{x}(\vec{y}_{0})}\;
\frac{\pp x^{j}}{\pp y^{l}}\bigg|_{\vec{y}=\vec{y}_{0}}\bigg)\;
(y^{l}-y^{l}_{0})+\ldots \;;  \\ \lb{s1_15d}
\mbox{fluctuation term}    &\Longrightarrow\;\;
\frac{\pp^{2}x^{i}}{\pp y^{k} \pp y^{l}}\bigg|_{\vec{y}=\vec{y}_{0}}
\frac{\pp\mscr{A}[\vec{x}]}{\pp x^{i}}\bigg|_{\vec{x}_{0}=\vec{x}(\vec{y}_{0})}+\frac{\pp x^{i}}{\pp y^{k}}\bigg|_{\vec{y}=\vec{y}_{0}}\;
\frac{\pp^{2}\mscr{A}[\vec{x}]}{\pp x^{i}\;\pp x^{j}}\bigg|_{\vec{x}_{0}=\vec{x}(\vec{y}_{0})}\;
\frac{\pp x^{j}}{\pp y^{l}}\bigg|_{\vec{y}=\vec{y}_{0}} =   \\ \notag \ph{\mbox{fluctuation term} } &=
\bigg(\frac{\pp\hat{J}_{\ph{i}k}^{i}}{\pp y^{l}}\bigg)_{\vec{y}=\vec{y}_{0}}\frac{\pp\mscr{A}[\vec{x}]}{\pp x^{i}}\bigg|_{\vec{x}_{0}=\vec{x}(\vec{y}_{0})}+
\bigl(\hat{J}_{\ph{i}k}^{i}\bigr)_{\vec{y}=\vec{y}_{0}}\;
\frac{\pp^{2}\mscr{A}[\vec{x}]}{\pp x^{i}\;\pp x^{j}}\bigg|_{\vec{x}_{0}=\vec{x}(\vec{y}_{0})}\;
\bigl(\hat{J}_{\ph{j}l}^{j}\bigr)_{\vec{y}=\vec{y}_{0}} \;.
\end{align}
\end{subequations}
In summary we can state that the classical solution, following from the first order variation, changes
under the transformation to Euclidean integration variables because this induces a transformation of
coordinates \(\vec{y}_{0}=\vec{y}(\vec{x}_{0})\) with the square root of the metric tensor. The second (and also higher n-th) order fluctuations are altered
by multiplications and by derivatives of the inverse square root of the metric tensor due to the chain rule for
the transformation \(\vec{x}=\vec{x}(\vec{y})\) aside from 'gauge' invariant transformations as in (\ref{s1_9a}-\ref{s1_9b})
\begin{subequations}
\begin{align}\notag \mbox{2nd }&\mbox{order fluctuations:} \\ \lb{s1_16a}
\mbox{Case 'I'=}&
\frac{\pp^{2}\mscr{A}[\vec{x}]}{\pp x^{i}\;\pp x^{j}}\bigg|_{\vec{x}=\vec{x}_{0}} \;;  \\  \lb{s1_16b}
\mbox{Case 'II'=}& \frac{\pp(\hat{g}^{-1/2}(\vec{y})\,)_{\ph{i}k}^{i}}{\pp y^{l}}\bigg|_{\vec{y}=\vec{y}_{0}}
\frac{\pp\mscr{A}[\vec{x}]}{\pp x^{i}}\bigg|_{\vec{x}_{0}=\vec{x}(\vec{y}_{0})} +
\bigl(\hat{g}^{-1/2}(\vec{y}_{0})\bigr)_{\ph{i}k}^{i}\;
\frac{\pp^{2}\mscr{A}[\vec{x}]}{\pp x^{i}\;\pp x^{j}}\bigg|_{\vec{x}_{0}=\vec{x}(\vec{y}_{0})}\;
\bigl(\hat{g}^{-1/2}(\vec{y}_{0})\bigr)^{j}_{\ph{j}l}\;.
\end{align}
\end{subequations}
In analogy we perform in section \ref{s5} transformations to Euclidean field degrees of freedom for the remaining
anomalous-doubled self-energy parts within the coset matrices. In a similar manner one does attain changes
for the classical field equations, according to the transformation of the field solution \(\vec{y}_{0}=\vec{y}(\vec{x}_{0})\)
and to the 'gauge' symmetry (\ref{s1_9a}-\ref{s1_9b})
in a 'transferred sense', but has to consider the inverse square root of the corresponding coset metric tensor
in order to conclude for the appropriate second and higher order fluctuation terms. A further possibility is given by a modified classical action,
according to
\begin{subequations}
\begin{align}\lb{s1_17a}
Z[\vec{x}] &=\int d[\vec{x}]\;\;\;\exp\bigl\{\im\:\mathsf{A}[\vec{x}]\bigr\} \;, \\  \lb{s1_17b}
\mathsf{A}[\vec{x}] &=-\frac{\im}{2}\;\mbox{tr}\bigl[\ln\bigl(\hat{g}(\vec{x})\bigr)\bigr]+\mscr{A}[\vec{x}] \;,
\end{align}
\end{subequations}
so that one allows for a further treatment of the nontrivial integration measure where one has to include the variations for the 'trace-log' term
of the metric tensor. This straightforward procedure of variations necessitates various kinds of Christoffel-symbols
of the metric tensor and also results in corrections, both in first and higher order variations, compared to cases 'I' and 'II'.
This straightforward kind of variation can be found in random log gases of random matrix theory \cite{loggas,Mehta}.

\subsection{A glimpse in advance for the case with locally Euclidean coset fields} \lb{s13}

In sections \ref{s31} to \ref{s33} we specify HSTs which range from the transformation with scalar, real-valued
self-energy variables to total self-energy matrices with anomalous terms in additional, off-diagonal blocks. However,
we extend the latter case with anomalous-doubled parts and perform a factorization of the self-energy matrix into
density-related self-energy matrices of a background functional and into coset matrices having generators of off-diagonal,
'Nambu' parts. This factorization is taken by a coset decomposition for exciton-related quasi-particles. Although the
various steps of section \ref{s4} and \ref{s5} are straightforward, we briefly list the resulting path integral of 'Euclidean'
coset matrices for an overview in advance with regard to the familiar kind of a nonlinear sigma model on a coset space
\(\mbox{SO}(\mfrak{N},\mfrak{N})\,/\,\mbox{U}(\mfrak{N})\otimes\mbox{U}(\mfrak{N})\)
\footnote{The dimension \(\mfrak{N}\) is given by \(\mfrak{N}=
(\mu=\mbox{'e','h'})\times(s=\uparrow,\downarrow)\times(\eta_{j}=\pm)\times(N=t_{N}/\Delta t)\times
(\mcal{N}_{x}=(L/\Delta x)^{d})=2\times 2\times 2\times N\times \mcal{N}_{x}\); 
a factor of two for the electron-hole, spin and contour time metric degrees of
freedom, respectively, times the number of time-like and discrete, spatial points defines the relevant dimension
\(\mfrak{N}\) within \(\mbox{SO}(\mfrak{N},\mfrak{N})\,/\,\mbox{U}(\mfrak{N})\otimes\mbox{U}(\mfrak{N})\;\).}
\begin{align}\lb{s1_18}
\ovv{Z[\hat{\mscr{J}}]} &\approx \int
d\bigl[\hat{\wt{a}}_{s\ppr s}^{(k)}(\Tetaxx{j\ppr}{\vec{x}\ppr};\Tetaxx{j}{\vec{x}})\bigr]\;\;\;\;
\mfrak{Z}\!\Bigl[\hat{\mathrm{T}}^{\boldsymbol{-1}}(\hat{\wt{a}});
\hat{\mathrm{T}}(\hat{\wt{a}});\hat{\mcal{H}}\Bigr]\;\times
\exp\biggl\{\frac{1}{2}\,\mcal{N}_{x}\sum_{\vec{x}}\sum_{j=0}^{2N+1}\sum_{\mu=e,h}^{s=\uparrow,\downarrow}
\;\times \\ \notag &\times\;
\TRAB\bigg[\ln\bigg(\boldsymbol{\hat{1}}-\bigl(\hat{\pp}_{(\ln-\breve{\mscr{H}})}\hat{\mscr{Z}}\bigr)^{b\ppr a\ppr}+
\hat{\mathrm{P}}\:\wt{\mscr{J}}\bigl[\hat{\mathrm{T}}^{-1}(\hat{\wt{a}}),\hat{\mathrm{T}}(\hat{\wt{a}})\bigr]\;
\Bigl(\breve{\mscr{H}}\bigl[\hat{\mfrak{H}},\langle\hat{\mfrak{s}}\rangle\bigr]\Bigr)^{\boldsymbol{-1}}\hat{\mathrm{P}}^{-1}
\bigg)\bigg]_{\mu,s;\mu,s}^{b=a}\hspace*{-0.9cm}(\Teta{j},\vec{x};\Teta{j},\vec{x})\biggr\}\;.
\end{align}
The independent field degrees of freedom are given by the matrices \(\hat{\wt{a}}(\nn{5};\nn{1})\) whose derivative or variational
increments '\(\sdelta\)' are given in terms of quaternion-valued matrix elements
\(\sdelta\hat{a}_{s_{5}s_{1}}^{(k)}(\Tetax{5};\Tetax{1})\) for off-diagonal elements and, 
aside from an additional phase factor (see second line in (\ref{s1_19a})),
by the field increments \(\sdelta\wt{a}_{ss}^{(2)}(\Tetaxx{j}{\vec{x}};\Tetaxx{j}{\vec{x}})=\sdelta(\,
|\wt{a}_{\mathsf{S}=\pm1}(\Tetaxx{j}{\vec{x}})|\:\exp\{\im\:\wt{\alpha}_{\mathsf{S}=\pm1}(\Tetaxx{j}{\vec{x}})\}\,)\) for the
particular elements along the main quaternion-diagonal with element \((\hat{\tau}_{2})_{\mu_{5}\mu_{1}}\). Apart from the contour
time \(\TT_{j}^{(\eta_{j}=\pm)}\) (\(j=0,\ldots,2N+1\)) and space vector \(\vec{x}\) dependence, we include the semiconductor-related,
two-band index \(\mu=\mbox{'e'},\mbox{'h'}\) and spin \(s=\uparrow,\downarrow,\:\mathsf{S}=2s=+1,-1\) degrees of freedom for the
independent, locally Euclidean, 'Nambu' fields \(\sdelta\hat{\wt{a}}(\nn{5};\nn{1})\); the already listed abbreviations
\(\nn{5},\nn{1},\ldots\) denote the total set of external and internal labels as the space-contour-time with additional band and spin indices,
respectively
\begin{subequations}
\begin{align}\lb{s1_19a}
\sdelta\hat{\wt{a}}(\nn{5};\nn{1}) &=-\sdelta\hat{\wt{a}}^{\boldsymbol{T}}(\nn{5};\nn{1})=\sum_{k=0}^{3}
(\hat{\tau}_{k})_{\mu_{5}\mu_{1}}\;\sdelta\hat{a}_{s_{5}s_{1}}^{(k)}(\Tetax{5};\Tetax{1})+  \\ \notag &+
(\hat{\tau}_{2})_{\mu_{5}\mu_{1}}\;\exp\bigg\{\im\int_{\TT_{0}^{(+)}}^{\Teta{j}}d\Teta{j\ppr}
\bigg(\frac{2\,|\wt{a}_{\mathsf{S}}(\Tetaxx{j\ppr}{\vec{x}})|}{\sinh\big(2\,|\wt{a}_{\mathsf{S}}(\Tetaxx{j\ppr}{\vec{x}})|\big)}-1\bigg)
\frac{\pp\wt{\alpha}_{\mathsf{S}}(\Tetaxx{j\ppr}{\vec{x}})}{\pp\Teta{j\ppr}}\bigg\}\;
\sdelta\wt{a}_{ss}^{(2)}(\Tetaxx{j}{\vec{x}};\Tetaxx{j}{\vec{x}})\;;  \\ \notag &
(\mbox{last term }(\Tetaxx{j}{\vec{x}}:=\Tetax{5}=\Tetax{1})\&(s:=s_{5}=s_{1}\;,\,\mathsf{S}=2s)\,)\;; \\  \lb{s1_19b}
\wt{a}_{ss}^{(2)}(\Tetaxx{j}{\vec{x}};\Tetaxx{j}{\vec{x}}) &=
|\wt{a}_{\mathsf{S}}(\Tetaxx{j}{\vec{x}})|\;\exp\{\im\:\wt{\alpha}_{\mathsf{S}}(\Tetaxx{j\ppr}{\vec{x}})\}\;; \\   \lb{s1_19c}
d\bigl[\hat{\wt{a}}_{s\ppr s}^{(k)}(\Tetaxx{j\ppr}{\vec{x}\ppr};\Tetaxx{j}{\vec{x}})\bigr] &=
 \prod_{s=\uparrow,\downarrow}^{\mathsf{S}=2s}
\prod_{j=1,\ldots,2N}^{\{\vec{x}\}}
\biggl(\frac{d\wt{a}_{ss}^{(2)*}(\Tetaxx{j}{\vec{x}};\Tetaxx{j}{\vec{x}})\wedge
d\wt{a}_{ss}^{(2)}(\Tetaxx{j}{\vec{x}};\Tetaxx{j}{\vec{x}})}{2\;\im}
\biggr) \times \\ \no &\times \prod_{k=0}^{3}
\Bigg(\prod_{s,s\ppr=\uparrow,\downarrow}^{\mathsf{S}=2s,\,\mathsf{S}\ppr=2s\ppr}\prod_{j_{1,2}=1,\ldots,2N}^{\{\vec{x}_{1,2}\}}
\bigl(\mbox{except : }(s\ppr=s,\,\mathsf{S}\ppr=\mathsf{S})\And(j_{1}=j_{2})\And(\vec{x}_{1}=\vec{x}_{2})\bigr)\;\;\times
\\ \no &\times
\frac{d\hat{a}_{s\ppr s}^{(k)*}(\Tetax{2};\Tetax{1})\wedge
d\hat{a}_{s\ppr s}^{(k)}(\Tetax{2};\Tetax{1})}{2\;\im}
\Bigg)^{\boldsymbol{1/2}}_{\mbox{.}}
\end{align}
\end{subequations}
In comparison to \cite{pop1,pop2},\cite{mie3,miesinh},\cite{BCSQCD}, we introduce a 
generalized gradient term \((\hat{\pp}_{(\ln-\breve{H})}\hat{\mscr{Z}}(\nn{3};\nn{2})\,)\) (\ref{s1_20}) within the
determinant of (\ref{s1_18}) which consists of the adjoint operator action of the sum of a one-particle operator and a saddle point solution
\(\langle\hat{\mfrak{s}}\rangle_{\nu,s\ppr;\mu,s}^{aa}(\Tetax{2};\Tetax{1})\) of the density-related background self-energy
\begin{align}\lb{s1_20}
\Bigl(\hat{\pp}_{\ln(-\breve{\mscr{H}})}\hat{\mscr{Z}}(\nn{3};\nn{2})\Bigr)^{ba}\hspace*{-0.4cm}(\nn{5};\nn{1}) &:=
\biggl[\bigg(\exp\Big\{\overrightarrow{\big[\ln(-\breve{\mscr{H}}[\hat{\mfrak{H}},\langle\hat{\mfrak{s}}\rangle])\;{\Large\boldsymbol{,}}
\ldots\big]_{-}}\Big\}-\hat{1}\bigg)\hat{\mscr{Z}}^{b\ppr a\ppr}(\nn{3};\nn{2})\biggr]^{ba}\hspace*{-0.4cm}(\nn{5};\nn{1})\;.
\end{align}
The coset matrix \(\hat{\mscr{Z}}^{ba}(\nn{5};\nn{1})\) with locally Euclidean, 'Nambu' field increments
\(\sdelta\hat{\wt{a}}(\nn{5};\nn{1})\), \(\sdelta\hat{\wt{a}}\pdag(\nn{5};\nn{1})\) in the off-diagonal blocks \(b\neq a\)
is outlined in (\ref{s1_21a}) with the block diagonal, density-related terms 
\(\sdelta\hat{\mscr{Y}}^{aa}(\nn{5};\nn{1})\) (\ref{s1_21b}) which depend on
the locally Euclidean fields of exciton-related quasi-particles
\begin{subequations}
\begin{align}\lb{s1_21a}
\sdelta\hat{\mscr{Z}}^{ba}(\nn{5};\nn{1}) &=
\left(\bea{cc} \sdelta\hat{\mscr{Y}}^{11}(\nn{5};\nn{1}) & \sdelta\hat{\wt{a}}(\nn{5};\nn{1}) \\
\sdelta\hat{\wt{a}}\pdag(\nn{5};\nn{1}) & \sdelta\hat{\mscr{Y}}^{22}(\nn{5};\nn{1}) \eea\right)^{ba}_{\mbox{;}}    \\  \lb{s1_21b}
\sdelta\hat{\mscr{Y}}^{11}(\nn{5};\nn{1}) &=-\sdelta\hat{\mscr{Y}}^{22,T}(\nn{5};\nn{1})=-\tfrac{1}{2}\sum_{k=0}^{3}
\bigl(\hat{\tau}_{k}\hat{\tau}_{2}\bigr)_{\mu_{5}\mu_{1}}\;\times \\  \notag &\times\Bigg\{
\bigg[\tanh\Bigl(\tfrac{|\wt{a}_{\mathsf{S}_{5}}(\Tetax{5})|+|\wt{a}_{\mathsf{S}_{1}}(\Tetax{1})|}{2}\Bigr)-
\tanh\Bigl(\tfrac{|\wt{a}_{\mathsf{S}_{5}}(\Tetax{5})|-|\wt{a}_{\mathsf{S}_{1}}(\Tetax{1})|}{2}\Bigr)\bigg]\times \\ \notag &\times
\exp\bigg\{-\im\int_{\TT_{0}^{(+)}}^{\Teta{j_{1}}}d\Teta{j\ppr}
\frac{2\,|\wt{a}_{\mathsf{S}_{1}}(\Tetaxx{j\ppr}{\vec{x}_{1}})|}{\sinh\big(2\,|\wt{a}_{\mathsf{S}_{1}}(\Tetaxx{j\ppr}{\vec{x}_{1}})|\big)}
\frac{\pp\wt{\alpha}_{\mathsf{S}_{1}}(\Tetaxx{j\ppr}{\vec{x}_{1}})}{\pp\Teta{j\ppr}}\bigg\}\;
\sdelta\hat{a}_{s_{5}s_{1}}^{(k)}(\Tetax{5};\Tetax{1}) +   \\  \notag  &-(-1)^{k}
\bigg[\tanh\Bigl(\tfrac{|\wt{a}_{\mathsf{S}_{5}}(\Tetax{5})|+|\wt{a}_{\mathsf{S}_{1}}(\Tetax{1})|}{2}\Bigr)+
\tanh\Bigl(\tfrac{|\wt{a}_{\mathsf{S}_{5}}(\Tetax{5})|-|\wt{a}_{\mathsf{S}_{1}}(\Tetax{1})|}{2}\Bigr)\bigg]\times \\ \notag &\times
\exp\bigg\{\im\int_{\TT_{0}^{(+)}}^{\Teta{j_{5}}}d\Teta{j\ppr}
\frac{2\,|\wt{a}_{\mathsf{S}_{5}}(\Tetaxx{j\ppr}{\vec{x}_{5}})|}{\sinh\big(2\,|\wt{a}_{\mathsf{S}_{5}}(\Tetaxx{j\ppr}{\vec{x}_{5}})|\big)}
\frac{\pp\wt{\alpha}_{\mathsf{S}_{5}}(\Tetaxx{j\ppr}{\vec{x}_{5}})}{\pp\Teta{j\ppr}}\bigg\}\;
\sdelta\hat{a}_{s_{5}s_{1}}^{(k)\dag}(\Tetax{5};\Tetax{1}) + \\  \notag &-\im\:\delta_{\mu_{5}\mu_{1}}\:\delta_{s_{5}s_{1}}\;
\tanh\bigl(|\wt{a}_{\mathsf{S}}(\Tetaxx{j}{\vec{x}})|\bigr)\;\;|\wt{a}_{\mathsf{S}}(\Tetaxx{j}{\vec{x}})|\;\;
\sdelta\wt{\alpha}_{\mathsf{S}}(\Tetaxx{j}{\vec{x}})\;;  \\ \notag &
(\mbox{last term }(\Tetaxx{j}{\vec{x}}:=\Tetax{5}=\Tetax{1})\&(s:=s_{5}=s_{1}\;,\,\mathsf{S}=2s)\,)\;;\;\;\;
(\mbox{matrix})\pdag=(\mbox{matrix})^{*,T}\;; \\  \notag &
(\mbox{cf. gauge fixing in eqs. (\ref{s5_15a},\ref{s5_15b}) and appendix \ref{sa}})\;.
\end{align}
\end{subequations}
The derived equations with various kinds of approximations can be applied for the investigation of self-induced transparency and
the area theorem in the case of ultrashort coherent transients propagating in matter with reduced absorption \cite{Hahn}-\cite{Galanin};
this can be further combined with holography and Fourier optics.

\section{Representation of second quantized Hamiltonians with coherent states} \lb{s2}

\subsection{The coherent state path integral for the time development} \lb{s21}

In this section we define the total Hamiltonian \(\hat{\mathrm{H}}_{\mbox{\scz tot.}}(\hat{\psi}\pdag,\hat{\psi};t)\)
of second quantized Fermi operators \(\hat{\psi}_{\mu,s}(\vec{x}_{1})\) (\ref{s2_1a}) and transform to a
coherent state path integral (\ref{s2_15})
which is used for the four different HST's in sections \ref{s31} to \ref{s33} and \ref{s4}.
We approximate the solid by a single pair of semiconductor-related electron-hole bands which are regarded by the indices \(\mu\),
\(\nu\), the opposite charges \(q_{e}\), \(q_{h}\) (\ref{s2_1b}) and further spin degrees of freedom
\(s,s\ppr=\uparrow,\downarrow\) (\ref{s2_1c}) with projection labels \(\alpha_{\mu}\), \(\beta_{\mu}\)
to be applied for the dipole moments. The corresponding anti-commutators of Fermi operators
\(\hat{\psi}_{\mu,s}(\vec{x}_{1})\), \(\hat{\psi}_{\nu,s\ppr}\pdag(\vec{x}_{2})\) are listed in (\ref{s2_1a})
with their various definitions (\ref{s2_1b},\ref{s2_1c}) for the single '(e)lectron', the single '(h)ole'
and the two spin degrees of freedom. The delta function \(\deltaN_{\vec{x}_{2},\vec{x}_{1}}\) denotes a Kronecker delta
\(\delta_{\vec{x}_{2},\vec{x}_{1}}\) weighted by
the total number \(\mcal{N}_{x}=(\tfrac{L}{\sdelta x})^{d}\) of space points in
'$d$' dimensions with system length '\(L\)' and spatial grid intervals '\(\sdelta x\)'.
The total number \(\mcal{N}_{x}=(\tfrac{L}{\sdelta x})^{d}\) of space points can be applied as a parameter of expansion
for the 'HST' transformed generating functions, provided that the density-related background fields do not vary considerably
on the spatial grid with \(\mcal{N}_{x}\) points in comparison to the 'Nambu'-related terms.
This parameter \(\mcal{N}_{x}=(\tfrac{L}{\sdelta x})^{d}\) of total space points is therefore similar to the parameter of
matrix dimensions within random matrix theories.
Spatial sums \(\sum_{\vec{x}}\ldots=\int_{L^{d}}(\tfrac{dx}{L})^{d}\ldots\) are normalized by the system
volume '\(L^{d}\)' to dimensionless integrals, whose overall dimensions are determined by the physical dimension
of the prevailing integrand, apart from the primed, unweighted case \(\sum_{\vec{x}}\ppr\ldots\) of pure summation
over numbered, spatial points. Double or multiple appearance of indices or spacetime points involve the standard convention
of summations, except for indices or spacetime labels set in parentheses
\begin{subequations}
\begin{align}\lb{s2_1a}
\boldsymbol{\{} \hat{\psi}_{\mu,s}(\vec{x}_{1})\;\text{\bf\Large,}\;\hat{\psi}_{\nu,s\ppr}\pdag(\vec{x}_{2})
\boldsymbol{\}_{+}} & =
\delta_{\mu\nu}\;\;\delta_{ss\ppr}\;\;\deltaN_{\vec{x}_{1},\vec{x}_{2}}\;; &
\boldsymbol{\{} \hat{\psi}_{\mu,s}(\vec{x}_{1})\;\text{\bf\Large,}\;\hat{\psi}_{\nu,s\ppr}(\vec{x}_{2})
\boldsymbol{\}_{+}}  &=\boldsymbol{\{} \hat{\psi}_{\mu,s}\pdag(\vec{x}_{1})\;\text{\bf\Large,}\;
\hat{\psi}_{\nu,s\ppr}\pdag(\vec{x}_{2})
\boldsymbol{\}_{+}} \equiv0 \;;  \\   \lb{s2_1b}
\mu,\;\nu & = \text{\scz'(e)lectron'},\;\text{\scz'(h)ole'}\;; &
q_{\mu},\; q_{\nu} \; :\; q_{e}=-1&,\; q_{h}=+1\;; \\   \lb{s2_1c}
s,\;s\ppr & =\uparrow,\downarrow\;; &
\alpha_{\mu},\;\beta_{\mu} \; : \;\alpha_{e}=+1&,\;\alpha_{h}=0\;;\;\;
\beta_{e}=0,\;\beta_{h}=+1 \;.
\end{align}
\end{subequations}
The total Hamiltonian \(\hat{\mathrm{H}}_{\mbox{\scz tot.}}(\hat{\psi}\pdag,\hat{\psi};t)\) (\ref{s2_2a}) is given in terms of a
bilinear, kinetic density part \(\hat{H}_{F}(\hat{\psi}\pdag,\hat{\psi})\) (\ref{s2_2b}) with general dispersion
\(\hat{\mscr{E}}_{\mu,s\ppr s}(\vec{x}_{2})\;\delta_{\vec{x}_{2},\vec{x}_{1}}^{(\mcal{N}_{x})}\), which can be
approximated by relations (\ref{s2_3a},\ref{s2_3b}), and also contains an anomalous part
\(\hat{H}_{E}(\hat{\psi}\pdag,\hat{\psi};t)\) (\ref{s2_2c}), which is driven by an electric field, coupled to a dipole
moment \(\vec{\mfrak{p}}_{\nu,s\ppr;\mu,s}\); hence, one only creates electron-hole pairs in correspondence with
charge conversation. Moreover, the general dipole moment \(\vec{\mfrak{p}}_{\nu,s\ppr;\mu,s}\) should be
constrained to the two parallel spin directions \(s=s\ppr=\uparrow,\downarrow\) in order to regard
the spin one-property of the electromagnetic field with the two possible spin components
\(2\,s=\mathsf{S}=2\,s\ppr=\mathsf{S}\ppr=\{+1,-1\}\).
We therefore summarize this electron-hole generation by a rather general, time-dependent,
driving dipole moment \(\hat{\mscr{D}}_{\nu,s\ppr;\mu,s}(\vec{x}_{2},\vec{x}_{1};t)\approx\delta_{s\ppr s}\:
\hat{\mscr{D}}_{\nu;\mu}(\vec{x}_{2},\vec{x}_{1};t)\) (\ref{s2_3c}) with the specific dipole
inter-band moment \(\vec{\mfrak{p}}_{\nu,s\ppr;\mu,s}(\vec{x}_{2},\vec{x}_{1};t)\approx \delta_{s\ppr s}\:
\vec{\mfrak{p}}_{\nu;\mu}(\vec{x}_{2},\vec{x}_{1};t)\),
a spatially dependent band gap \(E_{g}(\vec{x})\) and
an electric field of central frequency \(\omega\) and envelope \(\vec{E}_{0}(\vec{x};t)\).
According to various kinds of possible dipole moments, a rather general kind
is used with the type '\(\hat{\mscr{D}}_{\nu,s\ppr;\mu,s}(\vec{x}_{2},\vec{x}_{1};t)\)' and
projection labels \(\beta_{\nu}\), \(\alpha_{\mu}\) for inter-band transitions which may be approximated
by a constant dipole moment in the case of very ultrashort pulses \cite{Enz}. Apart from the Coulomb
interaction \(\hat{H}_{FF}(\hat{\psi}\pdag,\hat{\psi})\) (\ref{s2_2d}) of quartic Fermi operators and
Coulomb potential \(V(\vec{x}_{2},\vec{x}_{1})\), we also consider a random potential \(u(\vec{x},t)\)
with a Gaussian distribution (\ref{s2_3d},\ref{s2_3e}) for disorder and noise in the semiconductor-related
solid. After an ensemble average according to (\ref{s2_3d},\ref{s2_3e}), one thus also achieves an anti-hermitian,
dissipative, quartic interaction of Fermi fields with the spacetime variance \(f(\vec{x}_{2},t_{2};\vec{x}_{1},t_{1})\)
and energy scale '\(u_{0}\)'
\begin{subequations}
\begin{align}  \lb{s2_2a}
\hat{\mathrm{H}}_{\mbox{\scz tot.}}(\hat{\psi}\pdag,\hat{\psi};t) & =  \hat{H}_{F}(\hat{\psi}\pdag,\hat{\psi}) +
\hat{H}_{E}(\hat{\psi}\pdag,\hat{\psi};t) + \hat{H}_{FF}(\hat{\psi}\pdag,\hat{\psi}) +
\hat{H}_{D}(\hat{\psi}\pdag,\hat{\psi};t) \;; \\    \lb{s2_2b}
\hat{H}_{F}(\hat{\psi}\pdag,\hat{\psi}) & = \sum_{\vec{x}_{1,2}}^{\mu,s,s\ppr}
\hat{\psi}_{\mu,s\ppr}\pdag(\vec{x}_{2})\quad\hat{\mscr{E}}_{\mu,s\ppr s}(\vec{x}_{2})\;
\deltaN_{\vec{x}_{2},\vec{x}_{1}}\quad \psi_{\mu,s}(\vec{x}_{1})\;; \\    \lb{s2_2c}
\hat{H}_{E}(\hat{\psi}\pdag,\hat{\psi};t) & = \sum_{\vec{x}_{1,2}}^{\mu,\nu,s,s\ppr}
\bigg[\alpha_{\nu}\;\beta_{\mu}\;\hat{\psi}_{\nu,s\ppr}\pdag(\vec{x}_{2})\quad\;
\hat{\mscr{D}}_{\nu,s\ppr;\mu,s}\pdag(\vec{x}_{2},\vec{x}_{1};t)\quad\hat{\psi}_{\mu,s}\pdag(\vec{x}_{1})+
\\  \notag & \quad +
\beta_{\nu}\;\alpha_{\mu}\;\hat{\psi}_{\nu,s\ppr}(\vec{x}_{2})\quad\;
\hat{\mscr{D}}_{\nu,s\ppr;\mu,s}(\vec{x}_{2},\vec{x}_{1};t)\quad
\hat{\psi}_{\mu,s}(\vec{x}_{1})\bigg] \;; \\ \lb{s2_2d}
\hat{H}_{FF}(\hat{\psi}\pdag,\hat{\psi}) & = \sum_{\vec{x}_{1,2}}^{\mu,\nu,s,s\ppr}
\hat{\psi}_{\mu,s}\pdag(\vec{x}_{1})\;\hat{\psi}_{\nu,s\ppr}\pdag(\vec{x}_{2})\quad
q_{\mu}\;q_{\nu}\;\;V(\vec{x}_{2},\vec{x}_{1})\quad
\hat{\psi}_{\nu,s\ppr}(\vec{x}_{2})\;\hat{\psi}_{\mu,s}(\vec{x}_{1}) \;; \\  \lb{s2_2e}
\hat{H}_{D}(\hat{\psi}\pdag,\hat{\psi};t) & = \sum_{\vec{x},\mu,s}
\hat{\psi}_{\mu,s}\pdag(\vec{x})\quad q_{\mu}\;\;
u(\vec{x},t)\quad \psi_{\mu,s}(\vec{x}) \;.
\end{align}
\end{subequations}
\begin{subequations}
\begin{align} \lb{s2_3a}
\hat{\mscr{E}}_{\mu,s\ppr s}(\vec{x}_{2})\;\deltaN_{\vec{x}_{2},\vec{x}_{1}} & = \tfrac{1}{2}
\Big(\hat{M}_{\mu,s_{1}s\ppr}^{-1/2}(\vec{x}_{2})\tfrac{\hbar}{\im}\overrightarrow{\pp}_{\vec{x}_{2}}\Big)^{\!\!\dag}
\Big(\hat{M}_{\mu,s_{1}s}^{-1/2}(\vec{x}_{2})\tfrac{\hbar}{\im}\overrightarrow{\pp}_{\vec{x}_{2}}\Big)
\deltaN_{\vec{x}_{2},\vec{x}_{1}}  \;;  \\   \lb{s2_3b}
\hat{M}_{\mu,s\ppr s}^{\boldsymbol{-1}}(\vec{x}_{2})  &\approx
\tfrac{1}{m_{\mu}}\;\;\delta_{s\ppr s}\;\delta_{\vec{x}_{2},\vec{x}_{1}}+\ldots\;;\qquad
m_{e},\;m_{h}>0\;;   \\   \notag
\hat{M}_{\mu,s\ppr s}^{\boldsymbol{-1}}(\vec{x}_{2})  &: \mbox{generalized 'mass' metric tensor similar as in
general relativity theory}\;; \\   \lb{s2_3c}
\hat{\mscr{D}}_{\nu,s\ppr;\mu,s}(\vec{x}_{2},\vec{x}_{1};t) & \approx -\,\delta_{s\ppr s}\Bigl(
\vec{\mfrak{p}}_{\nu;\mu}\negthinspace\bigl(\vec{x}_{2},\vec{x}_{1};t\bigr)\boldsymbol{\cdot}
\vec{E}_{0}\bigl(\tfrac{1}{2}(\vec{x}_{2}+\vec{x}_{1});t\bigr)\Bigr)\quad
\cos\Bigl\{\tfrac{1}{\hbar}(\hbar\omega-
E_{g}\bigl(\tfrac{1}{2}(\vec{x}_{2}+\vec{x}_{1})\,\bigr)\cdot t\Bigr\}\;\;; \\ \notag
\hat{\mscr{D}}_{\nu,s\ppr;\mu,s}(\vec{x}_{2},\vec{x}_{1};t) & \approx -\,\delta_{s\ppr s}\Bigl(
\vec{\mfrak{p}}_{\nu;\mu}\negthinspace\bigl(\vec{x}_{2},\vec{x}_{1};t\bigr)\boldsymbol{\cdot}
\vec{E}_{0}\bigl(\tfrac{1}{2}(\vec{x}_{2}+\vec{x}_{1});t\bigr)\Bigr)\quad\;\tfrac{1}{2}\;
\underbrace{\exp\Bigl\{\tfrac{\im}{\hbar}(\hbar\omega-
E_{g}\bigl(\tfrac{1}{2}(\vec{x}_{2}+\vec{x}_{1})\,\bigr)\cdot t\Bigr\}}_{\mbox{\scz rotating wave
approximation}}\;\;;   \\   \notag
\vec{\mfrak{p}}_{\nu;\mu}(\vec{x}_{2},\vec{x}_{1};t) &\approx \beta_{\nu}\,\alpha_{\mu}\;
\deltaN_{\vec{x}_{2},\vec{x}_{1}}\;\vec{\mfrak{p}}_{\mbox{\scz h};\mbox{\scz e}}(\vec{x}_{1};t)\;;    \\  \lb{s2_3d}
\mscr{P}\bigl[u(\vec{x},t)\bigr] & = \mcal{N}_{\mscr{P}}\;\;\;
\exp\bigl\{-\tfrac{1}{2\;u_{0}^2}{\ts\sum_{\vec{x}_{1,2}}}\ppr\;u(\vec{x}_{2},t_{2})\;
f^{\boldsymbol{-1}}(\vec{x}_{2},t_{2};\vec{x}_{1},t_{1})\;u(\vec{x}_{1},t_{1})\bigr\}\;;  \\   \lb{s2_3e}
\ovv{u(\vec{x}_{1},t_{1})\;\;u(\vec{x}_{2},t_{2})} &= u_{0}^{2}\;\;f(\vec{x}_{1},t_{1};\vec{x}_{2},t_{2})\;\;\;.
\end{align}
\end{subequations}
We further extend the total Hamiltonian \(\hat{\mathrm{H}}_{\mbox{\scz tot.}}(\hat{\psi}\pdag,\hat{\psi};t)\) (\ref{s2_2a}) by a source
term '\(\hat{\mscr{J}}\)', defined in (\ref{s2_4},\ref{s2_9a},\ref{s2_10a})
for generating bilinear observables of Fermi fields, and
introduce the corresponding coherent state path integral (\ref{s2_4}) on a non-equilibrium time contour
which is taken into account by a 'plus' branch (\ref{s2_5a}) for forward propagation (lower line in (\ref{s2_4}))
and a 'minus' branch (\ref{s2_5b}) (upper line in (\ref{s2_4})) for backward propagation.
In order to include the changing sign in the exponent of the time-step development operators, a contour
metric \(\eta_{j}\) (\ref{s2_7})  is defined in correspondence with a contour time \(\TT_{\scrscr j}\) (\ref{s2_6})
which has equivalent values \(\TT_{\scrscr j}=\TT_{\scrscr 2N-j}=t_{\scrscr j}\) (\ref{s2_8a}), due
to the separation of the sign with metric \(\etan{j}=-\etan{2N-j}\), \((j\neq N)\). We accomplish a path integral
by using the anti-commuting coherent states (\ref{s2_8b}) with '(e)lectron'-, '(h)ole' and spin degrees
of freedom and further distinction of the prevailing plus-, minus-branch so that the labeling
\(\chi_{\mu,s}(\Teta{j},\vec{x})\) with an additional sign '\(^{(\etan{j})}\)' has to be applied
for distinguishing the Grassmann-valued coherent states between their plus-, minus-branch of propagation
\begin{align}\lb{s2_4}
Z[\hat{\mscr{J}}] & = \Big\langle0\Big|\exp\Big\{\tfrac{\im}{\hbar}\:\sdelta t\;
\hat{\mathrm{H}}_{\mbox{\scz tot.};\hat{\mscr{J}}}\bigl(\hat{\psi}\pdag,\hat{\psi};t_{1}\bigr)\Big\}\cdot\,\ldots\,\cdot
\exp\Big\{\tfrac{\im}{\hbar}\:\sdelta t\;
\hat{\mathrm{H}}_{\mbox{\scz tot.};\hat{\mscr{J}}}\bigl(\hat{\psi}\pdag,\hat{\psi};t_{N}\bigr)\Big\}\times \\ \notag  &\times\;
\exp\Big\{\negthickspace-\tfrac{\im}{\hbar}\:\sdelta t\;
\hat{\mathrm{H}}_{\mbox{\scz tot.};\hat{\mscr{J}}}\bigl(\hat{\psi}\pdag,\hat{\psi};t_{N}\bigr)\Big\}\cdot\,\ldots\,\cdot
\exp\Big\{\negthickspace-\tfrac{\im}{\hbar}\:\sdelta t\;
\hat{\mathrm{H}}_{\mbox{\scz tot.};\hat{\mscr{J}}}\bigl(\hat{\psi}\pdag,\hat{\psi};t_{1}\bigr)\Big\}\Big|0\Big\rangle\;;
\end{align}
\begin{subequations}
\begin{alignat}{9}\lb{s2_5a}
\text{Contour time}&: \text{'+' branch}&\;\; t_{\scrscr N}&>\ldots>&
t_{\scrscr j+1}\;\;&> \quad(t_{\scrscr j}=\TT_{\scrscr j})\quad &>\ldots &> \quad t_{\scrscr 1}
&> t_{\scrscr 0} \;; \\   \lb{s2_5b}
\text{Contour time}&: \text{'-' branch}&\;\; t_{\scrscr 0}&<\ldots<& \;t_{\scrscr N-(j+1)}&<
 (t_{{\scrscr N-j}}=\TT_{{\scrscr N+j}})\quad&<
\ldots &< t_{\scrscr N-1} &< (t_{\scrscr N});
\end{alignat}
\end{subequations}
\begin{alignat}{8}  \lb{s2_6}
\TT_{\scrscr 2N}&=t_{\scrscr0}\;;&\TT_{\scrscr 2N-1}&=t_{\scrscr 1}\;;&
\;\ldots\;;\TT_{\scrscr N+2}&=t_{\scrscr N-2}\;;&\TT_{\scrscr N+1}&=t_{\scrscr N-1}\;;&
\TT_{\scrscr N}&=t_{\scrscr N}\;;&\TT_{\scrscr N-1}&=t_{\scrscr N-1}\;;&\;\ldots\;;
\TT_{\scrscr 1}&=t_{\scrscr 1}\;; &\TT_{\scrscr 0}&=t_{\scrscr0}\;;\\ \lb{s2_7}
\eta_{\scrscr 2N}&={\scr-1}\;; &\eta_{\scrscr 2N-1}&={\scr-1}\;; &
\;\ldots\;;\eta_{\scrscr N+2}&={\scr-1}\;;
&\eta_{\scrscr N+1}&={\scr-1}\;;&\eta_{\scrscr N}&={\scr+1}\;;
&\eta_{\scrscr N-1}&={\scr+1}\;; &\;\ldots\;;
\eta_{\scrscr 1}&={\scr+1}\;; & \eta_{\scrscr 0}&={\scr+1}\;;
\end{alignat}
\begin{subequations}
\begin{align} \lb{s2_8a}
\TT_{\scrscr j} &= \TT_{\scrscr 2N-j} \;;&\quad \eta_{j}&=-\eta_{\scrscr 2N-j}\;;{\scr(0\leq j\leq N-1)}\;;
\;\;\eta_{\scrscr N}={\scr +1}\;;
\end{align}
\begin{align} \lb{s2_8b}&
\chi_{\mu,s}\bigl(\Teta{j},\vec{x}\bigr)\;;\;\chi_{\mu,s}\bigl(\Teta{j-1},\vec{x}\bigr)\;;\;
\chi_{\mu,s}\bigl(\Teta{j_{1}},\vec{x}\bigr)\;;\;\chi_{\mu,s}\bigl(\Teta{j_{1}-1},\vec{x}\bigr)\;;\;
\chi_{\mu,s}\bigl(\Teta{j_{2}},\vec{x}\bigr)\;;\;\chi_{\mu,s}\bigl(\Teta{j_{2}-1},\vec{x}\bigr)\;.
\end{align}
\end{subequations}
Due to the over-completeness relation of coherent states and the normal ordering of the total
Hamiltonian, we finally acquire the defining coherent state path integral (\ref{s2_9a},\ref{s2_9b})
for the time development
\begin{subequations}
\begin{align} \lb{s2_9a}
Z[\hat{\mscr{J}}] & = \Big\langle0\Big|\prod_{j=1}^{2N}\exp\Big\{\negthickspace-\tfrac{\im}{\hbar}\:\sdelta t\;
\eta_{j}\;\;
\hat{\mathrm{H}}_{\mbox{\scz tot.};\hat{\mscr{J}}}\bigl(\hat{\psi}\pdag,\hat{\psi};\TT_{j+(\eta_{j}-1)/2}\bigr)\Big\}
\Big|0\Big\rangle\;; \\     \lb{s2_9b}
Z[\hat{\mscr{J}}] & = \int\prod_{\vec{x},\mu,s}\prod_{j=0}^{2N}
\tfrac{d\chi_{\mu,s}^{*}\bigl(\Teta{j},\vec{x}\bigr)\;\;
d\chi_{\mu,s}\bigl(\Teta{j},\vec{x}\bigl)}{\mcal{N}_{x}}\;\;\times \\ \notag
&\times\exp\bigg\{\negthickspace-\sum_{\vec{x},\mu,s}\bigg(\sum_{j=0}^{2N}
\chi_{\mu,s}^{*}(\Teta{j},\vec{x})\;\;\chi_{\mu,s}(\Teta{j},\vec{x}) -
\sum_{j=1}^{2N}
\chi_{\mu,s}^{*}(\Teta{j},\vec{x})\;\;\chi_{\mu,s}(\Teta{j-1},\vec{x})\bigg)\bigg\}\times \\ \notag &\times
\prod_{j=1}^{2N}
\exp\bigg\{\negthickspace-\tfrac{\im}{\hbar}\:\sdelta t\;
\eta_{j}\;\;
\hat{\mathrm{H}}_{\mbox{\scz tot.};\hat{\mscr{J}}}\Bigl(\chi^{*}(\Teta{j}),\chi(\Teta{j-1});
\TT_{j+(\eta_{j}-1)/2}\Bigr)\bigg\}\;.
\end{align}
\end{subequations}

\subsection{Anomalous doubling of the one-particle part of the Hamiltonian} \lb{s22}

As one performs the additional ensemble average (\ref{s2_3d},\ref{s2_3e}), one attains the ensemble averaged
coherent state path integral \(\ovv{Z[\hat{\mscr{J}}]}\) (\ref{s2_10a}) with the two-body interaction
potential \(\hat{\mscr{V}}(\vec{x}_{2},\Teta{j_{2}};\vec{x}_{1},\Teta{j_{1}})\) (\ref{s2_10b}),
containing a dissipative noise part, and density pairs \(\mfrak{b}(\Teta{j_{1}},\vec{x}_{1})\) (\ref{s2_10c})
for the quartic interaction of fields. An anti-hermitian epsilon term (\(\ve_{+}>0\))
is added for analytic, convergent
properties of Green functions following after the integration over the Grassmann fields
\begin{subequations}
\begin{align}\lb{s2_10a}
\ovv{Z[\hat{\mscr{J}}]} &=\int\prod_{\vec{x},\mu,s}\prod_{j=0}^{2N}
\tfrac{d\chi_{\mu,s}^{*}\bigl(\Teta{j},\vec{x}\bigr)\;\;d\chi_{\mu,s}\bigl(\Teta{j},\vec{x}\bigr)}{\mcal{N}_{x}}
\;\;\times \\ \notag
&\times\exp\bigg\{\negthickspace-\sum_{\vec{x},\mu,s}\bigg(\sum_{j=0}^{2N}
\chi_{\mu,s}^{*}(\Teta{j},\vec{x})\;\;\chi_{\mu,s}(\Teta{j},\vec{x}) -
\sum_{j=1}^{2N}
\chi_{\mu,s}^{*}(\Teta{j},\vec{x})\;\;
\chi_{\mu,s}(\Teta{j-1},\vec{x})\bigg)\bigg\}\times \\ \notag &\times
\exp\bigg\{\negthickspace-
\sum_{\vec{x}_{1,2}}^{\mu,\nu,s,s\ppr}\sum_{j=1}^{2N}\eta_{j}\;\tfrac{\im}{\hbar}\;{\scrscr\Delta}t\Big[
\chi_{\nu,s\ppr}^{*}(\Teta{j},\vec{x}_{2})\;\delta_{\nu\mu}\;\;\Bigl(
\hat{\mscr{E}}_{\mu,s\ppr s}\bigl(\vec{x}_{2}\bigr)-\im\:\eta_{j}\:\delta_{s\ppr s}\:\ve_{+}\Bigr)\deltaN_{\vec{x}_{2},\vec{x}_{1}}\;
\chi_{\mu,s}(\Teta{j-1},\vec{x}_{1})+
\\ \notag &+ \alpha_{\nu}\,\beta_{\mu}\;\chi_{\nu,s\ppr}^{*}(\Teta{j},\vec{x}_{2})\quad\:
\hat{\mscr{D}}_{\nu,s\ppr;\mu,s}\pdag\bigl(\vec{x}_{2},\vec{x}_{1};\TT_{j+(\eta_{j}-1)/2}\bigr)\quad
\chi_{\mu,s}^{*}(\Teta{j},\vec{x}_{1}) + \\ \notag &+
\beta_{\nu}\,\alpha_{\mu}\;\chi_{\nu,s\ppr}(\Teta{j-1},\vec{x}_{2})\quad\:
\hat{\mscr{D}}_{\nu,s\ppr;\mu,s}\bigl(\vec{x}_{2},\vec{x}_{1};\TT_{j+(\eta_{j}-1)/2}\bigr)\quad
\chi_{\mu,s}(\Teta{j-1},\vec{x}_{1}) \Big]\bigg\}\times \\ \notag &\times
\exp\Big\{-\negthickspace\sum_{\vec{x}_{1,2}}\sum_{j_{1,2}=1}^{2N}
\mfrak{b}\bigl(\Teta{j_{2}},\vec{x}_{2}\bigr)\;\eta_{j_{2}}\quad
\Bigl(\im\tfrac{\sdelta t}{\hbar}\Bigr)\;\;
\hat{\mscr{V}}\bigl(\vec{x}_{2},\Teta{j_{2}};\vec{x}_{1},\Teta{j_{1}}\bigl)\quad\eta_{j_{1}}\;
\mfrak{b}\bigl(\Teta{j_{1}},\vec{x}_{1}\bigr)\Big\}\times \\ \notag &\times
\exp\Big\{-\sum_{\vec{x}_{1,2}}^{\mu_{1,2},s_{1,2}}\sum_{j_{1,2}=0}^{2N+1}
\Xi_{\mu_{2},s_{2}}^{\sharp,b}(\Teta{j_{2}},\vec{x}_{2})\;\eta_{j_{2}}\;\tfrac{1}{2}\;
\hat{\mscr{J}}_{\mu_{2},s_{2};\mu_{1},s_{1}}^{ba}(\Teta{j_{2}},\vec{x}_{2};
\Teta{j_{1}},\vec{x}_{1})\;\eta_{j_{1}}\;\Xi_{\mu_{1},s_{1}}^{a}(\Teta{j_{1}},\vec{x}_{1})\Big\} \;;\\ \lb{s2_10b} &
\hat{\mscr{V}}(\vec{x}_{2},\Teta{j_{2}};\vec{x}_{1},\Teta{j_{1}}) = \eta_{j_{1}}\;\delta_{j_{2},j_{1}}\;
\bigl(\delta_{\eta_{{j}_{2}},\eta_{{j}_{1}}}\bigr)\;
V(\vec{x}_{2},\vec{x}_{1})\;-\;
\tfrac{\im}{2}\;\tfrac{\sdelta t}{\hbar}\;\;u_{0}^{2}\;\;
f\bigl(\vec{x}_{2},\TT_{j_{2}};\vec{x}_{1},\TT_{j_{1}}\bigr);\;(j_{1},j_{2}=1,\ldots,2N)\;;   \\   \lb{s2_10c} &
\hspace*{0.9cm}\mfrak{b}\bigl(\Teta{j_{1}},\vec{x}_{1}\bigr)=\sum_{\mu_{1},s_{1}}
\chi_{\mu_{1},s_{1}}^{*}\bigl(\Teta{j_{1}},\vec{x}_{1}\bigr)\;q_{\mu_{1}}\;
\chi_{\mu_{1},s_{1}}(\Teta{j_{1}-1},\vec{x}_{1}\bigr)\;\;.
\end{align}
\end{subequations}
Note, that we have also specified the source term for generating observables in the last line of (\ref{s2_10a}).
However, aside from the density terms of \(\hat{H}_{F}(\hat{\psi}\pdag,\hat{\psi})\),
\(\hat{H}_{FF}(\hat{\psi}\pdag,\hat{\psi})\), and for \(\hat{H}_{D}(\hat{\psi}\pdag,\hat{\psi};t)\),
one has the anomalous operator \(\hat{H}_{E}(\hat{\psi}\pdag,\hat{\psi};t)\) with pairwise creation and
annihilation of electrons and holes so that an anomalous doubling has to be taken for the
anti-commuting fields. This is illustrated in relations (\ref{s2_11a}-\ref{s2_11g}) by introducing the anomalous
doubled fields \(\Xi_{\mu_{1},s_{1}}^{a}\!(\Teta{j_{1}},\vec{x}_{1})\) (\ref{s2_11a}). We emphasize the
{\it time step shift from '\(j_{1}-1\)' to '\(j_{1}\)'} between the upper
\(\chi_{\mu_{1},s_{1}}(\Teta{j_{1}-1},\vec{x}_{1})\) (\(a=1\)) and lower component
\(\chi_{\mu_{1},s_{1}}^{*}(\Teta{j_{1}},\vec{x}_{1})\) (\(a=2\)) of the anomalous-doubled field
\(\Xi_{\mu_{1},s_{1}}^{a}\!(\Teta{j_{1}},\vec{x}_{1})\) (\ref{s2_11a}), due to the normal ordering of
the Hamilton operators. Furthermore, we separately define extensions (\ref{s2_11b},\ref{s2_11c})
{\it without anomalous doubling} for the end
points '\(j_{1}=0\)' and '\(j_{1}=2N+1\)' in order to consider the exact, proper sequence of time steps for
the quantum problem (cf.\ Ref. \cite{mie3}). This extension
has also to be regarded for the hermitian conjugation '\(^{\sharp}\)' in (\ref{s2_11d}) with the peculiar
time step between upper '\(b=1\)' and lower component '\(b=2\)', but in reversed order compared to
(\ref{s2_11a}). The particular extension for the end points
'\(j_{2}=0\)' and '\(j_{2}=2N+1\)' is similarly given in Eqs. (\ref{s2_11e},\ref{s2_11f}), also
with the {\it absence of anomalous doubling} as in (\ref{s2_11b},\ref{s2_11c}). The hermitian conjugated field
\(\Xi_{\mu_{2},s_{2}}^{\sharp,b}\!(\Teta{j_{2}},\vec{x}_{2})\) (\ref{s2_11d}) also follows by multiplying
\(\Xi_{\mu_{1},s_{1}}^{a}\!\bigl(\Teta{j_{1}},\vec{x}_{1}\bigr)\) (\ref{s2_11a}) with the 'Nambu'-related Pauli matrix
\((\hat{\tau}_{1})^{ba}\) and a subsequent transposition so that one has property (\ref{s2_11g}) which is
used in removing the 'Nambu'-doubled anti-commuting fields by integration
\begin{subequations}
\begin{align} \lb{s2_11a}
\Xi_{\mu_{1},s_{1}}^{a}\!\bigl(\Teta{j_{1}},\vec{x}_{1}\bigr) &=
\left(\bea{c}\chi_{\mu_{1},s_{1}}(\Teta{j_{1}-1},\vec{x}_{1}) \\
\chi_{\mu_{1},s_{1}}^{*}(\Teta{j_{1}},\vec{x}_{1})\eea\right)^{a} \;; \\   \lb{s2_11b}
\Xi_{\mu_{1},s_{1}}^{a\equiv1}\!\!\Bigl(\TT_{(j_{1}=0)}^{\scrscr(\eta_{(j_{1}=0)}=+)},\vec{x}_{1}\Bigr) &=
\chi_{\mu_{1},s_{1}}^{*}\!\!\Bigl(\TT_{(j_{1}=0)}^{\scrscr(\eta_{(j_{1}=0)}=+)},\vec{x}_{1}\Bigr) \;; \\   \lb{s2_11c}
\Xi_{\mu_{1},s_{1}}^{a\equiv1}\!\!\Bigl(\TT_{(j_{1}=2N+1)}^{\scrscr(\eta_{(j_{1}=2N+1)}=-)},\vec{x}_{1}\Bigr) &=
\chi_{\mu_{1},s_{1}}\!\!\Bigl(\TT_{(j_{1}=2N)}^{\scrscr(\eta_{(j_{1}=2N)}=-)},\vec{x}_{1}\Bigr)  \;;  \\  \lb{s2_11d}
\Xi_{\mu_{2},s_{2}}^{\sharp,b}\!\!\Bigl(\Teta{j_{2}},\vec{x}_{2}\Bigr) &=
\Bigl(\bea{c}\chi_{\mu_{2},s_{2}}^{*}(\Teta{j_{2}},\vec{x}_{2})\;,\;
\chi_{\mu_{2},s_{2}}(\Teta{j_{2}-1},\vec{x}_{2})\eea\Bigr)^{b} \;;  \\   \lb{s2_11e}
\Xi_{\mu_{2},s_{2}}^{\sharp,b\equiv1}\!\!\Bigl(\TT_{(j_{2}=0)}^{\scrscr(\eta_{j_{2}=0}=+)},\vec{x}_{2}\Bigr) &=
\chi_{\mu_{2},s_{2}}^{*}\!\!\Bigl(\TT_{j_{2}=0}^{\scrscr(\eta_{j_{2}=0}=+)},\vec{x}_{2}\Bigr) \;; \\   \lb{s2_11f}
\Xi_{\mu_{2},s_{2}}^{\sharp,b\equiv1}\!\!\Bigl(\TT_{(j_{2}=2N+1)}^{\scrscr(\eta_{j_{2}=2N+1}=-)},\vec{x}_{2}\Bigr) &=
\chi_{\mu_{2},s_{2}}\!\!\Bigl(\TT_{j_{2}=2N}^{\scrscr(\eta_{j_{2}=2N}=-)},\vec{x}_{2}\Bigr)  \;; \\ \lb{s2_11g}
\Xi_{\mu_{2},s_{2}}^{\sharp,b}\!\!\Bigl(\Teta{j_{2}},\vec{x}_{2}\Bigr) &= \biggl(\bigl(\hat{\tau}_{1}\bigr)^{ba}\;
\Xi_{\mu_{2},s_{2}}^{a}\!\bigl(\Teta{j_{2}},\vec{x}_{2}\bigr)\biggr)^{\boldsymbol{T}}\;\;.
\end{align}
\end{subequations}
As one applies the anti-commuting fields
\(\Xi_{\mu_{1},s_{1}}^{a}\!(\Teta{j_{1}},\vec{x}_{1})\) (\ref{s2_11a}),
\(\Xi_{\mu_{2},s_{2}}^{\sharp,b}\!(\Teta{j_{2}},\vec{x}_{2})\) (\ref{s2_11d}) to the exponent in (\ref{s2_10a}),
one obtains relation (\ref{s2_12}) with doubled one-particle part
\(\hat{\mathsf{H}}_{\nu,s\ppr;\mu,s}^{ba}(\vec{x}_{2},\Teta{j_{2}};\vec{x}_{1},\Teta{j_{1}})\) (\ref{s2_14a}-\ref{s2_14d})
which has to be combined with the 'Nambu' metric tensors (\ref{s2_13a},\ref{s2_13b}) for the process
of anomalous doubling of anti-commuting fields (cf. Ref. \cite{mie3})
\begin{align}\lb{s2_12}&
\sum_{\vec{x},\mu,s}\bigg(\sum_{j=0}^{2N}
\chi_{\mu,s}^{*}(\Teta{j},\vec{k})\;\;\chi_{\mu,s}(\Teta{j},\vec{x}) -
\sum_{j=1}^{2N}
\chi_{\mu,s}^{*}(\Teta{j},\vec{x})\;\;
\chi_{\mu,s}(\Teta{j-1},\vec{x})\Big)+ \\ \notag &+
\sum_{\vec{x}_{1,2}}^{\mu,\nu,s,s\ppr}\sum_{j=1}^{2N}
\eta_{j}\;\tfrac{\im}{\hbar}\;{\scrscr\Delta}t\Big[
\chi_{\nu,s\ppr}^{*}(\Teta{j},\vec{x}_{2})\;\delta_{\nu\mu}
\Bigl(\hat{\mscr{E}}_{\mu,s\ppr s}(\vec{x}_{2})-\im\:\eta_{j}\:
\delta_{s\ppr s}\:\ve_{+}\Bigr)\deltaN_{\vec{x}_{2},\vec{x}_{1}}
\;\;\chi_{\mu,s}(\Teta{j-1},\vec{x}_{1})+
\\ \notag &+ \alpha_{\nu}\,\beta_{\mu}\;\chi_{\nu,s\ppr}^{*}(\Teta{j},\vec{x}_{2})\quad\:
\hat{\mscr{D}}_{\nu,s\ppr;\mu,s}\pdag\bigl(\vec{x}_{2},\vec{x}_{1};\TT_{j+(\eta_{j}-1)/2}\bigr)\quad
\chi_{\mu,s}^{*}(\Teta{j},\vec{x}_{1}) + \\ \notag &+
\beta_{\nu}\,\alpha_{\mu}\;\chi_{\nu,s\ppr}(\Teta{j-1},\vec{x}_{2})\quad\:
\hat{\mscr{D}}_{\nu,s\ppr;\mu,s}\bigl(\vec{x}_{2},\vec{x}_{1};\TT_{j+(\eta_{j}-1)/2}\bigr)\quad
\chi_{\mu,s}(\Teta{j-1},\vec{x}_{1}) \Big]+  \\ \notag &+
\sum_{\vec{x}_{1,2}}^{\mu_{1,2},s_{1,2}}\sum_{j_{1,2}=0}^{2N+1}
\Xi_{\mu_{2},s_{2}}^{\sharp,b}(\Teta{j_{2}},\vec{x}_{2})\;\eta_{j_{2}}\quad\tfrac{1}{2}\;
\hat{\mscr{J}}_{\mu_{2},s_{2};\mu_{1},s_{1}}^{ba}(\Teta{j_{2}},\vec{x}_{2};
\Teta{j_{1}},\vec{x}_{1})\quad\eta_{j_{1}}\;\Xi_{\mu_{1},s_{1}}^{a}(\Teta{j_{1}},\vec{x}_{1}) = \\ \notag &=\frac{1}{2}
\sum_{\vec{x}_{1,2}}^{\mu,\nu,s,s\ppr}\sum_{j_{1,2}=0}^{2N+1}
\Xi_{\nu,s\ppr}^{\sharp,b}(\Teta{j_{2}},\vec{x}_{2})\;\Big[\hat{\mathrm{I}}^{bb}\;
\hat{\mathsf{H}}_{\nu,s\ppr;\mu,s}^{ba}(\vec{x}_{2},\Teta{j_{2}};\vec{x}_{1},\Teta{j_{1}})\;
\hat{\mathrm{I}}^{aa}\;+\;  \\  \notag &\hspace*{3.6cm}+\;\eta_{j_{2}}\;
\hat{\mscr{J}}_{\nu,s\ppr;\mu,s}^{ba}(\Teta{j_{2}},\vec{x}_{2};\Teta{j_{1}},\vec{x}_{1})\;\eta_{j_{1}}\Big]\;
\Xi_{\mu,s}^{a}(\Teta{j_{1}},\vec{x}_{1}) \;;
\end{align}
\begin{subequations}
\begin{align}\lb{s2_13a}
\hat{\mathrm{I}}^{ab} &=\delta_{ab}\;\;\text{diag}\bigl\{\underbrace{\hat{1}}_{a=1}\;\boldsymbol{,}\;
\underbrace{\hat{\im}}_{a=2}\bigr\}\;;  \\ \lb{s2_13b}
\hat{\mathrm{S}}^{ab} &=\delta_{ab}\;\;
\text{diag}\bigl\{\underbrace{\hat{1}}_{a=1}\;\boldsymbol{,}\;\underbrace{-\hat{1}}_{a=2}\bigr\} =
\hat{\mathrm{I}}^{ac}\cdot\hat{\mathrm{I}}^{cb}\;;
\end{align}
\end{subequations}
\begin{subequations}
\begin{align} \lb{s2_14a}
\hat{\mathsf{H}}_{\nu,s\ppr;\mu,s}^{11}(\vec{x}_{2},\Teta{j_{2}};\vec{x}_{1},\Teta{j_{1}}) &=
\delta_{\nu\mu}\;\delta_{s\ppr s}\;\deltaN_{\vec{x}_{2},\vec{x}_{1}}\;\bigl(\delta_{j_{2}+1,j_{1}}-
\delta_{j_{2},j_{1}}\bigr)+    \\  \notag &+
\im\;\tfrac{\sdelta t}{\hbar}\;\;\delta_{j_{2},j_{1}}\;\eta_{j_{2}}\;\;
\delta_{\nu\mu}\Bigl(\hat{\mscr{E}}_{\mu,s\ppr s}(\vec{x}_{2})-\im\:\eta_{j_{2}}\:
\delta_{s\ppr s}\:\ve_{+}\Bigr)\deltaN_{\vec{x}_{2},\vec{x}_{1}} \;; \\  \lb{s2_14b}
\hat{\mathsf{H}}_{\nu,s\ppr;\mu,s}^{22}(\vec{x}_{2},\Teta{j_{2}};\vec{x}_{1},\Teta{j_{1}}) &=
\Bigl(\hat{\mathsf{H}}_{\nu,s\ppr;\mu,s}^{11}(\vec{x}_{2},\Teta{j_{2}};\vec{x}_{1},\Teta{j_{1}}) \Bigr)^{T}=
\delta_{\nu\mu}\;\delta_{s\ppr s}\;\deltaN_{\vec{x}_{2},\vec{x}_{1}}\;\bigl(\delta_{j_{2},j_{1}+1}-
\delta_{j_{2},j_{1}}\bigr)+    \\   \notag  &+
\im\;\tfrac{\sdelta t}{\hbar}\;\;\delta_{j_{2},j_{1}}\;\eta_{j_{1}}\;\;
\delta_{\nu\mu}\Bigl(\Bigl(\hat{\mscr{E}}_{\mu,s\ppr s}(\vec{x}_{2})-\im\:\eta_{j_{1}}\:
\delta_{s\ppr s}\:\ve_{+}\Bigr)\deltaN_{\vec{x}_{2},\vec{x}_{1}}\Bigr)^{T} \;; \\   \lb{s2_14c}
\hat{\mathsf{H}}_{\nu,s\ppr;\mu,s}^{12}(\vec{x}_{2},\Teta{j_{2}};\vec{x}_{1},\Teta{j_{1}}) &=
\tfrac{\sdelta t}{\hbar}\;\delta_{j_{2},j_{1}}\;\eta_{j_{2}}\;\;\alpha_{\nu}\;\beta_{\mu}\quad
\hat{\mscr{D}}_{\nu,s\ppr;\mu,s}\pdag\bigl(\vec{x}_{2},\vec{x}_{1};\TT_{j_{2}+(\eta_{j_{2}}-1)/2}\bigr) \;; \\  \lb{s2_14d}
\hat{\mathsf{H}}_{\nu,s\ppr;\mu,s}^{21}(\vec{x}_{2},\Teta{j_{2}};\vec{x}_{1},\Teta{j_{1}}) &=
\tfrac{\sdelta t}{\hbar}\;\delta_{j_{2},j_{1}}\;\eta_{j_{1}}\;\;\beta_{\nu}\;\alpha_{\mu}\quad
\hat{\mscr{D}}_{\nu,s\ppr;\mu,s}\bigl(\vec{x}_{2},\vec{x}_{1};\TT_{j_{1}+(\eta_{j_{1}}-1)/2}\bigr) \;.
\end{align}
\end{subequations}
The combination of (\ref{s2_10a}) with (\ref{s2_12}-\ref{s2_14d}) results into the path integral (\ref{s2_15})
which contains the one-particle part in an anomalous-doubled manner and the remaining interaction part
of quadratic, density related pairs \(\mfrak{b}(\Teta{j_{1}},\vec{x}_{1})\) (\ref{s2_10c}), each composed
of fermionic field pairs for density terms
\begin{align} \lb{s2_15}
\ovv{Z[\hat{\mscr{J}}]} &=\int\prod_{\vec{x},\mu,s}\prod_{j=0}^{2N}
\tfrac{d\chi_{\mu,s}^{*}\bigl(\Teta{j},\vec{x}\bigr)\;\;d\chi_{\mu,s}\bigl(\Teta{j},\vec{x}\bigr)}{\mcal{N}_{x}}
\;\;\times \\ \notag
&\times\exp\bigg\{\negthickspace-\frac{1}{2}
\sum_{\vec{x}_{1,2}}^{\mu,\nu,s,s\ppr}\sum_{j_{1,2}=0}^{2N+1}
\Xi_{\nu,s\ppr}^{\sharp,b}(\Teta{j_{2}},\vec{x}_{2})\;\Big[\hat{\mathrm{I}}^{bb}\;
\hat{\mathsf{H}}_{\nu,s\ppr;\mu,s}^{ba}(\vec{x}_{2},\Teta{j_{2}};\vec{x}_{1},\Teta{j_{1}})\;
\hat{\mathrm{I}}^{aa}\;+\;  \\  \notag &\hspace*{3.6cm}+\;\eta_{j_{2}}\;
\hat{\mscr{J}}_{\nu,s\ppr;\mu,s}^{ba}(\Teta{j_{2}},\vec{x}_{2};\Teta{j_{1}},\vec{x}_{1})\;\eta_{j_{1}}\Big]\;
\Xi_{\mu,s}^{a}(\Teta{j_{1}},\vec{x}_{1})\bigg\}\times \\ \notag &\times
\exp\bigg\{-\negthickspace\sum_{\vec{x}_{1,2}}\sum_{j_{1,2}=1}^{2N}
\mfrak{b}\bigl(\Teta{j_{2}},\vec{x}_{2}\bigr)\;\eta_{j_{2}}\quad
\Bigl(\im\tfrac{\sdelta t}{\hbar}\Bigr)\;\;
\hat{\mscr{V}}\bigl(\vec{x}_{2},\Teta{j_{2}};\vec{x}_{1},\Teta{j_{1}}\bigl)\quad\eta_{j_{1}}\;
\mfrak{b}\bigl(\Teta{j_{1}},\vec{x}_{1}\bigr)\bigg\}\;.
\end{align}

\section{The various kinds of Hubbard-Stratonovich transformations}\lb{s3}

\subsection{Auxiliary variables $\sigma(\Teta{j},\vec{x})$ as scalar, real-valued self-energy}\lb{s31}

The simplest possibility of a HST is given in (\ref{s3_1a}) for transforming from
densities \(\mfrak{b}(\Teta{j_{2}},\vec{x}_{2})\;\ldots\;
\mfrak{b}(\Teta{j_{1}},\vec{x}_{1})\) or quartic interaction of anti-commuting fields to a bilinear term
of fields with a single factor \(\mfrak{b}(\Teta{j_{1}},\vec{x}_{1})\) and auxiliary, real variable
\(\sigma(\Teta{j_{1}},\vec{x}_{1})\).
This is achieved by the additional Gaussian term (\ref{s3_1a}) of the real self-energy
variables whose 'variance' is determined in relation (\ref{s3_1b}) with
the 'inverted', ensemble-averaged potential
\(\hat{\mscr{V}}^{-1}(\vec{x}_{2},\Teta{j_{2}};\vec{x}_{1},\Teta{j_{1}})\).
The bilinear term of anti-commuting fields \(\mfrak{b}(\Teta{j_{2}},\vec{x}_{2})\)
with self-energy variable \(\sigma(\Teta{j_{1}},\vec{x}_{1})\) is converted to the anomalous-doubled form (\ref{s3_1c})
with Grassmann variables (\ref{s2_11a}-\ref{s2_11g}) and the 'Nambu' metric tensor
\(\hat{\mathrm{S}}^{ba}\) (\ref{s2_13b})
\begin{subequations}
\begin{align}\lb{s3_1a}&\hspace*{-0.9cm}
\exp\Big\{-\negthickspace\sum_{\vec{x}_{1,2}}\sum_{j_{1,2}=1}^{2N}
\mfrak{b}\bigl(\Teta{j_{2}},\vec{x}_{2}\bigr)\quad\eta_{j_{2}}\;\Bigl(\im\tfrac{\sdelta t}{\hbar}\Bigr)\;\;
\hat{\mscr{V}}\bigl(\vec{x}_{2},\Teta{j_{2}};\vec{x}_{1},\Teta{j_{1}}\bigl)\;\eta_{j_{1}}\quad
\mfrak{b}\bigl(\Teta{j_{1}},\vec{x}_{1}\bigr)\Big\} =  \\  \notag &=
\int d[\sigma(\Teta{j},\vec{x})]\;\;
\exp\Big\{\im\;\tfrac{\sdelta t}{\hbar}\sum_{j_{1,2}=1}^{2N}\sum_{\vec{x}_{1,2}}
\sigma(\Teta{j_{2}},\vec{x}_{2})\;\eta_{j_{2}}\;
\hat{\mscr{V}}^{\boldsymbol{-1}}\bigl(\vec{x}_{2},\Teta{j_{2}};\vec{x}_{1},\Teta{j_{1}}\bigr)\;\eta_{j_{1}}\;
\sigma(\Teta{j_{1}},\vec{x}_{1})\Big\} \times \\ \notag &\times
\exp\Big\{\negthickspace-\im\;\tfrac{\sdelta t}{\hbar}\sum_{j_{1,2}=1}^{2N}\sum_{\vec{x}_{1,2}}
\mfrak{b}(\Teta{j_{2}},\vec{x}_{2})\cdot\;\; 2\;
\deltaN_{\vec{x}_{2},\vec{x}_{1}}\;\delta_{j_{2},j_{1}}\;\etan{j_{1}}\;\cdot
\sigma(\Teta{j_{1}},\vec{x}_{1})\Big\}\;;    \\        \lb{s3_1b} &
\sum_{j_{4}=1}^{2N}\sum_{\vec{x}_{4}}\hat{\mscr{V}}(\vec{x}_{2},\Teta{j_{2}};\vec{x}_{4},\Teta{j_{4}})
\;\;\eta_{j_{4}}\;\;
\hat{\mscr{V}}^{\boldsymbol{-1}}(\vec{x}_{4},\Teta{j_{4}};\vec{x}_{1},\Teta{j_{1}}) =
\delta_{j_{2},j_{1}}\;\;\eta_{j_{1}}\;\;\deltaN_{\vec{x}_{2},\vec{x}_{1}}\;\;;  \\       \lb{s3_1c} &
\sum_{\vec{x}_{1,2}}\sum_{j_{1,2}=1}^{2N}
\mfrak{b}(\Teta{j_{2}},\vec{x}_{2})\cdot\;\; 2\;\deltaN_{\vec{x}_{2},\vec{x}_{1}}\;\delta_{j_{2},j_{1}}\;
\etan{j_{1}}\;\cdot\sigma(\Teta{j_{1}},\vec{x}_{1}) = \\ \notag &=
\sum_{\vec{x}_{1,2}}^{\mu,\nu,s,s\ppr}\sum_{j_{1,2}=1}^{2N}
\chi_{\nu,s\ppr}^{*}(\Teta{j_{2}},\vec{x}_{2})\quad 2\;
\delta_{\nu\mu}\;\delta_{s\ppr s}\;\deltaN_{\vec{x}_{2},\vec{x}_{1}}\;\delta_{j_{2},j_{1}}\;
\etan{j_{1}}\;\;q_{\mu}\;\;
\sigma(\Teta{j_{1}},\vec{x}_{1})\;\;\chi_{\mu,s}(\Teta{j_{1}-1},\vec{x}_{1}) \\ \notag &=
\frac{1}{2} \sum_{\vec{x}_{1,2}}^{\mu,\nu,s,s\ppr}\sum_{j_{1,2}=1}^{2N}
 \Xi_{\nu,s\ppr}^{\sharp,b}(\Teta{j_{2}},\vec{x}_{2})
\quad 2\;\delta_{\nu\mu}\;\delta_{s\ppr s}\;\deltaN_{\vec{x}_{2},\vec{x}_{1}}\;\delta_{j_{2},j_{1}}\;
\etan{j_{1}}\;\;q_{\mu}\;\;\hat{\mathrm{S}}^{ba}\quad
\sigma(\Teta{j_{1}},\vec{x}_{1})\;\;
\Xi_{\mu,s}^{a}(\Teta{j_{1}},\vec{x}_{1})\;\;.
\end{align}
\end{subequations}
As one applies the exponential form (\ref{s3_2}) of (\ref{s3_1c}), one finally acquires the coherent
state path integral (\ref{s3_3}) with only bilinear, anomalous-doubled anti-commuting fields (\ref{s2_11a}-\ref{s2_11g})
instead of the quartic interaction of Grassmann fields as in (\ref{s2_15})
\begin{align} \lb{s3_2}&
\exp\bigg\{-\negthickspace\sum_{\vec{x}_{1,2}}\sum_{j_{1,2}=1}^{2N}
\mfrak{b}\bigl(\Teta{j_{2}},\vec{x}_{2}\bigr)\quad\eta_{j_{2}}\;\Bigl(\im\tfrac{\sdelta t}{\hbar}\Bigr)\;\;
\hat{\mscr{V}}\bigl(\vec{x}_{2},\Teta{j_{2}};\vec{x}_{1},\Teta{j_{1}}\bigl)\;\eta_{j_{1}}\quad
\mfrak{b}\bigl(\Teta{j_{1}},\vec{x}_{1}\bigr)\bigg\} =  \\  \notag &=
\int d[\sigma(\Teta{j},\vec{x})]\;\;
\exp\bigg\{\im\;\tfrac{\sdelta t}{\hbar}\sum_{j_{1,2}=1}^{2N}\sum_{\vec{x}_{1,2}}
\sigma(\Teta{j_{2}},\vec{x}_{2})\;\eta_{j_{2}}\;
\hat{\mscr{V}}^{\boldsymbol{-1}}(\vec{x}_{2},\Teta{j_{2}};\vec{x}_{1},\Teta{j_{1}})\;\eta_{j_{1}}\;
\sigma(\Teta{j_{1}},\vec{x}_{1})\bigg\} \times \\ \notag &\times
\exp\bigg\{\negthickspace-\tfrac{1}{2}\sum_{\vec{x}_{1,2}}^{\mu,\nu,s,s\ppr}\sum_{j_{1,2}=1}^{2N}
 \Xi_{\nu,s\ppr}^{\sharp,b}(\Teta{j_{2}},\vec{x}_{2})
\; 2\;\delta_{\nu\mu}\;\delta_{s\ppr s}\;\deltaN_{\vec{x}_{2},\vec{x}_{1}}\;\delta_{j_{2}j_{1}}\;
\etan{j_{1}}\;\;q_{\mu}\;\;\hat{\mathrm{S}}^{ba}\;\bigl(\im\,\tfrac{\sdelta t}{\hbar}\bigr)\;
\sigma(\Teta{j_{1}},\vec{x}_{1})\;\;
\Xi_{\mu,s}^{a}(\Teta{j_{1}},\vec{x}_{1})\bigg\};
\end{align}
\begin{align} \lb{s3_3}
\ovv{Z[\hat{\mscr{J}}]} &= \int d[\sigma(\Teta{j},\vec{x})]\;\;
\exp\bigg\{\im\;\tfrac{\sdelta t}{\hbar}\sum_{j_{1,2}=1}^{2N}\sum_{\vec{x}_{1,2}}
\sigma(\Teta{j_{2}},\vec{x}_{2})\;\eta_{j_{2}}\;
\hat{\mscr{V}}^{\boldsymbol{-1}}(\vec{x}_{2},\Teta{j_{2}};\vec{x}_{1},\Teta{j_{1}})\;\eta_{j_{1}}\;
\sigma(\Teta{j_{1}},\vec{x}_{1})\bigg\} \times \\ \notag &\times
\int \prod_{\vec{x},\mu,s}\prod_{j=0}^{2N}
\tfrac{d\chi_{\mu,s}^{*}(\Teta{j},\vec{x})\;d\chi_{\mu,s}(\Teta{j},\vec{x})}{\mcal{N}_{x}}\;\;
\exp\bigg\{-\tfrac{1}{2}\sum_{\vec{x}_{1,2}}^{\mu,\nu,s,s\ppr}\sum_{j_{1,2}=0}^{2N+1}
\Xi_{\nu,s\ppr}^{\sharp,b}\bigl(\Teta{j_{2}},\vec{x}_{2}\bigr)\;\times \\ \notag &\times\;\bigg[\hat{\mathrm{I}}^{bb}\;
\hat{\mathsf{H}}_{\nu,s\ppr;\mu,s}^{ba}\bigl(\vec{x}_{2},\Teta{j_{2}};\vec{x}_{1},\Teta{j_{1}}\bigr)\;
\hat{\mathrm{I}}^{aa}\;
+\;\eta_{j_{2}}\;\;\hat{\mscr{J}}_{\nu,s\ppr;\mu,s}^{ba}\bigl(\Teta{j_{2}},\vec{x}_{2};\Teta{j_{1}},\vec{x}_{1}\bigr)\;\;
\eta_{j_{1}}\;+  \\   \notag &+\;
\delta_{\nu\mu}\;\delta_{s\ppr s}\;\eta_{j_{1}}\;2\;
\bigl(\im\tfrac{\sdelta t}{\hbar}\bigr)\;\deltaN_{\vec{x}_{2},\vec{x}_{1}}\;
\delta_{j_{2},j_{1}}\;q_{\mu}\;\hat{\mathrm{S}}^{ba}\;\;
\sigma\bigl(\Teta{j_{1}},\vec{x}_{1}\bigr)\bigg]_{\nu,s\ppr;\mu,s}^{ba}\hspace*{-0.8cm}
(\Teta{j_{2}},\vec{x}_{2};\Teta{j_{1}},\vec{x}_{1})\quad\Xi_{\mu,s}^{a}(\Teta{j_{1}},\vec{x}_{1})\bigg\}\;.
\end{align}
The bilinear Grassmann term in (\ref{s3_3}) is removed by the property of anti-commuting integrations
where one has to use the identity (\ref{s3_4}) for an anti-symmetric matrix \(\hat{M}\)
of even-integer dimensions as a one-particle part.
One can also state that the symmetric part of the matrix \(\hat{M}\) cancels in the exponent of (\ref{s3_4})
so that one is only left with the anti-symmetric part of \(\hat{M}\) in the resulting square root of the determinant
\((\det[\hat{M}])^{1/2}\)
\begin{align} \lb{s3_4}
\int d[\xi_{j}]\quad \exp\Bigl\{-\tfrac{1}{2}\sum_{i,j=1}^{2n}\xi_{i}^{T}\;\;\hat{M}_{ij}\;\;\xi_{j}\Bigr\}&=
\Big(\det\bigl[\hat{M}\bigr]\Big)^{1/2}\;;\quad \hat{M}=-\hat{M}^{T}\;.
\end{align}
As we define the matrix
\(\hat{M}_{I;\nu,s\ppr;\mu,s}^{(\hat{\mscr{J}})}(\Teta{j_{2}},\vec{x}_{2};\Teta{j_{1}},\vec{x}_{1})\) (\ref{s3_5a}) 
and use property (\ref{s2_11g}) between the hermitian conjugated fields
\(\Xi_{\mu_{2},s_{2}}^{\sharp,b}\!(\Teta{j_{2}},\vec{x}_{2})\) and their transposed form
\((\,(\hat{\tau}_{1})^{ba}\;\Xi_{\mu_{2},s_{2}}^{a}\!(\Teta{j_{2}},\vec{x}_{2})\,)^{\boldsymbol{T}}\),
we can reduce the path integral (\ref{s3_3}) to (\ref{s3_5b}) with the square root of the determinant as
exemplified in (\ref{s3_4}). Note that the Pauli matrix \((\hat{\tau}_{1})^{ba}\) appears as a redundant factor
within the determinant and can be disregarded due to the overall even dimensions of the matrix
\(\hat{M}_{I;\nu,s\ppr;\mu,s}^{(\hat{\mscr{J}})}(\Teta{j_{2}},\vec{x}_{2};\Teta{j_{1}},\vec{x}_{1})\)
so that the path integral (\ref{s3_5b}) is accomplished with the remaining, real- and even-valued self-energy
variables \(\sigma(\Teta{j_{1}},\vec{x}_{1})\)
\begin{subequations}
\begin{align}\notag
\hat{M}_{I;\nu,s\ppr;\mu,s}^{(\hat{\mscr{J}})}(\Teta{j_{2}},\vec{x}_{2};\Teta{j_{1}},\vec{x}_{1}) &=
\hat{\mathrm{I}}^{bb}\;
\hat{\mathsf{H}}_{\nu,s\ppr;\mu,s}^{ba}(\vec{x}_{2},\Teta{j_{2}};\vec{x}_{1},\Teta{j_{1}})\;
\hat{\mathrm{I}}^{aa}\;+\;\eta_{j_{2}}\;
\hat{\mscr{J}}_{\nu,s\ppr;\mu,s}^{ba}(\Teta{j_{2}},\vec{x}_{2};\Teta{j_{1}},\vec{x}_{1})\;\eta_{j_{1}}\;+\;
\\ \lb{s3_5a} &+
\delta_{\nu\mu}\:\delta_{s\ppr s}\:\eta_{j_{1}}\:2\bigl(\im\,\tfrac{\sdelta t}{\hbar}\bigr)\:
\deltaN_{\vec{x}_{2},\vec{x}_{1}}\:\delta_{j_{2},j_{1}}\:q_{\mu}\:\hat{\mathrm{S}}^{ba}\;
\sigma(\Teta{j_{1}},\vec{x}_{1})\;;  \\  \lb{s3_5b}
\ovv{Z[\hat{\mscr{J}}]} = \int d[\sigma(\Teta{j},\vec{x})]\;\;
\exp\bigg\{&\im\;\tfrac{\sdelta t}{\hbar}\sum_{j_{1,2}=1}^{2N}\sum_{\vec{x}_{1,2}}
\sigma(\Teta{j_{2}},\vec{x}_{2})\;\eta_{j_{2}}\;
\hat{\mscr{V}}^{-1}(\vec{x}_{2},\Teta{j_{2}};\vec{x}_{1},\Teta{j_{1}})\;\eta_{j_{1}}\;
\sigma(\Teta{j_{1}},\vec{x}_{1})\bigg\} \times \\ \notag &\times
\bigg(\mbox{DET}
\Big[\tfrac{1}{\mcal{N}_{x}}\hat{M}_{I;\nu,s\ppr;\mu,s}^{(\hat{\mscr{J}})}(\Teta{j_{2}},\vec{x}_{2};\Teta{j_{1}},\vec{x}_{1})\Big]\bigg)^{1/2}
\;;  \\   \lb{s3_5c}
0&\equiv \sigma(\Teta{j=0},\vec{x})=\sigma(\Teta{j=2N+1},\vec{x})\;\;\;.
\end{align}
\end{subequations}
The first order variation of the path integral (\ref{s3_5b}) with respect to \(\delta\sigma(\Teta{j_{1}},\vec{x}_{1})\)
results into the saddle point equation (\ref{s3_6}) where one has to scale the self-energy variable
\(\sigma_{0}(\Teta{j_{1}},\vec{x}_{1})\) by the value \(\mcal{N}_{x}^{-1/2}\) in order to achieve
proper solutions according to the total number of spatial points
\begin{align}\notag &
\frac{\sigma_{0}(\Teta{j_{1}},\vec{x}_{1})}{\mcal{N}_{x}^{1/2}} =-\tfrac{1}{2}\hspace*{-0.1cm}\sum_{{\scrscr j_{2}=1;}}^{{\scrscr 2N}}
\sum_{{\scrscr\vec{x}_{2};(a=1,2)}}^{\mu,s}\hspace*{-0.15cm}
\bigg[\hat{\mathrm{I}}^{b\ppr b\ppr}\,
\hat{\mathsf{H}}_{\mu_{2},s_{2};\mu_{1},s_{1}}^{b\ppr a\ppr}(\vec{x}_{4},\Teta{j_{4}};
\vec{x}_{3},\Teta{j_{3}})\,
\hat{\mathrm{I}}^{a\ppr a\ppr}\tfrac{1}{\mcal{N}_{x}^{3/2}}+
\delta_{\mu_{2}\mu_{1}}\:\delta_{s_{2}s_{1}}\:\eta_{j_{3}}\:2\bigl(\im\,\tfrac{\sdelta t}{\hbar}\bigr)\:
\delta_{j_{4},j_{3}}\:q_{\mu_{1}}\times \\ \lb{s3_6} &\times\,\hat{\mathrm{S}}^{b\ppr a\ppr}
\Big(\deltaN_{\vec{x}_{4},\vec{x}_{3}}\:\sigma_{0}\bigl(\Teta{j_{3}},\vec{x}_{3}\bigr)\Big/
\mcal{N}_{x}^{3/2}\Big)\bigg]_{\mu,s;\mu,s}^{\boldsymbol{-1};aa}\hspace*{-0.6cm}
\bigl(\Teta{j_{2}},\vec{x}_{2};\Teta{j_{2}},\vec{x}_{2}\bigr)\quad\eta_{j_{2}}\;q_{\mu}\;\hat{\mathrm{S}}^{aa}\;
\hat{\mscr{V}}(\vec{x}_{2},\Teta{j_{2}};\vec{x}_{1},\Teta{j_{1}})\;.
\end{align}
Solutions of (\ref{s3_6}) follow from iteration with \(\sigma_{0}(\Teta{j_{1}},\vec{x}_{1})/\mcal{N}_{x}^{1/2}\)
or continued fraction which involves the eigenvalue problem with the operator
\begin{align} \lb{s3_7} &
\sum_{j_{1}=0}^{2N+1}\sum_{\vec{x}_{1}}^{\mu,s}
\bigg[\hat{\mathrm{I}}^{bb}\;
\hat{\mathsf{H}}_{\nu,s\ppr;\mu,s}^{ba}(\vec{x}_{2},\Teta{j_{2}};\vec{x}_{1},\Teta{j_{1}})\;
\hat{\mathrm{I}}^{aa}\;+\; \\ \notag &+
\delta_{\nu\mu}\:\delta_{s\ppr s}\:\eta_{j_{1}}\:2\bigl(\im\,\tfrac{\sdelta t}{\hbar}\bigr)\:
\deltaN_{\vec{x}_{2},\vec{x}_{1}}\:\delta_{j_{2},j_{1}}\:q_{\mu}\:\hat{\mathrm{S}}^{ba}\;
\sigma_{0}(\Teta{j_{1}},\vec{x}_{1})\bigg]_{\nu,s\ppr;\mu,s}^{ba}\hspace*{-0.6cm}
(\Teta{j_{2}},\vec{x}_{2};\Teta{j_{1}},\vec{x}_{1})\quad
\Psi_{M;\mu,s}^{a}(\Teta{j_{1}},\vec{x}_{1}) = \\ \notag &=
\;E_{M}\;\;\Psi_{M;\nu,s\ppr}^{b}(\Teta{j_{2}},\vec{x}_{2})\;; \quad
(\mbox{eigenstate label $M$ with complex eigenvalue }E_{M}),
\end{align}
so that the first order variational equation (\ref{s3_6}) takes the form
\begin{align} \lb{s3_8}
\frac{\sigma_{0}(\Teta{j_{1}},\vec{x}_{1})}{\mcal{N}_{x}^{1/2}}=-\tfrac{1}{2}\sum_{j_{2}=1}^{2N}\sum_{\vec{x}_{2}}^{\mu,s}
\sum_{M\in\mbox{\scz eigenstates}}^{a=1,2}
\frac{\Psi_{M;\mu,s}^{a}(\Teta{j_{2}},\vec{x}_{2})\;\;
\Psi_{M;\mu,s}^{\boldsymbol{T},a}(\Teta{j_{2}},\vec{x}_{2})}{\bigl(E_{M}\big/\mcal{N}_{x}^{1/2}\bigr)}\quad
\eta_{j_{2}}\;q_{\mu}\;\hat{\mathrm{S}}^{aa}\;
\hat{\mscr{V}}(\vec{x}_{2},\Teta{j_{2}};\vec{x}_{1},\Teta{j_{1}}).
\end{align}
In general one could have problems to find proper solutions of (\ref{s3_7},\ref{s3_8}); however,
one can use a rotational or translational invariance in most cases so that a 2D or 3D problem simplifies
to radial equations with angular momenta labels or also wave-vectors. The iterated terms of (\ref{s3_6},\ref{s3_7})
yield a complex solution \(\sigma_{0}(\Teta{j},\vec{x})\) whose imaginary part has to comply with the
sign of the imaginary (\(\ve>0\))-terms for a proper, stable propagation or the correct analytic convergence
of Green functions. This implies a complex-valued eigenvalue problem (\ref{s3_7}) whose imaginary
eigenvalues \(E_{M}\) have also to correspond to the sign of the infinitesimal,
imaginary (\(\ve>0\))-term. If the imaginary parts of the solution \(\sigma_{0}(\Teta{j},\vec{x})\) or \(E_{M}\)
take opposite sign as the imaginary (\(\ve>0\))-term of the Green function, one acquires an instable range
of parameters for the semiconductor-related solid which may have a physical origin or may be related to
insufficient approximations with our simplified HST and auxiliary, scalar self-energy variable
\(\sigma(\Teta{j},\vec{x})\). This range of instable properties (also due to possible
artifacts of our HST in this section) can be improved by the more detailed kinds of HST's described in sections
\ref{s32} to \ref{s33}.

\subsection{Auxiliary matrix
$\hat{\Sigma}_{\nu,s\ppr;\mu,s}(\Teta{j_{2}},\vec{x}_{2};\Teta{j_{1}},\vec{x}_{1})$
as hermitian self-energy density} \lb{s32}

We extend the real self-energy variable \(\sigma(\Teta{j},\vec{x})\), replacing the
sum \(\sum_{\mu,s}\chi_{\mu,s}^{*}(\Teta{j},\vec{x})\:q_{\mu}\:\chi_{\mu,s}(\Teta{j-1},\vec{x})\), to a
self-energy matrix of fermionic densities which is specified as a dyadic product (\ref{s3_9a})
of \(\chi_{\mu_{1},s_{1}}(\Teta{j_{1}-1},\vec{x}_{1})\) with the complex conjugate
\(\chi_{\mu_{2},s_{2}}^{*}(\Teta{j_{2}},\vec{x}_{2})\). The self-energy density
\(\hat{\Sigma}_{\nu,s\ppr;\mu,s}(\Teta{j_{2}},\vec{x}_{2};\Teta{j_{1}},\vec{x}_{1})\) (\ref{s3_9b}) has a hermitian
symmetry where the hermitian conjugation involves the electron-hole and spin labels with inclusion
of the time contour and coordinate variables. As we apply the operation of dyadic products to the quartic
interaction of Fermi fields, one attains the interaction (\ref{s3_10})
in terms of two, solely density-related matrices from the dyadic product (\ref{s3_9a}),
which allow for a similar HST as in section \ref{s31}, but with matrices for densities in place of
scalar variables representing sums of fermionic densities
\begin{subequations}
\begin{align}\lb{s3_9a}
\hat{R}_{\mu_{1},s_{1};\mu_{2},s_{2}}\Big(\Teta{j_{1}},\vec{x}_{1};\Teta{j_{2}},\vec{x}_{2}\Big) & =
\chi_{\mu_{1},s_{1}}\bigl(\Teta{j_{1}-1},\vec{x}_{1}\bigr) \;\otimes \;
\chi_{\mu_{2},s_{2}}^{*}\bigl(\Teta{j_{2}},\vec{x}_{2}\bigr)\;;  \\          \lb{s3_9b}
\hat{\Sigma}_{\mu_{1},s_{1};\mu_{2},s_{2}}\Big(\Teta{j_{1}},\vec{x}_{1};\Teta{j_{2}},\vec{x}_{2}\Big) &=
\bigg(\hat{\Sigma}_{\mu_{1},s_{1};\mu_{2},s_{2}}\Big(\Teta{j_{1}},\vec{x}_{1};\Teta{j_{2}};\vec{x}_{2}
\Big)\bigg)\pdag\;\;;
\end{align}
\end{subequations}
\begin{align}\lb{s3_10}
&\sum_{\vec{x}_{1,2}}\sum_{j_{1,2}=1}^{2N}
\mfrak{b}\bigl(\Teta{j_{2}},\vec{x}_{2}\bigr)\;\eta_{j_{2}}\;\;\eta_{j_{1}}\;
\mfrak{b}\bigl(\Teta{j_{1}},\vec{x}_{1}\bigr)\quad
\hat{\mscr{V}}(\vec{x}_{2},\Teta{j_{2}};\vec{x}_{1},\Teta{j_{1}})  =
\sum_{\vec{x}_{1,2}}^{\mu_{1,2},s_{1,2}}\sum_{j_{1,2}=1}^{2N}
\hat{\mscr{V}}(\vec{x}_{2},\Teta{j_{2}};\vec{x}_{1},\Teta{j_{1}})\;\times \\ \notag &\times\;\;
\chi_{\mu_{2},s_{2}}^{*}\bigl(\Teta{j_{2}},\vec{x}_{2}\bigr)\;\;\eta_{j_{2}}\;q_{\mu_{2}}\;\;
\chi_{\mu_{2},s_{2}}\bigl(\Teta{j_{2}-1},\vec{x}_{2}\bigr)\;\;
\chi_{\mu_{1},s_{1}}^{*}\bigl(\Teta{j_{1}},\vec{x}_{1}\bigr)\;\;\eta_{j_{1}}\;q_{\mu_{1}}\;\;
\chi_{\mu_{1},s_{1}}\bigl(\Teta{j_{1}-1},\vec{x}_{1}\bigr) = \\ \notag &= -
\sum_{\vec{x}_{1,2}}^{\mu_{1,2},s_{1,2}}\sum_{j_{1,2}=1}^{2N}\;
 \hat{\mscr{V}}(\vec{x}_{2},\Teta{j_{2}};\vec{x}_{1},\Teta{j_{1}})\;\;\times \\ \notag &\times\;\bigg[
\eta_{j_{2}}\;q_{\mu_{2}}\;\hat{R}_{\mu_{2},s_{2};\mu_{1},s_{1}}\Big(\Teta{j_{2}},\vec{x}_{2};\Teta{j_{1}},
\vec{x}_{1}\Big) \;\eta_{j_{1}}\;q_{\mu_{1}}\;
\hat{R}_{\mu_{1},s_{1};\mu_{2},s_{2}}\Big(\Teta{j_{1}},\vec{x}_{1};\Teta{j_{2}},\vec{x}_{2}\Big)  \bigg]  \;\;.
\end{align}
One has to start from a similar Gaussian identity with a hermitian self-energy matrix of densities as
in the case of scalar self-energy variables \(\sigma(\Teta{j},\vec{x})\) in section \ref{s31}; we then transform
the quartic interaction of Grassmann fields to bilinear fields given as a hermitian dyadic product with
linear coupling to the hermitian self-energy matrix in a trace relation of internal and time contour,
coordinate variables
\begin{align} \lb{s3_11}
& \exp\bigg\{\negthickspace-\im\:\tfrac{\sdelta t}{\hbar}\sum_{\vec{x}_{1,2}}\sum_{j_{1,2}=1}^{2N}
\mfrak{b}\bigl(\Teta{j_{2}},\vec{x}_{2}\bigr)\;\eta_{j_{2}}\;\eta_{j_{1}}\;
\mfrak{b}\bigl(\Teta{j_{1}},\vec{x}_{1}\bigr)\;
\hat{\mscr{V}}(\vec{x}_{2},\Teta{j_{2}};\vec{x}_{1},\Teta{j_{1}}) \bigg\} = \\   \notag &=
\int d\bigl[\hat{\Sigma}_{\nu,s\ppr;\mu,s}(\Teta{j_{2}},\vec{x}_{2};\Teta{j_{1}},\vec{x}_{1})\bigr]\;\times\;
\exp\bigg\{\negthickspace-\im\:\tfrac{\sdelta t}{\hbar}
\sum_{\vec{x}_{1,2}}^{\mu_{1,2},s_{1,2}}\sum_{j_{1,2}=1}^{2N}
\hat{\mscr{V}}^{\boldsymbol{-1}}(\vec{x}_{2},\Teta{j_{2}};\vec{x}_{1},\Teta{j_{1}})\;\times  \\ \notag &\times
\bigg[\etan{j_{2}}\;\;q_{\mu_{2}}\;\;
\hat{\Sigma}_{\mu_{2},s_{2};\mu_{1},s_{1}}\Big(\Teta{j_{2}},\vec{x}_{2};\Teta{j_{1}},\vec{x}_{1}\Big)\;\;
\etan{j_{1}}\;\;q_{\mu_{1}}\;\;
\hat{\Sigma}_{\mu_{1},s_{1};\mu_{2},s_{2}}\Big(\Teta{j_{1}},\vec{x}_{1};\Teta{j_{2}},\vec{x}_{2}
\Big)\bigg]\bigg\}\times   \\   \notag &\times
\exp\bigg\{\negthinspace\im\:2\;\tfrac{\sdelta t}{\hbar}
\sum_{\vec{x}_{1,2}}^{\mu_{1,2},s_{1,2}}\sum_{j_{1,2}=1}^{2N}
\bigg[\etan{j_{2}}\;q_{\mu_{2}}\;
\hat{R}_{\mu_{2},s_{2};\mu_{1},s_{1}}\Big(\Teta{j_{2}},\vec{x}_{2};\Teta{j_{1}},\vec{x}_{1}\Big)\;
\etan{j_{1}}\;q_{\mu_{1}}\;
\hat{\Sigma}_{\mu_{1},s_{1};\mu_{2},s_{2}}\Big(\Teta{j_{1}},\vec{x}_{1};\Teta{j_{2}},\vec{x}_{2}
\Big)\bigg]\bigg\}_{\mbox{.}}
\end{align}
Since the one-particle part
\(\hat{\mathsf{H}}_{\nu,s\ppr;\mu,s}^{ba}(\vec{x}_{2},\Teta{j_{2}};\vec{x}_{1},\Teta{j_{1}})\)
(\ref{s2_14a}-\ref{s2_14d}) also contains anomalous terms with dipole moments, one has to perform the anomalous doubling
within the coupling of "(dyadic product)$\times$(hermitian self-energy matrix)"; hence, we have to
introduce two hermitian self-energy matrices
\(\hat{\Sigma}_{\nu,s\ppr;\mu,s}^{aa}(\Teta{j_{2}},\vec{x}_{2};\Teta{j_{1}},\vec{x}_{1})\), (\(a=1,2\))
which are related by transposition with opposite sign (\ref{s3_12a},\ref{s3_12b}).
One thus achieves relation (\ref{s3_12c}) with
the anomalous-doubled Fermi fields
\(\Xi_{\nu,s\ppr}^{\sharp,b}(\Teta{j_{2}},\vec{x}_{2})\ldots\Xi_{\mu,s}^{a}(\Teta{j_{1}},\vec{x}_{1})\)
and the two self-energy matrices which only comprise density terms without any 'Nambu' part
\begin{subequations}
\begin{align} \lb{s3_12a}&
\hat{\Sigma}_{\nu,s\ppr;\mu,s}^{aa}\!\bigl(\Teta{j_{2}},\vec{x}_{2};\Teta{j_{1}},\vec{x}_{1}\bigr)=
\bigg(\hat{\Sigma}_{\nu,s\ppr;\mu,s}^{aa}\!\bigl(\Teta{j_{2}},\vec{x}_{2};\Teta{j_{1}},\vec{x}_{1}\bigr)
\bigg)\pdag \;;  \\ \lb{s3_12b} &
\hat{\Sigma}_{\nu,s\ppr;\mu,s}^{22}\!\bigl(\Teta{j_{2}},\vec{x}_{2};\Teta{j_{1}},\vec{x}_{1}\bigr)=
\boldsymbol{-}
\bigg(\hat{\Sigma}_{\nu,s\ppr;\mu,s}^{11}\!\bigl(\Teta{j_{2}},\vec{x}_{2};\Teta{j_{1}},\vec{x}_{1}\bigr)
\bigg)^{\boldsymbol{T}} \;; \\ \lb{s3_12c}&
\exp\bigg\{\negthinspace\im\:2\;\tfrac{\sdelta t}{\hbar}
\sum_{\vec{x}_{1,2}}^{\mu_{1,2},s_{1,2}}\sum_{j_{1,2}=1}^{2N}
\bigg[\etan{j_{2}}\;\;q_{\mu_{2}}\;\;
\hat{R}_{\mu_{2},s_{2};\mu_{1},s_{1}}\Big(\Teta{j_{2}},\vec{x}_{2};\Teta{j_{1}},\vec{x}_{1}\Big)\;\;\times
\\  \notag &\times\;\;\etan{j_{1}}\;\;q_{\mu_{1}}\;\;
\hat{\Sigma}_{\mu_{1},s_{1};\mu_{2},s_{2}}\Big(\Teta{j_{1}},\vec{x}_{1};\Teta{j_{2}},\vec{x}_{2}\Big)\bigg]\bigg\} =
\exp\bigg\{\negthinspace-\im\:2\;\tfrac{\sdelta t}{\hbar}
\sum_{\vec{x}_{1,2}}^{\mu_{1,2},s_{1,2}}\sum_{j_{1,2}=1}^{2N}
\chi_{\mu_{1},s_{1}}^{*}\!\bigl(\Teta{j_{1}},\vec{x}_{1}\bigr)\;\;
\times   \\  \notag &\times\;\;\etan{j_{1}}\;\;q_{\mu_{1}}\;\;
\hat{\Sigma}_{\mu_{1},s_{1};\mu_{2},s_{2}}\Big(\Teta{j_{1}},\vec{x}_{1};\Teta{j_{2}},\vec{x}_{2}\Big)\;\;
\etan{j_{2}}\;\;q_{\mu_{2}}\;\;
\chi_{\mu_{2},j_{2}}\!\bigl(\Teta{j_{2}-1},\vec{x}_{2}\bigr)\bigg\} =
\exp\bigg\{-\negthickspace\tfrac{1}{2}
\sum_{\vec{x}_{1,2}}^{\mu,\nu,s,s\ppr}\sum_{j_{1,2}=1}^{2N}\times \\ \notag &\times\;\;
\Xi_{\nu,s\ppr}^{\sharp,a}\!\bigl(\Teta{j_{2}},\vec{x}_{2}\bigr)\;\;\etan{j_{2}}\;\;q_{\mu_{2}}\;\;
2\;\Bigl(\im\;\tfrac{\sdelta t}{\hbar}\Bigr)\;\;
\hat{\Sigma}_{\nu,s\ppr;\mu,s}^{aa}\!\bigl(\Teta{j_{2}},\vec{x}_{2};\Teta{j_{1}},\vec{x}_{1}\bigr)
\;\;\etan{j_{1}}\;\;q_{\mu_{1}}\;\;\Xi_{\mu,s}^{a}\!\bigl(\Teta{j_{1}},\vec{x}_{1}\bigr)\bigg\}\;.
\end{align}
\end{subequations}
After insertion of (\ref{s3_11}),(\ref{s3_12c}) into (\ref{s2_15}), we obtain the path integral (\ref{s3_13a}) whose 'Nambu'-doubled,
anti-commuting fields are removed by integration
according to the already given relation (\ref{s3_4}) in section \ref{s31}.
The resulting path integral (\ref{s3_13b}) only contains the hermitian self-energy density
\(\hat{\Sigma}_{\nu,s\ppr;\mu,s}(\Teta{j_{2}},\vec{x}_{2};\Teta{j_{1}},\vec{x}_{1})\) as remaining
quantum mechanical field degrees of freedom
\begin{subequations}
\begin{align} \lb{s3_13a}&
\ovv{Z[\hat{\mscr{J}}]} = \int
d\bigl[\hat{\Sigma}_{\nu,s\ppr;\mu,s}(\Teta{j_{2}},\vec{x}_{2};\Teta{j_{1}},\vec{x}_{1})\bigr]\;\;
\exp\bigg\{\negthickspace-\im\:\tfrac{\sdelta t}{\hbar}
\sum_{\vec{x}_{1,2}}^{\mu_{1,2},s_{1,2}}\sum_{j_{1,2}=1}^{2N}
\hat{\mscr{V}}^{\boldsymbol{-1}}(\vec{x}_{2},\Teta{j_{2}};\vec{x}_{1},\Teta{j_{1}})\;\;\times \\ \notag &\times
\bigg[\etan{j_{2}}\;\;q_{\mu_{2}}\;\;
\hat{\Sigma}_{\mu_{2},s_{2};\mu_{1},s_{1}}\Big(\Teta{j_{2}},\vec{x}_{2};\Teta{j_{1}},\vec{x}_{1}\Big)\;\;
\etan{j_{1}}\;\;q_{\mu_{1}}\;\;
\hat{\Sigma}_{\mu_{1},s_{1};\mu_{2},s_{2}}\Big(\Teta{j_{1}},\vec{x}_{1};\Teta{j_{2}},\vec{x}_{2}
\Big)\bigg]\bigg\}\;\times \\ \notag &\times\;
\int \prod_{\vec{x},\mu,s}\prod_{j=0}^{2N}
\tfrac{d\chi_{\mu,s}^{*}\bigl(\Teta{j},\vec{x}\bigr)\;\;d\chi_{\mu,s}\bigl(\Teta{j},\vec{x}\bigr)}{\mcal{N}_{x}}\;\times
\exp\bigg\{-\tfrac{1}{2}\sum_{\vec{x}_{1,2}}^{\mu,\nu,s,s\ppr}\sum_{j_{1,2}=0}^{2N+1}
\Xi_{\nu,s\ppr}^{\sharp,b}\bigl(\Teta{j_{2}},\vec{x}_{2}\bigr)\;\times \\ \notag &\times\;
\bigg[\hat{\mathrm{I}}^{bb}\;
\hat{\mathsf{H}}_{\nu,s\ppr;\mu,s}^{ba}\bigl(\vec{x}_{2},\Teta{j_{2}};
\vec{x}_{1},\Teta{j_{1}}\bigr)\;\hat{\mathrm{I}}^{aa}\;
+\;\eta_{j_{2}}\;\;\hat{\mscr{J}}_{\nu,s\ppr;\mu,s}^{ba}\bigl(\Teta{j_{2}},\vec{x}_{2};
\Teta{j_{1}},\vec{x}_{1}\bigr)\;\;\eta_{j_{1}}\;+ \;\etan{j_{2}}\;\;q_{\nu}\;\;
2\;\Bigl(\im\;\tfrac{\sdelta t}{\hbar}\Bigr)\;\;\times  \\ \notag &\times\;\;\delta_{ba}\;\;
\hat{\Sigma}_{\nu,s\ppr;\mu,s}^{aa}\!\bigl(\Teta{j_{2}},\vec{x}_{2};\Teta{j_{1}},\vec{x}_{1}\bigr)
\;\;\etan{j_{1}}\;\;q_{\mu}\bigg]_{\nu,s\ppr;\mu,s}^{ba}\hspace*{-0.75cm}
(\Teta{j_{2}},\vec{x}_{2};\Teta{j_{1}},\vec{x}_{1})\quad\Xi_{\mu,s}^{a}(\Teta{j_{1}},\vec{x}_{1})\bigg\}\;;
\end{align}
\begin{align} \lb{s3_13b}&
\ovv{Z[\hat{\mscr{J}}]} = \int
d\bigl[\hat{\Sigma}_{\nu,s\ppr;\mu,s}(\Teta{j_{2}},\vec{x}_{2};\Teta{j_{1}},\vec{x}_{1})\bigr]\;\;
\exp\bigg\{\negthickspace-\im\:\tfrac{\sdelta t}{\hbar}
\sum_{\vec{x}_{1,2}}^{\mu_{1,2},s_{1,2}}\sum_{j_{1,2}=1}^{2N}
\hat{\mscr{V}}^{\boldsymbol{-1}}(\vec{x}_{2},\Teta{j_{2}};\vec{x}_{1},\Teta{j_{1}})\;\;\times \\ \notag &\times
\bigg[\etan{j_{2}}\;\;q_{\mu_{2}}\;\;
\hat{\Sigma}_{\mu_{2},s_{2};\mu_{1},s_{1}}\Big(\Teta{j_{2}},\vec{x}_{2};\Teta{j_{1}},\vec{x}_{1}\Big)\;\;
\etan{j_{1}}\;\;q_{\mu_{1}}\;\;
\hat{\Sigma}_{\mu_{1},s_{1};\mu_{2},s_{2}}\Big(\Teta{j_{1}},\vec{x}_{1};\Teta{j_{2}},\vec{x}_{2}
\Big)\bigg]\bigg\}\;\times \\ \notag &\times\;
\mbox{DET}\bigg(\tfrac{1}{\mcal{N}_{x}}
\bigg[\hat{\mathrm{I}}^{bb}\;
\hat{\mathsf{H}}_{\nu,s\ppr;\mu,s}^{ba}\bigl(\vec{x}_{2},\Teta{j_{2}};
\vec{x}_{1},\Teta{j_{1}}\bigr)\;\hat{\mathrm{I}}^{aa}\;
+\;\eta_{j_{2}}\;\;\hat{\mscr{J}}_{\nu,s\ppr;\mu,s}^{ba}\bigl(\Teta{j_{2}},\vec{x}_{2};
\Teta{j_{1}},\vec{x}_{1}\bigr)\;\;\eta_{j_{1}}\;+   \\  \notag &+\;
\etan{j_{2}}\;\;q_{\nu}\;\;
2\;\Bigl(\im\;\tfrac{\sdelta t}{\hbar}\Bigr)\;\;\delta_{ba}\;\;
\hat{\Sigma}_{\nu,s\ppr;\mu,s}^{aa}\!\bigl(\Teta{j_{2}},\vec{x}_{2};\Teta{j_{1}},\vec{x}_{1}\bigr)
\;\;\etan{j_{1}}\;\;q_{\mu}\bigg]_{\nu,s\ppr;\mu,s}^{ba}\hspace*{-0.9cm}
(\Teta{j_{2}},\vec{x}_{2};\Teta{j_{1}},\vec{x}_{1})\bigg)^{\boldsymbol{1/2}}\;;  \\ \lb{s3_13c}&
\hat{\Sigma}_{\nu,s\ppr;\mu,s}(\Teta{j_{2}},\vec{x}_{2};\Teta{j_{1}},\vec{x}_{1})\equiv0\;\;\;
\mbox{ for }(j_{1}\mbox{\scz 'or'} j_{2}=0)\;\mbox{\scz 'or'}\;(j_{1}\mbox{\scz 'or'} j_{2}=2N+1)\;,(\mbox{cf. (\ref{s2_11b},\ref{s2_11f})}).
\end{align}
\end{subequations}
The final path integral (\ref{s3_13b}) consists of the matrix
\(\hat{M}_{II;\nu,s\ppr;\mu,s}^{(\hat{\mscr{J}})ba}(\Teta{j_{2}},\vec{x}_{2};\Teta{j_{1}},\vec{x}_{1})\) (\ref{s3_14})
whose anomalous terms \((b\neq a)\) are still restricted to those of the one-particle part
\(\hat{\mathsf{H}}_{\nu,s\ppr;\mu,s}^{b\neq a}(\vec{x}_{2},\Teta{j_{2}};\vec{x}_{1},\Teta{j_{1}})\).
In sections \ref{s33}, \ref{s4} we generalize to self-energy matrices which also comprise
the anomalous terms \(\hat{\Sigma}_{\nu,s\ppr;\mu,s}^{b\neq a}(\Teta{j_{2}},\vec{x}_{2};\Teta{j_{1}},\vec{x}_{1})\)
in the off-diagonal 'Nambu' blocks (\(b\neq a\))
\begin{align}\notag  &
\hat{M}_{II;\nu,s\ppr;\mu,s}^{(\hat{\mscr{J}})ba}(\Teta{j_{2}},\vec{x}_{2};\Teta{j_{1}},\vec{x}_{1}) =
\hat{\mathrm{I}}^{bb}\;
\hat{\mathsf{H}}_{\nu,s\ppr;\mu,s}^{ba}\bigl(\vec{x}_{2},\Teta{j_{2}};
\vec{x}_{1},\Teta{j_{1}}\bigr)\;\hat{\mathrm{I}}^{aa}\;
+\;\eta_{j_{2}}\;\;\hat{\mscr{J}}_{\nu,s\ppr;\mu,s}^{ba}\bigl(\Teta{j_{2}},\vec{x}_{2};
\Teta{j_{1}},\vec{x}_{1}\bigr)\;\;\eta_{j_{1}}\;+   \\  \lb{s3_14} &+\;
\etan{j_{2}}\;\;q_{\nu}\;\;
2\;\Bigl(\im\;\tfrac{\sdelta t}{\hbar}\Bigr)\;\;\delta_{ba}\;\;
\hat{\Sigma}_{\nu,s\ppr;\mu,s}^{aa}\!\bigl(\Teta{j_{2}},\vec{x}_{2};\Teta{j_{1}},\vec{x}_{1}\bigr)
\;\;\etan{j_{1}}\;\;q_{\mu}\;.
\end{align}
In analogy to section \ref{s31} we take the first order variation of (\ref{s3_13b})
so that one achieves the saddle point
equation (\ref{s3_15a}) in terms of the self-energy density
\(\hat{\Sigma}_{\nu,s\ppr;\mu,s}^{(0)aa}(\Teta{j_{2}},\vec{x}_{2};\Teta{j_{1}},\vec{x}_{1})\) whose solution
is also determined by continued fraction and a similar
eigenvalue problem (\ref{s3_15b},\ref{s3_15c}) of complex-valued matrices
in compliance with the imaginary (\(\ve>0\))-terms of stable propagating Green functions
(relations (\ref{s3_15a},\ref{s3_15c}) are free of any summations over spacetime coordinates and internal state labels,
aside from relation (\ref{s3_15b}) with the weighted spatial sum
\(\sum_{\vec{x}_{1}}\ldots=\tfrac{1}{\mcal{N}_{x}}\sum_{\vec{x}_{1}}\ppr\ldots\) and contour time step summations)
\begin{subequations}
\begin{align}\lb{s3_15a} &
\hat{\mscr{V}}^{\boldsymbol{-1}}(\vec{x}_{2},\Teta{j_{2}};\vec{x}_{1},\Teta{j_{1}})\;\;
\hat{\Sigma}_{\mu_{2},s_{2};\mu_{1},s_{1}}^{11,(0)}(\Teta{j_{2}},\vec{x}_{2};\Teta{j_{1}},\vec{x}_{1})\big/\mcal{N}_{x}=
\\ \notag &=\tfrac{1}{2}\,\mcal{N}_{x}\bigg[
\hat{M}_{II;\mu_{2},s_{2};\mu_{1},s_{1}}^{\boldsymbol{-1};(\hat{\mscr{J}}\equiv0),11}(
\Teta{j_{2}},\vec{x}_{2};\Teta{j_{1}},\vec{x}_{1}) -
\hat{M}_{II;\mu_{1},s_{1};\mu_{2},s_{2}}^{\boldsymbol{-1};(\hat{\mscr{J}}\equiv0),22}(
\Teta{j_{1}},\vec{x}_{1};\Teta{j_{2}},\vec{x}_{2})\bigg]\;;     \\   \lb{s3_15b}&
\sum_{j_{1}=0}^{2N+1}\sum_{\vec{x}_{1}}^{\mu,s}
\hat{M}_{II;\nu,s\ppr;\mu,s}^{(\hat{\mscr{J}}\equiv0)ba}(\Teta{j_{2}},\vec{x}_{2};\Teta{j_{1}},\vec{x}_{1})\quad
\Psi_{M;\mu,s}^{a}(\Teta{j_{1}},\vec{x}_{1}) \quad = \quad E_{M}\quad
\Psi_{M;\nu,s\ppr}^{b}(\Teta{j_{2}},\vec{x}_{2})\;; \\ \notag &
\quad(\mbox{eigenstate label $M$ with complex eigenvalue }E_{M})\;;   \\  \lb{s3_15c} &
\hat{\mscr{V}}^{\boldsymbol{-1}}(\vec{x}_{2},\Teta{j_{2}};\vec{x}_{1},\Teta{j_{1}})\;\;
\hat{\Sigma}_{\mu_{2},s_{2};\mu_{1},s_{1}}^{(0)}(\Teta{j_{2}},\vec{x}_{2};\Teta{j_{1}},\vec{x}_{1})\big/\mcal{N}_{x} =
\\  \notag  &=\tfrac{1}{2}\sum_{M\in\mbox{\scz eigenstates}}\bigg[
\frac{\Psi_{M;\mu_{2},s_{2}}^{(a=1)}(\Teta{j_{2}},\vec{x}_{2})\;\;
\Psi_{M;\mu_{1},s_{1}}^{\boldsymbol{T},(a=1)}(\Teta{j_{1}},\vec{x}_{1})}{E_{M}}-
\frac{\Psi_{M;\mu_{1},s_{1}}^{(a=2)}(\Teta{j_{1}},\vec{x}_{1})\;\;
\Psi_{M;\mu_{2},s_{2}}^{\boldsymbol{T},(a=2)}(\Teta{j_{2}},\vec{x}_{2})}{E_{M}}\bigg]\;.
\end{align}
\end{subequations}

\subsection{Self-energy
$\hat{\Sigma}_{\nu,s\ppr;\mu,s}^{ba}(\Teta{j_{2}},\vec{x}_{2};\Teta{j_{1}},\vec{x}_{1})$
of densities '$b=a$' and of anomalous parts '$b\neq a$'} \lb{s33}

Since the one-particle part \(\hat{\mathsf{H}}_{\nu,s\ppr;\mu,s}^{ba}(\vec{x}_{2},\Teta{j_{2}};\vec{x}_{1},\Teta{j_{1}})\)
contains anomalous terms, one has also to introduce 'Nambu' parts in the off-diagonal bloks (\(b\neq a\)) of
self-energy matrices. We therefore take the dyadic product (\ref{s3_16}) of the anomalous-doubled,
anti-commuting fields \(\Xi_{\mu_{1},s_{1}}^{a}(\Teta{j_{1}},\vec{x}_{1})\) and
\(\Xi_{\mu_{2},s_{2}}^{\sharp,b}(\Teta{j_{2}},\vec{x}_{2})\) and consider anomalous-doubled, fermionic
density pairs \(\mfrak{B}(\Teta{j_{1}},\vec{x}_{1})\) (\ref{s3_17}) with extended 'Nambu' metric tensor
\(\hat{\mathrm{S}}_{\mu}^{ba}\) (\ref{s3_18}) which implies an overall, additional factor \(\tfrac{1}{2}\). The combination of
(\ref{s3_16}-\ref{s3_18}) transforms the quartic interaction of Fermi fields to a trace relation (\ref{s3_19})
of density pairs
\(\hat{\mscr{R}}_{\mu_{1},s_{1};\mu_{2},s_{2}}^{ab}(\Teta{j_{1}},\vec{x}_{1};\Teta{j_{2}},\vec{x}_{2}) \)
 (\ref{s3_16}) with inclusion of 'Nambu' parts which involve the additional trace summation
(\(a,b=1,2\))
\begin{align}\lb{s3_16}&\hspace*{-2.5cm}
\hat{\mscr{R}}_{\mu_{1},s_{1};\mu_{2},s_{2}}^{ab}\Big(\Teta{j_{1}},\vec{x}_{1};\Teta{j_{2}},\vec{x}_{2}\Big)  =
\Xi_{\mu_{1},s_{1}}^{a}\!\bigl(\Teta{j_{1}},\vec{x}_{1}\bigr) \;\otimes \;
\Xi_{\mu_{2},s_{2}}^{\sharp,b}\bigl(\Teta{j_{2}},\vec{x}_{2}\bigr) = \\ \notag
&=\Biggl(\bea{c} \chi_{\mu_{1},s_{1}}\bigl(\Teta{j_{1}-1},\vec{x}_{1}\bigr) \\
\chi_{\mu_{1},s_{1}}^{*}\bigl(\Teta{j_{1}},\vec{x}_{1}\bigr) \eea\Biggr)^{a}\otimes
\Bigl(\chi_{\mu_{2},s_{2}}^{*}\bigl(\Teta{j_{2}},\vec{x}_{2}\bigr)\;\mbox{\bf\Large,}\;
\chi_{\mu_{2}}\bigl(\Teta{j_{2}-1},\vec{x}_{2}\bigr)\Bigr)^{b}\;;   \\    \lb{s3_17}
\mfrak{b}\bigl(\Teta{j_{1}},\vec{x}_{1}\bigr)&=\frac{1}{2}\;
\mfrak{B}\bigl(\Teta{j_{1}},\vec{x}_{1}\bigr)  =\frac{1}{2}\;\sum_{\vec{x}_{1},\mu,s}
\Xi_{\mu_{1},s_{1}}^{\sharp,a}\!\bigl(\Teta{j_{1}},\vec{x}_{1}\bigr)\;q_{\mu_{1}}\;
\hat{\mathrm{S}}^{aa}\;
\Xi_{\mu_{1},s_{1}}^{a}\bigl(\Teta{j_{1}},\vec{x}_{1}\bigr)\;;   \\   \lb{s3_18}
\hat{\mathrm{S}}_{\mu}^{ba} &=\;\hat{\mathrm{S}}^{ba}\;\;q_{\mu}\;\;\;;   \\   \lb{s3_19}
&\hspace*{-2.5cm}\sum_{\vec{x}_{1,2}}\sum_{j_{1,2}=1}^{2N}
\mfrak{b}\bigl(\Teta{j_{2}},\vec{x}_{2}\bigr)\;\eta_{j_{2}}\;\;\eta_{j_{1}}\;
\mfrak{b}\bigl(\Teta{j_{1}},\vec{x}_{1}\bigr)\quad
\hat{\mscr{V}}(\vec{x}_{2},\Teta{j_{2}};\vec{x}_{1},\Teta{j_{1}})  =
\sum_{\vec{x}_{1,2}}\sum_{j_{1,2}=1}^{2N}
\hat{\mscr{V}}(\vec{x}_{2},\Teta{j_{2}};\vec{x}_{1},\Teta{j_{1}})\;\times \\ \notag &\times\;\;
\frac{1}{4}\;\;\mfrak{B}\bigl(\Teta{j_{2}},\vec{x}_{2}\bigr)\;\eta_{j_{2}}\;\;\eta_{j_{1}}\;
\mfrak{B}\bigl(\Teta{j_{1}},\vec{x}_{1}\bigr)  =-\frac{1}{4}
\sum_{\vec{x}_{1,2}}^{\mu_{1,2},s_{1,2}}\sum_{j_{1,2}=1}^{2N}
\hat{\mscr{V}}(\vec{x}_{2},\Teta{j_{2}};\vec{x}_{1},\Teta{j_{1}})\;\times \\ \notag &\times
\TRAB\bigg[\hat{\mathrm{S}}_{\mu_{2}}^{bb}\:\eta_{j_{2}}\:
\hat{\mscr{R}}_{\mu_{2},s_{2};\mu_{1},s_{1}}^{ba}(\Teta{j_{2}},\vec{x}_{2};\Teta{j_{1}},\vec{x}_{1})\:
\hat{\mathrm{S}}_{\mu_{1}}^{aa}\:\eta_{j_{1}}\:
\hat{\mscr{R}}_{\mu_{1},s_{1};\mu_{2},s_{2}}^{ab}(\Teta{j_{1}},\vec{x}_{1};\Teta{j_{2}},\vec{x}_{2})
\bigg]_{\mbox{.}}
\end{align}
According to the dyadic product (\ref{s3_16}) of anomalous-doubled, anti-commuting fields, we take into
account 'Nambu' parts (\(b\neq a\)) of an overall, hermitian self-energy matrix
\(\hat{\Sigma}_{\mu_{1},s_{1};\mu_{2},s_{2}}^{ab}(\Teta{j_{1}},\vec{x}_{1};\Teta{j_{2}},\vec{x}_{2})\) (\ref{s3_20a})
whose diagonal blocks (\ref{s3_20b},\ref{s3_20c}) are related by opposite sign and transposition and whose
two, anti-symmetric 'Nambu' parts are related by hermitian conjugation (\ref{s3_20d},\ref{s3_20e},\ref{s3_20f})
\begin{subequations}
\begin{align}\lb{s3_20a}
\hat{\Sigma}_{\mu_{1},s_{1};\mu_{2},s_{2}}^{ab}\Big(\Teta{j_{1}},\vec{x}_{1};\Teta{j_{2}},\vec{x}_{1}\Big)
 &= \left(\bea{cc}
\hat{\Sigma}_{\mu_{1},s_{1};\mu_{2},s_{2}}^{11}\bigl(\Teta{j_{1}},\vec{x}_{1};\Teta{j_{2}},\vec{x}_{2}\bigr) &
\hat{\Sigma}_{\mu_{1},s_{1};\mu_{2},s_{2}}^{12}\bigl(\Teta{j_{1}},\vec{x}_{1};\Teta{j_{2}},\vec{x}_{2}\bigr)  \\
\hat{\Sigma}_{\mu_{1},s_{1};\mu_{2},s_{2}}^{21}\bigl(\Teta{j_{1}},\vec{x}_{1};\Teta{j_{2}},\vec{x}_{2}\bigr)  &
\hat{\Sigma}_{\mu_{1},s_{1};\mu_{2},s_{2}}^{22}\bigl(\Teta{j_{1}},\vec{x}_{1};\Teta{j_{2}},\vec{x}_{2}
\bigr) \eea\right)^{ab}_{\mbox{;}} \\ \lb{s3_20b}
\hat{\Sigma}_{\mu_{1},s_{1};\mu_{2},s_{2}}^{aa}\bigl(\Teta{j_{1}},\vec{x}_{1};\Teta{j_{2}},\vec{x}_{2}\bigr)
&= \biggl(
\hat{\Sigma}_{\mu_{1},s_{1};\mu_{2},s_{2}}^{aa}\bigl(\Teta{j_{1}},\vec{x}_{1};
\Teta{j_{2}},\vec{x}_{2}\bigr)\biggr)\pdag \;;  \\  \lb{s3_20c}
\hat{\Sigma}_{\mu_{1},s_{1};\mu_{2},s_{2}}^{22}\bigl(\Teta{j_{1}},\vec{x}_{1};
\Teta{j_{2}},\vec{x}_{2}\bigr) &= -\biggl(
\hat{\Sigma}_{\mu_{1},s_{1};\mu_{2},s_{2}}^{11}\bigl(\Teta{j_{1}},\vec{x}_{1};
\Teta{j_{2}},\vec{x}_{2}\bigr)\biggr)^{T} \;;  \\  \lb{s3_20d}
\hat{\Sigma}_{\mu_{1},s_{1};\mu_{2},s_{2}}^{21}\bigl(\Teta{j_{1}},\vec{x}_{1};\Teta{j_{2}},\vec{x}_{2}\bigr) &= \biggl(
\hat{\Sigma}_{\mu_{1},s_{1};\mu_{2},s_{2}}^{12}\bigl(\Teta{j_{1}},\vec{x}_{1};
\Teta{j_{2}},\vec{x}_{2}\bigr)\biggr)\pdag \;; \\   \lb{s3_20e}
\hat{\Sigma}_{\mu_{1},s_{1};\mu_{2},s_{2}}^{12}\bigl(\Teta{j_{1}},\vec{x}_{1};
\Teta{j_{2}},\vec{x}_{2}\bigr) &= -\biggl(
\hat{\Sigma}_{\mu_{1},s_{1};\mu_{2},s_{2}}^{12}\bigl(\Teta{j_{1}},\vec{x}_{1};
\Teta{j_{2}},\vec{x}_{1}\bigr)\biggr)^{T} \;;  \\   \lb{s3_20f}
\hat{\Sigma}_{\mu_{1},s_{1};\mu_{2},s_{2}}^{21}\bigl(\Teta{j_{1}},\vec{x}_{1};
\Teta{j_{2}},\vec{x}_{2}\bigr) &= -\biggl(
\hat{\Sigma}_{\mu_{1},s_{1};\mu_{2},s_{2}}^{21}\bigl(\Teta{j_{1}},\vec{x}_{1};
\Teta{j_{2}},\vec{x}_{1}\bigr)\biggr)^{T} \;;  \\  \lb{s3_20g}
\hat{\Sigma}_{\mu_{1},s_{1};\mu_{2},s_{2}}^{ab}\bigl(\Teta{j_{1}},\vec{x}_{1};
\Teta{j_{2}},\vec{x}_{2}\bigr) &\equiv 0 \;\;\;\;(\mbox{for }(j_{1}\mbox{\scz 'or'} j_{2}=0)\;\mbox{\scz 'or'}\;
(j_{1}\mbox{\scz 'or'} j_{2}=2N+1)\,)\;,(\mbox{cf. (\ref{s2_11b},\ref{s2_11f})})\;.
\end{align}
\end{subequations}
As one applies the dyadic product
\(\hat{\mscr{R}}_{\mu_{1},s_{1};\mu_{2},s_{2}}^{ab}(\Teta{j_{1}},\vec{x}_{1};\Teta{j_{2}},\vec{x}_{2})\)
(\ref{s3_16}-\ref{s3_19}) and the self-energy matrix (\ref{s3_20a}-\ref{s3_20f}) with the 'Nambu' parts in
the off-diagonal blocks (\(a\neq b\)) to the original, quartic interaction of Fermi fields, one attains the HST
(\ref{s3_21}) with Gaussian term of self-energy matrices and a phase factor with linear coupling between
dyadic product of doubled Fermi fields and the anomalous-doubled self-energy in the exponent
\begin{align}\lb{s3_21}
& \exp\bigg\{-\im\,\tfrac{\sdelta t}{\hbar}\sum_{\vec{x}_{1,2}}\sum_{j_{1,2}=1}^{2N}
\mfrak{b}\bigl(\Teta{j_{2}},\vec{x}_{2}\bigr)\;\eta_{j_{2}}\;\;\eta_{j_{1}}\;
\mfrak{b}\bigl(\Teta{j_{1}},\vec{x}_{1}\bigr)\quad
\hat{\mscr{V}}(\vec{x}_{2},\Teta{j_{2}};\vec{x}_{1},\Teta{j_{1}}) \bigg\} = \\ \notag &=
\exp\bigg\{\tfrac{\im}{4}\tfrac{\sdelta t}{\hbar}
\sum_{\vec{x}_{1,2}}^{\mu_{1,2},s_{1,2}}\sum_{j_{1,2}=1}^{2N}
\hat{\mscr{V}}(\vec{x}_{2},\Teta{j_{2}};\vec{x}_{1},\Teta{j_{1}}) \;\times
\TRAB\bigg[\hat{\mathrm{S}}_{\mu_{2}}^{bb}\;\eta_{j_{2}}\;
\hat{\mscr{R}}_{\mu_{2},s_{2};\mu_{1},s_{1}}^{ba}\Big(\Teta{j_{2}},\vec{x}_{2};\Teta{j_{1}},\vec{x}_{1}\Big)\;\times
\\  \notag &\times\;\hat{\mathrm{S}}_{\mu_{1}}^{aa}\;\eta_{j_{1}}\;
\hat{\mscr{R}}_{\mu_{1},s_{1};\mu_{2},s_{2}}^{ab}\Big(\Teta{j_{1}},\vec{x}_{1};\Teta{j_{2}},\vec{x}_{2}\Big)\bigg]\bigg\}\;=\;
\int d\bigl[\hat{\Sigma}_{\nu,s\ppr;\mu,s}^{ba}(\Teta{j_{2}},\vec{x}_{2};\Teta{j_{1}},\vec{x}_{1})\bigr]\;\times
\\ \notag &\times\;
\exp\bigg\{-\tfrac{\im}{4}\tfrac{\sdelta t}{\hbar}
\sum_{\vec{x}_{1,2}}^{\mu_{1,2},s_{1,2}}\sum_{j_{1,2}=1}^{2N}
\hat{\mscr{V}}^{\boldsymbol{-1}}(\vec{x}_{2},\Teta{j_{2}};\vec{x}_{1},\Teta{j_{1}}) \;\times
\TRAB\bigg[\hat{\mathrm{S}}_{\mu_{2}}^{bb}\;\eta_{j_{2}}\;
\hat{\Sigma}_{\mu_{2},s_{2};\mu_{1},s_{1}}^{ba}\Big(\Teta{j_{2}},\vec{x}_{2};\Teta{j_{1}},\vec{x}_{1}\Big)\;\times
\\ \notag &\times
\;\hat{\mathrm{S}}_{\mu_{1}}^{aa}\;\eta_{j_{1}}\;
\hat{\Sigma}_{\mu_{1},s_{1};\mu_{2},s_{2}}^{ab}\Big(\Teta{j_{1}},\vec{x}_{1};\Teta{j_{2}},\vec{x}_{2}\Big)\bigg]\bigg\}\;\times
\exp\bigg\{\tfrac{\im}{2}\tfrac{\sdelta t}{\hbar}
\sum_{\vec{x}_{1,2}}^{\mu_{1,2},s_{1,2}}\sum_{j_{1,2}=1}^{2N}\times \\  \notag &\times
\TRAB\bigg[\hat{\mathrm{S}}_{\mu_{2}}^{bb}\;\eta_{j_{2}}\;
\hat{\Sigma}_{\mu_{2},s_{2};\mu_{1},s_{1}}^{ba}\Big(\Teta{j_{2}},\vec{x}_{2};\Teta{j_{1}},\vec{x}_{1}\Big)
\;\hat{\mathrm{S}}_{\mu_{1}}^{aa}\;\eta_{j_{1}}\;
\hat{\mscr{R}}_{\mu_{1},s_{1};\mu_{2},s_{2}}^{ab}\Big(\Teta{j_{1}},\vec{x}_{1};
\Teta{j_{2}},\vec{x}_{2}\Big)\bigg]\bigg\}\;.
\end{align}
Insertion of the HST (\ref{s3_21}) into the path integral (\ref{s2_15}) results into relation (\ref{s3_22a})
with additional Gaussian factor of anomalous-doubled self-energy matrices, but with only bilinear,
anti-commuting fields. These are removed by integration according to relation (\ref{s3_4}) so that we finally
achieve path integral (\ref{s3_22b}) with the square root of the determinant and the 'Nambu'-doubled self-energy
matrix as remaining field degree of freedom
\begin{subequations}
\begin{align} \lb{s3_22a}&
\ovv{Z[\hat{\mscr{J}}]} = \int
d\bigl[\hat{\Sigma}_{\nu,s\ppr;\mu,s}^{ba}(\Teta{j_{2}},\vec{x}_{2};\Teta{j_{1}},\vec{x}_{1})\bigr]\;\;
\exp\bigg\{\negthickspace-\tfrac{\im}{4}\:\tfrac{\sdelta t}{\hbar}
\sum_{\vec{x}_{1,2}}^{\mu_{1,2},s_{1,2}}\sum_{j_{1,2}=1}^{2N}
\hat{\mscr{V}}^{\boldsymbol{-1}}(\vec{x}_{2},\Teta{j_{2}};\vec{x}_{1},\Teta{j_{1}})\;\;\times \\ \notag &\times
\TRAB\bigg[\hat{\mathrm{S}}_{\mu_{2}}^{bb}\;\;\etan{j_{2}}\;\;
\hat{\Sigma}_{\mu_{2},s_{2};\mu_{1},s_{1}}^{ba}\Big(\Teta{j_{2}},\vec{x}_{2};\Teta{j_{1}},\vec{x}_{1}\Big)\;\;
\hat{\mathrm{S}}_{\mu_{1}}^{aa}\;\;\etan{j_{1}}\;\;
\hat{\Sigma}_{\mu_{1},s_{1};\mu_{2},s_{2}}^{ab}\Big(\Teta{j_{1}},\vec{x}_{1};\Teta{j_{2}},\vec{x}_{2}
\Big)\bigg]\bigg\}\;\times \\ \notag &\times\;
\int \prod_{\vec{x},\mu,s}\prod_{j=0}^{2N}
\tfrac{d\chi_{\mu,s}^{*}\bigl(\Teta{j},\vec{x}\bigr)\;\;d\chi_{\mu,s}\bigl(\Teta{j},\vec{x}\bigr)}{\mcal{N}_{x}}\;\times
\exp\bigg\{-\tfrac{1}{2}\sum_{\vec{x}_{1,2}}^{\mu,\nu,s,s\ppr}\sum_{j_{1,2}=0}^{2N+1}
\Xi_{\nu,s\ppr}^{\sharp,b}\bigl(\Teta{j_{2}},\vec{x}_{2}\bigr)\;\times \\ \notag &\times\;
\bigg[\hat{\mathrm{I}}^{bb}\;
\hat{\mathsf{H}}_{\nu,s\ppr;\mu,s}^{ba}\bigl(\vec{x}_{2},\Teta{j_{2}};
\vec{x}_{1},\Teta{j_{1}}\bigr)\;\hat{\mathrm{I}}^{aa}\;
+\;\eta_{j_{2}}\;\;\hat{\mscr{J}}_{\nu,s\ppr;\mu,s}^{ba}\bigl(\Teta{j_{2}},\vec{x}_{2};
\Teta{j_{1}},\vec{x}_{1}\bigr)\;\;\eta_{j_{1}}\;+ \;\hat{\mathrm{S}}_{\nu}^{bb}\;\etan{j_{2}}\;\;
\Bigl(\im\;\tfrac{\sdelta t}{\hbar}\Bigr)\;\;\times  \\ \notag &\times\;\;
\hat{\Sigma}_{\nu,s\ppr;\mu,s}^{ba}\!\bigl(\Teta{j_{2}},\vec{x}_{2};\Teta{j_{1}},\vec{x}_{1}\bigr)
\;\;\hat{\mathrm{S}}_{\mu}^{aa}\;\etan{j_{1}}\bigg]_{\nu,s\ppr;\mu,s}^{ba}\hspace*{-0.9cm}
(\Teta{j_{2}},\vec{x}_{2};\Teta{j_{1}},\vec{x}_{1})\quad\Xi_{\mu,s}^{a}(\Teta{j_{1}},\vec{x}_{1})\bigg\}\;; \\   \lb{s3_22b}&
\ovv{Z[\hat{\mscr{J}}]} = \int
d\bigl[\hat{\Sigma}_{\nu,s\ppr;\mu,s}^{ba}(\Teta{j_{2}},\vec{x}_{2};\Teta{j_{1}},\vec{x}_{1})\bigr]\;\;
\exp\bigg\{\negthickspace-\tfrac{\im}{4}\:\tfrac{\sdelta t}{\hbar}
\sum_{\vec{x}_{1,2}}^{\mu_{1,2},s_{1,2}}\sum_{j_{1,2}=1}^{2N}
\hat{\mscr{V}}^{\boldsymbol{-1}}(\vec{x}_{2},\Teta{j_{2}};\vec{x}_{1},\Teta{j_{1}})\;\;\times \\ \notag &\times
\TRAB\bigg[\hat{\mathrm{S}}_{\mu_{2}}^{bb}\;\;\etan{j_{2}}\;\;
\hat{\Sigma}_{\mu_{2},s_{2};\mu_{1},s_{1}}^{ba}\Big(\Teta{j_{2}},\vec{x}_{2};\Teta{j_{1}},\vec{x}_{1}\Big)\;\;
\hat{\mathrm{S}}_{\mu_{1}}^{aa}\;\;\etan{j_{1}}\;\;
\hat{\Sigma}_{\mu_{1},s_{1};\mu_{2},s_{2}}^{ab}\Big(\Teta{j_{1}},\vec{x}_{1};\Teta{j_{2}},\vec{x}_{2}
\Big)\bigg]\bigg\}\;\times \\ \notag &\times\;
\mbox{DET}\bigg(\tfrac{1}{\mcal{N}_{x}}
\bigg[\hat{\mathrm{I}}^{bb}\;
\hat{\mathsf{H}}_{\nu,s\ppr;\mu,s}^{ba}\bigl(\vec{x}_{2},\Teta{j_{2}};
\vec{x}_{1},\Teta{j_{1}}\bigr)\;\hat{\mathrm{I}}^{aa}\;
+\;\eta_{j_{2}}\;\;\hat{\mscr{J}}_{\nu,s\ppr;\mu,s}^{ba}\bigl(\Teta{j_{2}},\vec{x}_{2};
\Teta{j_{1}},\vec{x}_{1}\bigr)\;\;\eta_{j_{1}}\;+   \\  \notag &+\;\hat{\mathrm{S}}_{\nu}^{bb}\;\;\etan{j_{2}}\;\;
\Bigl(\im\;\tfrac{\sdelta t}{\hbar}\Bigr)\;\;
\hat{\Sigma}_{\nu,s\ppr;\mu,s}^{ba}\!\bigl(\Teta{j_{2}},\vec{x}_{2};\Teta{j_{1}},\vec{x}_{1}\bigr)
\;\;\hat{\mathrm{S}}_{\mu}^{aa}\;\;\etan{j_{1}}\bigg]_{\nu,s\ppr;\mu,s}^{ba}\hspace*{-0.9cm}
(\Teta{j_{2}},\vec{x}_{2};\Teta{j_{1}},\vec{x}_{1})\bigg)^{\boldsymbol{1/2}}\;.
\end{align}
\end{subequations}
The saddle point equation of (\ref{s3_22b}) and its related eigenvalue problem follow in analogy to sections
\ref{s31}, \ref{s32} and are therefore listed in brevity in subsequent equations (\ref{s3_23a}-\ref{s3_23f})
(relations (\ref{s3_23b},\ref{s3_23c},\ref{s3_23e},\ref{s3_23f}) are free of any summations over space-contour-time
coordinates and internal state labels; note, however, that (\ref{s3_23d}) implies the scaled '\(\tfrac{1}{\mcal{N}_{x}}\)'
spatial and contour time step summations)
\begin{subequations}
\begin{align} \notag &
\hat{\mscr{M}}_{\nu,s\ppr;\mu,s}^{(\hat{\mscr{J}})ba}(\Teta{j_{2}},\vec{x}_{2};\Teta{j_{1}},\vec{x}_{1}) \;=\;
\hat{\mathrm{I}}^{bb}\;
\hat{\mathsf{H}}_{\nu,s\ppr;\mu,s}^{ba}\bigl(\vec{x}_{2},\Teta{j_{2}};
\vec{x}_{1},\Teta{j_{1}}\bigr)\;\hat{\mathrm{I}}^{aa}\;
+\;\eta_{j_{2}}\;\;\hat{\mscr{J}}_{\nu,s\ppr;\mu,s}^{ba}\bigl(\Teta{j_{2}},\vec{x}_{2};
\Teta{j_{1}},\vec{x}_{1}\bigr)\;\;\eta_{j_{1}}\;+   \\  \lb{s3_23a}   &+\;\hat{\mathrm{S}}_{\nu}^{bb}\;\;\etan{j_{2}}\;\;
\Bigl(\im\;\tfrac{\sdelta t}{\hbar}\Bigr)\;\;
\hat{\Sigma}_{\nu,s\ppr;\mu,s}^{ba}\!\bigl(\Teta{j_{2}},\vec{x}_{2};\Teta{j_{1}},\vec{x}_{1}\bigr)
\;\;\hat{\mathrm{S}}_{\mu}^{aa}\;\;\etan{j_{1}} \;;  \\  \lb{s3_23b} &
\hat{\mscr{V}}^{\boldsymbol{-1}}(\vec{x}_{2},\Teta{j_{2}};\vec{x}_{1},\Teta{j_{1}})\;\;
\hat{\Sigma}_{\mu_{2},s_{2};\mu_{1},s_{1}}^{(0)11}(\Teta{j_{2}},\vec{x}_{2};\Teta{j_{1}},\vec{x}_{1})\big/\mcal{N}_{x}=
\\  \notag &=\;\tfrac{1}{2}\,\mcal{N}_{x}\bigg[
\hat{\mscr{M}}_{\mu_{2},s_{2};
\mu_{1},s_{1}}^{\boldsymbol{-1};(\hat{\mscr{J}}\equiv0),11}(\Teta{j_{2}},\vec{x}_{2};\Teta{j_{1}},\vec{x}_{1}) -
\hat{\mscr{M}}_{\mu_{1},s_{1};
\mu_{2},s_{2}}^{\boldsymbol{-1};(\hat{\mscr{J}}\equiv0),22}(\Teta{j_{1}},\vec{x}_{1};\Teta{j_{2}},\vec{x}_{2})\bigg]\;;
\\ \lb{s3_23c}  &
\hat{\mscr{V}}^{\boldsymbol{-1}}(\vec{x}_{2},\Teta{j_{2}};\vec{x}_{1},\Teta{j_{1}})\;\;
\hat{\Sigma}_{\mu_{2},s_{2};\mu_{1},s_{1}}^{(0)b\neq a}(\Teta{j_{2}},\vec{x}_{2};
\Teta{j_{1}},\vec{x}_{1})\big/\mcal{N}_{x}=
\\  \notag &=\;\tfrac{1}{2}\,\mcal{N}_{x}\bigg[
\hat{\mscr{M}}_{\mu_{2},s_{2};
\mu_{1},s_{1}}^{\boldsymbol{-1};(\hat{\mscr{J}}\equiv0),b\neq a}(\Teta{j_{2}},\vec{x}_{2};\Teta{j_{1}},\vec{x}_{1}) -
\hat{\mscr{M}}_{\mu_{1},s_{1};
\mu_{2},s_{2}}^{\boldsymbol{-1};(\hat{\mscr{J}}\equiv0),b\neq a}(\Teta{j_{1}},
\vec{x}_{1};\Teta{j_{2}},\vec{x}_{2})\bigg]\;;
\\  \lb{s3_23d}   &
\sum_{j_{1}=0}^{2N+1}\sum_{\vec{x}_{1}}^{\mu,s}
\hat{\mscr{M}}_{\nu,s\ppr;\mu,s}^{(\hat{\mscr{J}}\equiv0)ba}(\Teta{j_{2}},\vec{x}_{2};\Teta{j_{1}},\vec{x}_{1})\quad
\Psi_{M;\mu,s}^{a}(\Teta{j_{1}},\vec{x}_{1}) \quad = \quad E_{M}\quad
\Psi_{M;\nu,s\ppr}^{b}(\Teta{j_{2}},\vec{x}_{2})\;; \\ \notag &
\quad(\mbox{eigenstate label $M$ with complex eigenvalue }E_{M})\;;   \\   \lb{s3_23e}  &
\hat{\mscr{V}}^{\boldsymbol{-1}}(\vec{x}_{2},\Teta{j_{2}};\vec{x}_{1},\Teta{j_{1}})\;\;
\hat{\Sigma}_{\mu_{2},s_{2};\mu_{1},s_{1}}^{(0)11}(\Teta{j_{2}},\vec{x}_{2};\Teta{j_{1}},\vec{x}_{1})\big/\mcal{N}_{x}=
\\  \notag &=\;\tfrac{1}{2}\sum_{M\in\mbox{\scz eigenstates}}\bigg[
\frac{\Psi_{M;\mu_{2},s_{2}}^{(a=1)}(\Teta{j_{2}},\vec{x}_{2})\;\;
\Psi_{M;\mu_{1},s_{1}}^{\boldsymbol{T},(a=1)}(\Teta{j_{1}},\vec{x}_{1})}{E_{M}}-\frac{
\Psi_{M;\mu_{1},s_{1}}^{(a=2)}(\Teta{j_{1}},\vec{x}_{1})\;\;
\Psi_{M;\mu_{2},s_{2}}^{\boldsymbol{T},(a=2)}(\Teta{j_{2}},\vec{x}_{2})}{E_{M}}\bigg]\;; \\  \lb{s3_23f} &
\hat{\mscr{V}}^{\boldsymbol{-1}}(\vec{x}_{2},\Teta{j_{2}};\vec{x}_{1},\Teta{j_{1}})\;\;
\hat{\Sigma}_{\mu_{2},s_{2};\mu_{1},s_{1}}^{(0)21}(\Teta{j_{2}},\vec{x}_{2};\Teta{j_{1}},\vec{x}_{1})\big/\mcal{N}_{x}=
\\  \notag &=\;\tfrac{1}{2}\sum_{M\in\mbox{\scz eigenstates}}\bigg[
\frac{\Psi_{M;\mu_{2},s_{2}}^{(b=2)}(\Teta{j_{2}},\vec{x}_{2})\;\;
\Psi_{M;\mu_{1},s_{1}}^{\boldsymbol{T},(a=1)}(\Teta{j_{1}},\vec{x}_{1})}{E_{M}}-\frac{
\Psi_{M;\mu_{1},s_{1}}^{(b=2)}(\Teta{j_{1}},\vec{x}_{1})\;\;
\Psi_{M;\mu_{2},s_{2}}^{\boldsymbol{T},(a=1)}(\Teta{j_{2}},\vec{x}_{2})}{E_{M}}\bigg]\;.
\end{align}
\end{subequations}

\section{Coset decomposition with density- and exciton-related parts} \lb{s4}

\subsection{SSB with 'hinge' fields and anomalous-doubled parts} \lb{s41}

In section \ref{s33} the density and 'Nambu' parts are regarded within a single, total matrix,
having the appropriate block structure (\ref{s3_20a}). In this section we aim at a further
factorization of the total self-energy into block diagonal, density-related parts and coset matrices
whose generators are composed of the anomalous-doubled field degrees of freedom. In order to
separate the coset matrices with 'Nambu' generators from the densities in a HST, one has to introduce
block diagonal self-energy densities
\(\hat{\mfrak{s}}_{\mu_{2},s_{2};\mu_{1},s_{1}}^{aa}(\Teta{j_{2}},\vec{x}_{2};\Teta{j_{1}},\vec{x}_{1})\)
(\ref{s4_1a},\ref{s4_1b}) in a SSB as invariant ground or vacuum state and a total self-energy
matrix \(\delta\wt{\Sigma}_{\mu_{2},s_{2};\mu_{1},s_{1}}^{ba}(\Teta{j_{2}},\vec{x}_{2};
\Teta{j_{1}},\vec{x}_{1})\) (\ref{s4_1c}) with anti-hermitian 'Nambu' parts for the factorization
with coset matrices (cf. section 3 in \cite{pop1}). The matrix
\(\delta\wt{\Sigma}_{\mu_{2},s_{2};\mu_{1},s_{1}}^{ba}(\Teta{j_{2}},\vec{x}_{2};
\Teta{j_{1}},\vec{x}_{1})\) (\ref{s4_1c}) has a similar structure (\ref{s4_1d}-\ref{s4_1g}) as
\(\hat{\Sigma}_{\mu_{2},s_{2};\mu_{1},s_{1}}^{ba}(\Teta{j_{2}},\vec{x}_{2};
\Teta{j_{1}},\vec{x}_{1})\) (\ref{s3_20a}) in relations (\ref{s3_20b}-\ref{s3_20f}) of section \ref{s33}, except from
the inclusion of the imaginary factor '\(\im\)' in the off-diagonal blocks for the anti-hermitian property which is required
for a factorization by a coset decomposition
\begin{subequations}
\begin{align}\lb{s4_1a}
\hat{\mfrak{s}}_{\mu_{2},s_{2};\mu_{1},s_{1}}^{aa}\bigl(\Teta{j_{2}},\vec{x}_{2};\Teta{j_{1}},\vec{x}_{1}\bigr) &=
\Big(\hat{\mfrak{s}}_{\mu_{2},s_{2};\mu_{1},s_{1}}^{aa}\bigl(\Teta{j_{2}},\vec{x}_{2};
\Teta{j_{1}},\vec{x}_{1}\bigr)\Big)\pdag\;;  \\ \lb{s4_1b}
\hat{\mfrak{s}}_{\mu_{2},s_{2};\mu_{1},s_{1}}^{22}\bigl(\Teta{j_{2}},\vec{x}_{2};\Teta{j_{1}},\vec{x}_{1}\bigr) &=
\boldsymbol{-}\Big(\hat{\mfrak{s}}_{\mu_{2},s_{2};\mu_{1},s_{1}}^{11}\bigl(\Teta{j_{2}},\vec{x}_{2};
\Teta{j_{1}},\vec{x}_{1}\bigr)\Big)^{\boldsymbol{T}} \;;  \\   \lb{s4_1c}
\delta\wt{\Sigma}_{\mu_{2},s_{2};\mu_{1},s_{1}}^{ba}\bigl(\Teta{j_{2}},\vec{x}_{2};
\Teta{j_{1}},\vec{x}_{1}\bigr) &=\biggl(\bea{cc}
\delta\hat{\Sigma}_{\mu_{2},s_{2};\mu_{1},s_{1}}^{11}\bigl(\Teta{j_{2}},\vec{x}_{2};
\Teta{j_{1}},\vec{x}_{1}\bigr)  & \im\;
\delta\hat{\Sigma}_{\mu_{2},s_{2};\mu_{1},s_{1}}^{12}\bigl(\Teta{j_{2}},\vec{x}_{2};
\Teta{j_{1}},\vec{x}_{1}\bigr) \\  \im\;
\delta\hat{\Sigma}_{\mu_{2},s_{2};\mu_{1},s_{1}}^{22}\bigl(\Teta{j_{2}},\vec{x}_{2};
\Teta{j_{1}},\vec{x}_{1}\bigr)  & \delta\hat{\Sigma}_{\mu_{2},s_{2};\mu_{1},s_{1}}^{22}\bigl(\Teta{j_{2}},\vec{x}_{2};
\Teta{j_{1}},\vec{x}_{1}\bigr) \eea\biggr)^{ba}_{\mbox{;}} \\  \lb{s4_1d}
\delta\hat{\Sigma}_{\mu_{2},s_{2};\mu_{1},s_{1}}^{aa}\bigl(\Teta{j_{2}},\vec{x}_{2};
\Teta{j_{1}},\vec{x}_{1}\bigr)  &=\Big(\delta\hat{\Sigma}_{\mu_{2},s_{2};\mu_{1},s_{1}}^{aa}\bigl(\Teta{j_{2}},\vec{x}_{2};
\Teta{j_{1}},\vec{x}_{1}\bigr)\Big)\pdag \;; \\   \lb{s4_1e}
\delta\hat{\Sigma}_{\mu_{2},s_{2};\mu_{1},s_{1}}^{22}\bigl(\Teta{j_{2}},\vec{x}_{2};
\Teta{j_{1}},\vec{x}_{1}\bigr)  &=\boldsymbol{-}
\Big(\delta\hat{\Sigma}_{\mu_{2},s_{2};\mu_{1},s_{1}}^{11}\bigl(\Teta{j_{2}},\vec{x}_{2};
\Teta{j_{1}},\vec{x}_{1}\bigr)\Big)^{\boldsymbol{T}} \;; \\  \lb{s4_1f}
\delta\hat{\Sigma}_{\mu_{2},s_{2};\mu_{1},s_{1}}^{21}\bigl(\Teta{j_{2}},\vec{x}_{2};
\Teta{j_{1}},\vec{x}_{1}\bigr)  &=
\Big(\delta\hat{\Sigma}_{\mu_{2},s_{2};\mu_{1},s_{1}}^{12}\bigl(\Teta{j_{2}},\vec{x}_{2};
\Teta{j_{1}},\vec{x}_{1}\bigr)\Big)\pdag \;; \\  \lb{s4_1g}
\delta\hat{\Sigma}_{\mu_{2},s_{2};\mu_{1},s_{1}}^{b\neq a}\bigl(\Teta{j_{2}},\vec{x}_{2};
\Teta{j_{1}},\vec{x}_{1}\bigr)  &=\boldsymbol{-}
\Big(\delta\hat{\Sigma}_{\mu_{2},s_{2};\mu_{1},s_{1}}^{b\neq a}\bigl(\Teta{j_{2}},\vec{x}_{2};
\Teta{j_{1}},\vec{x}_{1}\bigr)\Big)^{\boldsymbol{T}} \;.
\end{align}
\end{subequations}
The coset decomposition of (\ref{s4_1c}) needs further, block diagonal self-energy matrices of densities
or 'hinge' fields
\(\delta\hat{\Sigma}_{D;\mu_{2},s_{2};\mu_{1},s_{1}}^{aa}(\Teta{j_{2}},\vec{x}_{2};
\Teta{j_{1}},\vec{x}_{1})\)  with equivalent structure (\ref{s4_2a},\ref{s4_2b}) as
the matrices (\ref{s4_1a},\ref{s4_1b}) in order to perform the factorization (\ref{s4_3}) with the coset
matrices \(\hat{\mathrm{T}}_{\mu_{2},s_{2};\mu_{1},s_{1}}^{ba}(\Teta{j_{2}},\vec{x}_{2};
\Teta{j_{1}},\vec{x}_{1})\) (\ref{s4_4a}). The generators
\(\hat{\mathrm{Y}}_{\mu_{2},s_{2};\mu_{1},s_{1}}^{ba}(\Teta{j_{2}},\vec{x}_{2};
\Teta{j_{1}},\vec{x}_{1})\) (\ref{s4_4b}) of latter coset matrices consist of two anti-symmetric
sub-generators \(\hat{\mathrm{X}}_{\mu_{2},s_{2};\mu_{1},s_{1}}(\Teta{j_{2}},\vec{x}_{2};
\Teta{j_{1}},\vec{x}_{1})\), \(\hat{\mathrm{X}}_{\mu_{2},s_{2};\mu_{1},s_{1}}\pdag(\Teta{j_{2}},\vec{x}_{2};
\Teta{j_{1}},\vec{x}_{1})\) (\ref{s4_4c}) which describe the anomalous paired fer\-mi\-ons of a SSB
under inclusion of excitons\footnote{We apply the standard summation convention over double occurrence of
indices in (\ref{s4_3}) with {\it un-scaled} sums of discrete, spatial vectors \(\vec{x}_{3}\), \(\vec{x}_{4}\)
and with summations of time indices \(j_{3},j_{4}=1,\ldots,2N\) including summations of electron-, hole-bands
\(\mu_{3},\mu_{4}=\mbox{'e', 'h'}\) and of spin states \(s_{3},s_{4}=\uparrow,\downarrow\). This kind of
summation is also taken within the coset factorization in later steps of transformations.}
\begin{subequations}
\begin{align}\lb{s4_2a}
\delta\hat{\Sigma}_{D;\mu_{2},s_{2};\mu_{1},s_{1}}^{aa}\bigl(\Teta{j_{2}},\vec{x}_{2};\Teta{j_{1}},\vec{x}_{1}\bigr) &=
\Big(\delta\hat{\Sigma}_{D;\mu_{2},s_{2};\mu_{1},s_{1}}^{aa}\bigl(\Teta{j_{2}},\vec{x}_{2};
\Teta{j_{1}},\vec{x}_{1}\bigr)\Big)\pdag\;;  \\ \lb{s4_2b}
\delta\hat{\Sigma}_{D;\mu_{2},s_{2};\mu_{1},s_{1}}^{22}\bigl(\Teta{j_{2}},\vec{x}_{2};\Teta{j_{1}},\vec{x}_{1}\bigr) &=
\boldsymbol{-}\Big(\delta\hat{\Sigma}_{D;\mu_{2},s_{2};\mu_{1},s_{1}}^{11}\bigl(\Teta{j_{2}},\vec{x}_{2};
\Teta{j_{1}},\vec{x}_{1}\bigr)\Big)^{\boldsymbol{T}} \;;
\end{align}
\end{subequations}
\begin{align}   \lb{s4_3}
\delta\wt{\Sigma}_{\mu_{2},s_{2};\mu_{1},s_{1}}^{ba}\bigl(\Teta{j_{2}},\vec{x}_{2};
\Teta{j_{1}},\vec{x}_{1}\bigr) &= (\mbox{un-scaled, discrete sums }\vec{x}_{3,4},\,j_{3,4}=1,\ldots,2N,\,
\mu_{3,4}=e,h,\,s_{3,4}=\uparrow,\downarrow)= \\ \notag &\hspace*{-4.4cm}=
\hat{\mathrm{T}}_{\mu_{2},s_{2};\mu_{4},s_{4}}^{bb\ppr}\bigl(\Teta{j_{2}},\vec{x}_{2};
\Teta{j_{4}},\vec{x}_{4}\bigr)\;\;
\delta\hat{\Sigma}_{D;\mu_{4},s_{4};\mu_{3},s_{3}}^{b\ppr =a\ppr}\bigl(\Teta{j_{4}},\vec{x}_{4};
\Teta{j_{3}},\vec{x}_{3}\bigr)\;\;
\hat{\mathrm{T}}_{\mu_{3},s_{3};\mu_{1},s_{1}}^{\boldsymbol{-1};a\ppr a}\bigl(\Teta{j_{3}},\vec{x}_{3};
\Teta{j_{1}},\vec{x}_{1}\bigr)\;;
\end{align}
\begin{subequations}
\begin{align} \lb{s4_4a}
\hat{\mathrm{T}}_{\nu,s\ppr;\mu,s}^{ba}\bigl(\Teta{j_{2}},\vec{x}_{2};
\Teta{j_{1}},\vec{x}_{1}\bigr) &= \bigg[\exp\bigg\{-
\hat{\mathrm{Y}}_{\mu_{2},s_{2};\mu_{1},s_{1}}^{b\ppr a\ppr}\Bigl(\Teta{j_{4}},\vec{x}_{4};
\Teta{j_{3}},\vec{x}_{3}\Bigr)\bigg\}\bigg]_{\nu,s\ppr;\mu,s}^{ba}\hspace*{-0.9cm}\bigl(\Teta{j_{2}},\vec{x}_{2};
\Teta{j_{1}},\vec{x}_{1}\bigr) \;; \\  \lb{s4_4b}
\hat{\mathrm{Y}}_{\nu,s\ppr;\mu,s}^{ba}\bigl(\Teta{j_{2}},\vec{x}_{2};
\Teta{j_{1}},\vec{x}_{1}\bigr) &=\Biggl(\bea{cc}
0 & \hat{\mathrm{X}}_{\nu,s\ppr;\mu,s}\bigl(\Teta{j_{2}},\vec{x}_{2};
\Teta{j_{1}},\vec{x}_{1}\bigr)  \\ \hat{\mathrm{X}}_{\nu,s\ppr;\mu,s}\pdag\bigl(\Teta{j_{2}},\vec{x}_{2};
\Teta{j_{1}},\vec{x}_{1}\bigr) \eea\Biggr)^{ba} \;;  \\ \lb{s4_4c}
\hat{\mathrm{X}}_{\nu,s\ppr;\mu,s}\bigl(\Teta{j_{2}},\vec{x}_{2};
\Teta{j_{1}},\vec{x}_{1}\bigr) &=
\boldsymbol{-}\Bigl(\hat{\mathrm{X}}_{\nu,s\ppr;\mu,s}\bigl(\Teta{j_{2}},\vec{x}_{2};
\Teta{j_{1}},\vec{x}_{1}\bigr)\Bigr)^{\boldsymbol{T}}\;;\;\;(\mbox{with complex-valued entries}).
\end{align}
\end{subequations}
In analogy to \cite{mie3} we accomplish the HST (\ref{s4_5}) by one half with the densities
\(\hat{\mfrak{s}}_{\mu_{2},s_{2};\mu_{1},s_{1}}^{aa}(\Teta{j_{2}},\vec{x}_{2};\Teta{j_{1}},\vec{x}_{1})\)
(\ref{s4_1a}) and by one half with the anomalous parts of
\(\delta\wt{\Sigma}_{\mu_{2},s_{2};\mu_{1},s_{1}}^{b\neq a}(\Teta{j_{2}},\vec{x}_{2};
\Teta{j_{1}},\vec{x}_{1})\) ({\ref{s4_1c},\ref{s4_1f},\ref{s4_1g}). The 'hinge' fields
\(\delta\hat{\Sigma}_{\mu_{2},s_{2};\mu_{1},s_{1}}^{aa}(\Teta{j_{2}},\vec{x}_{2};
\Teta{j_{1}},\vec{x}_{1})\) (\ref{s4_1d},\ref{s4_1e})
or \(\delta\hat{\Sigma}_{D;\mu_{2},s_{2};\mu_{1},s_{1}}^{aa}(\Teta{j_{2}},\vec{x}_{2};
\Teta{j_{1}},\vec{x}_{1})\) (\ref{s4_2a},\ref{s4_2b})
only have the final effect of normalized Gaussian factors (cf. \cite{mie3}).
The true HST for the SSB to coset degrees of freedom is therefore given by the integrations of the densities
\(\hat{\mfrak{s}}_{\mu_{2},s_{2};\mu_{1},s_{1}}^{aa}(\Teta{j_{2}},\vec{x}_{2};\Teta{j_{1}},\vec{x}_{1})\)
(\ref{s4_1a}) and by the integrations of the anti-hermitian related 'Nambu' parts
of the total self-energy matrix, each contributing
in half for the transformation of the quartic interaction of Fermi fields
(\(\hat{\mathrm{S}}_{\mu}^{ba}=\hat{\mathrm{S}}^{ba}\;q_{\mu}\), cf.\ (\ref{s3_18}))
\begin{align} \notag &
\exp\bigg\{\hspace*{-0.2cm}-\im\;\tfrac{\sdelta t}{\hbar}\sum_{\vec{x}_{1,2}}\sum_{j_{1,2}=1}^{2N}
\mfrak{b}(\Teta{j_{2}},\vec{x}_{2})\;\eta_{j_{2}}\;\eta_{j_{1}}\;\mfrak{b}(\Teta{j_{1}},\vec{x}_{1})\;
\hat{\mscr{V}}(\vec{x}_{2},\Teta{j_{2}};\vec{x}_{1},\Teta{j_{1}})\bigg\}=  \\  \notag &=
\int d[\hat{\mfrak{s}}_{\nu,s\ppr;\mu,s}^{11}(\Teta{j_{2}},\vec{x}_{2};\Teta{j_{1}},\vec{x}_{1})]\;\times\;
\int d[\delta\wt{\Sigma}_{\nu,s\ppr;\mu,s}^{ba}(\Teta{j_{2}},\vec{x}_{2};
\Teta{j_{1}},\vec{x}_{1})]\;\times  \\ \notag &\times\;
\exp\bigg\{\hspace*{-0.1cm}-\tfrac{\im}{4}\tfrac{\sdelta t}{\hbar}\sum_{\vec{x}_{1,2}}\sum_{j_{1,2}=1}^{2N}
\hat{\mscr{V}}^{\boldsymbol{-1}}(\vec{x}_{2},\Teta{j_{2}};\vec{x}_{1},\Teta{j_{1}})
\bigg[
\hat{\mfrak{s}}_{\mu_{2},s_{2};\mu_{1},s_{1}}^{aa}(\Teta{j_{2}},\vec{x}_{2};\Teta{j_{1}},\vec{x}_{1})\,\times
\\  \lb{s4_5} &\times
\eta_{j_{1}}\;q_{\mu_{1}}\;
\hat{\mfrak{s}}_{\mu_{1},s_{1};\mu_{2},s_{2}}^{aa}(\Teta{j_{1}},\vec{x}_{1};\Teta{j_{2}},\vec{x}_{2})\;
\eta_{j_{2}}\;q_{\mu_{2}}\bigg]\bigg\} \times
\exp\bigg\{\hspace*{-0.1cm}\tfrac{\im}{4}\;\tfrac{\sdelta t}{\hbar}\sum_{\vec{x}_{1,2}}\sum_{j_{1,2}=1}^{2N}
\hat{\mscr{V}}^{\boldsymbol{-1}}(\vec{x}_{2},\Teta{j_{2}};\vec{x}_{1},\Teta{j_{1}})\;\times \\ \notag &\times
\TRAB\bigg[\hat{\mathrm{S}}_{\mu_{2}}^{bb}\:\eta_{j_{2}}\;
\delta\wt{\Sigma}_{\mu_{2},s_{2};\mu_{1},s_{1}}^{ba}(\Teta{j_{2}},\vec{x}_{2};
\Teta{j_{1}},\vec{x}_{1})\;\hat{\mathrm{S}}_{\mu_{1}}^{aa}\:\eta_{j_{1}}\;
\delta\wt{\Sigma}_{\mu_{1},s_{1};\mu_{2},s_{2}}^{ab}(\Teta{j_{1}},\vec{x}_{1};
\Teta{j_{2}},\vec{x}_{2})\bigg]\bigg\}\times \\ \notag &\times
\exp\bigg\{\hspace*{-0.1cm}-\tfrac{\im}{2}\;\tfrac{\sdelta t}{\hbar}
\sum_{\vec{x}_{1,2}}^{\mu_{1,2},s_{1,2}}\sum_{j_{1,2}=1}^{2N}
\Xi_{\mu_{2},s_{2}}^{\sharp,b}(\Teta{j_{2}},\vec{x}_{2})\;q_{\mu_{2}}\:\eta_{j_{2}}
\Big[\hat{\mfrak{s}}_{\mu_{2},s_{2};\mu_{1},s_{1}}^{b=a}(\Teta{j_{2}},\vec{x}_{2};\Teta{j_{1}},\vec{x}_{1})
+  \\  \notag &+\;\hat{\mathrm{I}}^{bb}\;
\delta\wt{\Sigma}_{\mu_{2},s_{2};\mu_{1},s_{1}}^{ba}(\Teta{j_{2}},\vec{x}_{2};
\Teta{j_{1}},\vec{x}_{1})\;\hat{\mathrm{I}}^{aa}\Big]q_{\mu_{1}}\:\eta_{j_{1}}\;
\Xi_{\mu_{1},s_{1}}^{a}(\Teta{j_{1}},\vec{x}_{1})\bigg\}_{\mbox{.}}
\end{align}
As we substitute (\ref{s4_5}) into the original path integral (\ref{s2_15}) and integrate according
to relation (\ref{s3_4}) with the doubled anti-commuting fields (\ref{s2_11a}-\ref{s2_11g}), we
attain the square root of the determinant and the Gaussian factors of the self-energy matrices which are
defined in (\ref{s4_1a}-\ref{s4_1g}). A shift of the off-diagonal, anomalous-doubled self-energy components
(\ref{s4_6b},\ref{s4_6c}) by the off-diagonal, driving dipole terms
\(\hat{\mathsf{H}}_{\nu,s\ppr;\mu,s}^{12}(\vec{x}_{2},\Teta{j_{2}};\vec{x}_{1},\Teta{j_{1}})\) (\ref{s2_14c}),
\(\hat{\mathsf{H}}_{\nu,s\ppr;\mu,s}^{21}(\vec{x}_{2},\Teta{j_{2}};\vec{x}_{1},\Teta{j_{1}})\) (\ref{s2_14d})
only leaves the block diagonal Hamiltonian parts (\ref{s2_14a},\ref{s2_14b}) within the determinant and
results into a modified Gaussian with a generating dipole term for the quadratic self-energy.
In order to distinguish between block diagonal \('b=a'\) and off-diagonal \('b\neq a'\) parts, we define in brevity
the two one-particle operators
\(\hat{\mfrak{H}}_{\nu,s\ppr;\mu,s}^{b=a}(\vec{x}_{2},\Teta{j_{2}};\vec{x}_{1},\Teta{j_{1}})\)  (\ref{s4_6d}) and
\(\hat{\mcal{H}}_{\nu,s\ppr;\mu,s}^{b\neq a}(\vec{x}_{2},\Teta{j_{2}};\vec{x}_{1},\Teta{j_{1}})\) (\ref{s4_6e})
for separate density-related and anomalous-related terms of the total one-particle operator
\(\hat{\mathsf{H}}_{\nu,s\ppr;\mu,s}^{ba}(\vec{x}_{2},\Teta{j_{2}};\vec{x}_{1},\Teta{j_{1}})\) (\ref{s2_14a}-\ref{s2_14d})
\begin{subequations}
\begin{align} \lb{s4_6a}&
\ovv{Z[\hat{\mscr{J}}]} =
\int d[\hat{\mfrak{s}}_{\nu,s\ppr;\mu,s}^{11}(\Teta{j_{2}},\vec{x}_{2};\Teta{j_{1}},\vec{x}_{1})]\;\times
\exp\bigg\{\hspace*{-0.1cm}-\tfrac{\im}{4}\tfrac{\sdelta t}{\hbar}
\sum_{\vec{x}_{1,2}}^{\mu_{1,2},s_{1,2}}\sum_{j_{1,2}=1}^{2N}
\hat{\mscr{V}}^{\boldsymbol{-1}}(\vec{x}_{2},\Teta{j_{2}};\vec{x}_{1},\Teta{j_{1}})\;\times \\ \notag &\times
\bigg[\eta_{j_{2}}\;q_{\mu_{2}}\;
\hat{\mfrak{s}}_{\mu_{2},s_{2};\mu_{1},s_{1}}^{aa}(\Teta{j_{2}},\vec{x}_{2};\Teta{j_{1}},\vec{x}_{1})\;
\eta_{j_{1}}\;q_{\mu_{1}}\;
\hat{\mfrak{s}}_{\mu_{1},s_{1};\mu_{2},s_{2}}^{aa}(\Teta{j_{1}},\vec{x}_{1};\Teta{j_{2}},\vec{x}_{2})
\bigg]\bigg\} \times \\  \notag &\int
d\bigl[\delta\wt{\Sigma}_{\nu,s\ppr;\mu,s}^{ba}(\Teta{j_{2}},\vec{x}_{2};\Teta{j_{1}},\vec{x}_{1})\bigr]\;
\exp\bigg\{\hspace*{-0.1cm}\tfrac{\im}{4}\;\tfrac{\sdelta t}{\hbar}
\sum_{\vec{x}_{1,2}}^{\mu_{1,2},s_{1,2}}\sum_{j_{1,2}=1}^{2N}
\hat{\mscr{V}}^{\boldsymbol{-1}}(\vec{x}_{2},\Teta{j_{2}};\vec{x}_{1},\Teta{j_{1}})\;\times \\ \notag &\times
\TRAB\bigg[\hat{\mathrm{S}}_{\mu_{2}}^{bb}\:\eta_{j_{2}}\;
\delta\wt{\Sigma}_{\mu_{2},s_{2};\mu_{1},s_{1}}^{ba}(\Teta{j_{2}},\vec{x}_{2};
\Teta{j_{1}},\vec{x}_{1})\;\hat{\mathrm{S}}_{\mu_{1}}^{aa}\:\eta_{j_{1}}\;
\delta\wt{\Sigma}_{\mu_{1},s_{1};\mu_{2},s_{2}}^{ab}(\Teta{j_{1}},\vec{x}_{1};
\Teta{j_{2}},\vec{x}_{2}) \bigg]\bigg\}  \\ \notag &\times\;
\mbox{DET}\bigg(\tfrac{1}{\mcal{N}_{x}}
\bigg[\hat{\mathrm{I}}^{bb}\;
\hat{\mathsf{H}}_{\nu,s\ppr;\mu,s}^{ba}\bigl(\vec{x}_{2},\Teta{j_{2}};
\vec{x}_{1},\Teta{j_{1}}\bigr)\;\hat{\mathrm{I}}^{aa}\;
+\;\eta_{j_{2}}\;\;\hat{\mscr{J}}_{\nu,s\ppr;\mu,s}^{ba}\bigl(\Teta{j_{2}},\vec{x}_{2};
\Teta{j_{1}},\vec{x}_{1}\bigr)\;\;\eta_{j_{1}}\;+   \\  \notag &+\;
q_{\nu}\:\etan{j_{2}}\:
\Bigl(\im\;\tfrac{\sdelta t}{\hbar}\Bigr)\:
\Big(\hat{\mfrak{s}}_{\nu,s\ppr;\mu,s}^{b=a}\!\bigl(\Teta{j_{2}},\vec{x}_{2};\Teta{j_{1}},\vec{x}_{1}\bigr)+
\hat{\mathrm{I}}^{bb}\:
\delta\wt{\Sigma}_{\nu,s\ppr;\mu,s}^{ba}\!\bigl(\Teta{j_{2}},\vec{x}_{2};\Teta{j_{1}},\vec{x}_{1}\bigr)
\:\hat{\mathrm{I}}^{aa}\Big)\:q_{\mu}\:\etan{j_{1}}\bigg]_{\nu,s\ppr;\mu,s}^{ba}\bigg)^{\boldsymbol{1/2}}\;.
\end{align}
\begin{align} \lb{s4_6b} &
\delta\wt{\Sigma}_{\nu,s\ppr;\mu,s}^{12}(\Teta{j_{2}},\vec{x}_{2};\Teta{j_{1}},\vec{x}_{1}) \rightarrow
\delta\wt{\Sigma}_{\nu,s\ppr;\mu,s}^{12}(\Teta{j_{2}},\vec{x}_{2};\Teta{j_{1}},\vec{x}_{1}) + \\ \notag & \hspace*{5.5cm}
+q_{\nu}\;\eta_{j_{2}}\;\Bigl(\im\tfrac{\hbar}{\Delta t}\Bigr)\;
\hat{\mathsf{H}}_{\nu,s\ppr;\mu,s}^{12}(\vec{x}_{2},\Teta{j_{2}};\vec{x}_{1},\Teta{j_{1}})\;
q_{\mu}\;\eta_{j_{1}} \;;  \\ \lb{s4_6c} &
\delta\wt{\Sigma}_{\nu,s\ppr;\mu,s}^{21}(\Teta{j_{2}},\vec{x}_{2};\Teta{j_{1}},\vec{x}_{1}) \rightarrow
\delta\wt{\Sigma}_{\nu,s\ppr;\mu,s}^{21}(\Teta{j_{2}},\vec{x}_{2};\Teta{j_{1}},\vec{x}_{1}) + \\ \notag & \hspace*{5.5cm}
+q_{\nu}\;\eta_{j_{2}}\;\Bigl(\im\tfrac{\hbar}{\Delta t}\Bigr)\;
\hat{\mathsf{H}}_{\nu,s\ppr;\mu,s}^{21}(\vec{x}_{2},\Teta{j_{2}};\vec{x}_{1},\Teta{j_{1}})\;
q_{\mu}\;\eta_{j_{1}} \;; \\  \lb{s4_6d} &
\hat{\mfrak{H}}_{\nu,s\ppr;\mu,s}^{b=a}(\vec{x}_{2},\Teta{j_{2}};\vec{x}_{1},\Teta{j_{1}})=
\hat{\mathsf{H}}_{\nu,s\ppr;\mu,s}^{b=a}(\vec{x}_{2},\Teta{j_{2}};\vec{x}_{1},\Teta{j_{1}}) \;; \\ \lb{s4_6e} &
\hat{\mcal{H}}_{\nu,s\ppr;\mu,s}^{b\neq a}(\vec{x}_{2},\Teta{j_{2}};\vec{x}_{1},\Teta{j_{1}})=
\hat{\mathsf{H}}_{\nu,s\ppr;\mu,s}^{b\neq a}(\vec{x}_{2},\Teta{j_{2}};\vec{x}_{1},\Teta{j_{1}})\;.
\end{align}
\end{subequations}
The coset decomposition follows from the factorization (\ref{s4_3}-\ref{s4_4c}) where one has to consider
the integration measure from the coset metric tensor and the Vandermonde determinants  of eigenvalues \cite{Mehta}
in order to replace the Euclidean integration variables
\(d[\delta\wt{\Sigma}_{\nu,s\ppr;\mu,s}^{ba}(\Teta{j_{2}},\vec{x}_{2};\Teta{j_{1}},\vec{x}_{1})]\) \cite{pop1,pop2,mie3}.
A particular simple change of integration measure arises from (\ref{s4_7}) as one diagonalizes the sub-generators
\(\hat{\mathrm{X}}_{\mu_{2},s_{2};\mu_{1},s_{1}}(\Teta{j_{2}},\vec{x}_{2};
\Teta{j_{1}},\vec{x}_{1})\), \(\hat{\mathrm{X}}_{\mu_{2},s_{2};\mu_{1},s_{1}}\pdag(\Teta{j_{2}},\vec{x}_{2};
\Teta{j_{1}},\vec{x}_{1})\) (\ref{s4_4c}) within the total generator
\(\hat{\mathrm{Y}}_{\mu_{2},s_{2};\mu_{1},s_{1}}^{ba}(\Teta{j_{2}},\vec{x}_{2};
\Teta{j_{1}},\vec{x}_{1})\) (\ref{s4_4b})
of coset matrices \(\hat{\mathrm{T}}_{\mu_{2},s_{2};\mu_{1},s_{1}}^{ba}(\Teta{j_{2}},\vec{x}_{2};
\Teta{j_{1}},\vec{x}_{1})\) (\ref{s4_4a}) with eigenvalue matrix
\(\hat{\mathrm{Y}}_{D;\nu,s;\mu,s}^{ba}(\Teta{j},\vec{x})\) (\ref{s4_8a}-\ref{s4_8c}) and eigenvectors
\(\hat{\mathrm{P}}_{\nu,s\ppr;\mu,s}^{aa}(\Teta{j_{2}},\vec{x}_{2};\Teta{j_{1}},\vec{x}_{1})\)
(\ref{s4_9a}-\ref{s4_9f}). We regard the anti-symmetry
of \(\hat{\mathrm{X}}_{\mu_{2},s_{2};\mu_{1},s_{1}}(\Teta{j_{2}},\vec{x}_{2};
\Teta{j_{1}},\vec{x}_{1})\) and of its hermitian conjugate
\(\hat{\mathrm{X}}_{\mu_{2},s_{2};\mu_{1},s_{1}}\pdag(\Teta{j_{2}},\vec{x}_{2};
\Teta{j_{1}},\vec{x}_{1})\) (\ref{s4_4c}) by quaternion
eigenvalues \(\ovv{f}_{\mathsf{S}}(\Teta{j},\vec{x})\) (\(\mathsf{S}=\{+1,-1\}\))
with Pauli matrix \((\hat{\tau}_{2})_{\nu\mu}\) and block diagonal 'eigenvectors'
\(\hat{\mathrm{P}}_{\nu,s\ppr;\mu,s}^{aa}(\Teta{j_{2}},\vec{x}_{2};\Teta{j_{1}},\vec{x}_{1})\) (\ref{s4_9a}-\ref{s4_9f}).
Furthermore, we use quaternion matrix elements
\(\hat{f}_{D;s\ppr s}^{(k)}(\Teta{j_{2}},\vec{x}_{2};\Teta{j_{1}},\vec{x}_{1})\) (\ref{s4_8c})
for the sub-generators \(\hat{\mathrm{X}}_{\mu_{2},s_{2};\mu_{1},s_{1}}(\Teta{j_{2}},\vec{x}_{2};
\Teta{j_{1}},\vec{x}_{1})\), \(\hat{\mathrm{X}}_{\mu_{2},s_{2};\mu_{1},s_{1}}\pdag(\Teta{j_{2}},\vec{x}_{2};
\Teta{j_{1}},\vec{x}_{1})\) (\ref{s4_4c}) of
\(\hat{\mathrm{Y}}_{\mu_{2},s_{2};\mu_{1},s_{1}}^{ba}(\Teta{j_{2}},\vec{x}_{2};
\Teta{j_{1}},\vec{x}_{1})\) (\ref{s4_4b}). The correct number of independent field degrees of freedom is maintained
by setting the quaternion-diagonal elements
\(\hat{\mscr{G}}_{D;\nu,s;\mu,s}(\Teta{j},\vec{x};\Teta{j},\vec{x})\) (\ref{s4_9e}) as part of
the total generator (\ref{s4_9d}) for the 'eigenvector matrices'
\(\hat{\mathrm{P}}_{\nu,s\ppr;\mu,s}^{aa}(\Teta{j_{2}},\vec{x}_{2};\Teta{j_{1}},\vec{x}_{1})\) (\(a=1,2\))
(\ref{s4_9c},\ref{s4_9f}) identical to zero; these \(2\times2\) quaternion degrees of freedom along the main diagonal
are already contained within the anti-symmetric, quaternion eigenvalues
\(\hat{\mathrm{X}}_{D;\nu,s;\mu,s}(\Teta{j},\vec{x})\) (\ref{s4_8b}) with
Pauli matrix \((\hat{\tau}_{2})_{\nu\mu}\)
\begin{align} \lb{s4_7}
\hat{\mathrm{Y}}_{\nu,s\ppr;\mu,s}^{ba}(\Teta{j_{2}},\vec{x}_{2};\Teta{j_{1}},\vec{x}_{1}) &=
\hat{\mathrm{P}}_{\nu,s\ppr;\mu_{2},s_{1}}^{\boldsymbol{-1};bb}(\Teta{j_{2}},\vec{x}_{2};\Teta{j},\vec{x})\;\;
\hat{\mathrm{Y}}_{D;\mu_{2},s_{1};\mu_{1},s_{1}}^{ba}(\Teta{j},\vec{x})\;\;
\hat{\mathrm{P}}_{\mu_{1},s_{1};\mu,s}^{aa}(\Teta{j},\vec{x};\Teta{j_{1}},\vec{x}_{1})\;;
\end{align}
\begin{subequations}
\begin{align}\lb{s4_8a}
\hat{\mathrm{Y}}_{D;\mu_{2},s;\mu_{1},s}^{ba}(\Teta{j},\vec{x}) &=
\biggl(\bea{cc} 0 & \hat{\mathrm{X}}_{D;\mu_{2},s;\mu_{1},s}(\Teta{j},\vec{x}) \\
\hat{\mathrm{X}}_{D;\mu_{2},s;\mu_{1},s}\pdag(\Teta{j},\vec{x}) & 0 \eea\biggr)^{ba} \;; \\ \lb{s4_8b}
\hat{\mathrm{X}}_{D;\mu_{2},s;\mu_{1},s}(\Teta{j},\vec{x}) &=(\hat{\tau}_{2})_{\mu_{2}\mu_{1}}\;\;
\ovv{f}_{\mathsf{S}}(\Teta{j},\vec{x}) \;;(\mathsf{S}=2s=\{+1,-1\}\;\mbox{\scz for }s=\uparrow,\downarrow);  \\ \notag
\hat{\mathrm{X}}_{\nu,s\ppr;\mu,s}(\Teta{j_{2}},\vec{x}_{2};\Teta{j_{1}},\vec{x}_{1}) &=
\sum_{k=0}^{3} (\hat{\tau}_{k})_{\nu\mu}\;\;
\hat{f}_{D;s\ppr s}^{(k)}(\Teta{j_{2}},\vec{x}_{2};\Teta{j_{1}},\vec{x}_{1})\;;\;\;\;
(\hat{f}_{D;ss}^{(k=0,1,3)}(\Teta{j},\vec{x};\Teta{j},\vec{x})\equiv0)\;;  \\  \lb{s4_8c}
\hat{f}_{D;ss}^{(2)}(\Teta{j},\vec{x};\Teta{j},\vec{x}) &=
\ovv{f}_{\mathsf{S}}(\Teta{j},\vec{x}) = \bigl|\ovv{f}_{\mathsf{S}}(\Teta{j},\vec{x})\bigr|\;\;
\exp\bigl\{\im\,\phi_{\mathsf{S}}(\Teta{j},\vec{x})\bigr\}\in\mathbb{C}\;;
\end{align}
\end{subequations}
\begin{subequations}
\begin{align}\lb{s4_9a}
\hat{\mathrm{P}}_{\nu,s\ppr;\mu,s}^{aa}(\Teta{j_{2}},\vec{x}_{2};\Teta{j_{1}},\vec{x}_{1}) &=
\biggl(\bea{cc} \hat{\mathrm{P}}_{\nu,s\ppr;\mu,s}^{11}(\Teta{j_{2}},\vec{x}_{2};\Teta{j_{1}},\vec{x}_{1}) & 0 \\
0 & \hat{\mathrm{P}}_{\nu,s\ppr;\mu,s}^{22}(\Teta{j_{2}},\vec{x}_{2};\Teta{j_{1}},\vec{x}_{1}) \eea\biggr)_{\mbox{;}}
\\  \lb{s4_9b} \hat{\mathrm{P}}_{\nu,s\ppr;\mu,s}^{22}(\Teta{j_{2}},\vec{x}_{2};\Teta{j_{1}},\vec{x}_{1}) &=
\Bigl(\hat{\mathrm{P}}_{\nu,s\ppr;\mu,s}^{11;\boldsymbol{-1}}(\Teta{j_{2}},\vec{x}_{2};\Teta{j_{1}},\vec{x}_{1})
\Bigr)^{\boldsymbol{T}}\;;  \\  \lb{s4_9c}
\hat{\mathrm{P}}_{\nu,s\ppr;\mu,s}^{11}(\Teta{j_{2}},\vec{x}_{2};\Teta{j_{1}},\vec{x}_{1}) &=
\biggl[\exp\Bigl\{\im\:\hat{\mscr{G}}_{D;\mu_{4},s_{4};\mu_{3},s_{3}}\bigl(\Teta{j_{4}},\vec{x}_{4};
\Teta{j_{3}},\vec{x}_{3}\bigr)\Bigr\}\biggr]_{\nu,s\ppr;\mu,s}\hspace*{-0.9cm}(\Teta{j_{2}},\vec{x}_{2};
\Teta{j_{1}},\vec{x}_{1})\;\;\;; \\  \lb{s4_9d}
\hat{\mscr{G}}_{D;\nu,s\ppr;\mu,s}(\Teta{j_{2}},\vec{x}_{2};\Teta{j_{1}},\vec{x}_{1}) &=
\Bigl(\hat{\mscr{G}}_{D;\nu,s\ppr;\mu,s}(\Teta{j_{2}},\vec{x}_{2};\Teta{j_{1}},\vec{x}_{1})\Bigr)\pdag \;;
\\  \lb{s4_9e} \hat{\mscr{G}}_{D;\nu,s;\mu,s}(\Teta{j},\vec{x};\Teta{j},\vec{x}) &\equiv 0\;\;;\;\;\;
(\mu,\nu=\mbox{\scz (e)lectron, (h)ole})\;;  \\ \lb{s4_9f}
\hat{\mathrm{P}}_{\nu,s\ppr;\mu,s}^{22}(\Teta{j_{2}},\vec{x}_{2};\Teta{j_{1}},\vec{x}_{1}) &=
\biggl[\exp\Bigl\{-\im\:\hat{\mscr{G}}_{D;\mu_{4},s_{4};\mu_{3},s_{3}}^{\boldsymbol{T}}
\bigl(\Teta{j_{4}},\vec{x}_{4};\Teta{j_{3}},\vec{x}_{3}\bigr)\Bigr\}\biggr]_{\nu,s\ppr;\mu,s}
\hspace*{-0.9cm}(\Teta{j_{2}},\vec{x}_{2};\Teta{j_{1}},\vec{x}_{1})\;\;.
\end{align}
\end{subequations}
The chosen kind of eigenvalues (\ref{s4_7}-\ref{s4_8b}) and parametrization with quaternion matrices
\((\hat{\tau}_{k})_{\nu\mu}\) (\ref{s4_8c}) takes into account the strong Coulomb attraction
between electrons and holes resulting into exciton quasi-particles of neutral charge. According
from a rigorous, mathematical point of view, one could also prefer eigenvalues
\(\hat{\mathrm{X}}_{D;\mu,s_{2};\mu,s_{1}}(\Teta{j},\vec{x})=(\hat{\tau}_{2})_{s_{2}s_{1}}\:
\ovv{f}_{\mu}(\Teta{j},\vec{x})\)
with separate eigenvalue parameters \(\ovv{f}_{\mu}(\Teta{j},\vec{x})\) for the respective
electron- '$\mu=\mbox{e}$' and hole-band '$\mu=\mbox{h}$' which are composed of opposite spin pairs due to
\((\hat{\tau}_{2})_{s_{2}s_{1}}\) for separate degrees of freedom of
"electron$\uparrow$-electron$\downarrow$" and "hole$\uparrow$-hole$\downarrow$" pairs.
This latter kind of parametrization of eigenvalues is more appropriate for superconductors with Cooper-pairs
of equal kind of fermions caused by an attractive interaction, e.g.\ by phonons. However, corresponding
to the strong Coulomb interaction in the presented case defined within section \ref{s2}, the excitons are
certainly the prevailing, appropriate kind of quasi-particle formation for the eigenvalues
\(\hat{\mathrm{X}}_{D;\mu_{2},s;\mu_{1},s}(\Teta{j},\vec{x})=
(\hat{\tau}_{2})_{\mu_{2}\mu_{1}}\:\ovv{f}_{\mathsf{S}}(\Teta{j},\vec{x})\) with parallel spins of electron-hole
pairs leading to the two possible, total spin values
\(\mathsf{S}=2\, s=\{+1,-1\}\) for \(s=\{\uparrow=+\tfrac{1}{2}\,,\,\downarrow=-\tfrac{1}{2}\}\).
One might be astonished about the missing, total spin value \(\mathsf{S}=0\) of electron-hole pairs,
having opposite spin directions, within our chosen parametrization of eigenvalues; however,
the direct generation of electron-hole pairs with opposite spins is negligible in comparison to the case with
parallel spin directions according to the massless, spin-one property of the electromagnetic field with the only
two possible spin components \(\mathsf{S}=\{+1,-1\}\), but with lacking spin zero component (cf.\ (\ref{s2_3c})).

Further factorization of the block diagonal self-energy densities
\(\delta\hat{\Sigma}_{D;\nu,s\ppr;\mu,s}^{aa}(\Teta{j_{2}},\vec{x}_{2};\Teta{j_{1}},\vec{x}_{1})\) (\ref{s4_2a},\ref{s4_2b})
(as 'hinge' fields for a SSB) is specified in relations (\ref{s4_10a}-\ref{s4_10f}) with the anomalous-doubled eigenvalues
\(\delta\hat{\Lambda}_{\mu,s}^{a}(\Teta{j},\vec{x})=
\{+\delta\hat{\lambda}_{\mu,s}(\Teta{j},\vec{x})\,;\,-\delta\hat{\lambda}_{\mu,s}(\Teta{j},\vec{x})\}\) (\ref{s4_10b})
and 'rotation' matrices \(\hat{\mathrm{Q}}_{\nu,s\ppr;\mu,s}^{11}(\Teta{j_{2}},\vec{x}_{2};\Teta{j_{1}},\vec{x}_{1})\)
(\ref{s4_10c}), \(\hat{\mathrm{Q}}_{\nu,s\ppr;\mu,s}^{22}(\Teta{j_{2}},\vec{x}_{2};\Teta{j_{1}},\vec{x}_{1})\)
(\ref{s4_10d}) so that we regard the hermiticity of
\(\delta\hat{\Sigma}_{D;\nu,s\ppr;\mu,s}^{aa}(\Teta{j_{2}},\vec{x}_{2};\Teta{j_{1}},\vec{x}_{1})\) (\ref{s4_2a})
and their relation (\ref{s4_2b}) of transposition. The generator
\(\hat{\mscr{F}}_{D;\nu,s\ppr;\mu,s}(\Teta{j_{2}},\vec{x}_{2};\Teta{j_{1}},\vec{x}_{1})\) (\ref{s4_10c},\ref{s4_10d})
of \(\hat{\mathrm{Q}}_{\nu,s\ppr;\mu,s}^{aa}(\Teta{j_{2}},\vec{x}_{2};\Teta{j_{1}},\vec{x}_{1})\)
consists of a hermitian matrix (\ref{s4_10e}), but
with vanishing diagonal elements (\ref{s4_10f}) in order to achieve the proper number of independent fields
as in \(\delta\hat{\Sigma}_{D;\nu,s\ppr;\mu,s}^{aa}(\Teta{j_{2}},\vec{x}_{2};\Teta{j_{1}},\vec{x}_{1})\)
\begin{subequations}
\begin{align} \lb{s4_10a}
\delta\hat{\Sigma}_{D;\nu,s\ppr;\mu,s}^{aa}(\Teta{j_{2}},\vec{x}_{2};\Teta{j_{1}},\vec{x}_{1})
&=   \\ \notag =\hat{\mathrm{Q}}_{\nu,s\ppr;\mu_{3},s_{3}}^{aa;\boldsymbol{-1}}(\Teta{j_{2}},\vec{x}_{2};\Teta{j_{3}},\vec{x}_{3})
&\ph{\times}\hspace*{-0.3cm}
\biggl(\bea{cc} \delta\hat{\lambda}_{\mu_{3},s_{3}}(\Teta{j_{3}},\vec{x}_{3}) &  \\ &
-\delta\hat{\lambda}_{\mu_{3},s_{3}}(\Teta{j_{3}},\vec{x}_{3}) \eea\biggr)^{aa}
\hat{\mathrm{Q}}_{\mu_{3},s_{3};\mu,s}^{aa}(\Teta{j_{3}},\vec{x}_{3};\Teta{j_{1}},\vec{x}_{1}) \;;  \\  \lb{s4_10b}
\delta\hat{\Lambda}_{\mu,s}^{a}(\Teta{j},\vec{x}) &=
\Bigl\{\underbrace{+\delta\hat{\lambda}_{\mu,s}(\Teta{j},\vec{x})}_{a=1}\,;\,
\underbrace{-\delta\hat{\lambda}_{\mu,s}(\Teta{j},\vec{x})}_{a=2}\Bigr\}  \;;  \\ \lb{s4_10c}
\hat{\mathrm{Q}}_{\nu,s\ppr;\mu,s}^{11}(\Teta{j_{2}},\vec{x}_{2};\Teta{j_{1}},\vec{x}_{1})  &=
\biggl[\exp\Bigl\{\im\:\hat{\mscr{F}}_{D;\mu_{4},s_{4};\mu_{3},s_{3}}\bigl(\Teta{j_{4}},\vec{x}_{4};
\Teta{j_{3}},\vec{x}_{3}\bigr)\Bigr\}\biggr]_{\nu,s\ppr;\mu,s}\hspace*{-0.9cm}(\Teta{j_{2}},\vec{x}_{2};
\Teta{j_{1}},\vec{x}_{1})   \;;   \\  \lb{s4_10d}
\hat{\mathrm{Q}}_{\nu,s\ppr;\mu,s}^{22}(\Teta{j_{2}},\vec{x}_{2};\Teta{j_{1}},\vec{x}_{1})  &=
\Bigl(\hat{\mathrm{Q}}_{\nu,s\ppr;\mu,s}^{11;\boldsymbol{-1}}(\Teta{j_{2}},\vec{x}_{2};\Teta{j_{1}},\vec{x}_{1})
\Bigr)^{\boldsymbol{T}} = \\ \notag & =
\biggl[\exp\Bigl\{-\im\:\hat{\mscr{F}}_{D;\mu_{4},s_{4};\mu_{3},s_{3}}^{\boldsymbol{T}}\bigl(\Teta{j_{4}},\vec{x}_{4};
\Teta{j_{3}},\vec{x}_{3}\bigr)\Bigr\}\biggr]_{\nu,s\ppr;\mu,s}\hspace*{-0.9cm}(\Teta{j_{2}},\vec{x}_{2};
\Teta{j_{1}},\vec{x}_{1})   \;;  \\  \lb{s4_10e}
\hat{\mscr{F}}_{D;\nu,s\ppr;\mu,s}\bigl(\Teta{j_{2}},\vec{x}_{2};\Teta{j_{1}},\vec{x}_{1}\bigr)  &=
\Bigl(\hat{\mscr{F}}_{D;\nu,s\ppr;\mu,s}\bigl(\Teta{j_{2}},\vec{x}_{2};\Teta{j_{1}},\vec{x}_{1}\bigr)\Bigr)\pdag  \;;   \\  \lb{s4_10f}
\hat{\mscr{F}}_{D;\mu;s;\mu,s}(\Teta{j},\vec{x};\Teta{j},\vec{x}) &\equiv 0 \;\;; \;\;
(\mu=\mbox{\scz e,h};\,s=\uparrow,\downarrow)\;.
\end{align}
\end{subequations}
In terms of the above parameters (\ref{s4_7}-\ref{s4_9f}) and(\ref{s4_10a}-\ref{s4_10f}),
it is straightforward to derive the Jacobi determinant for the change
of integration measure from Euclidean integration variables
\(d[\delta\wt{\Sigma}_{\nu,s\ppr;\mu,s}^{ba}(\Teta{j_{2}},\vec{x}_{2};\Teta{j_{1}},\vec{x}_{1})]\) to the
coset part \(d[(\hat{\mathrm{T}}^{-1}\:d\hat{\mathrm{T}})_{\nu,s\ppr;\mu,s}^{ba}(\Teta{j_{2}},\vec{x}_{2};
\Teta{j_{1}},\vec{x}_{1})]\) under inclusion of the block diagonal self-energy densities
\(d[\delta\hat{\Sigma}_{D;\nu,s\ppr;\mu,s}^{aa}(\Teta{j_{2}},\vec{x}_{2};\Teta{j_{1}},\vec{x}_{1})]\) and
Vandermonde eigenvalue polynomial \(\mbox{Poly}(\delta\hat{\lambda}_{\mu,s}(\Teta{j},\vec{x})\,)\).
According to appendix A of Ref. \cite{pop1}, we can state the transformation of integration variables
\begin{subequations}
\begin{align} \lb{s4_11a}
d\bigl[\delta\wt{\Sigma}_{\nu,s\ppr;\mu,s}^{ba}(\Teta{j_{2}},\vec{x}_{2};\Teta{j_{1}},\vec{x}_{1})\bigr] &\stackrel{\wedge}{=}
\biggl(\prod_{s,s\ppr=\uparrow,\downarrow}^{\mu,\nu=e,h}
\prod_{j_{1,2}=1,\ldots,2N}^{\{\vec{x}_{1,2}\}}
d\bigl[\delta\wt{\Sigma}_{\nu,s\ppr;\mu,s}^{ba}(\Teta{j_{2}},\vec{x}_{2};\Teta{j_{1}},\vec{x}_{1})\bigr]\biggr)= \\ \no =
d\bigl[\delta\hat{\Sigma}_{D;\nu,s\ppr;\mu,s}^{ba}(\Teta{j_{2}},\vec{x}_{2};\Teta{j_{1}},\vec{x}_{1})] &\times
\mbox{Poly}\bigl(\delta\hat{\lambda}_{\mu,s}(\Teta{j},\vec{x})\,\bigr) \;\times\;
d\bigl[(\hat{\mathrm{T}}^{-1}\:d\hat{\mathrm{T}})_{\nu,s\ppr;\mu,s}^{ba}(\Teta{j_{2}},\vec{x}_{2};
\Teta{j_{1}},\vec{x}_{1})\bigr] \;;    \\  \lb{s4_11b}
\mbox{Poly}\bigl(\delta\hat{\lambda}_{\mu,s}(\Teta{j},\vec{x})\bigr) &=
\prod_{s=\uparrow,\downarrow}\prod_{j=1,\ldots,2N}^{\{\vec{x}\}}
\Bigl(\delta\lambda_{\mbox{\scz e},s}(\Teta{j},\vec{x})+
\delta\lambda_{\mbox{\scz h},s}(\Teta{j},\vec{x})\Bigr)^{\boldsymbol{2}}\times
\\ \no &\hspace*{-2.8cm}\times \Bigg(
\prod_{s,s\ppr=\uparrow,\downarrow}\prod_{j_{1,2}=1,\ldots,2N}^{\{\vec{x}_{1,2}\}}
\bigl(\mbox{except : }(s\ppr=s)\And(j_{1}=j_{2})\And(\vec{x}_{1}=\vec{x}_{2})\bigr)\;\;\times
\\ \no &\hspace*{-2.8cm}\times
\Bigl(\delta\lambda_{\mbox{\scz e},s}(\Teta{j_{1}},\vec{x}_{1})+
\delta\lambda_{\mbox{\scz e},s\ppr}(\Teta{j_{2}},\vec{x}_{2})\Bigr)^{\boldsymbol{2}}\;\times\;
\Bigl(\delta\lambda_{\mbox{\scz h},s}(\Teta{j_{1}},\vec{x}_{1})+
\delta\lambda_{\mbox{\scz h},s\ppr}(\Teta{j_{2}},\vec{x}_{2})\Bigr)^{\boldsymbol{2}}\;\;\times
\\   \no &\hspace*{-2.8cm}\times
\Bigl(\delta\lambda_{\mbox{\scz h},s}(\Teta{j_{1}},\vec{x}_{1})+
\delta\lambda_{\mbox{\scz e},s\ppr}(\Teta{j_{2}},\vec{x}_{2})\Bigr)^{\boldsymbol{2}}\;\times \;
\Bigl(\delta\lambda_{\mbox{\scz e},s}(\Teta{j_{1}},\vec{x}_{1})+
\delta\lambda_{\mbox{\scz h},s\ppr}(\Teta{j_{2}},\vec{x}_{2})
\Bigr)^{\boldsymbol{2}}\;\;\Bigg)^{\boldsymbol{1/2}}_{\mbox{.}}
\end{align}
\end{subequations}
\begin{align} \lb{s4_12}
d\bigl[(\hat{\mathrm{T}}^{-1}\:d\hat{\mathrm{T}})_{\nu,s\ppr;\mu,s}^{ba}(\Teta{j_{2}},\vec{x}_{2};
\Teta{j_{1}},\vec{x}_{1})\bigr] &= \\ \notag &\hspace*{-4.6cm}= \prod_{s=\uparrow,\downarrow}^{\mathsf{S}=2s}
\prod_{j=1,\ldots,2N}^{\{\vec{x}\}}
\biggl(\frac{d\hat{f}_{D;ss}^{(2)*}(\Teta{j},\vec{x};\Teta{j},\vec{x})\wedge
d\hat{f}_{D;ss}^{(2)}(\Teta{j},\vec{x};\Teta{j},\vec{x})}{2\;\im}\times
\frac{\sinh\bigl(2\:\bigl|\ovv{f}_{\mathsf{S}}(\Teta{j},\vec{x})\bigr|\bigr)}{\bigl|\ovv{f}_{\mathsf{S}}(\Teta{j},\vec{x})\bigr|}
\biggr) \times \\ \no &\hspace*{-4.6cm}\times \prod_{k=0}^{3}
\Bigg(\prod_{s,s\ppr=\uparrow,\downarrow}^{\mathsf{S}=2s,\,\mathsf{S}\ppr=2s\ppr}\prod_{j_{1,2}=1,\ldots,2N}^{\{\vec{x}_{1,2}\}}
\bigl(\mbox{except : }(s\ppr=s,\,\mathsf{S}\ppr=\mathsf{S})\And(j_{1}=j_{2})\And(\vec{x}_{1}=\vec{x}_{2})\bigr)\;\;\times
\\ \no &\hspace*{-4.6cm}\times
\frac{d\hat{f}_{D;s\ppr s}^{(k)*}(\Teta{j_{2}},\vec{x}_{2};\Teta{j_{1}},\vec{x}_{1})\wedge
d\hat{f}_{D;s\ppr s}^{(k)}(\Teta{j_{2}},\vec{x}_{2};\Teta{j_{1}},\vec{x}_{1})}{2\;\im}\times \\ \no &\hspace*{-4.6cm}\times\,4\,
\bigg|\frac{\sinh\bigl(\bigl|\ovv{f}_{\mathsf{S}\ppr}(\Teta{j_{2}},\vec{x}_{2})\bigr|+
\bigl|\ovv{f}_{\mathsf{S}}(\Teta{j_{1}},\vec{x}_{1})\bigr|\bigr)}{\bigl|\ovv{f}_{\mathsf{S}\ppr}(\Teta{j_{2}},\vec{x}_{2})\bigr|+
\bigl|\ovv{f}_{\mathsf{S}}(\Teta{j_{1}},\vec{x}_{1})\bigr|}\bigg|\times
\bigg|\frac{\sinh\bigl(\bigl|\ovv{f}_{\mathsf{S}\ppr}(\Teta{j_{2}},\vec{x}_{2})\bigr|-
\bigl|\ovv{f}_{\mathsf{S}}(\Teta{j_{1}},\vec{x}_{1})\bigr|\bigr)}{\bigl|\ovv{f}_{\mathsf{S}\ppr}(\Teta{j_{2}},\vec{x}_{2})\bigr|-
\bigl|\ovv{f}_{\mathsf{S}}(\Teta{j_{1}},\vec{x}_{1})\bigr|}\bigg|\Bigg)^{\boldsymbol{1/2}}_{\mbox{.}}
\end{align}
As we combine the change of integration measure (\ref{s4_11a}-\ref{s4_12}) with the coset decomposition
(\ref{s4_2a}-\ref{s4_4c}) and its eigenvalue parametrization (\ref{s4_7}-\ref{s4_9f}) for the total,
anomalous-doubled self-energy
\(\delta\wt{\Sigma}_{\nu,s\ppr;\mu,s}^{ba}(\Teta{j_{2}},\vec{x}_{2};\Teta{j_{1}},\vec{x}_{1})\) and also perform
the projection onto the coset field degrees of freedom, we finally obtain the path integral (\ref{s4_13})
replacing (\ref{s4_6a}) with Euclidean self-energy variables
\begin{align} \lb{s4_13} &
\ovv{Z[\hat{\mscr{J}}]} =
\int d[\hat{\mfrak{s}}_{\nu,s\ppr;\mu,s}^{11}(\Teta{j_{2}},\vec{x}_{2};\Teta{j_{1}},\vec{x}_{1})]\;\times
\exp\bigg\{\hspace*{-0.1cm}-\tfrac{\im}{4}\tfrac{\sdelta t}{\hbar}
\sum_{\vec{x}_{1,2}}^{\mu_{1,2},s_{1,2}}\sum_{j_{1,2}=1}^{2N}
\hat{\mscr{V}}^{\boldsymbol{-1}}(\vec{x}_{2},\Teta{j_{2}};\vec{x}_{1},\Teta{j_{1}})\;\times \\ \notag &\times
\bigg[\eta_{j_{2}}\;q_{\mu_{2}}\;
\hat{\mfrak{s}}_{\mu_{2},s_{2};\mu_{1},s_{1}}^{aa}(\Teta{j_{2}},\vec{x}_{2};\Teta{j_{1}},\vec{x}_{1})\;
\eta_{j_{1}}\;q_{\mu_{1}}\;
\hat{\mfrak{s}}_{\mu_{1},s_{1};\mu_{2},s_{2}}^{aa}(\Teta{j_{1}},\vec{x}_{1};\Teta{j_{2}},\vec{x}_{2})
\bigg]\bigg\} \times \\  \notag &\int
d\bigl[\bigl(\hat{\mathrm{T}}^{\boldsymbol{-1}}\:
d\hat{\mathrm{T}}\bigr)_{\nu,s\ppr;\mu,s}^{ba}(\Teta{j_{2}},\vec{x}_{2};\Teta{j_{1}},\vec{x}_{1})\bigr]\;\;\;\;
\mfrak{Z}\!\Bigl[\hat{\mathrm{T}}^{\boldsymbol{-1}};
\hat{\mathrm{T}};\hat{\mcal{H}}\Bigr]\;\times   \\ \notag &\times\;
\mbox{DET}\bigg(\tfrac{1}{\mcal{N}_{x}}
\bigg[\breve{\mscr{H}}\bigl[\hat{\mfrak{H}},\hat{\mfrak{s}}\bigr] +
\sdeltaT\breve{\mscr{H}}\bigl[\hat{\mfrak{H}},\hat{\mfrak{s}};\hat{\mathrm{T}}^{-1},\hat{\mathrm{T}}\bigr]+
\wt{\mscr{J}}\bigl[\hat{\mathrm{T}}^{-1},\hat{\mathrm{T}}\bigr]\bigg]_{\nu,s\ppr;\mu,s}^{ba}
\hspace*{-0.9cm}(\Teta{j_{2}},\vec{x}_{2};\Teta{j_{1}},\vec{x}_{1})\bigg)^{\boldsymbol{1/2}}\;.
\end{align}
Apart from the Gaussian factor with the density-related self-energy
\(\hat{\mfrak{s}}_{\nu,s\ppr;\mu,s}^{11}(\Teta{j_{2}},\vec{x}_{2};\Teta{j_{1}},\vec{x}_{1})\), the path
integral (\ref{s4_13}) comprises the functional \(\mfrak{Z}\![\hat{\mathrm{T}}^{\boldsymbol{-1}};
\hat{\mathrm{T}};\hat{\mcal{H}}]\) (\ref{s4_15}) with the driving interaction
\(\hat{\mcal{H}}_{\nu,s\ppr;\mu,s}^{b\neq a}(\vec{x}_{2},\Teta{j_{2}};\vec{x}_{1},\Teta{j_{1}})\)
(\ref{s2_14c},\ref{s2_14d},\ref{s4_6e}), creating exciton parts, and the square root of the determinant;
this determinant has a density-related
part (\ref{s4_14a}), which is completely block diagonal in 'Nambu' indices '\(b=a\)', and a gradient term
(\ref{s4_14b}) with the coset matrices, which is weighted relatively to the density part (\ref{s4_14a}).
Furthermore, the source term (\ref{s4_14c}) allows to track the original observables, given as combinations of
even-numbered anti-commuting fields, to the presented case (\ref{s4_13}) with even-valued self-energy matrices
\begin{subequations}
\begin{align}\lb{s4_14a}
\breve{\mscr{H}}\bigl[\hat{\mfrak{H}},\hat{\mfrak{s}}\bigr]_{\nu,s\ppr;\mu,s}^{b=a}
(\Teta{j_{2}},\vec{x}_{2};\Teta{j_{1}},\vec{x}_{1}) &=
\hat{\mfrak{H}}_{\nu,s\ppr;\mu,s}^{b=a}(\vec{x}_{2},\Teta{j_{2}};\vec{x}_{1},\Teta{j_{1}}) + \\ \notag &+
\hat{\mathrm{S}}^{b=a}\:q_{\nu}\:\eta_{j_{2}}\:\bigl(\im\tfrac{\sdelta t}{\hbar}\bigr)\;
\hat{\mfrak{s}}_{\nu,s\ppr;\mu,s}^{b=a}(\Teta{j_{2}},\vec{x}_{2};\Teta{j_{1}},\vec{x}_{1})\;q_{\mu}\:\eta_{j_{1}}\;;
\\  \notag  \breve{\mscr{H}}\bigl[\hat{\mfrak{H}},\hat{\mfrak{s}}\bigr]_{\nu,s\ppr;\mu,s}^{22}
(\Teta{j_{2}},\vec{x}_{2};\Teta{j_{1}},\vec{x}_{1}) &=\bigg(
\breve{\mscr{H}}\bigl[\hat{\mfrak{H}},\hat{\mfrak{s}}\bigr]_{\nu,s\ppr;\mu,s}^{11}
(\Teta{j_{2}},\vec{x}_{2};\Teta{j_{1}},\vec{x}_{1}) \bigg)^{\boldsymbol{T}} \;;
\\  \lb{s4_14b} \sdeltaT\breve{\mscr{H}}\bigl[\hat{\mfrak{H}},\hat{\mfrak{s}};
\hat{\mathrm{T}}^{-1},\hat{\mathrm{T}}\bigr]_{\nu,s\ppr;\mu,s}^{ba}(\Teta{j_{2}},\vec{x}_{2};\Teta{j_{1}},\vec{x}_{1})
&=\biggl[\hat{\mathrm{T}}^{\boldsymbol{-1}}\:\breve{\mscr{H}}\bigl[\hat{\mfrak{H}},\hat{\mfrak{s}}\bigr]\:
\hat{\mathrm{T}}-
\breve{\mscr{H}}\bigl[\hat{\mfrak{H}},
\hat{\mfrak{s}}\bigr]\biggr]_{\nu,s\ppr;\mu,s}^{ba}\hspace*{-0.9cm}(\Teta{j_{2}},\vec{x}_{2};
\Teta{j_{1}},\vec{x}_{1}) =    \\ \notag
\lefteqn{\hspace*{-6.4cm}=\hat{\mathrm{T}}_{\nu,s\ppr;\mu_{4},s_{4}}^{\boldsymbol{-1};bb\ppr}(\Teta{j_{2}},\vec{x}_{2};
\Teta{j_{4}},\vec{x}_{4})\;\;\breve{\mscr{H}}\Bigl[\hat{\mfrak{H}},\hat{\mfrak{s}}\Bigr]_{\mu_{4},s_{4};
\mu_{3},s_{3}}^{b\ppr= a\ppr}\hspace*{-1.2cm}(\Teta{j_{4}},\vec{x}_{4};\Teta{j_{3}},\vec{x}_{3})\;\;
\hat{\mathrm{T}}_{\mu_{3},s_{3};\mu,s}^{a\ppr a}(\Teta{j_{3}},\vec{x}_{3};\Teta{j_{1}},\vec{x}_{1}) +}
\\ \notag \lefteqn{\hspace*{-6.4cm}-\breve{\mscr{H}}\Bigl[\hat{\mfrak{H}},\hat{\mfrak{s}}\Bigr]_{\nu,s\ppr;
\mu,s}^{b=a}\hspace*{-0.6cm}(\Teta{j_{2}},\vec{x}_{2};\Teta{j_{1}},\vec{x}_{1})\;;}  \\  \lb{s4_14c}\lefteqn{\hspace*{-7.6cm}
\wt{\mscr{J}}\Bigl[\hat{\mathrm{T}}^{-1},\hat{\mathrm{T}}\Bigr]_{\nu,s\ppr;\mu,s}^{ba}
\hspace*{-0.9cm}(\Teta{j_{2}},\vec{x}_{2};\Teta{j_{1}},\vec{x}_{1}) =
\hat{\mathrm{T}}_{\nu,s\ppr;\mu_{4},s_{4}}^{\boldsymbol{-1};bb\ppr}(\Teta{j_{2}},\vec{x}_{2};
\Teta{j_{4}},\vec{x}_{4})\;\times} \\  \notag \lefteqn{\hspace*{-4.6cm}\times\;
\eta_{j_{4}}\;\hat{\mathrm{I}}^{\boldsymbol{-1};b\ppr b\ppr}\;\tfrac{1}{\mcal{N}_{x}}\,
\hat{\mscr{J}}_{\mu_{4},s_{4};\mu_{3},s_{3}}^{b\ppr a\ppr}(\Teta{j_{4}},\vec{x}_{4};\Teta{j_{3}},
\vec{x}_{3})\;\hat{\mathrm{I}}^{\boldsymbol{-1};a\ppr a\ppr}\;\eta_{j_{3}}\;
\hat{\mathrm{T}}_{\mu_{3},s_{3};\mu,s}^{a\ppr a}(\Teta{j_{3}},\vec{x}_{3};\Teta{j_{1}},\vec{x}_{1})\;;}
\end{align}
\end{subequations}
\begin{align}\lb{s4_15}
\mfrak{Z}\!\Bigl[\hat{\mathrm{T}}^{\boldsymbol{-1}};
\hat{\mathrm{T}};\hat{\mcal{H}}\Bigr] &= \int d\bigl[\delta\hat{\Sigma}_{D;\nu,s\ppr;\mu,s}^{11}(\Teta{j_{2}},\vec{x}_{2};
\Teta{j_{1}},\vec{x}_{1})\bigr]\;\;
\mbox{Poly}\bigl(\delta\hat{\lambda}_{\mu,s}(\Teta{j},\vec{x})\,\bigr)\;\;\times\;
\exp\biggl\{\tfrac{\im}{4}\,\tfrac{\sdelta t}{\hbar}\sum_{j_{1,2}=1}^{2N} \times \\ \notag &\hspace*{-2.3cm}\times\;
\hat{\mscr{V}}^{-1}(\vec{x}_{2},\Teta{j_{2}};\vec{x}_{1},\Teta{j_{1}})\times
\TRAB\bigg[\hat{\mathrm{S}}_{\mu_{2}}^{bb}\:\etan{j_{2}}\;
\hat{\mathrm{T}}_{\mu_{2},s_{2};\mu_{4},s_{4}}^{bb\ppr}(\Teta{j_{2}},\vec{x}_{2};\Teta{j_{4}},
\vec{x}_{4})\;\delta\hat{\Sigma}_{D;\mu_{4},s_{4};\mu_{3},s_{3}}^{b\ppr=a\ppr}(\Teta{j_{4}},
\vec{x}_{4};\Teta{j_{3}},\vec{x}_{3}) \;\times \\ \notag &\times\;
\hat{\mathrm{T}}_{\mu_{3},s_{3};\mu_{1},s_{1}}^{\boldsymbol{-1};a\ppr a}(\Teta{j_{3}},\vec{x}_{3};
\Teta{j_{1}},\vec{x}_{1})\;\hat{\mathrm{S}}_{\mu_{1}}^{aa}\;\etan{j_{1}}\;
\hat{\mathrm{T}}_{\mu_{1},s_{1};\mu_{6},s_{6}}^{aa\pppr}
(\Teta{j_{1}},\vec{x}_{1};\Teta{j_{6}},
\vec{x}_{6})\;\times \\ \notag &\times\;
\delta\hat{\Sigma}_{D;\mu_{6},s_{6};\mu_{5},s_{5}}^{a\pppr=b\pppr}
(\Teta{j_{6}},\vec{x}_{6};\Teta{j_{5}},\vec{x}_{5}) \;
\hat{\mathrm{T}}_{\mu_{5},s_{5};\mu_{2},s_{2}}^{\boldsymbol{-1};b\pppr b}(\Teta{j_{5}},\vec{x}_{5};
\Teta{j_{2}},\vec{x}_{2})\biggr]\biggr\} \;\;\times\;
\exp\biggl\{-\tfrac{1}{2}\sum_{j_{1,2}=1}^{2N} \times \\ \notag &\hspace*{-2.3cm}\times\;
\hat{\mscr{V}}^{-1}(\vec{x}_{2},\Teta{j_{2}};\vec{x}_{1},\Teta{j_{1}})\times
\TRAB\bigg[\hat{\mathrm{S}}^{bb}\;
\hat{\mathrm{T}}_{\mu_{2},s_{2};\mu_{4},s_{4}}^{bb\ppr}(\Teta{j_{2}},\vec{x}_{2};\Teta{j_{4}},
\vec{x}_{4})\;\delta\hat{\Sigma}_{D;\mu_{4},s_{4};\mu_{3},s_{3}}^{b\ppr=a\ppr}(\Teta{j_{4}},
\vec{x}_{4};\Teta{j_{3}},\vec{x}_{3}) \;\times \\ \notag &\times\;
\hat{\mathrm{T}}_{\mu_{3},s_{3};\mu_{1},s_{1}}^{\boldsymbol{-1};a\ppr a}(\Teta{j_{3}},\vec{x}_{3};
\Teta{j_{1}},\vec{x}_{1})\;\hat{\mathrm{S}}^{aa}\;
\hat{\mcal{H}}_{\mu_{1},s_{1};\mu_{2},s_{2}}^{a\neq b}(\vec{x}_{1},\Teta{j_{1}};\vec{x}_{2},\Teta{j_{2}})
\biggr]\biggr\}\;.
\end{align}

\subsection{Saddle point approximation with various kinds of gradient and $1/\mcal{N}_{x}$ expansions} \lb{s42}

The factorization into coset and density-related field degrees of freedom with corresponding factorization of
the integration measure allows to separate the total path integral (\ref{s4_13});
corresponding to this coset decomposition, the anomalous-doubled
fields of the generator (\ref{s4_7}-\ref{s4_9f}) for the coset matrices are regarded to be 'immersed' in
density-related self-energies of a background functional (\ref{s4_17}). This 'background averaging' (\ref{s4_17}) of the
path integral (\ref{s4_16}) for the coset degrees of freedom is denoted by
\(\langle(\ldots)\rangle_{\boldsymbol{\hat{\mfrak{s}}^{aa}}}\)
\begin{align}\lb{s4_16}
\ovv{Z[\hat{\mscr{J}}]} &=\bigg\langle \int
d\bigl[\bigl(\hat{\mathrm{T}}^{\boldsymbol{-1}}\:
d\hat{\mathrm{T}}\bigr)_{\nu,s\ppr;\mu,s}^{ba}(\Teta{j_{2}},\vec{x}_{2};\Teta{j_{1}},\vec{x}_{1})\bigr]\;\;\;\;
\mfrak{Z}\!\Bigl[\hat{\mathrm{T}}^{\boldsymbol{-1}};
\hat{\mathrm{T}};\hat{\mcal{H}}\Bigr]\;\times   \\ \notag &\times
\mbox{DET}\bigg(
\bigg[\boldsymbol{\hat{1}}+
\sdeltaT\breve{\mscr{H}}\bigl[\hat{\mfrak{H}},\hat{\mfrak{s}};\hat{\mathrm{T}}^{-1},\hat{\mathrm{T}}\bigr]\;
\Bigl(\breve{\mscr{H}}\bigl[\hat{\mfrak{H}},\hat{\mfrak{s}}\bigr]\Bigr)^{\boldsymbol{-1}}+
\wt{\mscr{J}}\bigl[\hat{\mathrm{T}}^{-1},\hat{\mathrm{T}}\bigr]\;
\Bigl(\breve{\mscr{H}}\bigl[\hat{\mfrak{H}},\hat{\mfrak{s}}\bigr]\Bigr)^{\boldsymbol{-1}}\;
\bigg]_{\nu,s\ppr;\mu,s}^{ba}
\hspace*{-0.9cm}(\Teta{j_{2}},\vec{x}_{2};\Teta{j_{1}},\vec{x}_{1})\bigg)^{\boldsymbol{1/2}}
\bigg\rangle_{\!\!\boldsymbol{\hat{\mfrak{s}}^{aa}}} \hspace*{-0.4cm}=  \\  \notag &=
\bigg\langle \int
d\bigl[\bigl(\hat{\mathrm{T}}^{\boldsymbol{-1}}\:
d\hat{\mathrm{T}}\bigr)_{\nu,s\ppr;\mu,s}^{ba}(\Teta{j_{2}},\vec{x}_{2};\Teta{j_{1}},\vec{x}_{1})\bigr]\;\;\;\;
\mfrak{Z}\!\Bigl[\hat{\mathrm{T}}^{\boldsymbol{-1}};
\hat{\mathrm{T}};\hat{\mcal{H}}\Bigr]\;\times
\exp\biggl\{\frac{1}{2}\,\mcal{N}_{x}\sum_{\vec{x}}\sum_{j=0}^{2N+1}\sum_{\mu=e,h}^{s=\uparrow,\downarrow}
\;\times \\ \notag &\times\;
\TRAB\bigg[\ln\bigg( \boldsymbol{\hat{1}}+
\sdeltaT\breve{\mscr{H}}\bigl[\hat{\mfrak{H}},\hat{\mfrak{s}};\hat{\mathrm{T}}^{-1},\hat{\mathrm{T}}\bigr]\;
\Bigl(\breve{\mscr{H}}\bigl[\hat{\mfrak{H}},\hat{\mfrak{s}}\bigr]\Bigr)^{\boldsymbol{-1}}+
\wt{\mscr{J}}\bigl[\hat{\mathrm{T}}^{-1},\hat{\mathrm{T}}\bigr]\;
\Bigl(\breve{\mscr{H}}\bigl[\hat{\mfrak{H}},\hat{\mfrak{s}}\bigr]\Bigr)^{\boldsymbol{-1}}
\bigg)\bigg]_{\mu,s;\mu,s}^{b=a}
\hspace*{-0.9cm}(\Teta{j},\vec{x};\Teta{j},\vec{x})\biggr\}
\bigg\rangle_{\!\!\boldsymbol{\hat{\mfrak{s}}^{aa}}} \;;
\end{align}
\begin{align}\lb{s4_17}
&\bigg\langle \Bigl(\mbox{path integral of coset field degrees of freedom}\Bigr) \bigg\rangle_{\boldsymbol{\hat{\mfrak{s}}^{aa}}} =
\int d\bigl[\hat{\mfrak{s}}_{\nu,s\ppr;\mu,s}^{11}(\Teta{j_{2}},\vec{x}_{2};\Teta{j_{1}},\vec{x}_{1})\bigr]\;\;\times \\ \notag &\times
\;\mbox{DET}\Bigl(\tfrac{1}{\mcal{N}_{x}}\breve{\mscr{H}}\bigl[\hat{\mfrak{H}},\hat{\mfrak{s}}\bigr]^{aa}\Bigr)^{\boldsymbol{1/2}}\;\times
\exp\biggl\{-\tfrac{\im}{4}\,\tfrac{\sdelta t}{\hbar}\sum_{\vec{x}_{1,2}}^{a=1,2}
\sum_{\mu_{1,2}=e,h}^{s_{1,2}=\uparrow,\downarrow}\sum_{j_{1,2}=1}^{2N}
\hat{\mscr{V}}^{-1}(\vec{x}_{2},\Teta{j_{2}};\vec{x}_{1},\Teta{j_{1}})\;\times \\ \no &\times\;
\Big[\etan{j_{2}}\;q_{\mu_{2}}\;\hat{\mfrak{s}}_{\mu_{2},s_{2};\mu_{1},s_{1}}^{aa}(\Teta{j_{2}},\vec{x}_{2};
\Teta{j_{1}},\vec{x}_{1})\;\etan{j_{1}}\;q_{\mu_{1}}\;
\hat{\mfrak{s}}_{\mu_{1},s_{1};\mu_{2},s_{2}}^{aa}(\Teta{j_{1}},\vec{x}_{1};\Teta{j_{2}},\vec{x}_{2})
\Big]\biggr\}\;\times
\\ \notag &\times \;\Bigl(\mbox{path integral of coset field degrees of freedom}\Bigr) =  \\ \notag & =
\int \!\!d\bigl[\hat{\mfrak{s}}_{\nu,s\ppr;\mu,s}^{11}(\Teta{j_{2}},\vec{x}_{2};\Teta{j_{1}},\vec{x}_{1})\bigr]
\;\mbox{Det}\Bigl(\tfrac{1}{\mcal{N}_{x}}\breve{\mscr{H}}\bigl[\hat{\mfrak{H}},\hat{\mfrak{s}}\bigr]^{11}\Bigr)\;
\exp\biggl\{\!-\tfrac{\im}{2}\,\tfrac{\sdelta t}{\hbar}\sum_{\vec{x}_{1,2}}
\sum_{\mu_{1,2}=e,h}^{s_{1,2}=\uparrow,\downarrow}\sum_{j_{1,2}=1}^{2N}\!\!
\hat{\mscr{V}}^{-1}(\vec{x}_{2},\Teta{j_{2}};\vec{x}_{1},\Teta{j_{1}})\,\times \\ \no &\times\;
\Big[\etan{j_{2}}\;q_{\mu_{2}}\;\hat{\mfrak{s}}_{\mu_{2},s_{2};\mu_{1},s_{1}}^{11}(\Teta{j_{2}},\vec{x}_{2};
\Teta{j_{1}},\vec{x}_{1})\;\etan{j_{1}}\;q_{\mu_{1}}\;
\hat{\mfrak{s}}_{\mu_{1},s_{1};\mu_{2},s_{2}}^{11}(\Teta{j_{1}},\vec{x}_{1};\Teta{j_{2}},\vec{x}_{2})
\Big]\biggr\}\times
\\ \notag &\times \;\Bigl(\mbox{path integral of coset field degrees of freedom}\Bigr) \;.
\end{align}
Moreover, it is possible to derive from the background functional (\ref{s4_17}) of the self-energy densities
a saddle point equation (\ref{s4_18a}) ; in consequence one can approximate the path integral (\ref{s4_16}) of the
anomalous-doubled field variables by replacing the density-related field variables
\(\hat{\mfrak{s}}_{\nu,s\ppr;\mu,s}^{aa}(\Teta{j_{2}},\vec{x}_{2};\Teta{j_{1}},\vec{x}_{1})\) with definite,
fixed functions \(\langle\hat{\mfrak{s}}\rangle_{\nu,s\ppr;\mu,s}^{aa}(\Teta{j_{2}},\vec{x}_{2};\Teta{j_{1}},\vec{x}_{1})\).
The latter saddle point solution results from a similar eigenvalue problem (\ref{s4_18b}) as in section \ref{s32}, but
without the off-diagonal term of the driving interaction which has been shifted into the functional
\(\mfrak{Z}\![\hat{\mathrm{T}}^{-1};\hat{\mathrm{T}};\hat{\mcal{H}}]\) (\ref{s4_15}), so that the computation of the saddle
point equation can be reduced to the '11' block of densities
\begin{subequations}
\begin{align} \lb{s4_18a}&
\hat{\mscr{V}}^{-1}(\vec{x}_{2},\Teta{j_{2}};\vec{x}_{1},\Teta{j_{1}})\;
\langle\hat{\mfrak{s}}\rangle_{\mu_{2},s_{2};\mu_{1},s_{1}}^{11}(\Teta{j_{2}},\vec{x}_{2};\Teta{j_{1}},\vec{x}_{1})\Big/\mcal{N}_{x}=
\mcal{N}_{x}\Bigl[\hat{\mfrak{H}}_{\mu_{4},s_{4};\mu_{3},s_{3}}^{11}(\vec{x}_{4},\Teta{j_{4}};\vec{x}_{3},\Teta{j_{3}})+ \\ \notag &+
\eta_{j_{4}}\;q_{\mu_{4}}\;\bigl(\im\tfrac{\Delta t}{\hbar}\bigr)\;
\langle\hat{\mfrak{s}}\rangle_{\mu_{4},s_{4};\mu_{3},s_{3}}^{11}(\Teta{j_{4}},\vec{x}_{4};\Teta{j_{3}},\vec{x}_{3})\;
\eta_{j_{3}}\;q_{\mu_{3}}\Bigr]_{\mu_{2},s_{2};\mu_{1},s_{1}}^{\boldsymbol{-1};11}
\hspace*{-0.9cm}(\Teta{j_{2}},\vec{x}_{2};\Teta{j_{1}},\vec{x}_{1})\;; \\ \lb{s4_18b}&
\sum_{j_{1}=0}^{2N+1}\sum_{\vec{x}_{1}}^{\mu,s}\Big[
\hat{\mfrak{H}}_{\nu,s\ppr;\mu,s}^{11}(\vec{x}_{2},\Teta{j_{2}};\vec{x}_{1},\Teta{j_{1}})+ \\ \notag &+
\eta_{j_{2}}\;q_{\nu}\;\bigl(\im\tfrac{\Delta t}{\hbar}\bigr)\;
\langle\hat{\mfrak{s}}\rangle_{\nu,s\ppr;\mu,s}^{11}(\Teta{j_{2}},\vec{x}_{2};\Teta{j_{1}},\vec{x}_{1})\;
\eta_{j_{1}}\;q_{\mu}\Big]\psi_{M;\mu,s}(\Teta{j_{1}},\vec{x}_{1})=E_{M}\;\psi_{M;\nu,s\ppr}(\Teta{j_{2}},\vec{x}_{2})\;; \\ \notag &
\quad(\mbox{eigenstate label $M$ with complex eigenvalue }E_{M})\;;   \\  \lb{s4_18c} &
\hat{\mscr{V}}^{\boldsymbol{-1}}(\vec{x}_{2},\Teta{j_{2}};\vec{x}_{1},\Teta{j_{1}})\;
\langle\hat{\mfrak{s}}\rangle_{\mu_{2},s_{2};\mu_{1},s_{1}}^{11}(\Teta{j_{2}},\vec{x}_{2};\Teta{j_{1}},\vec{x}_{1})\big/\mcal{N}_{x} =
\hspace*{-0.3cm}\sum_{M\in\mbox{\scz eigenstates}}\hspace*{-0.66cm}
\frac{\psi_{M;\mu_{2},s_{2}}(\Teta{j_{2}},\vec{x}_{2})\;
\psi_{M;\mu_{1},s_{1}}^{\boldsymbol{T}}(\Teta{j_{1}},\vec{x}_{1})}{E_{M}}_{\mbox{.}}
\end{align}
\end{subequations}
The saddle point approximation of (\ref{s4_16}) with (\ref{s4_18a}-\ref{s4_18c}) even leads to a further simplification
as one considers the relation between the block diagonal propagator
\((\breve{\mscr{H}}[\hat{\mfrak{H}},\langle\hat{\mfrak{s}}\rangle])^{\boldsymbol{-1};aa}\)
and the self-energy density \(\langle\hat{\mfrak{s}}\rangle^{aa}\) as consequence of (\ref{s4_18a}-\ref{s4_18c})
\begin{align}\lb{s4_19}
\mcal{N}_{x}\biggl(\breve{\mscr{H}}\bigl[\hat{\mfrak{H}},\langle\hat{\mfrak{s}}\rangle\bigr]\biggr)_{\nu,s\ppr;\mu,s}^{\boldsymbol{-1};b=a}
\hspace*{-0.9cm}(\Teta{j_{2}},\vec{x}_{2};\Teta{j_{1}},\vec{x}_{1}) &=
\hat{\mscr{V}}^{-1}(\vec{x}_{2},\Teta{j_{2}};\vec{x}_{1},\Teta{j_{1}})\;\;\hat{\mathrm{S}}^{b=a}\;\;
\langle\hat{\mfrak{s}}\rangle_{\nu,s\ppr;\mu,s}^{b=a}(\Teta{j_{2}},\vec{x}_{2};\Teta{j_{1}},\vec{x}_{1})\Big/\mcal{N}_{x}\;.
\end{align}
After insertion of the saddle point solution \(\langle\hat{\mfrak{s}}\rangle\) from (\ref{s4_18a}-\ref{s4_18c}) with
inclusion of (\ref{s4_19}), we can remove the block diagonal propagator
\((\breve{\mscr{H}}[\hat{\mfrak{H}},\langle\hat{\mfrak{s}}\rangle])^{\boldsymbol{-1};aa}\) of purely density-related terms and
obtain the path integral (\ref{s4_20a},\ref{s4_20b}) with the gradient term (\ref{s4_20c},\ref{s4_20d})
which simplifies the gradient expansion of the
trace-log functional. In this manner one has succeeded into a path integral of exciton related coset matrices
\begin{subequations}
\begin{align}\lb{s4_20a}
\ovv{Z[\hat{\mscr{J}}]} &\approx \int
d\bigl[\bigl(\hat{\mathrm{T}}^{\boldsymbol{-1}}\:
d\hat{\mathrm{T}}\bigr)_{\nu,s\ppr;\mu,s}^{ba}(\Teta{j_{2}},\vec{x}_{2};\Teta{j_{1}},\vec{x}_{1})\bigr]\;\;\;\;
\mfrak{Z}\!\Bigl[\hat{\mathrm{T}}^{\boldsymbol{-1}};
\hat{\mathrm{T}};\hat{\mcal{H}}\Bigr]\;\times
\exp\biggl\{\frac{1}{2}\,\mcal{N}_{x}\sum_{\vec{x}}\sum_{j=0}^{2N+1}\sum_{\mu=e,h}^{s=\uparrow,\downarrow}
\;\times \\ \notag &\times\;
\TRAB\bigg[\ln\bigg(\boldsymbol{\hat{1}}+
\sdeltaT\breve{\mscr{H}}\bigl[\hat{\mfrak{H}},\langle\hat{\mfrak{s}}\rangle;\hat{\mathrm{T}}^{-1},\hat{\mathrm{T}}\bigr]\;
\Bigl(\breve{\mscr{H}}\bigl[\hat{\mfrak{H}},\langle\hat{\mfrak{s}}\rangle\bigr]\Bigr)^{\boldsymbol{-1}}+
\wt{\mscr{J}}\bigl[\hat{\mathrm{T}}^{-1},\hat{\mathrm{T}}\bigr]\;
\Bigl(\breve{\mscr{H}}\bigl[\hat{\mfrak{H}},\langle\hat{\mfrak{s}}\rangle\bigr]\Bigr)^{\boldsymbol{-1}}
\bigg)\bigg]_{\mu,s;\mu,s}^{b=a}\hspace*{-0.9cm}(\Teta{j},\vec{x};\Teta{j},\vec{x})\biggr\}\;;  \\ \notag
  \ovv{Z[\hat{\mscr{J}}]} &\approx
\int d\bigl[\bigl(\hat{\mathrm{T}}^{\boldsymbol{-1}}\:
d\hat{\mathrm{T}}\bigr)_{\nu,s\ppr;\mu,s}^{ba}(\Teta{j_{2}},\vec{x}_{2};\Teta{j_{1}},\vec{x}_{1})\bigr]\;\;\;\;
\mfrak{Z}\!\Bigl[\hat{\mathrm{T}}^{\boldsymbol{-1}};
\hat{\mathrm{T}};\hat{\mcal{H}}\Bigr]\;\times
\exp\biggl\{\frac{1}{2}\,\mcal{N}_{x}\sum_{\vec{x}}\sum_{j=0}^{2N+1}\sum_{\mu=e,h}^{s=\uparrow,\downarrow}
\;\TRAB\bigg[\ln\bigg(\boldsymbol{\hat{1}}+  \\ \lb{s4_20b} &\hspace*{-0.9cm}+
\sdeltaT\breve{\mscr{H}}\bigl[\hat{\mfrak{H}},\langle\hat{\mfrak{s}}\rangle;\hat{\mathrm{T}}^{-1},\hat{\mathrm{T}}\bigr]\;
\Bigl(\hat{\mscr{V}}^{-1}\;\hat{\mathrm{S}}^{aa}\;\langle\hat{\mfrak{s}}\rangle^{aa}/\mcal{N}_{x}^{\,2}\Bigr)+
\wt{\mscr{J}}\bigl[\hat{\mathrm{T}}^{-1},\hat{\mathrm{T}}\bigr]\;
\Bigl(\hat{\mscr{V}}^{-1}\;\hat{\mathrm{S}}^{aa}\;\langle\hat{\mfrak{s}}\rangle^{aa}/\mcal{N}_{x}^{\,2}\Bigr)
\bigg)\bigg]_{\mu,s;\mu,s}^{b=a}\hspace*{-0.9cm}(\Teta{j},\vec{x};\Teta{j},\vec{x})\biggr\}  \;;
\end{align}
\begin{align}\lb{s4_20c}
\sdeltaT\breve{\mscr{H}}
\bigl[\hat{\mfrak{H}},\langle\hat{\mfrak{s}}\rangle;\hat{\mathrm{T}}^{-1},\hat{\mathrm{T}}\bigr]\;
\Bigl(\breve{\mscr{H}}\bigl[\hat{\mfrak{H}},\langle\hat{\mfrak{s}}\rangle\bigr]\Bigr)^{\boldsymbol{-1}}
&= \biggl[\hat{\mathrm{T}}^{-1}\;\;\breve{\mscr{H}}\bigl[\hat{\mfrak{H}},\langle\hat{\mfrak{s}}\rangle\bigr]\;\;\hat{\mathrm{T}}\;
\Bigl(\breve{\mscr{H}}\bigl[\hat{\mfrak{H}},\langle\hat{\mfrak{s}}\rangle\bigr]\Bigr)^{\boldsymbol{-1}}
-\boldsymbol{\hat{1}}\biggr]    \\  \notag  &=
\biggl[\biggl(\exp\Bigl\{\overrightarrow{\bigl[\hat{\mathrm{Y}}\boldsymbol{,}\,\ldots\bigr]_{-}}\Bigr\}\,
\breve{\mscr{H}}\bigl[\hat{\mfrak{H}},\langle\hat{\mfrak{s}}\rangle\bigr]\biggr)\;
\Bigl(\breve{\mscr{H}}\bigl[\hat{\mfrak{H}},\langle\hat{\mfrak{s}}\rangle\bigr]\Bigr)^{\boldsymbol{-1}}
-\boldsymbol{\hat{1}}\biggr] \;; \\ \lb{s4_20d} \sdeltaT\breve{\mscr{H}}
\bigl[\hat{\mfrak{H}},\langle\hat{\mfrak{s}}\rangle;\hat{\mathrm{T}}^{-1},\hat{\mathrm{T}}\bigr]\;
\Bigl(\breve{\mscr{H}}\bigl[\hat{\mfrak{H}},\langle\hat{\mfrak{s}}\rangle\bigr]\Bigr)^{\boldsymbol{-1}} &\approx
\biggl[\biggl(\exp\Bigl\{\overrightarrow{\bigl[\hat{\mathrm{Y}}\boldsymbol{,}\,\ldots\bigr]_{-}}\Bigr\}\,
\breve{\mscr{H}}\bigl[\hat{\mfrak{H}},\langle\hat{\mfrak{s}}\rangle\bigr]\biggr)\;
\Bigl(\hat{\mscr{V}}^{-1}\;\hat{\mathrm{S}}^{aa}\;\langle\hat{\mfrak{s}}\rangle^{aa}/\mcal{N}_{x}^{\,2}\Bigr)
-\boldsymbol{\hat{1}}\biggr]\;.
\end{align}
\end{subequations}
There are various possibilities to expand the determinants in (\ref{s4_20a},\ref{s4_20b}). One possibility refers to the parameter
\(\mcal{N}_{x}=(L/\sdelta x)^{d}\) of total space points which allows to consider the self-energy
\(\langle\hat{\mfrak{s}}\rangle^{aa}_{\nu,s\ppr;\mu,s}(\Teta{j_{2}},\vec{x}_{2};\Teta{j_{1}},\vec{x}_{1})\) as a background
density provided that its spatial variation is negligible compared to that of the coset matrices. A further possibility follows from
performing the total trace for the determinants in (\ref{s4_20a},\ref{s4_20b}) in a momentum basis of plane waves with a gradient
expansion of spatial derivative or momentum {\it operators} \cite{BCSQCD}; in this manner one can derive effective Lagrangians
of coset matrices with a minimum number of derivatives and effective coupling coefficients consisting of the saddle point solution.
One of the guiding principles for such a gradient expansion has to include 'Derrick's theorem' \cite{raja} which restricts the
expansion to stable configurations under a minimum number of derivatives of coset matrices. Moreover, one can introduce a
generalized gradient operator in the two following kinds
\begin{align} \lb{s4_21}
\hat{\pp}_{\breve{\mscr{H}}}\bigl(\mbox{matrix}\bigr)^{ba} &:=
\biggl[\breve{\mscr{H}}\bigl[\hat{\mfrak{H}},\langle\hat{\mfrak{s}}\rangle\bigr]^{bb}\;\bigl(\mbox{matrix}\bigr)^{ba}\;
\breve{\mscr{H}}\bigl[\hat{\mfrak{H}},\langle\hat{\mfrak{s}}\rangle\bigr]^{\boldsymbol{-1};aa}-
\bigl(\mbox{matrix}\bigr)^{ba}\biggr]\;;  \\  \lb{s4_22}
\hat{\pp}_{\ln(-\breve{\mscr{H}})}\bigl(\mbox{matrix}\bigr)^{ba} &:=
\biggl[\bigg(\exp\Big\{\overrightarrow{\big[\ln(-\breve{\mscr{H}}[\hat{\mfrak{H}},\langle\hat{\mfrak{s}}\rangle])\;{\Large\boldsymbol{,}}
\ldots\big]_{-}}\Big\}-\hat{1}\bigg)\bigl(\mbox{matrix}\bigr)^{b\ppr a\ppr}\biggr]^{ba}\;,
\end{align}
where the exponentiation \(-\breve{\mscr{H}}[\hat{\mfrak{H}},\langle\hat{\mfrak{s}}\rangle]=
\exp\{\ln(-\breve{\mscr{H}}[\hat{\mfrak{H}},\langle\hat{\mfrak{s}}\rangle])\}\) allows for a further possible expansion in terms of
\([\ln(-\breve{\mscr{H}}[\hat{\mfrak{H}},\langle\hat{\mfrak{s}}\rangle])\;{\large\boldsymbol{,}}
\ldots]_{-}\). Since the one-particle operators \(\hat{\mfrak{H}}_{\nu,s\ppr;\mu,s}^{aa}(\vec{x}_{2},\Teta{j_{2}};\vec{x}_{1},\Teta{j_{1}})\)
contain the total unit operator (in fact '\(-\hat{1}\)', cf.\ (\ref{s2_14a},\ref{s2_14b}), therefore
\(\hat{\pp}_{\ln(\boldsymbol{-}\breve{\mscr{H}})}\) instead of \(\hat{\pp}_{\ln(\breve{\mscr{H}})}\) in (\ref{s4_22})), one also achieves
reasonable approximations and simplifications from the expansion of the logarithm by using
\(\ln(1-x)=-x-\tfrac{1}{2}x^{2}-\tfrac{1}{3}x^{3}-\ldots\)
\begin{align} \notag
\ln\bigl(-\breve{\mscr{H}}[\hat{\mfrak{H}},\langle\hat{\mfrak{s}}\rangle] \bigr)&=
\ln\bigg(-\hat{\mfrak{H}}_{\nu,s\ppr;\mu,s}^{11}(\vec{x}_{2},\Teta{j_{2}};\vec{x}_{1},\Teta{j_{1}})-
q_{\nu}\:\eta_{j_{2}}\:\bigl(\im\,\tfrac{\sdelta t}{\hbar}\bigr)\:
\langle\hat{\mfrak{s}}\rangle_{\nu,s\ppr;\mu,s}^{11}(\Teta{j_{2}},\vec{x}_{2};\Teta{j_{1}},\vec{x}_{1})\;q_{\mu}\:\eta_{j_{1}}\bigg)=
\\ \lb{s4_23} &\hspace*{-2.8cm}= \ln\bigg[\delta_{\nu\mu}\,\delta_{s\ppr s}\,\deltaN_{\vec{x}_{2},\vec{x}_{1}}\,\delta_{j_{2},j_{1}}-\bigg(
\delta_{\nu\mu}\,\delta_{s\ppr s}\,\deltaN_{\vec{x}_{2},\vec{x}_{1}}\,\delta_{j_{2}+1,j_{1}}+
\im\,\tfrac{\sdelta t}{\hbar}\Big[\delta_{j_{2},j_{1}}\,\eta_{j_{2}}\,\delta_{\nu\mu}
\big(\hat{\mscr{E}}_{\mu,s\ppr s}(\vec{x}_{2})-\im\,\eta_{j_{2}}\,\delta_{s\ppr s}\,\ve_{+}\big)\deltaN_{\vec{x}_{2},\vec{x}_{1}}+
\\ \notag &\hspace*{4.6cm}+ q_{\nu}\:\eta_{j_{2}}\:
\langle\hat{\mfrak{s}}\rangle_{\nu,s\ppr;\mu,s}^{11}(\Teta{j_{2}},\vec{x}_{2};\Teta{j_{1}},\vec{x}_{1})\;q_{\mu}\:\eta_{j_{1}}\Big]\bigg)
\bigg]\;.
\end{align}
However, the appropriate expansion or approximation of the determinant in (\ref{s4_20a},\ref{s4_20b}) relies on the prevailing phenomenon
which may refer to universal fluctuations of a 'Random Matrix Theory', an effective expansion with second and fourth order gradients
for Skyrme-like Lagrangians or extraction of relevant terms for a renormalizable theory. Therefore, 'ultimately' valid expressions
from expansions of the determinant are hardly accessible or universally applicable so that the prevailing approximation of terms has to be
adapted to the corresponding physical phenomenon which has originally been defined as a coherent state path integral (\ref{s2_9a}-\ref{s2_10c})
\cite{Negele}, a Dyson equation or further diagrammatic expressions \cite{Martin}-\cite{Sadovs}.

\section{Transformation to Euclidean path integration variables} \lb{s5}

\subsection{Transformation with the square root of the coset metric tensor} \lb{s51}

In the following we abbreviate the contour time variable \(\Teta{j_{i}}\) and the spatial vector \(\vec{x}_{i}\) by \(\Tetax{i}\)
and also combine the total set of variables with electron-hole and spin indices into respective numbers \(\nn{1},\,\nn{2},\,\ldots\)
for brevity
\begin{subequations}
\begin{align} \lb{s5_1a}
\Teta{j_{i}}\,,\,\vec{x}_{i} &\stackrel{\wedge}{=}\;\Tetax{i}\;; \\ \lb{s5_1b}
\nn{1}\;;\;\nn{2}\;;\;\ldots &\stackrel{\wedge}{=}\;'(\Teta{j_{1}}\,,\,\vec{x}_{1}\,,\,\mu_{1}\,,\,s_{1})'\;;
\;'(\Teta{j_{2}}\,,\,\vec{x}_{2}\,,\,\mu_{2}\,,\,s_{2})'\;;\;\ldots\;.
\end{align}
\end{subequations}
Section \ref{s12} contains a description of a nontrivial, multi-dimensional integral with the square root of a metric tensor as the Jacobi matrix.
In order to convey these transformations to coset spaces as \(\mbox{SO}(\mfrak{N},\mfrak{N})\,/\,\mbox{U}(\mfrak{N})\), we identify the
matrix combination \(\hat{\mathrm{T}}^{-1}(\sdelta\hat{\mathrm{T}})\) as the analogous square root of the metric tensor as in section \ref{s12}.
The derivative symbol '\(\sdelta\)'  can denote a partial derivative of space '\(\pp_{\vec{x}}=\tfrac{\pp}{\pp \vec{x}}\)' or time '\(\pp/\Teta{j}\)', 
a total derivative '\(d\)' with respect to time slices for the contour time ordering of path integration variables or 
a variation symbol '\(\delta\)' for deriving classical equations in the
exponentials on the condition that the integration measure has been transformed to Euclidean variables as in this section.
The derivative symbol '\(\sdelta\)' can even refer to the matrix operations (\ref{s4_21}-\ref{s4_23}) for generalized derivatives
\((\hat{\pp}_{\breve{\mscr{H}}}\hat{\mathrm{T}})\), \((\hat{\pp}_{\ln(-\breve{\mscr{H}})}\hat{\mathrm{T}})\)
of the coset matrix (\ref{s4_4a}-\ref{s4_4c}). We achieve the transformation to
Euclidean variables by using the property that '\(\hat{\mathrm{T}}^{-1}(\sdelta\hat{\mathrm{T}})\)' can be expressed as elements of the
\(\mbox{so}(\mfrak{N},\mfrak{N})\) Lie algebra. This can be illustrated by consideration of a Lie group \(\hat{U}(\varphi^{i})\) of matrices with
Lie algebra generators \(\hat{t}_{i}\) and parameters \(\varphi^{i}\). As we take the variation \(\sdelta\hat{U}(\varphi^{i})=
\hat{U}(\varphi^{i}+\sdelta\varphi^{i})-\hat{U}(\varphi^{i})\) of \(\hat{U}(\varphi^{i})\), we can apply the closed Lie algebra of
generators \(\hat{t}_{i}\) in order to relate \(\hat{U}(\varphi^{i}+\sdelta\varphi^{i})\) to a rotation of  \(\hat{U}(\varphi^{i})\) by another
Lie group element \(\hat{U}(\sdelta f^{j}(\varphi^{i},\sdelta\varphi^{i})\,)\) with function increments
\(\sdelta f^{j}(\varphi^{i},\sdelta\varphi^{i})\,)\) to be determined. Nevertheless, we can specify \(\hat{U}^{-1}(\varphi^{i})\;
(\sdelta\hat{U}(\varphi^{i})\,)\) in terms of the generators \(\hat{t}_{j}\)
\begin{subequations}
\begin{align}\lb{s5_2a}
\hat{U}(\varphi^{i}) &= \exp\bigl\{\im\;\varphi^{i}\;\hat{t}_{i}\bigr\} \;; \\ \lb{s5_2b}
\hat{U}(\varphi^{i}+\sdelta\varphi^{i}) &=\exp\bigl\{\im\;(\varphi^{i}+\sdelta\varphi^{i})\;\hat{t}_{i}\bigr\}=
\hat{U}(\varphi^{i}) \;\hat{U}\bigl(\sdelta f^{j}(\varphi^{i},\sdelta\varphi^{i})\bigr) =
\exp\bigl\{\im\;\varphi^{i}\;\hat{t}_{i}\bigr\}\;\;\exp\bigl\{\im\;\sdelta f^{j}(\varphi^{i},\sdelta\varphi^{i})\;\hat{t}_{j}\bigr\}\;;
\\ \lb{s5_2c} \hat{U}^{-1}(\varphi^{i})\,(\sdelta\hat{U}(\varphi^{i})\,) &=
\hat{U}^{-1}(\varphi)\;\bigl(\hat{U}(\varphi^{i}+\sdelta\varphi^{i})-\hat{U}(\varphi)\bigr) =
\im\;\sdelta f^{j}(\varphi^{i},\sdelta\varphi^{i})\;\hat{t}_{j}=\im\;\sdelta\varphi^{k}\;\frac{\pp f^{j}(\varphi^{i})}{\pp\varphi^{k}}\;
\hat{t}_{j}\;.
\end{align}
\end{subequations}
Application of (\ref{s5_2a}-\ref{s5_2c}) to \(\hat{\mathrm{T}}^{-1}\,(\sdelta\hat{\mathrm{T}})\) combined with an additional gauge
transformation \(\hat{\mathrm{P}}\ldots\hat{\mathrm{P}}^{-1}\) therefore gives rise to the \(\mbox{so}(\mfrak{N},\mfrak{N})\) Lie algebra
elements \(\sdelta\hat{\mscr{Z}}^{ba}(\nn{5};\nn{1})\) (\ref{s5_3a})
with the block diagonal, density-related parts \(\sdelta\hat{\mscr{Y}}^{11}(\nn{5};\nn{1})\),
\(\sdelta\hat{\mscr{Y}}^{22}(\nn{5};\nn{1})\) (\ref{s5_3d},\ref{s5_3e}) as dependent variables
and off-diagonal, anomalous-doubled parts \(\sdelta\hat{\mscr{X}}(\nn{5};\nn{1})\),
\(\sdelta\hat{\mscr{X}}\pdag(\nn{5};\nn{1})\) (\ref{s5_3b},\ref{s5_3c}) as the independent variables
which we also define by quaternion-valued matrix elements \(\sdelta\hat{a}_{s_{5}s_{1}}^{(k)}(\Tetax{5};\Tetax{1})\).
The block diagonal density parts consist of anti-hermitian sub-matrices \(\im\;\sdelta\hat{g}(\nn{5};\nn{1})\) which are functions of the
coset variables \(\sdelta\hat{\mscr{X}}(\nn{5};\nn{1})\),
\(\sdelta\hat{\mscr{X}}\pdag(\nn{5};\nn{1})\) or their quaternion-valued correspondents
\(\sdelta\hat{a}_{s_{5}s_{1}}^{(k)}(\Tetax{5};\Tetax{1})\), \(\sdelta\hat{a}_{s_{5}s_{1}}^{(k)*}(\Tetax{5};\Tetax{1})\)
\begin{subequations}
\begin{align}\lb{s5_3a}
\sdelta\hat{\mscr{Z}}^{ba}(\nn{5};\nn{1}) &=
-\Bigl(\hat{\mathrm{P}}(\nn{5};\nn{4})\;\hat{\mathrm{T}}^{-1}(\nn{4};\nn{3})\;\bigl(\sdelta\hat{\mathrm{T}}(\nn{3};\nn{2})\bigr)\;
\hat{\mathrm{P}}^{-1}(\nn{2};\nn{1})\Bigr)^{ba} =
\left(\bea{cc} \sdelta\hat{\mscr{Y}}^{11}(\nn{5};\nn{1}) & \sdelta\hat{\mscr{X}}(\nn{5};\nn{1}) \\
\sdelta\hat{\mscr{X}}\pdag(\nn{5};\nn{1}) & \sdelta\hat{\mscr{Y}}^{22}(\nn{5};\nn{1}) \eea\right)^{ba}_{\mbox{;}} \\  \lb{s5_3b}
\sdelta\hat{\mscr{X}}(\nn{5};\nn{1}) &=\sdelta\hat{\mscr{X}}_{\mu_{5},s_{5};\mu_{1},s_{1}}(\Tetax{5};\Tetax{1}) =
\sdelta\hat{a}(\nn{5};\nn{1})=\sum_{k=0}^{3}\bigl(\hat{\tau}_{k}\bigr)_{\mu_{5}\mu_{1}}\;
\sdelta\hat{a}_{s_{5}s_{1}}^{(k)}(\Tetax{5};\Tetax{1})\;\;; \\   \lb{s5_3c}
\sdelta\hat{a}(\nn{5};\nn{1}) &= -\sdelta\hat{a}^{T}(\nn{5};\nn{1})\;;  \\  \lb{s5_3d}
\sdelta\hat{\mscr{Y}}^{11}(\nn{5};\nn{1}) &=-\sdelta\hat{\mscr{Y}}^{22;T}(\nn{5};\nn{1}) =
\im\;\sdelta\hat{g}(\nn{5};\nn{1}) \;;  \\  \lb{s5_3e}
\sdelta\hat{g}(\nn{5};\nn{1}) &=\sdelta\hat{g}\pdag(\nn{5};\nn{1}) \;\;\;.
\end{align}
\end{subequations}
In order to calculate the dependence of \(\sdelta\hat{a}(\nn{5};\nn{1})\) on the anomalous-doubled fields within the generator
\(\hat{\mathrm{Y}}(\nn{5};\nn{1})\) (\ref{s4_7}-\ref{s4_9f}) of
coset matrices \(\hat{\mathrm{T}}\) (\ref{s4_4a}-\ref{s4_4c}), we use the property
\begin{align}\lb{s5_4}
\exp\big\{\hat{B}\big\}\;\delta\Big(\exp\big\{-\hat{B}\big\}\Big) &= -\int_{0}^{1}dv\;
\exp\big\{v\;\hat{B}\big\}\;\;\delta\hat{B}\;\;\exp\big\{-v\;\hat{B}\big\} \;,
\end{align}
and perform the derivative symbol '\(\sdelta\)' of \(\hat{\mathrm{P}}\;\hat{\mathrm{T}}^{-1}\;(\sdelta\hat{\mathrm{T}})\;\hat{\mathrm{P}}^{-1}\)
with its additional gauge transformation so that the eigenvalue decomposition of \(\hat{\mathrm{Y}}(\nn{4};\nn{1})\) into
\(\hat{\mathrm{P}}^{-1}(\nn{4};\nn{3})\;\hat{\mathrm{Y}}_{D}(\nn{3};\nn{2})\;\hat{\mathrm{P}}(\nn{2};\nn{1})\) simplifies the
expression considerably
\begin{align}\lb{s5_5}
-\sdelta\hat{\mscr{Z}}^{ba}(\nn{5};\nn{1}) &=\Bigl(\hat{\mathrm{P}}(\nn{5};\nn{4})\;
\hat{\mathrm{T}}^{-1}(\nn{4};\nn{3})\;\bigl(\sdelta\hat{\mathrm{T}}(\nn{3};\nn{2})\bigr)\;\hat{\mathrm{P}}^{-1}(\nn{2};\nn{1})\Bigr)^{ba} \\
\notag &= \Bigl(\hat{\mathrm{P}}(\nn{5};\nn{4})\;\exp\{\hat{\mathrm{Y}}(\nn{\wt{4}};\nn{\wt{3}})\}\;
\bigl(\sdelta\exp\{-\hat{\mathrm{Y}}(\nn{\wt{3}};\nn{\wt{2}})\}\bigr)\;\hat{\mathrm{P}}^{-1}(\nn{2};\nn{1})\Bigr)^{ba} \\ \notag &=
\Bigl(\hat{\mathrm{P}}(\nn{5};\nn{4})\;\bigl(\exp\{\hat{\mathrm{P}}^{-1}\;\hat{\mathrm{Y}}_{D}\;\hat{\mathrm{P}}\}(\nn{4};\nn{3})\bigr)\;
\bigl(\sdelta\exp\{-\hat{\mathrm{P}}^{-1}\;\hat{\mathrm{Y}}_{D}\;\hat{\mathrm{P}}\}(\nn{3};\nn{2})\bigr)\;
\hat{\mathrm{P}}^{-1}(\nn{2};\nn{1})\Bigr)^{ba} \\ \notag &=
-\int_{0}^{1}dv\;\;\bigl(\exp\{v\;\hat{\mathrm{Y}}_{D}\}(\nn{5};\nn{4})\bigr)\;\;
\bigl(\sdelta\hat{\mathrm{Y}}\ppr(\nn{4};\nn{3})\bigr)\;\;\bigl(\exp\{-v\;\hat{\mathrm{Y}}_{D}\}(\nn{3};\nn{1})\bigr)\;\;.
\end{align}
Note, however, that we do not directly transform from \((\sdelta\hat{\mathrm{Y}}(\nn{5};\nn{1})\,)\) (\ref{s5_6a})
to \(\sdelta\hat{\mscr{Z}}^{ba}(\nn{5};\nn{1})\)
in (\ref{s5_5}), but from rotated variables \((\sdelta\hat{\mathrm{Y}}\ppr(\nn{4};\nn{1})\,)=\hat{\mathrm{P}}(\nn{4};\nn{3})\;
(\sdelta\hat{\mathrm{Y}}(\nn{3};\nn{2})\,)\;\hat{\mathrm{P}}^{-1}(\nn{2};\nn{1})\) (\ref{s5_6b})
to \(\sdelta\hat{\mscr{Z}}^{ba}(\nn{5};\nn{1})\)
which does not alter the integration measure due to the invariance of the trace (\ref{s5_6c})
\begin{subequations}
\begin{align}\lb{s5_6a}
\sdelta\hat{\mathrm{Y}}(\nn{4};\nn{1}) &=
\left(\bea{cc} 0 & \sdelta\hat{\mathrm{X}}(\nn{4};\nn{1}) \\ \sdelta\hat{\mathrm{X}}\pdag(\nn{4};\nn{1}) & 0 \eea\right)^{ba}\;; \hspace*{0.6cm}
\sdelta\hat{\mathrm{X}}(\nn{4};\nn{1}) = \sum_{k=0}^{3}\bigl(\hat{\tau}_{k}\bigr)_{\mu_{4}\mu_{1}}\;\;
\sdelta f_{D;s_{4}s_{1}}^{(k)}(\Tetax{4};\Tetax{1}) \;;  \\   \lb{s5_6b}
\sdelta\hat{\mathrm{Y}}\ppr(\nn{4};\nn{1}) &= \hat{\mathrm{P}}(\nn{4};\nn{3})\;\;
\sdelta\hat{\mathrm{Y}}(\nn{3};\nn{2})\;\;\hat{\mathrm{P}}^{-1}(\nn{2};\nn{1})\;\;\;;   \hspace*{0.6cm}
\sdelta\hat{\mathrm{X}}\ppr(\nn{4};\nn{1}) = \sum_{k=0}^{3}\bigl(\hat{\tau}_{k}\bigr)_{\mu_{4}\mu_{1}}\;\;
\sdelta f_{D;s_{4}s_{1}}^{\bpr(k)}(\Tetax{4};\Tetax{1}) \;;  \\   \lb{s5_6c}  &
\TRAB\Bigl[ \bigl(\sdelta\hat{\mathrm{Y}}\ppr(\nn{3};\nn{2})\bigr)\;\bigl(\sdelta\hat{\mathrm{Y}}\ppr(\nn{2};\nn{1})\bigr)\Bigr]
=\TRAB\Bigl[ \bigl(\sdelta\hat{\mathrm{Y}}(\nn{3};\nn{2})\bigr)\;\bigl(\sdelta\hat{\mathrm{Y}}(\nn{2};\nn{1})\bigr)\Bigr]\;\;\;.
\end{align}
\end{subequations}
It remains to compute the exponential \((\exp\{\pm v\:\hat{\mathrm{Y}}_{D}\}(\nn{5};\nn{4})\,)\) of the quaternion-eigenvalue matrix
\(\hat{\mathrm{Y}}_{D}(\nn{3};\nn{2})\) (\ref{s4_7}-\ref{s4_9f}) with anti-symmetric element \((\hat{\tau}_{2})_{\nu\mu}\) and
to integrate over \(v\in[0,1]\) in (\ref{s5_5})
in order to relate the 'rotated' independent fields \((\sdelta\hat{\mathrm{Y}}\ppr(\nn{3};\nn{2})\,)\) (\ref{s5_6b}) or
\(\sdelta f_{D;s_{3}s_{2}}^{\bpr(k)}(\Tetax{3};\Tetax{2})\) to \(\sdelta\hat{\mscr{Z}}^{ba}(\nn{5};\nn{1})\). This has to be accomplished
for the various parts \(a,b=1,2\) of \(\sdelta\hat{\mscr{Z}}^{ba}(\nn{5};\nn{1})\)  (\ref{s5_3a}-\ref{s5_3e})
where the off-diagonal parts comprise the independent
integration variables \(\sdelta\hat{a}_{\mu_{5},s_{5};\mu_{1},s_{1}}(\Tetax{5};\Tetax{1})\) (\ref{s5_3b},\ref{s5_3c})
and where the dependent, block diagonal
parts are to be determined by corresponding off-diagonal entries. We emphasize that the eigenvalue decomposition
\(\hat{\mathrm{Y}}(\nn{4};\nn{1})=\hat{\mathrm{P}}^{-1}(\nn{4};\nn{3})\;\hat{\mathrm{Y}}_{D}(\nn{3};\nn{2})\;\hat{\mathrm{P}}(\nn{2};\nn{1})\)
(\ref{s4_7}-\ref{s4_9f}) diagonalizes the coset metric tensor in analogy to section \ref{s12} in a 'transferred sense'
and yields straightforward, simplified expressions.
Therefore, we simply list the various parts, following from (\ref{s5_5}), and specify the various variables \(\sdelta\hat{a}(\nn{5};\nn{1})\),
\(\sdelta\hat{a}_{s_{5}s_{1}}^{(k)}(\Tetax{5};\Tetax{1})\) (\ref{s5_3b},\ref{s5_3c})
in terms of \(\sdelta\hat{f}_{D;s_{5}s_{1}}^{\bpr(k)}(\Tetax{5};\Tetax{1})\) (\ref{s5_6b}) or
\(\sdelta\hat{f}_{D;s_{5}s_{1}}^{(k)}(\Tetax{5};\Tetax{1})\) (\ref{s5_6a})
and compute the respective change of integration measure so that we can attain
Euclidean integration variables \(\sdelta\hat{a}_{s_{5}s_{1}}^{(k)}(\Tetax{5};\Tetax{1})\)
under consideration of the integration measure (\ref{s4_12})
for \(\sdelta\hat{f}_{D;s_{5}s_{1}}^{\bpr(k)}(\Tetax{5};\Tetax{1})\) or \(\sdelta\hat{f}_{D;s_{5}s_{1}}^{(k)}(\Tetax{5};\Tetax{1})\).
The diagonal elements for the anomalous-doubled (\ref{s5_3b},\ref{s5_3c}) and block-diagonal terms (\ref{s5_3d},\ref{s5_3e})
are given in (\ref{s5_7a}-\ref{s5_7c}) and (\ref{s5_9}) with the respective change of integration
measure (\ref{s5_8}) which cancels the corresponding term in (\ref{s4_12}) and results in Euclidean path integration fields
\(d\hat{a}_{ss}^{(2)}(\Tetaxx{j}{\vec{x}};\Tetaxx{j}{\vec{x}})\,\wedge\,d\hat{a}_{ss}^{(2)*}(\Tetaxx{j}{\vec{x}};\Tetaxx{j}{\vec{x}})\)
\begin{subequations}
\begin{align}\notag&
-\Bigl(\hat{\mathrm{P}}\;\hat{\mathrm{T}}^{-1}\;\bigl(\sdelta\hat{\mathrm{T}}\bigr)\;\hat{\mathrm{P}}^{-1}\Bigr)_{\mu_{5},s;\mu_{1},s}^{12}
\hspace*{-0.6cm}(\Tetaxx{j}{\vec{x}};\Tetaxx{j}{\vec{x}}) =-
\Bigl(\hat{\mathrm{P}}\;\hat{\mathrm{T}}^{-1}\;\bigl(\sdelta\hat{\mathrm{T}}\bigr)\;\hat{\mathrm{P}}^{-1}\Bigr)_{\mu_{5},s;\mu_{1},s}^{21,\dag}
\hspace*{-0.6cm}(\Tetaxx{j}{\vec{x}};\Tetaxx{j}{\vec{x}})= 
\sdelta\hat{a}_{\mu_{5},s;\mu_{1},s}(\Tetaxx{j}{\vec{x}};\Tetaxx{j}{\vec{x}})= \\  \lb{s5_7a} &=
\bigl(\hat{\tau}_{2}\bigr)_{\mu_{5}\mu_{1}}\;\;\sdelta\hat{a}_{ss}^{(2)}(\Tetaxx{j}{\vec{x}};\Tetaxx{j}{\vec{x}})=
\bigl(\hat{\tau}_{2}\bigr)_{\mu_{5}\mu_{1}}\;\bigg[\bigg(\tfrac{1}{2}+\tfrac{\sinh\big(2\:|\ovv{f}_{\mathsf{S}}(\Tetaxx{j}{\vec{x}})|\big)}{4\;
|\ovv{f}_{\mathsf{S}}(\Tetaxx{j}{\vec{x}})|}\bigg)\;\sdelta\hat{f}_{D;ss}^{\bpr(2)}(\Tetaxx{j}{\vec{x}};\Tetaxx{j}{\vec{x}})  +  \\  \notag &\hspace*{0.9cm}+
\bigg(\tfrac{1}{2}-\tfrac{\sinh\big(2\:|\ovv{f}_{\mathsf{S}}(\Tetaxx{j}{\vec{x}})|\big)}{4\;|\ovv{f}_{\mathsf{S}}(\Tetaxx{j}{\vec{x}})|}\bigg)\;
\exp\{\im\:2\:\phi_{\mathsf{S}}(\Tetaxx{j}{\vec{x}})\}\;\sdelta\hat{f}_{D;ss}^{\bpr(2)*}(\Tetaxx{j}{\vec{x}};\Tetaxx{j}{\vec{x}})\bigg]\;;
\\  \lb{s5_7b}   &\left(\bea{c}
\sdelta\hat{a}_{ss}^{(2)}(\Tetaxx{j}{\vec{x}};\Tetaxx{j}{\vec{x}}) \\
\sdelta\hat{a}_{ss}^{(2)*}(\Tetaxx{j}{\vec{x}};\Tetaxx{j}{\vec{x}})  \eea\right) =  \\ \notag &=
\left(\bea{cc}
\bigg(\tfrac{1}{2}+\tfrac{\sinh\big(2\;|\ovv{f}_{\mathsf{S}}(\Tetaxx{j}{\vec{x}})|\big)}{4\;
|\ovv{f}_{\mathsf{S}}(\Tetaxx{j}{\vec{x}})|}\bigg)  &
e^{\im\;2\phi_{\mathsf{S}}(\Tetaxx{j}{\vec{x}})}\;
\bigg(\tfrac{1}{2}-\tfrac{\sinh\big(2\;|\ovv{f}_{\mathsf{S}}(\Tetaxx{j}{\vec{x}})|\big)}{4\;
|\ovv{f}_{\mathsf{S}}(\Tetaxx{j}{\vec{x}})|}\bigg)
\\ e^{-\im\;2\phi_{\mathsf{S}}(\Tetaxx{j}{\vec{x}})}\;\bigg(\tfrac{1}{2}-
\tfrac{\sinh\big(2\;|\ovv{f}_{\mathsf{S}}(\Tetaxx{j}{\vec{x}})|\big)}{4\;
|\ovv{f}_{\mathsf{S}}(\Tetaxx{j}{\vec{x}})|}\bigg) &
\bigg(\tfrac{1}{2}+\tfrac{\sinh\big(2\;|\ovv{f}_{\mathsf{S}}(\Tetaxx{j}{\vec{x}})|\big)}{4\;
|\ovv{f}_{\mathsf{S}}(\Tetaxx{j}{\vec{x}})|}\bigg) \eea\right)
\left(\bea{c} \sdelta\hat{f}_{D;ss}^{\bpr(2)}(\Tetaxx{j}{\vec{x}};\Tetaxx{j}{\vec{x}}) \\
\sdelta\hat{f}_{D;ss}^{\bpr(2)*}(\Tetaxx{j}{\vec{x}};\Tetaxx{j}{\vec{x}}) \eea\right)_{\mbox{;}}
 \\  \lb{s5_7c} &
\left(\bea{c} \sdelta\hat{f}_{D;ss}^{\bpr(2)}(\Tetaxx{j}{\vec{x}};\Tetaxx{j}{\vec{x}}) \\
\sdelta\hat{f}_{D;ss}^{\bpr(2)*}(\Tetaxx{j}{\vec{x}};\Tetaxx{j}{\vec{x}}) \eea\right)=  \\ \notag &=
\left(\bea{cc}
\bigg(\tfrac{1}{2}+\tfrac{|\ovv{f}_{\mathsf{S}}(\Tetaxx{j}{\vec{x}})|}{\sinh\big(2\;
|\ovv{f}_{\mathsf{S}}(\Tetaxx{j}{\vec{x}})|\big)}\bigg)  &
e^{\im\;2\phi_{\mathsf{S}}(\Tetaxx{j}{\vec{x}})}\;\bigg(\tfrac{1}{2}-
\tfrac{|\ovv{f}_{\mathsf{S}}(\Tetaxx{j}{\vec{x}})|}{\sinh\big(2\;
|\ovv{f}_{\mathsf{S}}(\Tetaxx{j}{\vec{x}})|\big)}\bigg)
\\ e^{-\im\;2\phi_{\mathsf{S}}(\Tetaxx{j}{\vec{x}})}\;\bigg(\tfrac{1}{2}-
\tfrac{|\ovv{f}_{\mathsf{S}}(\Tetaxx{j}{\vec{x}})|}{\sinh\big(2\;
|\ovv{f}_{\mathsf{S}}(\Tetaxx{j}{\vec{x}})|\big)}\bigg) &
\bigg(\tfrac{1}{2}+\tfrac{|\ovv{f}_{\mathsf{S}}(\Tetaxx{j}{\vec{x}}|}{\sinh\big(2\;
|\ovv{f}_{\mathsf{S}}(\Tetaxx{j}{\vec{x}})|\big)}\bigg) \eea\right)
\left(\bea{c}
\sdelta\hat{a}_{ss}^{(2)}(\Tetaxx{j}{\vec{x}};\Tetaxx{j}{\vec{x}}) \\
\sdelta\hat{a}_{ss}^{(2)*}(\Tetaxx{j}{\vec{x}};\Tetaxx{j}{\vec{x}})  \eea\right)_{\mbox{;}}
\end{align}
\end{subequations}
\begin{align}\lb{s5_8}
d\hat{f}_{D;ss}^{\bpr(2)}(\Tetaxx{j}{\vec{x}};\Tetaxx{j}{\vec{x}})\wedge d\hat{f}_{D;ss}^{\bpr(2)*}(\Tetaxx{j}{\vec{x}};\Tetaxx{j}{\vec{x}}) &=
d\hat{f}_{D;ss}^{(2)}(\Tetaxx{j}{\vec{x}};\Tetaxx{j}{\vec{x}})\wedge d\hat{f}_{D;ss}^{(2)*}(\Tetaxx{j}{\vec{x}};\Tetaxx{j}{\vec{x}}) = \\  \notag &=
d\hat{a}_{ss}^{(2)}(\Tetaxx{j}{\vec{x}};\Tetaxx{j}{\vec{x}})\wedge d\hat{a}_{ss}^{(2)*}(\Tetaxx{j}{\vec{x}};\Tetaxx{j}{\vec{x}})\;\;
\tfrac{2\;|\ovv{f}_{\mathsf{S}}(\Tetaxx{j}{\vec{x}})|}{\sinh\big(2\:|\ovv{f}_{\mathsf{S}}(\Tetaxx{j}{\vec{x}})|\big)} \;;
\end{align}
\begin{align}\lb{s5_9}&
-\Bigl(\hat{\mathrm{P}}\;\hat{\mathrm{T}}^{-1}\;\bigl(\sdelta\hat{\mathrm{T}}\bigr)\;\hat{\mathrm{P}}^{-1}\Bigr)_{\mu_{5},s;\mu_{1},s}^{11}
\hspace*{-0.6cm}(\Tetaxx{j}{\vec{x}};\Tetaxx{j}{\vec{x}}) =
\Bigl(\hat{\mathrm{P}}\;\hat{\mathrm{T}}^{-1}\;\bigl(\sdelta\hat{\mathrm{T}}\bigr)\;\hat{\mathrm{P}}^{-1}\Bigr)_{\mu_{5},s;\mu_{1},s}^{22,T}
\hspace*{-0.6cm}(\Tetaxx{j}{\vec{x}};\Tetaxx{j}{\vec{x}})= 
\im\,\sdelta\hat{g}_{\mu_{5},s;\mu_{1},s}(\Tetaxx{j}{\vec{x}};\Tetaxx{j}{\vec{x}})=
\\  \notag &=
\delta_{\mu_{5}\mu_{1}}\;\tfrac{\big(\sinh\big(|\ovv{f}_{\mathsf{S}}(\Tetaxx{j}{\vec{x}}|\big)\,\big)^{2}}{2\;
|\ovv{f}_{\mathsf{S}}(\Tetaxx{j}{\vec{x}})|}\;
\Big[\sdelta\hat{f}_{D;ss}^{\bpr(2)*}(\Tetaxx{j}{\vec{x}};\Tetaxx{j}{\vec{x}})\;
\exp\{\im\;\phi_{\mathsf{S}}(\Tetaxx{j}{\vec{x}})\}-
\sdelta\hat{f}_{D;ss}^{\bpr(2)}(\Tetaxx{j}{\vec{x}};\Tetaxx{j}{\vec{x}})\;
\exp\{-\im\;\phi_{\mathsf{S}}(\Tetaxx{j}{\vec{x}})\}\Big] = \\ \notag &=-\delta_{\mu_{5}\mu_{1}}\;
\tfrac{1}{2}\;\tanh(|\ovv{f}_{\mathsf{S}}(\Tetaxx{j}{\vec{x}})|)\,
\bigg[\sdelta\hat{a}_{ss}^{(2)}(\Tetaxx{j}{\vec{x}};\Tetaxx{j}{\vec{x}})\;
\exp\{-\im\:\phi_{\mathsf{S}}(\Tetaxx{j}{\vec{x}})\}-\sdelta\hat{a}_{ss}^{(2)*}(\Tetaxx{j}{\vec{x}};\Tetaxx{j}{\vec{x}})\;
\exp\{\im\:\phi_{\mathsf{S}}(\Tetaxx{j}{\vec{x}})\}\bigg]\;.
\end{align}
Similarly, we calculate the off-diagonal matrix elements for anomalous-doubled (\ref{s5_3b},\ref{s5_3c}) and
block diagonal parts (\ref{s5_3d},\ref{s5_3e}). Expressions are abbreviated
by coefficients \(A(\nn{5};\nn{1})\), \(B(\nn{5};\nn{1})\), \(C(\nn{5};\nn{1})\), \(D(\nn{5};\nn{1})\) in order to attain corresponding results
from calculating and analyzing matrix terms in eqs. (\ref{s5_3a}-\ref{s5_3e},\ref{s5_5},\ref{s5_6a}-\ref{s5_6c}). Note that we again
obtain the result of eqs. (\ref{s5_7a}-\ref{s5_9}) with the diagonal elements, as one takes the limit process \(\nn{5}\rightarrow\nn{1}\)
of the spacetime labels \(\Tetax{5}\rightarrow\Tetax{1}\) with identical spins \(s_{5}=s_{1}\) and \(2\times2\) quaternion elements
\(\mu_{5},\mu_{1}=\mbox{\scz'e','h'}\) 
\begin{subequations}
\begin{align}\lb{s5_10a}&
-\Bigl(\hat{\mathrm{P}}\;\hat{\mathrm{T}}^{-1}\;\bigl(\sdelta\hat{\mathrm{T}}\bigr)\;\hat{\mathrm{P}}^{-1}\Bigr)_{\mu_{5},s_{5};\mu_{1},s_{1}}^{12}
\hspace*{-0.4cm}(\Tetax{5};\Tetax{1}) \stackrel{\nn{5}\neq\nn{1}}{=}
-\Bigl(\hat{\mathrm{P}}\;\hat{\mathrm{T}}^{-1}\;\bigl(\sdelta\hat{\mathrm{T}}\bigr)\;\hat{\mathrm{P}}^{-1}\Bigr)_{\mu_{5},s_{5};\mu_{1},s_{1}}^{21,\dag}
\hspace*{-0.4cm}(\Tetax{5};\Tetax{1}) =  \\ \notag &\stackrel{\nn{5}\neq\nn{1}}{=}
\sum_{k=0}^{3}\bigl(\hat{\tau}_{k}\bigr)_{\mu_{5}\mu_{1}}\;\sdelta\hat{a}_{s_{5}s_{1}}^{(k)}(\Tetax{5};\Tetax{1})\;;  \\  \lb{s5_10b} &
A(\nn{5};\nn{1}) =\tfrac{\big|\ovv{f}_{\mathsf{S}_{5}}(\Tetax{5})\big|\;\cosh\big(|\ovv{f}_{\mathsf{S}_{1}}(\Tetax{1})|\,\big)\;
\sinh\big(|\ovv{f}_{\mathsf{S}_{5}}(\Tetax{5})|\,\big)-\big|\ovv{f}_{\mathsf{S}_{1}}(\Tetax{1})\big|\;\cosh\big(|\ovv{f}_{\mathsf{S}_{5}}(\Tetax{5})|\,\big)\;
\sinh\big(|\ovv{f}_{\mathsf{S}_{1}}(\Tetax{1})|\,\big)}{\big|\ovv{f}_{\mathsf{S}_{5}}(\Tetax{5})\big|^{2}-\big|\ovv{f}_{\mathsf{S}_{1}}(\Tetax{1})\big|^{2}}\;; \\ \lb{s5_10c} &
B(\nn{5};\nn{1}) =\tfrac{\big|\ovv{f}_{\mathsf{S}_{5}}(\Tetax{5})\big|\;\cosh\big(|\ovv{f}_{\mathsf{S}_{5}}(\Tetax{5})|\,\big)\;
\sinh\big(|\ovv{f}_{\mathsf{S}_{1}}(\Tetax{1})|\,\big)-\big|\ovv{f}_{\mathsf{S}_{1}}(\Tetax{1})\big|\;\cosh\big(|\ovv{f}_{\mathsf{S}_{1}}(\Tetax{1})|\,\big)\;
\sinh\big(|\ovv{f}_{\mathsf{S}_{5}}(\Tetax{5})|\,\big)}{\big|\ovv{f}_{\mathsf{S}_{5}}(\Tetax{5})\big|^{2}-\big|\ovv{f}_{\mathsf{S}_{1}}(\Tetax{1})\big|^{2}}\;;  \\   \lb{s5_10d}&
\left(\bea{c} \sdelta\hat{a}_{s_{5}s_{1}}^{(0)}(\Tetax{5};\Tetax{1}) \\ \sdelta\hat{a}_{s_{5}s_{1}}^{(0)*}(\Tetax{5};\Tetax{1}) \eea\right) = \\ \notag
&=\left(\bea{cc} A(\nn{5};\nn{1}) & \hspace*{-1.6cm}\exp\big\{\im\big(\phi_{\mathsf{S}_{5}}\big(\Tetax{5}\big)+
\phi_{\mathsf{S}_{1}}\big(\Tetax{1}\big)\big)\big\}\;B(\nn{5};\nn{1}) \\
\exp\big\{\!\!\!-\!\im\big(\phi_{\mathsf{S}_{5}}\big(\Tetax{5}\big)+
\phi_{\mathsf{S}_{1}}\big(\Tetax{1}\big)\big)\big\}\;B(\nn{5};\nn{1}) \hspace*{-1.6cm}& A(\nn{5};\nn{1}) \eea\right)
\left(\bea{c} \sdelta\hat{f}_{D;s_{5}s_{1}}^{\bpr(0)}(\Tetax{5};\Tetax{1}) \\ \sdelta\hat{f}_{D;s_{5}s_{1}}^{\bpr(0)*}(\Tetax{5};\Tetax{1})
\eea\right)_{\!\!\!\mbox{;}}     \\    \lb{s5_10e}  &
\left(\bea{c} \sdelta\hat{f}_{D;s_{5}s_{1}}^{\bpr(0)}(\Tetax{5};\Tetax{1}) \\ \sdelta\hat{f}_{D;s_{5}s_{1}}^{\bpr(0)*}(\Tetax{5};\Tetax{1})
\eea\right)= \frac{1}{A^{2}(\nn{5};\nn{1})-B^{2}(\nn{5};\nn{1})} \\ \notag &=
\left(\bea{cc} A(\nn{5};\nn{1}) & \hspace*{-1.6cm}-\exp\big\{\im\big(\phi_{\mathsf{S}_{5}}\big(\Tetax{5}\big)+
\phi_{\mathsf{S}_{1}}\big(\Tetax{1}\big)\big)\big\}\;B(\nn{5};\nn{1}) \\
-\exp\big\{\!\!\!-\!\im\big(\phi_{\mathsf{S}_{5}}\big(\Tetax{5}\big)+
\phi_{\mathsf{S}_{1}}\big(\Tetax{1}\big)\big)\big\}\;B(\nn{5};\nn{1}) \hspace*{-1.6cm}& A(\nn{5};\nn{1}) \eea\right)
\left(\bea{c} \sdelta\hat{a}_{s_{5}s_{1}}^{(0)}(\Tetax{5};\Tetax{1}) \\ \sdelta\hat{a}_{s_{5}s_{1}}^{(0)*}(\Tetax{5};\Tetax{1}) \eea\right) _{\!\!\!\mbox{;}} \\  \lb{s5_10f} &
\left(\bea{c} \sdelta\hat{a}_{s_{5}s_{1}}^{(k)}(\Tetax{5};\Tetax{1}) \\ \sdelta\hat{a}_{s_{5}s_{1}}^{(k)*}(\Tetax{5};\Tetax{1}) \eea\right)
\stackrel{\boldsymbol{k=1,2,3}}{=} \\ \notag &=
\left(\bea{cc} A(\nn{5};\nn{1}) & \hspace*{-1.8cm}-\exp\big\{\im\big(\phi_{\mathsf{S}_{5}}\big(\Tetax{5}\big)+
\phi_{\mathsf{S}_{1}}\big(\Tetax{1}\big)\big)\big\}\;B(\nn{5};\nn{1}) \\
-\exp\big\{\!\!\!-\!\im\big(\phi_{\mathsf{S}_{5}}\big(\Tetax{5}\big)+
\phi_{\mathsf{S}_{1}}\big(\Tetax{1}\big)\big)\big\}\;B(\nn{5};\nn{1}) \hspace*{-1.8cm}& A(\nn{5};\nn{1}) \eea\right)
\left(\bea{c} \sdelta\hat{f}_{D;s_{5}s_{1}}^{\bpr(k)}(\Tetax{5};\Tetax{1}) \\ \sdelta\hat{f}_{D;s_{5}s_{1}}^{\bpr(k)*}(\Tetax{5};\Tetax{1})
\eea\right)_{\!\!\!\mbox{;}}   \\   \lb{s5_10g} &
\left(\bea{c} \sdelta\hat{f}_{D;s_{5}s_{1}}^{\bpr(k)}(\Tetax{5};\Tetax{1}) \\ \sdelta\hat{f}_{D;s_{5}s_{1}}^{\bpr(k)*}(\Tetax{5};\Tetax{1})
\eea\right)\stackrel{\boldsymbol{k=1,2,3}}{=}  \frac{1}{A^{2}(\nn{5};\nn{1})-B^{2}(\nn{5};\nn{1})} \\ \notag &=
\left(\bea{cc} A(\nn{5};\nn{1}) & \hspace*{-1.6cm}\exp\big\{\im\big(\phi_{\mathsf{S}_{5}}\big(\Tetax{5}\big)+
\phi_{\mathsf{S}_{1}}\big(\Tetax{1}\big)\big)\big\}\;B(\nn{5};\nn{1}) \\
\exp\big\{\!\!\!-\!\im\big(\phi_{\mathsf{S}_{5}}\big(\Tetax{5}\big)+
\phi_{\mathsf{S}_{1}}\big(\Tetax{1}\big)\big)\big\}\;B(\nn{5};\nn{1}) \hspace*{-1.6cm}& A(\nn{5};\nn{1}) \eea\right)
\left(\bea{c} \sdelta\hat{a}_{s_{5}s_{1}}^{(k)}(\Tetax{5};\Tetax{1}) \\ \sdelta\hat{a}_{s_{5}s_{1}}^{(k)*}(\Tetax{5};\Tetax{1}) \eea\right) _{\!\!\!\mbox{;}} \\ \lb{s5_10h} &
\frac{A(\nn{5};\nn{1})}{A^{2}(\nn{5};\nn{1})-B^{2}(\nn{5};\nn{1})} =\tfrac{1}{2}\;
\bigg(\tfrac{\big|\ovv{f}_{\mathsf{S}_{5}}(\Tetax{5})\big|-\big|\ovv{f}_{\mathsf{S}_{1}}(\Tetax{1})\big|}{\sinh\big(
|\ovv{f}_{\mathsf{S}_{5}}(\Tetax{5})|-|\ovv{f}_{\mathsf{S}_{1}}(\Tetax{1})|\big)}+
\tfrac{\big|\ovv{f}_{\mathsf{S}_{5}}(\Tetax{5})\big|+\big|\ovv{f}_{\mathsf{S}_{1}}(\Tetax{1})\big|}{\sinh\big(
|\ovv{f}_{\mathsf{S}_{5}}(\Tetax{5})|+|\ovv{f}_{\mathsf{S}_{1}}(\Tetax{1})|\big)}\bigg)_{\mbox{;}}  \\  \lb{s5_10i} &
\frac{B(\nn{5};\nn{1})}{A^{2}(\nn{5};\nn{1})-B^{2}(\nn{5};\nn{1})} =\tfrac{1}{2}\;
\bigg(\tfrac{\big|\ovv{f}_{\mathsf{S}_{5}}(\Tetax{5})\big|-\big|\ovv{f}_{\mathsf{S}_{1}}(\Tetax{1})\big|}{\sinh\big(
|\ovv{f}_{\mathsf{S}_{5}}(\Tetax{5})|-|\ovv{f}_{\mathsf{S}_{1}}(\Tetax{1})|\big)}-
\tfrac{\big|\ovv{f}_{\mathsf{S}_{5}}(\Tetax{5})\big|+\big|\ovv{f}_{\mathsf{S}_{1}}(\Tetax{1})\big|}{\sinh\big(
|\ovv{f}_{\mathsf{S}_{5}}(\Tetax{5})|+|\ovv{f}_{\mathsf{S}_{1}}(\Tetax{1})|\big)}\bigg)_{\mbox{;}}
\end{align}
\end{subequations}
\begin{align}\lb{s5_11}
d\hat{f}_{D;s_{5}s_{1}}^{\bpr(k)}(\Tetax{5};\Tetax{1})\wedge d\hat{f}_{D;s_{5}s_{1}}^{\bpr(k)*}(\Tetax{5};\Tetax{1})&=
d\hat{f}_{D;s_{5}s_{1}}^{(k)}(\Tetax{5};\Tetax{1})\wedge d\hat{f}_{D;s_{5}s_{1}}^{(k)*}(\Tetax{5};\Tetax{1})= \\ \notag &\hspace*{-8.2cm}=
d\hat{a}_{s_{5}s_{1}}^{(k)}(\Tetax{5};\Tetax{1})\wedge d\hat{a}_{s_{5}s_{1}}^{(k)*}(\Tetax{5};\Tetax{1})\;
\tfrac{|\ovv{f}_{\mathsf{S}_{5}}(\Tetax{5})|+|\ovv{f}_{\mathsf{S}_{1}}(\Tetax{1})|}{\sinh(
|\ovv{f}_{\mathsf{S}_{5}}(\Tetax{5})|+|\ovv{f}_{\mathsf{S}_{1}}(\Tetax{1})|)}\;
\tfrac{|\ovv{f}_{\mathsf{S}_{5}}(\Tetax{5})|-|\ovv{f}_{\mathsf{S}_{1}}(\Tetax{1})|}{\sinh(
|\ovv{f}_{\mathsf{S}_{5}}(\Tetax{5})|-|\ovv{f}_{\mathsf{S}_{1}}(\Tetax{1})|)}\;;
\end{align}
\begin{subequations} 
\begin{align} \lb{s5_12a} &
-\Big(\hat{\mathrm{P}}\;\hat{\mathrm{T}}^{-1}\;\big(\sdelta\hat{\mathrm{T}}\big)\;\hat{\mathrm{{P}}}^{-1}\Big)_{
\mu_{5},s_{5};\mu_{1},s_{1}}^{11}\hspace*{-0.4cm}(\Tetax{5};\Tetax{1})
 \stackrel{\nn{5}\neq \nn{1}}{=}  \Big(\hat{\mathrm{P}}\;\hat{\mathrm{T}}^{-1}\;\big(\sdelta\hat{\mathrm{T}}\big)\;
\hat{\mathrm{P}}^{-1}\Big)_{\mu_{5},s_{5};\mu_{1},s_{1}}^{22,T}\hspace*{-0.4cm}(\Tetax{5};\Tetax{1})
\stackrel{\nn{5}\neq \nn{1}}{=} \hspace*{-0.2cm} \im\,\sdelta\hat{g}(\nn{5};\nn{1})  =
\\ \notag &=
-\big(\tau_{2}\big)_{\mu_{5}\mu_{1}}\;
\Big[e^{-\im\;\phi_{\mathsf{S}_{1}}(\Tetax{1})}\;C(\nn{5};\nn{1})\;
\sdelta\hat{f}_{D;s_{5}s_{1}}^{\bpr(0)}(\Tetax{5};\Tetax{1})+
e^{\im\;\phi_{\mathsf{S}_{5}}(\Tetax{5})}\;D(\nn{5};\nn{1})\;
\sdelta\hat{f}_{D;s_{5}s_{1}}^{\bpr(0)*}(\Tetax{5};\Tetax{1})\Big]+ \\ \notag &-
\sum_{k=1,2,3}\hspace*{-0.3cm}\big(\hat{m}_{k}\big)_{\mu_{5}\mu_{1}}
\Big[e^{-\im\;\phi_{\mathsf{S}_{1}}(\Tetax{1})}\;C(\nn{5};\nn{1})\;
\sdelta\hat{f}_{D;s_{5}s_{1}}^{\bpr(k)}(\Tetax{5};\Tetax{1})+
e^{\im\;\phi_{\mathsf{S}_{5}}(\Tetax{5})}\;D(\nn{5};\nn{1})\;
\sdelta\hat{f}_{D;s_{5}s_{1}}^{\bpr(k)*}(\Tetax{5};\Tetax{1})\Big]_{\mbox{;}} \\  & \lb{s5_12b}
\big(\hat{m}_{1}\big)_{\mu_{5}\mu_{1}}=\im\;\big(\hat{\tau}_{3}\big)_{\mu_{5}\mu_{1}}\;;\hspace*{0.5cm}
\big(\hat{m}_{2}\big)_{\mu_{5}\mu_{1}}=\big(\hat{\tau}_{0}\big)_{\mu_{5}\mu_{1}}\;;\hspace*{0.5cm}
\big(\hat{m}_{3}\big)_{\mu_{5}\mu_{1}}=-\im\;\big(\hat{\tau}_{1}\big)_{\mu_{5}\mu_{1}}\;;  \\   \lb{s5_12c} &
C(\nn{5};\nn{1})=   \\   \notag &= \tfrac{-\big|\ovv{f}_{\mathsf{S}_{1}}(\Tetax{1})\big|+\big|\ovv{f}_{\mathsf{S}_{1}}(\Tetax{1})\big|
\;\cosh\big(|\ovv{f}_{\mathsf{S}_{1}}(\Tetax{1})|\big)\;
\cosh\big(|\ovv{f}_{\mathsf{S}_{5}}(\Tetax{5})|\big)-\big|\ovv{f}_{\mathsf{S}_{5}}(\Tetax{5})\big|
\;\sinh\big(|\ovv{f}_{\mathsf{S}_{1}}(\Tetax{1})|\big)\;
\sinh\big(|\ovv{f}_{\mathsf{S}_{5}}(\Tetax{5})|\big)}{\big|\ovv{f}_{\mathsf{S}_{1}}(\Tetax{1})\big|^{2}-
\big|\ovv{f}_{\mathsf{S}_{5}}(\Tetax{5})\big|^{2}} \;; \\ \lb{s5_12d}&
D(\nn{5};\nn{1})=C(\nn{1};\nn{5})=  \\   \notag &=\tfrac{-\big|\ovv{f}_{\mathsf{S}_{5}}(\Tetax{5})\big|+\big|\ovv{f}_{\mathsf{S}_{5}}(\Tetax{5})\big|
\;\cosh\big(|\ovv{f}_{\mathsf{S}_{5}}(\Tetax{5})|\big)\;
\cosh\big(|\ovv{f}_{\mathsf{S}_{1}}(\Tetax{1})|\big)-\big|\ovv{f}_{\mathsf{S}_{1}}(\Tetax{1})\big|
\;\sinh\big(|\ovv{f}_{\mathsf{S}_{5}}(\Tetax{5})|\big)\;
\sinh\big(|\ovv{f}_{\mathsf{S}_{1}}(\Tetax{1})|\big)}{\big|\ovv{f}_{\mathsf{S}_{5}}(\Tetax{5})\big|^{2}-
\big|\ovv{f}_{\mathsf{S}_{1}}(\Tetax{1})\big|^{2}} \;;    \\  \lb{s5_12e}&
-\Big(\hat{\mathrm{P}}\;\hat{\mathrm{T}}^{-1}\;\big(\sdelta\hat{\mathrm{T}}\big)\;\hat{\mathrm{{P}}}^{-1}\Big)_{
\mu_{5},s_{5};\mu_{1},s_{1}}^{11}\hspace*{-0.4cm}(\Tetax{5};\Tetax{1})
 \stackrel{\nn{5}\neq \nn{1}}{=}  \Big(\hat{\mathrm{P}}\;\hat{\mathrm{T}}^{-1}\;\big(\sdelta\hat{\mathrm{T}}\big)\;
\hat{\mathrm{P}}^{-1}\Big)_{\mu_{5},s_{5};\mu_{1},s_{1}}^{22,T}\hspace*{-0.4cm}(\Tetax{5};\Tetax{1})
\stackrel{\nn{5}\neq \nn{1}}{=}   \\  \notag &= \im\,\sdelta\hat{g}_{\mu_{5},s_{5};\mu_{1},s_{1}}  =
-\big(\tau_{2}\big)_{\mu_{5}\mu_{1}}\;\times   \\     \notag &\times\bigg[
\tfrac{e^{-\im\;\phi_{\mathsf{S}_{1}}(\Tetax{1})}\;
\sinh\big(|\ovv{f}_{\mathsf{S}_{1}}(\Tetax{1})|\big)\;\sdelta\hat{a}_{s_{5}s_{1}}^{(0)}(\Tetax{5};\Tetax{1})+
e^{\im\;\phi_{\mathsf{S}_{5}}(\Tetax{5})}\;\sinh\big(|\ovv{f}_{\mathsf{S}_{5}}(\Tetax{5})|\big)\;
\sdelta\hat{a}_{s_{5}s_{1}}^{(0)*}(\Tetax{5};\Tetax{1})\big)}{
\cosh\big(|\ovv{f}_{\mathsf{S}_{5}}(\Tetax{5})|\big)+
\cosh\big(|\ovv{f}_{\mathsf{S}_{1}}(\Tetax{1})|\big)} \bigg] +   \\  \notag &-\hspace*{-0.2cm}\sum_{k=1,2,3}\hspace*{-0.2cm}
(\!\hat{m}_{k}\!)_{\mu_{5}\mu_{1}}\!\bigg[
\tfrac{e^{-\im\;\phi_{\mathsf{S}_{1}}(\Tetax{1})}\,
\sinh\big(|\ovv{f}_{\mathsf{S}_{1}}(\Tetax{1})|\big)\,
\sdelta\hat{a}_{s_{5}s_{1}}^{(k)}(\Tetax{5};\Tetax{1})-
e^{\im\;\phi_{\mathsf{S}_{5}}(\Tetax{5})}\,
\sinh\big(|\ovv{f}_{\mathsf{S}_{5}}(\Tetax{5})|\big)\,\sdelta\hat{a}_{s_{5}s_{1}}^{(k)*}(\Tetax{5};\Tetax{1})}{
\cosh\big(|\ovv{f}_{\mathsf{S}_{5}}(\Tetax{5})|\big)+
\cosh\big(|\ovv{f}_{\mathsf{S}_{1}}(\Tetax{1}|\big)} \bigg] \\ \notag &=
-\tfrac{1}{2}\,(\hat{\tau}_{2})_{\mu_{5}\mu_{1}}\Big\{
\tanh\Big(\tfrac{|\ovv{f}_{\mathsf{S}_{5}}(\Tetax{5})|+|\ovv{f}_{\mathsf{S}_{1}}(\Tetax{1})|}{2}\Big)\;\times \\ \notag &\times
\Big[e^{-\im\;\phi_{\mathsf{S}_{1}}(\Tetax{1})}\;\sdelta\hat{a}_{s_{5}s_{1}}^{(0)}(\Tetax{5};\Tetax{1})+
e^{\im\;\phi_{\mathsf{S}_{5}}(\Tetax{5})}\;\sdelta\hat{a}_{s_{5}s_{1}}^{(0)*}(\Tetax{5};\Tetax{1})\Big] +  \\ \notag &-
\tanh\Big(\tfrac{|\ovv{f}_{\mathsf{S}_{5}}(\Tetax{5})|-|\ovv{f}_{\mathsf{S}_{1}}(\Tetax{1})|}{2}\Big)\;
\Big[e^{-\im\;\phi_{\mathsf{S}_{1}}(\Tetax{1})}\;\sdelta\hat{a}_{s_{5}s_{1}}^{(0)}(\Tetax{5};\Tetax{1})-
e^{\im\;\phi_{\mathsf{S}_{5}}(\Tetax{5})}\;
\sdelta\hat{a}_{s_{5}s_{1}}^{(0)*}(\Tetax{5};\Tetax{1})\Big]\Big\} +  \\ \notag &-
\tfrac{1}{2}\sum_{k=1,2,3}\hspace*{-0.3cm}
(\hat{m}_{k})_{\mu_{5}\mu_{1}}\;\Big\{\tanh\Big(\tfrac{|\ovv{f}_{\mathsf{S}_{5}}(\Tetax{5})|+
|\ovv{f}_{\mathsf{S}_{1}}(\Tetax{1})|}{2}\Big)\;\times  \\  \notag &\times
\Big[e^{-\im\;\phi_{\mathsf{S}_{1}}(\Tetax{1})}\;\sdelta\hat{a}_{s_{5}s_{1}}^{(k)}(\Tetax{5};\Tetax{1})-
e^{\im\;\phi_{\mathsf{S}_{5}}(\Tetax{5})}\;
\sdelta\hat{a}_{s_{5}s_{1}}^{(k)*}(\Tetax{5};\Tetax{1})\Big] +  \\ \notag &-
\tanh\Big(\tfrac{|\ovv{f}_{\mathsf{S}_{5}}(\Tetax{5})|-|\ovv{f}_{\mathsf{S}_{1}}(\Tetax{1})|}{2}\Big)\;
\Big[e^{-\im\;\phi_{\mathsf{S}_{1}}(\Tetax{1})}\;\sdelta\hat{a}_{s_{5}s_{1}}^{(k)}(\Tetax{5};\Tetax{1})+
e^{\im\;\phi_{\mathsf{S}_{5}}(\Tetax{5})}\;
\sdelta\hat{a}_{s_{5}s_{1}}^{(k)*}(\Tetax{5};\Tetax{1})\Big] \Big\}  \\ \notag &=
-\tfrac{1}{2} \sum_{k=0}^{3}(\hat{\tau}_{k}\hat{\tau}_{2})_{\mu_{5}\mu_{1}}
\Big\{\tanh\Big(\tfrac{|\ovv{f}_{\mathsf{S}_{5}}(\Tetax{5})|+|\ovv{f}_{\mathsf{S}_{1}}(\Tetax{1})|}{2}\Big)\;
\Big[e^{-\im\;\phi_{\mathsf{S}_{1}}(\Tetax{1})}\;
\sdelta\hat{a}_{s_{5}s_{1}}^{(k)}(\Tetax{5};\Tetax{1})+     \\   \notag   &-(-1)^{k}\,
e^{\im\;\phi_{\mathsf{S}_{5}}(\Tetax{5})}\;
\sdelta\hat{a}_{s_{5}s_{1}}^{(k)\dag}(\Tetax{5};\Tetax{1})\Big] -
\tanh\Big(\tfrac{|\ovv{f}_{\mathsf{S}_{5}}(\Tetax{5})|-|\ovv{f}_{\mathsf{S}_{1}}(\Tetax{1})|}{2}\Big)\;\times \\ \notag &\times
\Big[e^{-\im\;\phi_{\mathsf{S}_{1}}(\Tetax{1})}\;
\sdelta\hat{a}_{s_{5}s_{1}}^{(k)}(\Tetax{5};\Tetax{1})+(-1)^{k}\,
e^{\im\;\phi_{\mathsf{S}_{5}}(\Tetax{5})}\;
\sdelta\hat{a}_{s_{5}s_{1}}^{(k)\dag}(\Tetax{5};\Tetax{1})\Big]\Big\}_{\mbox{;}}  \;\;
(\mbox{matrix})\pdag=(\mbox{matrix})^{*,T}\;.
\end{align}
\end{subequations}

\subsection{Eigenvalues of coset generators and their transformed, Euclidean correspondents}\lb{s52}

In the previous section \ref{s51} we have accomplished various relations (\ref{s5_7a}-\ref{s5_8},\ref{s5_10a}-\ref{s5_11}) 
between the rotated variables
\((\sdelta\hat{\mathrm{Y}}\ppr(\nn{4};\nn{1})\,)=\hat{\mathrm{P}}(\nn{4};\nn{3})\;
(\sdelta\hat{\mathrm{Y}}(\nn{3};\nn{2})\,)\;\hat{\mathrm{P}}^{-1}(\nn{2};\nn{1})\) or their quaternion
representation with \(\sdelta\hat{\mathrm{X}}\ppr(\nn{4};\nn{1})=\sum_{k=0}^{3}(\hat{\tau}_{k})_{\mu_{4}\mu_{1}}
\sdelta\hat{f}_{D;s_{4}s_{1}}^{\bpr(k)}(\Tetax{4};\Tetax{1})\) and the Euclidean path integration variables
\(\sdelta\hat{a}_{s_{5}s_{1}}^{(k)}(\Tetax{5};\Tetax{1})\); aside from the anomalous-doubled fields, we have also
transformed the block diagonal density parts (\ref{s5_9},\ref{s5_12a}-\ref{s5_12e}) as dependent variables on the off-diagonal terms. The
eigenvalue terms \(\hat{\mathrm{Y}}_{D}(\nn{3};\nn{2})\) or \(\ovv{f}_{\mathsf{S}}(\Tetaxx{j}{\vec{x}})\)
of \(\hat{\mathrm{Y}}(\nn{4};\nn{1})\) (\ref{s4_7}-\ref{s4_9f}) have diagonalized the coset metric tensor and have simplified the
resulting transformations between \(\sdelta\hat{a}_{s_{5}s_{1}}^{(k)}(\Tetax{5};\Tetax{1})\) and
\(\sdelta\hat{f}_{D;s_{5}s_{1}}^{\bpr(k)}(\Tetax{5};\Tetax{1})\); therefore, it remains to determine the
exciton-related eigenvalues \(\ovv{f}_{\mathsf{S}}(\Tetaxx{j}{\vec{x}})\) (\ref{s4_8b},\ref{s4_8c}) with absolute value
\(|\ovv{f}_{\mathsf{S}}(\Tetaxx{j}{\vec{x}})|\) and phase \(\phi_{\mathsf{S}}(\Tetaxx{j}{\vec{x}})\)
in terms of the Euclidean fields \(d\hat{a}_{s_{5}s_{1}}^{(k)}(\Tetax{5};\Tetax{1})\). Since the coherent state
path integral has been derived from a contour time ordering of the time development operator, one has to apply
the infinitesimal integration increment '\(d\)' of the contour time in order to relate the Euclidean fields
\(d\hat{a}(\nn{5};\nn{1})\) to the eigenvalues \(d\ovv{f}_{\mathsf{S}}(\Tetaxx{j}{\vec{x}})\)
\begin{align}\lb{s5_13}
|\ovv{f}_{\mathsf{S}}(\Tetaxx{j}{\vec{x}})|- |\ovv{f}_{\mathsf{S}}(\mscr{T}_{0,\vec{x}}^{\scrscr(+)})| &=
\int_{\mscr{T}_{0}^{(+)}}^{\Teta{j}}\hspace*{-0.37cm}d\Teta{j\ppr}\;
\frac{\pp|\ovv{f}_{\mathsf{S}}(\Tetaxx{j\ppr}{\vec{x}})|}{\pp \Teta{j\ppr}}\;;\hspace*{0.1cm}
\phi_{\mathsf{S}}(\Tetaxx{j}{\vec{x}})-\phi_{\mathsf{S}}(\mscr{T}_{0,\vec{x}}^{\scrscr(+)}) =
\int_{\mscr{T}_{0}^{(+)}}^{\Teta{j}}\hspace*{-0.37cm}d\Teta{j\ppr}\;
\frac{\pp\phi_{\mathsf{S}}(\Tetaxx{j\ppr}{\vec{x}})}{\pp \Teta{j\ppr}}\;;  \\  \lb{s5_14}
\hat{g}_{\mu,s;\nu,s}(\Tetaxx{j}{\vec{x}};\Tetaxx{j}{\vec{x}})&- \hat{g}_{\mu,s;\nu,s}(\mscr{T}_{0,\vec{x}}^{(+)};\mscr{T}_{0,\vec{x}}^{(+)}) 
=\delta_{\mu\nu}\int_{\mscr{T}_{0}^{(+)}}^{\Teta{j}}\hspace*{-0.37cm}d\Teta{j\ppr}\;
\frac{\pp\hat{g}_{\mu,s;\nu,s}(\Tetaxx{j\ppr}{\vec{x}};\Tetaxx{j\ppr}{\vec{x}})}{\pp \Teta{j\ppr}}\;.
\end{align}
As we reconsider the relation (\ref{s5_5}) for '\(\sdelta\)'='\(d\)' of the contour time and expand
\(d\hat{\mathrm{Y}}=(d\hat{\mathrm{P}}^{-1}\:\hat{\mathrm{Y}}_{D}\:\hat{\mathrm{P}})\), we note the
additional commutator \([\hat{\mathrm{Y}}_{D}\,{\large\boldsymbol{,}}\,(d\hat{\mathrm{P}})\hat{\mathrm{P}}^{-1}]_{\boldsymbol{-}}\)
whose resulting diagonal quaternion elements in the off-diagonal parts can be chosen to vanish due to the
gauge invariance of the integration measure with respect to \(\hat{\mathrm{P}}\ldots\hat{\mathrm{P}}^{-1}\)
(cf.\ (\ref{s5_6c}), compare appendix \ref{sa})
\begin{subequations}
\begin{align}\notag
-\big(d\hat{\mscr{Z}}^{ab}(\nn{5};\nn{1})\big)&=
\Big(\hat{\mathrm{P}}\;\hat{\mathrm{T}}^{-1}\;\big(d\hat{\mathrm{T}}\big)\;\hat{\mathrm{P}}^{-1}\Big)^{ab}\hspace*{-0.3cm}(\nn{5};\nn{1})
 = -\int_{0}^{1}dv\;\;\Big(e^{v\;\hat{\mathrm{Y}}_{D}}\;\hat{\mathrm{P}}\;\Big(d\;\hat{\mathrm{P}}^{-1}\;
\hat{\mathrm{Y}}_{D}\;\hat{\mathrm{P}}\Big)\;\hat{\mathrm{P}}^{-1}\;
e^{-v\;\hat{\mathrm{Y}}_{D}} \Big)^{ab}\hspace*{-0.3cm}(\nn{5};\nn{1}) \\ \lb{s5_15a} &=-\int_{0}^{1}dv\;\;
\Big(e^{v\;\hat{\mathrm{Y}}_{D}}\;
\bigg(\big(d\hat{\mathrm{Y}}_{D}\big)+
\Big[\hat{\mathrm{Y}}_{D}\;{\large\boldsymbol{,}}\;\big(d\hat{\mathrm{P}}\big)\;\hat{\mathrm{P}}^{-1}\Big]_{\boldsymbol{-}}\bigg)\;
e^{-v\;\hat{\mathrm{Y}}_{D}}\Big)^{ab}\hspace*{-0.3cm}(\nn{5};\nn{1})\;;   \\ \lb{s5_15b}  0&\equiv
\Big(\big(d\hat{\mathrm{P}}\big)\;\hat{\mathrm{P}}^{-1}\Big)_{\mu_{5},s;\mu_{1},s}\hspace*{-1.2cm}(\Tetaxx{j}{\vec{x}};
\Tetaxx{j}{\vec{x}}) \;;\hspace*{0.5cm}
\Big(\Big[\hat{\mathrm{Y}}_{D}\:{\large\boldsymbol{,}}\:\big(d\hat{\mathrm{P}})\,\hat{\mathrm{P}}^{-1}\Big]_{\boldsymbol{-}}\Big)_{\mu_{5},s;\mu_{1},s}\hspace*{-1.2cm}(\Tetaxx{j}{\vec{x}};\Tetaxx{j}{\vec{x}})\equiv 0 \;.
\end{align}
\end{subequations}
The chosen gauge (\ref{s5_15b}) simplifies (\ref{s5_15a}) and leads to the direct transformation (\ref{s5_16a}) between the diagonal elements of the Euclidean
path integration fields \(d\hat{\mscr{Z}}^{ab}(\nn{5};\nn{1})\) and the eigenvalues \(d\hat{\mathrm{Y}}_{D}\)
where the integration of \(v\in[0,1]\) and the calculation of \(\exp\{\pm v\,\hat{\mathrm{Y}}_{D}\}\)
becomes straightforward due to the restriction to diagonal matrix elements
\begin{subequations}
\begin{align}\lb{s5_16a}
d\hat{\mscr{Z}}_{\mu,s;\nu,s}^{ab}(\Tetaxx{j}{\vec{x}};\Tetaxx{j}{\vec{x}})&=\int_{0}^{1}dv\;\;
\Big(\exp\{v\;\hat{\mathrm{Y}}_{D}\}\;\;\big(d\hat{\mathrm{Y}}_{D}\big)\;\;
\exp\{-v\;\hat{\mathrm{Y}}_{D}\}\Big)_{\mu,s;\nu,s}^{ab}\hspace*{-0.8cm}(\Tetaxx{j}{\vec{x}};\Tetaxx{j}{\vec{x}} ) \\ \lb{s5_16b}
d\hat{\mscr{Z}}_{\mu,s;\nu,s}^{12}(\Tetaxx{j}{\vec{x}};\Tetaxx{j}{\vec{x}}) &=
(\tau_{2})_{\mu\nu}\;d\hat{a}_{ss}^{(2)}(\Tetaxx{j}{\vec{x}};\Tetaxx{j}{\vec{x}})\;;  \\ \lb{s5_16c}
d\hat{\mscr{Z}}_{\mu,s;\nu,s}^{21}(\Tetaxx{j}{\vec{x}};\Tetaxx{j}{\vec{x}}) &=
(\tau_{2})_{\mu\nu}\;d\hat{a}_{ss}^{(2)*}(\Tetaxx{j}{\vec{x}};\Tetaxx{j}{\vec{x}}) \;;   \\  \lb{s5_16d}
d\hat{\mscr{Z}}_{\mu,s;\nu,s}^{11}(\Tetaxx{j}{\vec{x}};\Tetaxx{j}{\vec{x}}) &= \im\;\delta_{\mu\nu}\;
d\hat{g}_{\mu,s;\nu,s}(\Tetaxx{j}{\vec{x}};\Tetaxx{j}{\vec{x}}) \;;  \\  \lb{s5_16e}
d\hat{\mscr{Z}}_{\mu,s;\nu,s}^{22}(\Tetaxx{j}{\vec{x}};\Tetaxx{j}{\vec{x}}) &= -\im\;\delta_{\mu\nu}\;
d\hat{g}_{\mu,s;\nu,s}(\Tetaxx{j}{\vec{x}};\Tetaxx{j}{\vec{x}})\;.
\end{align}
\end{subequations}
After computation of (\ref{s5_16a}), we achieve for the various diagonal elements of (\ref{s5_16b}-\ref{s5_16e}) following relations
\begin{subequations}
\begin{align}\lb{s5_17a}
d\hat{a}_{ss}^{(2)}(\Tetaxx{j}{\vec{x}};\Tetaxx{j}{\vec{x}})&= d\big(|\ovv{a}_{\mathsf{S}}(\Tetaxx{j}{\vec{x}})|\;
\exp\{\im\;\alpha_{\mathsf{S}}(\Tetaxx{j}{\vec{x}})\}\big) =  \\  \notag &\hspace*{-1.9cm}=
d\ovv{f}_{\mathsf{S}}^{*}(\Tetaxx{j}{\vec{x}})\;\exp\{\im\;2\phi_{\mathsf{S}}(\Tetaxx{j}{\vec{x}})\}\;
\bigg(\tfrac{1}{2}-\tfrac{\sinh\big(2\;|\ovv{f}_{\mathsf{S}}(\Tetaxx{j}{\vec{x}})|\big)}{4\;|\ovv{f}_{\mathsf{S}}(\Tetaxx{j}{\vec{x}})|}\bigg)+
d\ovv{f}_{\mathsf{S}}(\Tetaxx{j}{\vec{x}})\;
\bigg(\tfrac{1}{2}+
\tfrac{\sinh\big(2\;|\ovv{f}_{\mathsf{S}}(\Tetaxx{j}{\vec{x}})|\big)}{4\;|\ovv{f}_{\mathsf{S}}(\Tetaxx{j}{\vec{x}})|}\bigg)  \;;  \\  \notag
\im\,d\hat{g}_{\mu,s;\mu,s}(\Tetaxx{j}{\vec{x}};\Tetaxx{j}{\vec{x}}) 
&= -\Big[\big(d\ovv{f}_{\mathsf{S}}(\Tetaxx{j}{\vec{x}})\big)\,
\exp\{-\im\;\phi_{\mathsf{S}}(\Tetaxx{j}{\vec{x}})\}-
\big(d\ovv{f}_{\mathsf{S}}^{*}(\Tetaxx{j}{\vec{x}})\big)\,
\exp\{\im\;\phi_{\mathsf{S}}(\Tetaxx{j}{\vec{x}})\}\Big]\,
\tfrac{\Big(\sinh\big(|\ovv{f}_{\mathsf{S}}(\Tetaxx{j}{\vec{x}})|\big)\Big)^{2}}{2\;
\big|\ovv{f}_{\mathsf{S}}(\Tetaxx{j}{\vec{x}})\big|}; \\  \lb{s5_17b} 
\hat{g}_{\mu,s;\mu,s}(\Tetaxx{j}{\vec{x}};\Tetaxx{j}{\vec{x}}) &\in\mathbb{R}\;;
\end{align}
\end{subequations}
\begin{subequations}
\begin{align}\lb{s5_18a}\mbox{and}\hspace*{0.6cm}
d\hat{a}_{ss}^{(2)}(\Tetaxx{j}{\vec{x}};\Tetaxx{j}{\vec{x}})&= d\big(|\ovv{a}_{\mathsf{S}}(\Tetaxx{j}{\vec{x}})|\;
\exp\{\im\;\alpha_{\mathsf{S}}(\Tetaxx{j}{\vec{x}})\}\big) = \\  \notag &=
\exp\{\im\;\phi_{\mathsf{S}}(\Tetaxx{j}{\vec{x}})\}\;
\Big[\big(d|\ovv{f}_{\mathsf{S}}(\Tetaxx{j}{\vec{x}})|\big)+\im\;\frac{\sinh\big(2\;|\ovv{f}_{\mathsf{S}}(\Tetaxx{j}{\vec{x}})|\big)}{2}\;\;
\big(d\phi_{\mathsf{S}}(\Tetaxx{j}{\vec{x}})\big)\Big]  \;;  \\   \lb{s5_18b}
d\hat{g}_{\mu,s;\mu,s}(\Tetaxx{j}{\vec{x}};\Tetaxx{j}{\vec{x}}) 
 &= -\Big(\sinh\big(|\ovv{f}_{\mathsf{S}}(\Tetaxx{j}{\vec{x}})|\big)\Big)^{2}\;\;
d\phi_{\mathsf{S}}(\Tetaxx{j}{\vec{x}})\;;
\end{align}
\end{subequations}
where we have also introduced the separation of the exciton-related, diagonal elements
\(d\hat{a}_{ss}^{(2)}(\Tetaxx{j}{\vec{x}};\Tetaxx{j}{\vec{x}})=
d(|\ovv{a}_{\mathsf{S}}(\Tetaxx{j}{\vec{x}})|\:\exp\{\im\,\alpha_{\mathsf{S}}(\Tetaxx{j}{\vec{x}})\})\)
into absolute value \(|\ovv{a}_{\mathsf{S}}(\Tetaxx{j}{\vec{x}})|\) and phase
\(\alpha_{\mathsf{S}}(\Tetaxx{j}{\vec{x}})\). However, in order to compare real and imaginary parts of the
transformations (\ref{s5_17a}-\ref{s5_18b}), we perform a phase rotation from
\(d\hat{a}_{ss}^{(2)}(\Tetaxx{j}{\vec{x}};\Tetaxx{j}{\vec{x}})\) to
\(d\wt{a}_{ss}^{(2)}(\Tetaxx{j}{\vec{x}};\Tetaxx{j}{\vec{x}})\) with
absolute value \(|\wt{a}_{\mathsf{S}}(\Tetaxx{j}{\vec{x}})|\) and
corresponding phase \(\wt{\alpha}_{\mathsf{S}}(\Tetaxx{j}{\vec{x}})\)
\begin{align}\lb{s5_19}
\wt{a}_{ss}^{(2)}(\Tetaxx{j}{\vec{x}};\Tetaxx{j}{\vec{x}}) &= |\wt{a}_{\mathsf{S}}(\Tetaxx{j}{\vec{x}})|\;\;
\exp\{\im\;\wt{\alpha}_{\mathsf{S}}(\Tetaxx{j}{\vec{x}})\}\;;  \\  \notag
\exp\{-\im\;\wt{\alpha}_{\mathsf{S}}(\Tetaxx{j}{\vec{x}})\}\;\;d\wt{a}_{ss}^{(2)}(\Tetaxx{j}{\vec{x}};\Tetaxx{j}{\vec{x}})&=
\exp\{-\im\;\phi_{\mathsf{S}}(\Tetaxx{j}{\vec{x}})\}\;\;d\hat{a}_{ss}^{(2)}(\Tetaxx{j}{\vec{x}};\Tetaxx{j}{\vec{x}})\;\Longrightarrow \\  \notag &\hspace*{-2.8cm}
\Longrightarrow\;\;
d\wt{a}_{ss}^{(2)*}(\Tetaxx{j}{\vec{x}};\Tetaxx{j}{\vec{x}})\wedge d\wt{a}_{ss}^{(2)}(\Tetaxx{j}{\vec{x}};\Tetaxx{j}{\vec{x}})=
d\hat{a}_{rr}^{(2)*}(\Tetaxx{j}{\vec{x}};\Tetaxx{j}{\vec{x}})\wedge d\hat{a}_{rr}^{(2)}(\Tetaxx{j}{\vec{x}};\Tetaxx{j}{\vec{x}})\;.
\end{align}
Accordingly, we can replace the diagonal, Euclidean, pair condensate integration variables of eqs. (\ref{s5_17a}-\ref{s5_18b}) by
\(d\wt{a}_{ss}^{(2)}(\Tetaxx{j}{\vec{x}};\Tetaxx{j}{\vec{x}})\), (+c.c.) and obtain new relations between
the coset eigenvalues \(\ovv{f}_{\mathsf{S}}(\Tetaxx{j}{\vec{x}})=|\ovv{f}_{\mathsf{S}}(\Tetaxx{j}{\vec{x}})|\:
\exp\{\im\,\phi_{\mathsf{S}}(\Tetaxx{j}{\vec{x}})\}\) and the new Euclidean elements
\(\wt{a}_{ss}^{(2)}(\Tetaxx{j}{\vec{x}};\Tetaxx{j}{\vec{x}})=|\wt{a}_{\mathsf{S}}(\Tetaxx{j}{\vec{x}})|\:
\exp\{\im\,\wt{\alpha}_{\mathsf{S}}(\Tetaxx{j}{\vec{x}})\}\) (\ref{s5_19}). Note, that the absolute values transform as
state variables whereas the phase transformations are path dependent and ,therefore, can only result from
contour time integrals
\begin{subequations}
\begin{align}\lb{s5_20a}
d|\wt{a}_{\mathsf{S}}(\Tetaxx{j}{\vec{x}})|+
\im\,|\wt{a}_{\mathsf{S}}(\Tetaxx{j}{\vec{x}})|\;\;d\wt{\alpha}_{\mathsf{S}}(\Tetaxx{j}{\vec{x}}) &=
d|\ovv{f}_{\mathsf{S}}(\Tetaxx{j}{\vec{x}})| +\im\,
\tfrac{\sinh\big(2\;|\ovv{f}_{\mathsf{S}}(\Tetaxx{j}{\vec{x}})|\big)}{2}\;\;d\phi_{\mathsf{S}}(\Tetaxx{j}{\vec{x}})  \;;  \\ \lb{s5_20b}
|\ovv{f}_{\mathsf{S}}(\Tetaxx{j}{\vec{x}})|-\underbrace{|\ovv{f}_{\mathsf{S}}(\mscr{T}_{0,\vec{x}}^{(+)})|}_{=0}&=
|\wt{a}_{\mathsf{S}}(\Tetaxx{j}{\vec{x}})|-\underbrace{|\wt{a}_{\mathsf{S}}(\mscr{T}_{0,\vec{x}}^{(+)})|}_{=0} \;; \\  \lb{s5_20c}
\phi_{\mathsf{S}}(\Tetaxx{j}{\vec{x}})-\underbrace{\phi_{\mathsf{S}}(\mscr{T}_{0,\vec{x}}^{(0)})}_{=0} =
\int_{\mscr{T}_{0}^{(+)}}^{\Teta{j}}\hspace*{-0.37cm}d\Teta{j\ppr}\;&
\frac{\pp\phi_{\mathsf{S}}(\Tetaxx{j\ppr}{\vec{x}})}{\pp \Teta{j\ppr}}  =
\int_{\mscr{T}_{0}^{(+)}}^{\Teta{j}}\hspace*{-0.37cm}d\Teta{j\ppr}\;
\tfrac{2\;|\wt{a}_{\mathsf{S}}(\Tetaxx{j\ppr}{\vec{x}})|}{\sinh\big(2\;
|\wt{a}_{\mathsf{S}}(\Tetaxx{j\ppr}{\vec{x}})|\big)}\;\;
\tfrac{\pp\wt{\alpha}_{\mathsf{S}}(\Tetaxx{j\ppr}{\vec{x}})}{\pp \Teta{j\ppr}}  \;;  \\  \lb{s5_20d}
\hat{g}_{\mu,s;\mu,s}(\Tetaxx{j}{\vec{x}};\Tetaxx{j}{\vec{x}})-
\underbrace{\hat{g}_{\mu,s;\mu,s}(\mscr{T}_{0,\vec{x}}^{(+)};\mscr{T}_{0,\vec{x}}^{(+)})}_{=0}
&=-\int_{\mscr{T}_{0}^{(+)}}^{\Teta{j}}\hspace*{-0.37cm}d\Teta{j\ppr}\;
\tanh\big(|\wt{a}_{\mathsf{S}}(\Tetaxx{j\ppr}{\vec{x}})|\big)\;\;|\wt{a}_{\mathsf{S}}(\Tetaxx{j\ppr}{\vec{x}})|\;\;
\tfrac{\pp\wt{\alpha}_{\mathsf{S}}(\Tetaxx{j\ppr}{\vec{x}})}{\pp \Teta{j\ppr}}  \;.
\end{align}
\end{subequations}
Consequently, we have the final relations between absolute values (\ref{s5_21a}) and the phases (\ref{s5_21b})
where the Euclidean field variables \(\wt{a}_{ss}^{(2)}(\Tetaxx{j}{\vec{x}};\Tetaxx{j}{\vec{x}})=
|\wt{a}_{\mathsf{S}}(\Tetaxx{j}{\vec{x}})|\:\exp\{\im\,\wt{\alpha}_{\mathsf{S}}(\Tetaxx{j}{\vec{x}})\}\)
have to replace the original diagonal elements
\(\hat{a}_{ss}^{(2)}(\Tetaxx{j}{\vec{x}};\Tetaxx{j}{\vec{x}})\) by subsequent relation (\ref{s5_21c}),
due to the inclusion of the phase rotation (\ref{s5_19})
\begin{subequations}
\begin{align} \lb{s5_21a}
|\ovv{f}_{\mathsf{S}}(\Tetaxx{j}{\vec{x}})| &= |\wt{a}_{\mathsf{S}}(\Tetaxx{j}{\vec{x}})| \;;  \\   \lb{s5_21b}   
\phi_{\mathsf{S}}(\Tetaxx{j}{\vec{x}}) &=
\int_{\mscr{T}_{0}^{(+)}}^{\Teta{j}}\hspace*{-0.37cm}d\Teta{j\ppr}\;
\frac{\pp\phi_{\mathsf{S}}(\Tetaxx{j\ppr}{\vec{x}})}{\pp \Teta{j\ppr}}  =
\int_{\mscr{T}_{0}^{(+)}}^{\Teta{j}}\hspace*{-0.37cm}d\Teta{j\ppr}\;
\tfrac{2\;|\wt{a}_{\mathsf{S}}(\Tetaxx{j\ppr}{\vec{x}})|}{\sinh\big(2\;
|\wt{a}_{\mathsf{S}}(\Tetaxx{j\ppr}{\vec{x}})|\big)}\;
\tfrac{\pp\wt{\alpha}_{\mathsf{S}}(\Tetaxx{j\ppr}{\vec{x}})}{\pp \Teta{j\ppr}}  \;;  \\  \lb{s5_21c}
d\hat{a}_{ss}^{(2)}(\Tetaxx{j}{\vec{x}};\Tetaxx{j}{\vec{x}})&=
\exp\{\im(\phi_{\mathsf{S}}(\Tetaxx{j}{\vec{x}})-\wt{\alpha}_{\mathsf{S}}(\Tetaxx{j}{\vec{x}})\,)\}\;\;
d\wt{a}_{ss}^{(2)}(\Tetaxx{j}{\vec{x}};\Tetaxx{j}{\vec{x}})  =  \\ \notag &=
\exp\bigg\{\im\int_{\mscr{T}_{0}^{(+)}}^{\Teta{j}}\hspace*{-0.37cm}d\Teta{j\ppr}\;
\Big(\tfrac{2\;|\wt{a}_{\mathsf{S}}(\Tetaxx{j\ppr}{\vec{x}})|}{\sinh\big(2\;
|\wt{a}_{\mathsf{S}}(\Tetaxx{j\ppr}{\vec{x}})|\big)}-1\Big)\;
\tfrac{\pp\wt{\alpha}_{\mathsf{S}}(\Tetaxx{j\ppr}{\vec{x}})}{\pp \Teta{j\ppr}}\bigg\}\;
d\wt{a}_{ss}^{(2)}(\Tetaxx{j}{\vec{x}};\Tetaxx{j}{\vec{x}})\;.
\end{align}
\end{subequations}

\subsection{Nonlinear sigma model with Euclidean coset fields
$\mbox{SO}(\mfrak{N},\mfrak{N})\,/\,\mbox{U}(\mfrak{N})\otimes\mbox{U}(\mfrak{N})$} \lb{s53}

It remains to collect and summarize the transformation to Euclidean field degrees of freedom and their relation
to the eigenvalues of the coset generators following from sections \ref{s51}, \ref{s52}. According to these
transformations, the final nonlinear sigma model, following from (\ref{s4_20a}-\ref{s4_20d}) is determined to be
\begin{align}\lb{s5_22}
\ovv{Z[\hat{\mscr{J}}]} &\approx \int
d\bigl[\hat{\wt{a}}_{s\ppr s}^{(k)}(\Tetaxx{j\ppr}{\vec{x}\ppr};\Tetaxx{j}{\vec{x}})\bigr]\;\;\;\;
\mfrak{Z}\!\Bigl[\hat{\mathrm{T}}^{\boldsymbol{-1}}(\hat{\wt{a}});
\hat{\mathrm{T}}(\hat{\wt{a}});\hat{\mcal{H}}\Bigr]\;\times
\exp\biggl\{\frac{1}{2}\,\mcal{N}_{x}\sum_{\vec{x}}\sum_{j=0}^{2N+1}\sum_{\mu=e,h}^{s=\uparrow,\downarrow}
\;\times \\ \notag &\times\;
\TRAB\bigg[\ln\bigg(\boldsymbol{\hat{1}}-\bigl(\hat{\pp}_{(\ln-\breve{\mscr{H}})}\hat{\mscr{Z}}\bigr)^{b\ppr a\ppr}+
\hat{\mathrm{P}}\:\wt{\mscr{J}}\bigl[\hat{\mathrm{T}}^{-1}(\hat{\wt{a}}),\hat{\mathrm{T}}(\hat{\wt{a}})\bigr]\;
\Bigl(\breve{\mscr{H}}\bigl[\hat{\mfrak{H}},\langle\hat{\mfrak{s}}\rangle\bigr]\Bigr)^{\boldsymbol{-1}}\hat{\mathrm{P}}^{-1}
\bigg)\bigg]_{\mu,s;\mu,s}^{b=a}\hspace*{-0.9cm}(\Teta{j},\vec{x};\Teta{j},\vec{x})\biggr\}\;,
\end{align}
with the sub-functional \(\mfrak{Z}[\hat{\mathrm{T}}^{-1}(\hat{\wt{a}});\hat{\mathrm{T}}^{-1}(\hat{\wt{a}});\hat{\mcal{H}}]\)
(\ref{s4_15}) containing the driving laser field for creation of exciton quasi-particles in the off-diagonal,
anomalous parts
\begin{align}\lb{s5_23}
\mfrak{Z}\!\Bigl[\hat{\mathrm{T}}^{\boldsymbol{-1}};
\hat{\mathrm{T}};\hat{\mcal{H}}\Bigr] &= \int d\bigl[\delta\hat{\Sigma}_{D;\nu,s\ppr;\mu,s}^{11}(\Teta{j_{2}},\vec{x}_{2};
\Teta{j_{1}},\vec{x}_{1})\bigr]\;\;
\mbox{Poly}\bigl(\delta\hat{\lambda}_{\mu,s}(\Teta{j},\vec{x})\,\bigr)\;\;\times\;
\exp\biggl\{\tfrac{\im}{4}\,\tfrac{\sdelta t}{\hbar}\sum_{j_{1,2}=1}^{2N} \times \\ \notag &\hspace*{-2.3cm}\times\;
\hat{\mscr{V}}^{-1}(\vec{x}_{2},\TT{\scrscr j_{2}};\vec{x}_{1},\TT{\scrscr j_{1}})\times
\TRAB\bigg[\hat{\mathrm{S}}_{\mu_{2}}^{bb}\:\etan{j_{2}}\;
\hat{\mathrm{T}}_{\mu_{2},s_{2};\mu_{4},s_{4}}^{bb\ppr}(\Teta{j_{2}},\vec{x}_{2};\Teta{j_{4}},
\vec{x}_{4})\;\delta\hat{\Sigma}_{D;\mu_{4},s_{4};\mu_{3},s_{3}}^{b\ppr=a\ppr}(\Teta{j_{4}},
\vec{x}_{4};\Teta{j_{3}},\vec{x}_{3}) \;\times \\ \notag &\times\;
\hat{\mathrm{T}}_{\mu_{3},s_{3};\mu_{1},s_{1}}^{\boldsymbol{-1};a\ppr a}(\Teta{j_{3}},\vec{x}_{3};
\Teta{j_{1}},\vec{x}_{1})\;\hat{\mathrm{S}}_{\mu_{1}}^{aa}\;\etan{j_{1}}\;
\hat{\mathrm{T}}_{\mu_{1},s_{1};\mu_{6},s_{6}}^{aa\pppr}
(\Teta{j_{1}},\vec{x}_{1};\Teta{j_{6}},
\vec{x}_{6})\;\times \\ \notag &\times\;
\delta\hat{\Sigma}_{D;\mu_{6},s_{6};\mu_{5},s_{5}}^{a\pppr=b\pppr}
(\Teta{j_{6}},\vec{x}_{6};\Teta{j_{5}},\vec{x}_{5}) \;
\hat{\mathrm{T}}_{\mu_{5},s_{5};\mu_{2},s_{2}}^{\boldsymbol{-1};b\pppr b}(\Teta{j_{5}},\vec{x}_{5};
\Teta{j_{2}},\vec{x}_{2})\biggr]\biggr\} \;\;\times\;
\exp\biggl\{-\tfrac{1}{2}\sum_{j_{1,2}=1}^{2N} \times \\ \notag &\hspace*{-2.3cm}\times\;
\hat{\mscr{V}}^{-1}(\vec{x}_{2},\TT{\scrscr j_{2}};\vec{x}_{1},\TT{\scrscr j_{1}})\times
\TRAB\bigg[\hat{\mathrm{S}}^{bb}\;
\hat{\mathrm{T}}_{\mu_{2},s_{2};\mu_{4},s_{4}}^{bb\ppr}(\Teta{j_{2}},\vec{x}_{2};\Teta{j_{4}},
\vec{x}_{4})\;\delta\hat{\Sigma}_{D;\mu_{4},s_{4};\mu_{3},s_{3}}^{b\ppr=a\ppr}(\Teta{j_{4}},
\vec{x}_{4};\Teta{j_{3}},\vec{x}_{3}) \;\times \\ \notag &\times\;
\hat{\mathrm{T}}_{\mu_{3},s_{3};\mu_{1},s_{1}}^{\boldsymbol{-1};a\ppr a}(\Teta{j_{3}},\vec{x}_{3};
\Teta{j_{1}},\vec{x}_{1})\;\hat{\mathrm{S}}^{aa}\;
\hat{\mcal{H}}_{\mu_{1},s_{1};\mu_{2},s_{2}}^{a\neq b}(\vec{x}_{1},\Teta{j_{1}};\vec{x}_{2},\Teta{j_{2}})
\biggr]\biggr\}\;.
\end{align}
The increments of Euclidean path integration fields \(\sdelta\hat{\wt{a}}(\nn{5};\nn{1})\) (\ref{s5_3a}-\ref{s5_3e}) are specified by
quaternion matrix elements \(\sdelta\hat{a}_{s_{5}s_{1}}^{(k)}(\Tetax{5};\Tetax{1})\) with the exception of the
diagonal elements \(\sdelta\wt{a}_{ss}^{(2)}(\Tetaxx{j}{\vec{x}};\Tetaxx{j}{\vec{x}})\)
which incorporate a contour time dependent phase correction (\ref{s5_19}), denoted by the tilde '\(\wt{\ph{A}}\)'
in \(\sdelta\hat{\wt{a}}(\nn{5};\nn{1})\)
\begin{subequations}
\begin{align}\lb{s5_24a}
\sdelta\hat{\wt{a}}(\nn{5};\nn{1}) &=-\sdelta\hat{\wt{a}}^{\boldsymbol{T}}(\nn{5};\nn{1})=\sum_{k=0}^{3}
(\hat{\tau}_{k})_{\mu_{5}\mu_{1}}\;\sdelta\hat{a}_{s_{5}s_{1}}^{(k)}(\Tetax{5};\Tetax{1})+  \\ \notag &+
(\hat{\tau}_{2})_{\mu_{5}\mu_{1}}\;\exp\bigg\{\im\int_{\TT_{0}^{(+)}}^{\Teta{j}}d\Teta{j\ppr}
\bigg(\frac{2\,|\wt{a}_{\mathsf{S}}(\Tetaxx{j\ppr}{\vec{x}})|}{\sinh\big(2\,|\wt{a}_{\mathsf{S}}(\Tetaxx{j\ppr}{\vec{x}})|\big)}-1\bigg)
\frac{\pp\wt{\alpha}_{\mathsf{S}}(\Tetaxx{j\ppr}{\vec{x}})}{\pp\Teta{j\ppr}}\bigg\}\;
\sdelta\wt{a}_{ss}^{(2)}(\Tetaxx{j}{\vec{x}};\Tetaxx{j}{\vec{x}})\;;  \\ \notag &
(\mbox{last term }(\Tetaxx{j}{\vec{x}}:=\Tetax{5}=\Tetax{1})\&(s:=s_{5}=s_{1}\;,\,\mathsf{S}=2s)\,)\;; \\    \lb{s5_24b}
\wt{a}_{ss}^{(2)}(\Tetaxx{j}{\vec{x}};\Tetaxx{j}{\vec{x}}) &=
|\wt{a}_{\mathsf{S}}(\Tetaxx{j}{\vec{x}})|\;\exp\{\im\:\wt{\alpha}_{\mathsf{S}}(\Tetaxx{j\ppr}{\vec{x}})\}\;; \\   \lb{s5_24c}
d\bigl[\hat{\wt{a}}_{s\ppr s}^{(k)}(\Tetaxx{j\ppr}{\vec{x}\ppr};\Tetaxx{j}{\vec{x}})\bigr] &=
 \prod_{s=\uparrow,\downarrow}^{\mathsf{S}=2s}
\prod_{j=1,\ldots,2N}^{\{\vec{x}\}}
\biggl(\frac{d\wt{a}_{ss}^{(2)*}(\Tetaxx{j}{\vec{x}};\Tetaxx{j}{\vec{x}})\wedge
d\wt{a}_{ss}^{(2)}(\Tetaxx{j}{\vec{x}};\Tetaxx{j}{\vec{x}})}{2\;\im}
\biggr) \times \\ \no &\times \prod_{k=0}^{3}
\Bigg(\prod_{s,s\ppr=\uparrow,\downarrow}^{\mathsf{S}=2s,\,\mathsf{S}\ppr=2s\ppr}\prod_{j_{1,2}=1,\ldots,2N}^{\{\vec{x}_{1,2}\}}
\bigl(\mbox{except : }(s\ppr=s,\,\mathsf{S}\ppr=\mathsf{S})\And(j_{1}=j_{2})\And(\vec{x}_{1}=\vec{x}_{2})\bigr)\;\;\times
\\ \no &\times
\frac{d\hat{a}_{s\ppr s}^{(k)*}(\Tetax{2};\Tetax{1})\wedge
d\hat{a}_{s\ppr s}^{(k)}(\Tetax{2};\Tetax{1})}{2\;\im}
\Bigg)^{\boldsymbol{1/2}}_{\mbox{.}}
\end{align}
\end{subequations}
Aside from the source field \(\wt{\mscr{J}}[\hat{\mathrm{T}}^{-1}(\hat{\wt{a}}),\hat{\mathrm{T}}^{-1}(\hat{\wt{a}})]\)
for generating the observables,
\begin{align}\lb{s5_25}  &
\wt{\mscr{J}}\Bigl[\hat{\mathrm{T}}^{-1},\hat{\mathrm{T}}\Bigr]_{\nu,s\ppr;\mu,s}^{ba}
\hspace*{-0.9cm}(\Teta{j_{2}},\vec{x}_{2};\Teta{j_{1}},\vec{x}_{1}) =
\hat{\mathrm{T}}_{\nu,s\ppr;\mu_{4},s_{4}}^{\boldsymbol{-1};bb\ppr}(\Teta{j_{2}},\vec{x}_{2};
\Teta{j_{4}},\vec{x}_{4})\;\times    \\  \notag  &\times\;
\eta_{j_{4}}\;\hat{\mathrm{I}}^{\boldsymbol{-1};b\ppr b\ppr}\;
\hat{\mscr{J}}_{\mu_{4},s_{4};\mu_{3},s_{3}}^{b\ppr a\ppr}(\Teta{j_{4}},\vec{x}_{4};\Teta{j_{3}},
\vec{x}_{3})\;\hat{\mathrm{I}}^{\boldsymbol{-1};a\ppr a\ppr}\;\eta_{j_{3}}\;
\hat{\mathrm{T}}_{\mu_{3},s_{3};\mu,s}^{a\ppr a}(\Teta{j_{3}},\vec{x}_{3};\Teta{j_{1}},\vec{x}_{1})\;,
\end{align}
we have introduced generalized gradient operators (\ref{s4_21},\ref{s4_22}), (\ref{s5_26}-\ref{s5_28}) 
with the adjoint action of the logarithm of the density-related one-particle operator (\ref{s4_23})
\begin{align}\lb{s5_26}
\hat{\pp}_{\breve{\mscr{H}}}\bigl(\mbox{matrix}\bigr)^{ba} &:=
\biggl[\breve{\mscr{H}}\bigl[\hat{\mfrak{H}},\langle\hat{\mfrak{s}}\rangle\bigr]^{bb}\;\bigl(\mbox{matrix}\bigr)^{ba}\;
\breve{\mscr{H}}\bigl[\hat{\mfrak{H}},\langle\hat{\mfrak{s}}\rangle\bigr]^{\boldsymbol{-1};aa}-
\bigl(\mbox{matrix}\bigr)^{ba}\biggr]\;;  \\     \lb{s5_27}
\hat{\pp}_{\ln(-\breve{\mscr{H}})}\bigl(\mbox{matrix}\bigr)^{ba} &:=
\biggl[\bigg(\exp\Big\{\overrightarrow{\big[\ln(-\breve{\mscr{H}}[\hat{\mfrak{H}},\langle\hat{\mfrak{s}}\rangle])\;{\Large\boldsymbol{,}}
\ldots\big]_{-}}\Big\}-\hat{1}\bigg)\bigl(\mbox{matrix}\bigr)^{b\ppr a\ppr}\biggr]^{ba}\;;   \\     \lb{s5_28}
\Bigl(\hat{\pp}_{\ln(-\breve{\mscr{H}})}\hat{\mscr{Z}}(\nn{3};\nn{2})\Bigr)^{ba}\hspace*{-0.4cm}(\nn{5};\nn{1}) &:=
\biggl[\bigg(\exp\Big\{\overrightarrow{\big[\ln(-\breve{\mscr{H}}[\hat{\mfrak{H}},\langle\hat{\mfrak{s}}\rangle])\;{\Large\boldsymbol{,}}
\ldots\big]_{-}}\Big\}-\hat{1}\bigg)\hat{\mscr{Z}}^{b\ppr a\ppr}(\nn{3};\nn{2})\biggr]^{ba}\hspace*{-0.4cm}(\nn{5};\nn{1})\;.
\end{align}
The logarithm of the density-related, block diagonal operator \(\breve{\mscr{H}}[\hat{\mfrak{H}},\langle\hat{\mfrak{s}}\rangle]\)
allows to simplify the action of the adjoint operator \(\mathsf{ad}_{\ln(-\breve{\mscr{H}})}\) in (\ref{s5_27},\ref{s5_28}) by expanding the logarithm
around the unit value with the saddle point solution \(\langle\hat{\mfrak{s}}\rangle_{\nu,s\ppr;\mu,s}^{aa}(\Tetax{2};\Tetax{1})\)
and kinetic energy operator of the one-particle part
\begin{subequations}
\begin{align}\notag
\ln\bigl(-\breve{\mscr{H}}[\hat{\mfrak{H}},\langle\hat{\mfrak{s}}\rangle] \bigr)&=
\ln\bigg(-\hat{\mfrak{H}}_{\nu,s\ppr;\mu,s}^{11}(\vec{x}_{2},\Teta{j_{2}};\vec{x}_{1},\Teta{j_{1}})-
q_{\nu}\:\eta_{j_{2}}\:\bigl(\im\,\tfrac{\sdelta t}{\hbar}\bigr)\:
\langle\hat{\mfrak{s}}\rangle_{\nu,s\ppr;\mu,s}^{11}(\Teta{j_{2}},\vec{x}_{2};\Teta{j_{1}},\vec{x}_{1})\;q_{\mu}\:\eta_{j_{1}}\bigg)=
\\ \notag &\hspace*{-2.8cm}= \ln\bigg[\delta_{\nu\mu}\,\delta_{s\ppr s}\,\deltaN_{\vec{x}_{2},\vec{x}_{1}}\,\delta_{j_{2},j_{1}}-\bigg(
\delta_{\nu\mu}\,\delta_{s\ppr s}\,\deltaN_{\vec{x}_{2},\vec{x}_{1}}\,\delta_{j_{2}+1,j_{1}}+
\im\,\tfrac{\sdelta t}{\hbar}\Big[\delta_{j_{2},j_{1}}\,\eta_{j_{2}}\,\delta_{\nu\mu}
\big(\hat{\mscr{E}}_{\mu,s\ppr s}(\vec{x}_{2})-\im\,\eta_{j_{2}}\,\delta_{s\ppr s}\,\ve_{+}\big)\deltaN_{\vec{x}_{2},\vec{x}_{1}}+
\\ \lb{s5_29a}    &\hspace*{3.7cm}+ q_{\nu}\:\eta_{j_{2}}\:
\langle\hat{\mfrak{s}}\rangle_{\nu,s\ppr;\mu,s}^{11}(\Teta{j_{2}},\vec{x}_{2};\Teta{j_{1}},\vec{x}_{1})\;q_{\mu}\:\eta_{j_{1}}\Big]\bigg)
\bigg]\;;    \\      \lb{s5_29b}
\hat{\mscr{V}}^{-1}(\vec{x}_{2},\TT_{\scrscr j_{2}};\vec{x}_{1},\TT_{\scrscr j_{1}})\;&
\langle\hat{\mfrak{s}}\rangle_{\mu_{2},s_{2};\mu_{1},s_{1}}^{11}(\Teta{j_{2}},\vec{x}_{2};\Teta{j_{1}},\vec{x}_{1})\Big/\mcal{N}_{x}=
\mcal{N}_{x}\Bigl[\hat{\mfrak{H}}_{\mu_{4},s_{4};\mu_{3},s_{3}}^{11}(\vec{x}_{4},\Teta{j_{4}};\vec{x}_{3},\Teta{j_{3}})+ \\ \notag &\hspace*{-2.8cm}+
\eta_{j_{4}}\;q_{\mu_{4}}\;\bigl(\im\tfrac{\Delta t}{\hbar}\bigr)\;
\langle\hat{\mfrak{s}}\rangle_{\mu_{4},s_{4};\mu_{3},s_{3}}^{11}(\Teta{j_{4}},\vec{x}_{4};\Teta{j_{3}},\vec{x}_{3})\;
\eta_{j_{3}}\;q_{\mu_{3}}\Bigr]_{\mu_{2},s_{2};\mu_{1},s_{1}}^{\boldsymbol{-1};11}
\hspace*{-0.9cm}(\Teta{j_{2}},\vec{x}_{2};\Teta{j_{1}},\vec{x}_{1})\;; \\      \lb{s5_29c}   &\hspace*{-2.8cm}
\sum_{j_{1}=0}^{2N+1}\sum_{\vec{x}_{1}}^{\mu,s}\Big[
\hat{\mfrak{H}}_{\nu,s\ppr;\mu,s}^{11}(\vec{x}_{2},\Teta{j_{2}};\vec{x}_{1},\Teta{j_{1}})+ \\ \notag &\hspace*{-2.8cm}+
\eta_{j_{2}}\;q_{\nu}\;\bigl(\im\tfrac{\Delta t}{\hbar}\bigr)\;
\langle\hat{\mfrak{s}}\rangle_{\nu,s\ppr;\mu,s}^{11}(\Teta{j_{2}},\vec{x}_{2};\Teta{j_{1}},\vec{x}_{1})\;
\eta_{j_{1}}\;q_{\mu}\Big]\psi_{M;\mu,s}(\Teta{j_{1}},\vec{x}_{1})=E_{M}\;\psi_{M;\nu,s\ppr}(\Teta{j_{2}},\vec{x}_{2})\;; \\   \lb{s5_29d}
&\hspace*{-2.8cm}
\mcal{N}_{x}\biggl(\breve{\mscr{H}}\bigl[\hat{\mfrak{H}},\langle\hat{\mfrak{s}}\rangle\bigr]\biggr)_{\nu,s\ppr;\mu,s}^{\boldsymbol{-1};b=a}
\hspace*{-0.9cm}(\Teta{j_{2}},\vec{x}_{2};\Teta{j_{1}},\vec{x}_{1}) =
\hat{\mscr{V}}^{-1}(\vec{x}_{2},\TT_{\scrscr j_{2}};\vec{x}_{1},\TT_{\scrscr j_{1}})\;\;\tfrac{\hat{\mathrm{S}}^{b=a}}{\mcal{N}_{x}}\;\;
\langle\hat{\mfrak{s}}\rangle_{\nu,s\ppr;\mu,s}^{b=a}(\Teta{j_{2}},\vec{x}_{2};\Teta{j_{1}},\vec{x}_{1})\;.
\end{align}
\end{subequations}
The coset matrix \(\hat{\mscr{Z}}^{ba}(\nn{5};\nn{1})\) (\ref{s5_3a}-\ref{s5_3e}), (\ref{s5_30a}-\ref{s5_30c})
consists of the anomalous-doubled, exciton-related, anti-symmetric
matrices \(\sdelta\hat{\wt{a}}(\nn{5};\nn{1})\), \(\sdelta\hat{\wt{a}}\pdag(\nn{5};\nn{1})\) as the independent field degrees of
freedom which prescribe the dependent, block diagonal, density-related parts (\ref{s5_30d},\ref{s5_30e})
\begin{subequations}
\begin{align}\lb{s5_30a}
\sdelta\hat{\mscr{Z}}^{ba}(\nn{5};\nn{1}) &=
-\Bigl(\hat{\mathrm{P}}(\nn{5};\nn{4})\;\hat{\mathrm{T}}^{-1}(\nn{4};\nn{3})\;\bigl(\sdelta\hat{\mathrm{T}}(\nn{3};\nn{2})\bigr)\;
\hat{\mathrm{P}}^{-1}(\nn{2};\nn{1})\Bigr)^{ba} =
\left(\bea{cc} \sdelta\hat{\mscr{Y}}^{11}(\nn{5};\nn{1}) & \sdelta\hat{\mscr{X}}(\nn{5};\nn{1}) \\
\sdelta\hat{\mscr{X}}\pdag(\nn{5};\nn{1}) & \sdelta\hat{\mscr{Y}}^{22}(\nn{5};\nn{1}) \eea\right)^{ba}_{\mbox{;}} \\ \lb{s5_30b}
\sdelta\hat{\mscr{X}}(\nn{5};\nn{1}) &=\sdelta\hat{\mscr{X}}_{\mu_{5},s_{5};\mu_{1},s_{1}}(\Tetax{5};\Tetax{1}) =
\sdelta\hat{\wt{a}}(\nn{5};\nn{1})=\sum_{k=0}^{3}\bigl(\hat{\tau}_{k}\bigr)_{\mu_{5}\mu_{1}}\;
\sdelta\hat{\wt{a}}_{s_{5}s_{1}}^{(k)}(\Tetax{5};\Tetax{1})\;\;; \\    \lb{s5_30c}
\sdelta\hat{\wt{a}}(\nn{5};\nn{1}) &= -\sdelta\hat{\wt{a}}^{T}(\nn{5};\nn{1})\;;  \\    \lb{s5_30d}
\sdelta\hat{\mscr{Y}}^{11}(\nn{5};\nn{1}) &=-\sdelta\hat{\mscr{Y}}^{22;T}(\nn{5};\nn{1}) =
\im\;\sdelta\hat{g}(\nn{5};\nn{1}) \;;  \hspace*{0.5cm}
\sdelta\hat{g}(\nn{5};\nn{1}) =\sdelta\hat{g}\pdag(\nn{5};\nn{1}) \;\;\;;   \\    \lb{s5_30e}
\sdelta\hat{\mscr{Y}}^{11}(\nn{5};\nn{1}) &=-\sdelta\hat{\mscr{Y}}^{22,T}(\nn{5};\nn{1})=-\tfrac{1}{2}\sum_{k=0}^{3}
\bigl(\hat{\tau}_{k}\hat{\tau}_{2}\bigr)_{\mu_{5}\mu_{1}}\;\times \\  \notag &\times\Bigg\{
\bigg[\tanh\Bigl(\tfrac{|\wt{a}_{\mathsf{S}_{5}}(\Tetax{5})|+|\wt{a}_{\mathsf{S}_{1}}(\Tetax{1})|}{2}\Bigr)-
\tanh\Bigl(\tfrac{|\wt{a}_{\mathsf{S}_{5}}(\Tetax{5})|-|\wt{a}_{\mathsf{S}_{1}}(\Tetax{1})|}{2}\Bigr)\bigg]\times \\ \notag &\times
\exp\bigg\{-\im\int_{\TT_{0}^{(+)}}^{\Teta{j_{1}}}d\Teta{j\ppr}
\frac{2\,|\wt{a}_{\mathsf{S}_{1}}(\Tetaxx{j\ppr}{\vec{x}_{1}})|}{\sinh\big(2\,|\wt{a}_{\mathsf{S}_{1}}(\Tetaxx{j\ppr}{\vec{x}_{1}})|\big)}
\frac{\pp\wt{\alpha}_{\mathsf{S}_{1}}(\Tetaxx{j\ppr}{\vec{x}_{1}})}{\pp\Teta{j\ppr}}\bigg\}\;
\sdelta\hat{a}_{s_{5}s_{1}}^{(k)}(\Tetax{5};\Tetax{1}) +   \\  \notag  &-(-1)^{k}
\bigg[\tanh\Bigl(\tfrac{|\wt{a}_{\mathsf{S}_{5}}(\Tetax{5})|+|\wt{a}_{\mathsf{S}_{1}}(\Tetax{1})|}{2}\Bigr)+
\tanh\Bigl(\tfrac{|\wt{a}_{\mathsf{S}_{5}}(\Tetax{5})|-|\wt{a}_{\mathsf{S}_{1}}(\Tetax{1})|}{2}\Bigr)\bigg]\times \\ \notag &\times
\exp\bigg\{\im\int_{\TT_{0}^{(+)}}^{\Teta{j_{5}}}d\Teta{j\ppr}
\frac{2\,|\wt{a}_{\mathsf{S}_{5}}(\Tetaxx{j\ppr}{\vec{x}_{5}})|}{\sinh\big(2\,|\wt{a}_{\mathsf{S}_{5}}(\Tetaxx{j\ppr}{\vec{x}_{5}})|\big)}
\frac{\pp\wt{\alpha}_{\mathsf{S}_{5}}(\Tetaxx{j\ppr}{\vec{x}_{5}})}{\pp\Teta{j\ppr}}\bigg\}\;
\sdelta\hat{a}_{s_{5}s_{1}}^{(k)\dag}(\Tetax{5};\Tetax{1}) + \\  \notag &-\im\:\delta_{\mu_{5}\mu_{1}}\:\delta_{s_{5}s_{1}}\;
\tanh\bigl(|\wt{a}_{\mathsf{S}}(\Tetaxx{j}{\vec{x}})|\bigr)\;|\wt{a}_{\mathsf{S}}(\Tetaxx{j}{\vec{x}})|\;
\sdelta\wt{\alpha}_{\mathsf{S}}(\Tetaxx{j}{\vec{x}})\;;  \\ \notag &
(\mbox{last term }(\Tetaxx{j}{\vec{x}}:=\Tetax{5}=\Tetax{1})\&(s:=s_{5}=s_{1}\;,\,\mathsf{S}=2s)\,)\;; 
\;\;\;\;(\mbox{matrix})\pdag=(\mbox{matrix})^{*,T}\;; \\  \notag &
(\mbox{cf. gauge fixing in eqs. (\ref{s5_15a},\ref{s5_15b}) and appendix \ref{sa}})\;.
\end{align}
\end{subequations}
The transformations to Euclidean fields yields corrections for the derivation of classical field equations by first order variation of the
anomalous field parts \(\sdelta\hat{\wt{a}}(\nn{5};\nn{1})\) and becomes even more important for higher order variation of
fluctuation terms (cf.\ section \ref{s12}). The derivations to Euclidean path integration fields within sections \ref{s51}, \ref{s52}
therefore hint at the importance of the correct coset integration measure, orginally included in the factorization of the total self-energy
\(\delta\wt{\Sigma}_{\nu,s\ppr;\mu,s}^{ba}(\Tetax{2};\Tetax{1})\) into block diagonal, density-related self-energies as hinge fields
of a SSB and into anomalous-related, off-diagonal parts by using a coset decomposition.

\section{Summary and conclusion}\lb{s6}

\subsection{Observable quantities by differentiating with the generating source term}\lb{s61}

In sections \ref{s31} to \ref{s33} we have performed the three possible Gaussian transformations of increasing complexity in order to
convert the quartic interaction of Fermi fields to even-valued self-energy variables and matrices. These path integration fields,
following from the prevailing HST (\ref{s3_1a}-\ref{s3_3}), (\ref{s3_9a}-\ref{s3_12c}), (\ref{s3_16}-\ref{s3_21}) 
of sections \ref{s31} to \ref{s33}, only take values in completely Euclidean spaces and flat
integration measures so that the calculation of observables results from differentiation of the corresponding HST-transformed
generating function (\ref{s3_5a}-\ref{s3_5c}), (\ref{s3_13a}-\ref{s3_14}), (\ref{s3_22a}-\ref{s3_22b}) 
with respect to the source field \(\hat{\mscr{J}}_{\nu,s\ppr;\mu,s}^{ba}(\Tetaxx{j\ppr}{\vec{x}\ppr};\Tetaxx{j}{\vec{x}})\)
(\ref{s2_10a}). The latter source field allows to track the original observables, as combinations of anti-commuting fields in even number, to
corresponding Green functions with the prevailing self-energy variable or matrix which can be restricted by the computation of
a saddle point equation (\ref{s3_6}-\ref{s3_8}), (\ref{s3_15a}-\ref{s3_15c}), (\ref{s3_23c}-\ref{s3_23f}).

We outline the calculation for the more profound SSB case with the factorization into density- and anomalous-related parts by a
coset decomposition according to sections \ref{s4} and \ref{s5}. The density- and Nambu-related parts follow straightforwardly as one
tracks the original observables in terms of fermionic coherent states to the independent, anomalous-related, locally Euclidean coset
fields \(\hat{\wt{a}}(\nn{5};\nn{1})\) (\ref{s5_30a}-\ref{s5_30e}) by differentiation with 
respect to \(\hat{\mscr{J}}_{\nu,s\ppr;\mu,s}^{ba}(\Tetaxx{j\ppr}{\vec{x}\ppr};\Tetaxx{j}{\vec{x}})\).
This leads to the following three terms (\(j=1,\ldots,2N\))
\begin{subequations}
\begin{align}\lb{s6_1a}
\Big\langle\tfrac{1}{\mcal{N}_{x}}\,\chi_{\nu,s\ppr}^{*}(\Tetaxx{j\ppr}{\vec{x}\ppr})\;\chi_{\mu,s}(\Tetaxx{j-1}{\vec{x}})
\Big\rangle_{\mbox{\scz\bf eq.(\ref{s2_15})}}&=
-\mcal{N}_{x}\sum_{a,b=1,2}\hat{\mathrm{S}}^{ba}
\frac{\pp\ovv{Z[\hat{\mscr{J}}]}}{\pp\hat{\mscr{J}}_{\nu,s\ppr;\mu,s}^{ba}(\Tetaxx{j\ppr}{\vec{x}\ppr};\Tetaxx{j}{\vec{x}})}= \\ \notag &\hspace*{-5.9cm}=
-\mcal{N}_{x}\sum_{a,b=1,2}\bigg\langle\hat{\mathrm{S}}^{ba}\bigg(\hat{\mathrm{I}}^{-1}\,\eta\,\hat{\mathrm{T}}(\hat{\wt{a}})\:\breve{\mscr{H}}^{-1}\,
\hat{\mathrm{P}}^{-1}\Big(\hat{1}+\hat{\mscr{Z}}-\breve{\mscr{H}}\:\hat{\mscr{Z}}\:\breve{\mscr{H}}^{-1}\Big)^{\boldsymbol{-1}}\hat{\mathrm{P}}\:
\hat{\mathrm{T}}^{-1}\!(\hat{\wt{a}})\:\eta\,\hat{\mathrm{I}}^{-1}\bigg)_{\mu,s;\nu,s\ppr}^{\boldsymbol{-1};ab}
\hspace*{-0.9cm}(\Tetaxx{j}{\vec{x}};\Tetaxx{j\ppr}{\vec{x}\ppr})
\bigg\rangle_{\mbox{\scz\bf eq.(\ref{s5_22})}}\;; \\   \lb{s6_1b}
\Big\langle\tfrac{1}{\mcal{N}_{x}}\,\chi_{\nu,s\ppr}(\Tetaxx{j\ppr-1}{\vec{x}\ppr})\;\chi_{\mu,s}(\Tetaxx{j-1}{\vec{x}})
\Big\rangle_{\mbox{\scz\bf eq.(\ref{s2_15})}}&=
-2\mcal{N}_{x}
\frac{\pp\ovv{Z[\hat{\mscr{J}}]}}{\pp\hat{\mscr{J}}_{\nu,s\ppr;\mu,s}^{21}(\Tetaxx{j\ppr}{\vec{x}\ppr};\Tetaxx{j}{\vec{x}})}= \\ \notag &\hspace*{-5.9cm}=
-2\mcal{N}_{x}\bigg\langle\bigg(\hat{\mathrm{I}}^{-1}\,\eta\,\hat{\mathrm{T}}(\hat{\wt{a}})\:\breve{\mscr{H}}^{-1}\,
\hat{\mathrm{P}}^{-1}\Big(\hat{1}+\hat{\mscr{Z}}-\breve{\mscr{H}}\:\hat{\mscr{Z}}\:\breve{\mscr{H}}^{-1}\Big)^{\boldsymbol{-1}}\hat{\mathrm{P}}\:
\hat{\mathrm{T}}^{-1}\!(\hat{\wt{a}})\:\eta\,\hat{\mathrm{I}}^{-1}\bigg)_{\mu,s;\nu,s\ppr}^{\boldsymbol{-1};12}\hspace*{-0.9cm}(\Tetaxx{j}{\vec{x}};\Tetaxx{j\ppr}{\vec{x}\ppr})
\bigg\rangle_{\mbox{\scz\bf eq.(\ref{s5_22})}}\;; \\   \lb{s6_1c}
\Big\langle\tfrac{1}{\mcal{N}_{x}}\,\chi_{\nu,s\ppr}^{*}(\Tetaxx{j\ppr}{\vec{x}\ppr})\;\chi_{\mu,s}^{*}(\Tetaxx{j}{\vec{x}})
\Big\rangle_{\mbox{\scz\bf eq.(\ref{s2_15})}}&=
-2\mcal{N}_{x}
\frac{\pp\ovv{Z[\hat{\mscr{J}}]}}{\pp\hat{\mscr{J}}_{\nu,s\ppr;\mu,s}^{12}(\Tetaxx{j\ppr}{\vec{x}\ppr};\Tetaxx{j}{\vec{x}})}= \\ \notag &\hspace*{-5.9cm}=
-2\mcal{N}_{x}\bigg\langle\bigg(\hat{\mathrm{I}}^{-1}\,\eta\,\hat{\mathrm{T}}(\hat{\wt{a}})\:\breve{\mscr{H}}^{-1}\,
\hat{\mathrm{P}}^{-1}\Big(\hat{1}+\hat{\mscr{Z}}-\breve{\mscr{H}}\:\hat{\mscr{Z}}\:\breve{\mscr{H}}^{-1}\Big)^{\boldsymbol{-1}}\hat{\mathrm{P}}\:
\hat{\mathrm{T}}^{-1}\!(\hat{\wt{a}})\:\eta\,\hat{\mathrm{I}}^{-1}\bigg)_{\mu,s;\nu,s\ppr}^{\boldsymbol{-1};21}\hspace*{-0.9cm}(\Tetaxx{j}{\vec{x}};\Tetaxx{j\ppr}{\vec{x}\ppr})
\bigg\rangle_{\mbox{\scz\bf eq.(\ref{s5_22})}}\;;   \\   \lb{s6_1d}
\breve{\mscr{H}}\bigl[\hat{\mfrak{H}},\hat{\mfrak{s}}\bigr]_{\nu,s\ppr;\mu,s}^{b=a}
(\Teta{j_{2}},\vec{x}_{2};\Teta{j_{1}},\vec{x}_{1}) &=
\hat{\mfrak{H}}_{\nu,s\ppr;\mu,s}^{b=a}(\vec{x}_{2},\Teta{j_{2}};\vec{x}_{1},\Teta{j_{1}}) + \\ \notag &+
\hat{\mathrm{S}}^{b=a}\:q_{\nu}\:\eta_{j_{2}}\:\bigl(\im\tfrac{\sdelta t}{\hbar}\bigr)\;
\hat{\mfrak{s}}_{\nu,s\ppr;\mu,s}^{b=a}(\Teta{j_{2}},\vec{x}_{2};\Teta{j_{1}},\vec{x}_{1})\;q_{\mu}\:\eta_{j_{1}}\;;
\\  \notag  \breve{\mscr{H}}\bigl[\hat{\mfrak{H}},\hat{\mfrak{s}}\bigr]_{\nu,s\ppr;\mu,s}^{22}
(\Teta{j_{2}},\vec{x}_{2};\Teta{j_{1}},\vec{x}_{1}) &=\bigg(
\breve{\mscr{H}}\bigl[\hat{\mfrak{H}},\hat{\mfrak{s}}\bigr]_{\nu,s\ppr;\mu,s}^{11}
(\Teta{j_{2}},\vec{x}_{2};\Teta{j_{1}},\vec{x}_{1}) \bigg)^{\boldsymbol{T}} \;;
\end{align}
\end{subequations}
Applying relations of section \ref{s4} and \ref{s5}, we can reduce and simplify the inverse of combined anomalous- and density-related
parts in (\ref{s6_1a}-\ref{s6_1c}) so that one is reminded of the astonishing result that the exciton quasi-particles in the off-diagonal blocks of
\(\hat{\mscr{Z}}^{ba}(\nn{5};\nn{1})\) are multiplied by the one-particle operator \(\breve{\mscr{H}}[\hat{\mfrak{H}},\langle\hat{\mfrak{s}}\rangle]\)
in order to achieve the true exciton field
\begin{align}\lb{s6_2}
\bigg(\hat{\mathrm{I}}^{-1}\,\eta\,\hat{\mathrm{T}}(\hat{\wt{a}})\:\breve{\mscr{H}}^{-1}\,
\hat{\mathrm{P}}^{-1}\Big(\hat{1}+\hat{\mscr{Z}}-\breve{\mscr{H}}\:\hat{\mscr{Z}}\:\breve{\mscr{H}}^{-1}\Big)^{\boldsymbol{-1}}\hat{\mathrm{P}}\:
\hat{\mathrm{T}}^{-1}\!(\hat{\wt{a}})\:\eta\,\hat{\mathrm{I}}^{-1}\bigg)_{\mu,s;\nu,s\ppr}^{\boldsymbol{-1};ab}\hspace*{-0.9cm}(\Tetaxx{j}{\vec{x}};\Tetaxx{j\ppr}{\vec{x}\ppr})&=
\\ \notag &\hspace{-11.5cm}=\bigg(\hat{\mathrm{I}}\,\eta\,\hat{\mathrm{P}}^{-1}\bigg[
\bigg(\hat{1}-\left[\left(\exp\{\overrightarrow{[\ln(-\breve{\mscr{H}}){\large\boldsymbol{,}}\ldots]_{-}}\}-\hat{1}\right)\hat{\mscr{Z}}\right]\bigg)
\hat{\mathrm{P}}\:\breve{\mscr{H}}\:\hat{\mathrm{P}}^{-1}
\exp\{-\overleftarrow{[\ldots{\large\boldsymbol{,}}\hat{\mathrm{Y}}_{D}]_{-}}\}\bigg]\hat{\mathrm{P}}\,\eta\,\hat{\mathrm{I}}
\bigg)_{\mu,s;\nu,s\ppr}^{\boldsymbol{-1};ab}\hspace*{-0.9cm}(\Tetaxx{j}{\vec{x}};\Tetaxx{j\ppr}{\vec{x}\ppr})\;.
\end{align}
In consequence, the 'Nambu'-doubled, locally Euclidean fields \(\sdelta\hat{\wt{a}}(\nn{5};\nn{1})\)
have to be regarded as exciton fields relatively to the density-related parts. Further restriction to solely diagonal elements
\(\wt{a}_{ss}^{(2)}(\Tetaxx{j}{\vec{x}};\Tetaxx{j}{\vec{x}})\) of \(\hat{\wt{a}}(\nn{5};\nn{1})\) (\ref{s5_24a},\ref{s5_24b},
\ref{s5_30a}-\ref{s5_30e})
allows to derive and investigate classical
soliton equations from first and higher order variations which are determined by the prevailing coset integration measure (cf.\ section \ref{s12}).
Therefore, we have pointed out and intensively examined the symmetries of the corresponding self-energies throughout the paper
under inclusion of the precise time contour step ordering of the original time development operator. 
The derived equations of section \ref{s53} for the observables (\ref{s6_1a}-\ref{s6_2}) can be used to study self-induced transparency of
ultrashort coherent transients and the area theorem in combination with holography and selective Fourier optics for possible means of
reduced absorption within matter \cite{Hahn}-\cite{Galanin}. Possible extensions of the presented approaches of sections \ref{s31}-\ref{s33}
and \ref{s4}-\ref{s5} may include the quantization of the electromagnetic field which can also be represented in terms of coherent states so
that corresponding HSTs lead to analogous self-energies as for the quartic interaction of fermionic fields.

\newpage

\begin{appendix}
\section{Gauge fixing for 'Nambu'-fields within the coset integration measure}\lb{sa}

In section \ref{s52} with eqs. (\ref{s5_15a},\ref{s5_15b}) we have to require that the quaternionic diagonal, anti-symmetric
matrix elements have to vanish in the gauge combination
\((\,(d\hat{\mathrm{P}})\:\hat{\mathrm{P}}^{-1})_{\mu_{5},s;\mu_{1},s}(\Tetaxx{j}{\vec{x}};\Tetaxx{j}{\vec{x}})\) (\ref{s5_15b})
of the block diagonal matrices \(\hat{\mathrm{P}}_{\mu_{5},s_{5};\mu_{1},s_{1}}^{aa}(\Tetax{5};\Tetax{1})\) (\ref{s4_7}-\ref{s4_9f})
with their contour time derivatives \('d'\)
\begin{align}\lb{sa_1}
0&\stackrel{!}{=} \Big((d\hat{\mathrm{P}})\,\hat{\mathrm{P}}^{-1}\Big)_{\mu_{5},s;\mu_{1},s}^{aa}\hspace*{-0.6cm}
(\Tetaxx{j}{\vec{x}};\Tetaxx{j}{\vec{x}})\;\;\;.
\end{align}
This can be accomplished by a gauge transformation (\ref{sa_2}) of \(\hat{\mathrm{P}}(\nn{5};\nn{1})\) with
a quaternion diagonal matrix \(\hat{\mathrm{P}}_{D;\mu_{5},s;\mu_{1},s}^{aa}(\Tetaxx{j}{\vec{x}};\Tetaxx{j}{\vec{x}})\) 
(\ref{sa_3},\ref{sa_4}) which has only non-vanishing matrix elements 
\(\hat{\mscr{G}}_{D;\mu_{5},s;\mu_{1},s}(\Tetaxx{j}{\vec{x}};\Tetaxx{j}{\vec{x}})\neq 0\)
along the \(2\times 2\) diagonals, just in opposite to \(\hat{\mathrm{P}}^{aa}(\nn{5};\nn{1})\) (\ref{s4_9a}-\ref{s4_9f}).
These hermitian \(2\times 2\) matrix elements  \(\hat{\mscr{G}}_{D;\mu_{5},s;\mu_{1},s}(\Tetaxx{j}{\vec{x}};\Tetaxx{j}{\vec{x}})\) 
(\ref{sa_5}) along the main diagonal have to depend on the
off-diagonal parameters of the ladder operators in \(\hat{\mathrm{P}}^{11}(\nn{5};\nn{1})\), 
\(\hat{\mathrm{P}}^{22}(\nn{5};\nn{1})\) and have to be chosen with suitable dependence in such a manner 
that the block diagonal, gauge transformed matrices 
\(\hat{\mscr{P}}^{aa}(\nn{5};\nn{1})=\hat{\mathrm{P}}_{D}^{aa}(\nn{5};\nn{4})\;\hat{\mathrm{P}}^{aa}(\nn{4};\nn{1})\)
(\ref{sa_2}) fulfill the property of
\((\,(d\hat{\mscr{P}})\,\hat{\mscr{P}}^{-1})_{\mu_{5},s;\mu_{1},s}^{aa}(\Tetaxx{j}{\vec{x}};\Tetaxx{j}{\vec{x}})\equiv0\) (\ref{sa_1}). 
One has to take into account the quaternion algebra in order to achieve (\ref{sa_7}) for
diagonal elements referring to the quaternion matrix elements with anti-symmetric element \((\hat{\tau}_{2})_{\mu_{5}\mu_{1}}\)
of exciton quasi-particles 
\begin{align} \lb{sa_2}
\hat{\mathrm{P}}_{\mu_{5},s_{5};\mu_{1},s_{1}}^{aa}(\Tetax{5};\Tetax{1}) &\rightarrow\;
\hat{\mscr{P}}_{\mu_{5},s_{5};\mu_{1},s_{1}}^{aa}(\Tetax{5};\Tetax{1}) \;;  \\   \notag 
\hat{\mscr{P}}_{\mu_{5},s_{5};\mu_{1},s_{1}}^{aa}(\Tetax{5};\Tetax{1}) &=\sum_{\nu=\mbox{\scz 'e','h'}}
\hat{\mathrm{P}}_{D;\mu_{5},s_{5};\nu,s_{5}}^{aa}(\Tetax{5};\Tetax{5}) \;\;
\hat{\mathrm{P}}_{\nu,s_{5};\mu_{1},s_{1}}^{aa}(\Tetax{5};\Tetax{1})\;;   \\  \lb{sa_3} 
\hat{\mathrm{P}}_{D;\mu_{5},s;\mu_{1},s}^{11}(\Tetaxx{j}{\vec{x}};\Tetaxx{j}{\vec{x}})&=\exp\Big\{\im\,
\hat{\mscr{G}}_{D;\mu_{5}\ppr,s\ppr;\mu_{1}\ppr,s\ppr}(\Tetaxx{j\ppr}{\vec{x}\ppr};\Tetaxx{j\ppr}{\vec{x}\ppr})
\Big\}_{\mu_{5},s;\mu_{1},s}\hspace*{-0.9cm}(\Tetaxx{j}{\vec{x}};\Tetaxx{j}{\vec{x}})\;;   \\  \lb{sa_4}
\hat{\mathrm{P}}_{D;\mu_{5},s;\mu_{1},s}^{22}(\Tetaxx{j}{\vec{x}};\Tetaxx{j}{\vec{x}})&=\exp\Big\{-\im\,
\hat{\mscr{G}}_{D;\mu_{5}\ppr,s\ppr;\mu_{1}\ppr,s\ppr}^{\boldsymbol{T}}(\Tetaxx{j\ppr}{\vec{x}\ppr};\Tetaxx{j\ppr}{\vec{x}\ppr})
\Big\}_{\mu_{5},s;\mu_{1},s}\hspace*{-0.9cm}(\Tetaxx{j}{\vec{x}};\Tetaxx{j}{\vec{x}})\;;   \\  \lb{sa_5}
\hat{\mscr{G}}_{D;\mu_{5},s;\mu_{1},s}(\Tetaxx{j}{\vec{x}};\Tetaxx{j}{\vec{x}})&=
\hat{\mscr{G}}_{D}\Big(\hat{\mscr{G}}_{D}(\nn{3};\nn{2});\nn{3}\neq \nn{2}\Big)_{\mu_{5},s;\mu_{1},s}\hspace*{-1.2cm}
(\Tetaxx{j}{\vec{x}};\Tetaxx{j}{\vec{x}}) \;; \\ \lb{sa_6}
\Big(\big(d\hat{\mscr{P}}^{aa}\big)\;\hat{\mscr{P}}^{-1;aa}\Big)(\nn{5};\nn{1}) &=
\hat{\mathrm{P}}_{D}^{aa}(\nn{5};\nn{4})\;\;
d\hat{\mathrm{P}}^{aa}(\nn{4};\nn{3})\;\hat{\mathrm{P}}^{-1;aa}(\nn{3};\nn{2})\;\;
\hat{\mathrm{P}}_{D}^{-1;aa}(\nn{2};\nn{1}) + \\  \notag &\hspace*{1.6cm}+
d\hat{\mathrm{P}}_{D}^{aa}(\nn{5};\nn{4})\;\;\hat{\mathrm{P}}_{D}^{-1;aa}(\nn{4};\nn{1})\;;   \\  \lb{sa_7}&\hspace*{-3.6cm}
\sum_{\mu_{4}=\mbox{\scz 'e','h'}}
\hat{\mathrm{P}}_{D;\mu_{5},s;\mu_{4},s}^{-1;aa}(\Tetax{5};\Tetax{5})\;\;
d\hat{P}_{D;\mu_{4},s;\mu_{1},s}^{aa}(\Tetax{5};\Tetax{5}) =  \\  \notag &=- \sum_{j_{4}=1}^{2N}
\sum_{\vec{x}_{4}}^{\mu_{4},s_{4}}\,\hspace*{-0.15cm}\ppr\;
d\hat{\mathrm{P}}_{\mu_{5},s;\mu_{4},s_{4}}^{aa}(\Tetax{5};\Tetax{4})\;\;
\hat{\mathrm{P}}_{\mu_{4},s_{4};\mu_{1},s}^{-1;aa}(\Tetax{4};\Tetax{5}) \;.
\end{align}
Note that there is no summation over the contour time and coordinate space labels '\(\Tetax{5}\)' and the spin label '\({\scr s_{5}}\)'
in relations (\ref{sa_2},\ref{sa_7}), due to the 'diagonal' property of \(\hat{\mathrm{P}}_{D;\mu_{5},s;\mu_{1},s}^{aa}(\Tetax{5};\Tetax{5})\)
with respect to these labels, apart from the electron-hole labels '\({\scr\mu_{5},\mu_{1}}=\mbox{\scz 'e','h'}\)' which restrict the degrees of freedom to the 
anti-symmetric, quaternion element \((\hat{\tau}_{2})_{\mu_{5}\mu_{1}}\). A corresponding description holds for relation (\ref{sa_6})
with the quaternion diagonal matrix elements \(\hat{\mathrm{P}}_{D;\nu,s;\mu,s}^{aa}(\Tetaxx{j}{\vec{x}};\Tetaxx{j}{\vec{x}})\),
having only the quaternion elements  \((\hat{\tau}_{2})_{\mu_{5}\mu_{4}}\),  \((\hat{\tau}_{2})_{\mu_{2}\mu_{1}}\),
 \((\hat{\tau}_{2})_{\mu_{5}\mu_{4}}\),  \((\hat{\tau}_{2})_{\mu_{4}\mu_{1}}\) for respective gauge matrices
\(\hat{\mathrm{P}}_{D}^{aa}(\nn{5};\nn{4})\), \(\hat{\mathrm{P}}_{D}^{-1;aa}(\nn{2};\nn{1})\), \(d\hat{\mathrm{P}}_{D}^{aa}(\nn{5};\nn{4})\),
\(\hat{\mathrm{P}}_{D}^{-1;aa}(\nn{4};\nn{1})\).

\end{appendix}

% Set the ending of a LaTeX document
\end{document}